\documentclass[11pt, a4paper]{article}
\pdfoutput=1
\usepackage{jheparxiv}
\usepackage[latin1]{inputenc}
\usepackage{amsmath}
\usepackage{amsfonts}
\usepackage{amssymb}
\usepackage{latexsym}
\usepackage{mathrsfs}
\usepackage{graphicx}
\usepackage{color}
\usepackage{twistor}
\usepackage[all]{xy}

\renewcommand{\d}{\mathrm{d}}

\title{Twistor actions for gauge theory and gravity}

\author{Tim Adamo}

\affiliation{The Mathematical Institute, University of Oxford \\
	24-29 St.~Giles', Oxford OX1 3LB, United Kingdom}
\affiliation{Department of Applied Mathematics \& Theoretical Physics, University of Cambridge \\
        Wilberforce Road, Cambridge CB3 0WA, United Kingdom}
	
\emailAdd{t.adamo@damtp.cam.ac.uk}

\abstract{This is a review of recent developments in the study of perturbative gauge theory and gravity using action functionals on twistor space.  It is intended to provide a user-friendly introduction to twistor actions, geared towards researchers or graduate students interested in learning something about the utility, prospects, and shortcomings of this approach.  For those already familiar with the twistor approach, it should provide a condensed overview of the literature as well as several novel results of potential interest.  This work is based primarily upon the author's D.Phil. thesis.

We first consider four-dimensional, maximally supersymmetric Yang-Mills theory as a gauge theory in twistor space.  We focus on the perturbation theory associated to this action, which in an axial gauge leads to the MHV formalism.  This allows us to efficiently compute scattering amplitudes at tree-level (and beyond) in twistor space.  Other gauge theory observables such as local operators and null polygonal Wilson loops can also be formulated twistorially, leading to proofs for several correspondences between correlation functions and Wilson loops, as well as a recursive formula for computing mixed Wilson loop / local operator correlators.  We then apply the twistor action approach to general relativity, using the on-shell equivalence between conformal and Einstein gravity.  This can be extended to $\cN=4$ supersymmetry. The perturbation theory of the twistor action leads to formulae for the MHV amplitude with and without cosmological constant, yields a candidate for the Einstein twistor action, and induces a MHV formalism on twistor space.  Appendices include discussion of super-connections and Coulomb branch regularization on twistor space.}

\begin{document}

\maketitle


\section{Introduction}
\label{Chapter1}

Twistor theory, as first outlined by Roger Penrose in the late 1960s \cite{Penrose:1967wn}, is a program which relates physical objects on (in general, complex) Minkowski space-time to geometrical data in complex projective spaces called \emph{twistor spaces}.  This general picture of representing physics by complex geometry is captured by three of the classic results in twistor theory, each of which is an equivalence: between zero-rest-mass fields on space-time and cohomology on twistor space (the Penrose transform) \cite{Penrose:1969ae}; between Yang-Mills instantons on space-time and holomorphic vector bundles over twistor space (the Ward Correspondence) \cite{Ward:1977ta}; and between self-dual four-manifolds and integrable complex structures on twistor space (the non-linear graviton) \cite{Penrose:1976jq}.

Since its inception, twistor theory has provided proofs for theorems in classical general relativity \cite{Newman:1976gc, Hansen:1978jz, Adamo:2009vu}; informed the study of integrable systems \cite{Mason:1996rf, Hitchin:1999, Mason:2003}; and been of utility in a wide array of mathematical and physical applications (e.g., \cite{Salamon:1982, Neitzke:2007ke, Alexandrov:2012bu}).  Despite these advances, twistor theory had fallen well short of its initial ambitions, namely: to serve as a unifying mathematical framework for describing both quantum field theory and gravity.  By the early 2000s, this failure could be captured by two fundamental problems which had proven insurmountable in the preceding decades: the `googly problem', and the inability to make meaningful contact with quantum field theory.

The first of these captures the difficulty of dealing with arbitrarily curved space-times in twistor theory.  The non-linear graviton construction indicates that traditional twistor methods can be applied to any four-manifold whose anti-self-dual Weyl curvature vanishes; but for real Lorentzian space-times, the only example of such a space-time is Minkowski space itself.  Hence, it seemed impossible for any progress to be made if twistors could not be adapted to the most basic of situations in general relativity.  This issue is also present at the level of gauge theory: the Ward correspondence treats only self-dual gauge bundles on space-time.  Much effort was dedicated to overcoming this problem starting in the late 1970s, but diminishing returns soon turned this arena of research into a no-man's land.

The second problem is equally fundamental: in spite of constructs such as the Penrose transform and Ward correspondence, there was no clear proposal for how twistor theory could be used to make contact with basic questions in quantum field theory.  In particular, how could twistors be used to compute physical observables like scattering amplitudes or cross-sections in a gauge theory?  Once again, for a theory aiming to provide a mathematical formalism for both gravity and quantum field theory, this was a rather embarrassing problem.  While Hodges' twistor diagram formalism did make some progress in this area \cite{Hodges:1980hn}, twistor theory had by-and-large failed to make an impact on the study of quantum field theory.

This state of affairs changed dramatically in 2003/4, when Witten discovered that scattering amplitudes in (planar) maximally supersymmetric ($\cN=4$) super-Yang-Mills theory could be computed, at least at tree-level, via a topological B string theory in twistor space \cite{Witten:2003nn}.  This not only provided an answer to the second fundamental question plaguing twistor theory, but also gave a perturbative solution to the googly problem.  In the twistor-string setting, the anti-self-dual interactions of the theory are accounted for by $D1$-instantons in the target space.\footnote{Or alternatively, by a sum over worldsheet instantons in a heterotic formulation of twistor-string theory \cite{Mason:2007zv}.}  

Twistor-string theory has spurred an impressive list of advances in our understanding of gauge theory in general, and the planar sector of $\cN=4$ super-Yang-Mills in particular.  It has led to the development of efficient techniques for computing scattering amplitudes which are non-obvious from the space-time Lagrangian (such as the MHV formalism \cite{Cachazo:2004kj} and BCFW recursion \cite{Britto:2005fq}), and this in turn has influenced the computation of real processes in QCD which are measured at particle colliders (e.g., \cite{Berger:2008ag}).  It has motivated the study of dual conformal symmetry and the discovery of an infinite dimensional symmetry algebra associated with the scattering amplitudes of $\cN=4$ super-Yang-Mills \cite{Drummond:2008vq}.  Furthermore, it is an important influence behind emergent space-time proposals such as the Grassmannian formalism of Arkani-Hamed and collaborators \cite{ArkaniHamed:2009dn, ArkaniHamed:2012nw}, as well as numerous other insights and advances.

While Witten's theory correctly describes planar gauge theory at tree level \cite{Skinner:2010cz, Dolan:2011za}, it is not without its problems.  The most glaring is that the gravitational degrees of freedom in the string theory correspond to conformal super-gravity, a theory which is widely believed to be non-physical \cite{Berkovits:2004jj}.  Since these conformal gravity modes will run in loops, gauge theory scattering amplitudes calculated by twistor-string will be contaminated beyond tree-level.  

Recently, Skinner proposed a new twistor-string theory which eliminates the modes of conformal gravity and, in the flat-space limit, correctly produces the tree-level S-matrix of $\cN=8$ supergravity from the worldsheet theory \cite{Skinner:2013xp}.  The key difference between this theory and all previous twistor-string theories is the addition of \emph{worldsheet} supersymmetry; anomaly cancellation conditions then uniquely restrict to a twistor space which manifests the maximal $\cN=8$ gravitational supersymmetry.  While Skinner's model undoubtedly represents an incredible breakthrough for the twistor approach, there are still many facets of the theory which are not properly understood: it is unclear how- or if- the theory describes the analogues of scattering amplitudes for non-flat backgrounds (i.e., gauged supergravity on anti-de Sitter space), and while anomaly-free for any genus worldsheet, it is not known if the theory correctly computes scattering amplitudes beyond tree level (even at the level of a loop integrand).      

The most successful solution to the puzzle of studying gauge theory without gravity has been the \emph{twistor action} proposal of Mason \cite{Mason:2005zm}.  This approaches gauge theory via an action functional on twistor space which is the classical generating functional for the gauge theory degrees of freedom in twistor-string theory, completely eliminating gravity from the picture!  This means that physical observables can be studied to all orders in perturbation theory using the twistor action.  On the gravitational side, Mason also found a twistor action functional for conformal gravity, and the existence of Skinner's twistor-string hints that an action for Einstein gravity itself should also exist.

\medskip

In this review, we study twistor actions as theories in their own right.  That is, we consider the twistor action as the primary object (rather than the space-time theory) and attempt to study the basic structures of gauge theory and gravity from an intrinsically twistorial point of view.  As we shall see, asking rather basic questions about the twistor action (e.g., `What are its Feynman rules?') leads to surprisingly interesting answers (e.g., a derivation of the MHV formalism).  Furthermore, studying basic physical observables (such as correlation functions and Wilson loops) in twistor space allows us to prove powerful statements about space-time physics.

Much of our presentation will focus on results which have already appeared in published form elsewhere; our aim is to provide a coherent and self-contained explanation of these findings which (hopefully) also incorporates a novel presentation.  However, there are many results included here which have not been published before.  These range from technical lemmas which may catch the eye of twistor theorists, to more general findings which may interest researchers interested in scattering amplitudes and the ways in which they can be studied using twistor theory. 

Section \ref{Chapter2} contains review material pertaining to flat-space twistor theory, some basic calculational tools, and $\cN=4$ super-Yang-Mills.  In particular, we discuss the twistor correspondence between points in (complex) Minkowski space and linearly embedded Riemann spheres in twistor space, and introduce the concepts of Penrose transform and Ward correspondence.  We also set out a calculus of distributional forms which will be used throughout the paper, and discuss some salient features of $\cN=4$ super-Yang-Mills theory (SYM).  The reader who is already acquainted with these issues could skim this section in order to progress more quickly.  

Section \ref{Chapter3} deals with the twistor action of $\cN=4$ SYM (first given in \cite{Boels:2006ir}).  We discuss the gauge freedom and perturbation theory of this action, and demonstrate how it can be used to arrive at a twistorial derivation of the MHV formalism.  This also leads to a natural method for computing the scattering amplitudes of $\cN=4$ SYM on twistor space itself \cite{Adamo:2011cb} which manifests superconformal symmetry, and should be contrasted against the \emph{momentum twistor} approach of \cite{Bullimore:2010pj}, which computes the integrand of a scattering amplitude divided by a tree-level MHV factor and manifests \emph{dual} superconformal symmetry.  The twistor action thereby allows us to compute the entire tree-level S-matrix of the gauge theory, and we also discuss the prospects for computing loop-level amplitudes in this formalism.

In Section \ref{Chapter4}, we consider other natural observables in gauge theory from a twistor perspective: correlation functions involving local operators and null polygonal Wilson loops.  We show that these operators have an algebro-geometric formulation in twistor space, and their expectation values can be computed using the Feynman rules of the twistor action developed in the preceding section.  This allows us to provide proofs (at the level of the loop integrand) for the supersymmetric correlation function / Wilson loop correspondence \cite{Adamo:2011dq} as well as several conjectures regarding mixed Wilson loop / local operator correlators \cite{Adamo:2011cd}.  Additionally, we can build on the BCFW deformation of the Wilson loop in twistor space \cite{Bullimore:2011ni} to derive novel recursion relations for these mixed operators.

We switch our focus from gauge theory to gravity in Section \ref{Chapter5}, beginning with a review of the basic result in twistor theory for curved space-times: the non-linear graviton construction.  We then discuss the embedding of Einstein gravity into conformal gravity on an asymptotically de Sitter background \cite{Maldacena:2011mk}.  Using the Plebanski formalism, we can state this embedding precisely at the level of generating functionals for the MHV amplitudes of the two theories.  On twistor space, we introduce the twistor action for conformal gravity and its minimal $\cN=4$ supersymmetric extension, and reduce its degrees of freedom to those of Einstein gravity.  This not only gives a twistorial expression for the MHV generating functionals, but also produces a candidate twistor action for Einstein gravity \cite{Adamo:2013tja}.  As an interesting curiosity, we also discuss the possibility of formulating twistor actions for \emph{non-minimal} $\cN=4$ conformal supergravity, such as the theory which arises from the gravitational degrees of freedom in the Berkovits-Witten twistor-string \cite{Berkovits:2004jj}.

Section \ref{Chapter6} is dedicated to studying the perturbation theory associated with the conformal gravity twistor action reduced to Einstein states, as well as the Einstein twistor action itself.  We show that the vertices for both of these actions correspond to the MHV amplitudes of Einstein gravity (with cosmological constant); this is accomplished by translating the iterative solution of an integral equation determining the scattering background into a diagram calculus on the Riemann sphere.  We show that the resulting formulae are gauge invariant and limit onto Hodges' formula for the MHV amplitude \cite{Hodges:2012ym} when the cosmological constant is sent to zero.  We then discuss the propagator structure on twistor space, arguing that it induces a MHV formalism for Einstein gravity.  We conclude by providing an additional formula for the MHV amplitude which is based on BCFW recursion.

Section \ref{Chapter7} concludes with a discussion of open questions and future directions related to this work.  Appendices \ref{Appendix1} and \ref{Appendix2} provide some results which are useful supplements to our discussions.  Appendix \ref{Appendix1} presents some results concerning superconnections in $\cN=4$ super-Yang-Mills theory, while Appendix \ref{Appendix2} defines a Coulomb branch twistor action and derives the massive MHV formalism on twistor space.     

\subsection{Advice for the Reader}

This review is adapted from the author's D.Phil. thesis, and a word of warning may prove useful to the reader.  In particular, the reader may find that the degree of precision varies substantially throughout the text: some (rather minor) results are proved explicitly while others are simply outlined or referenced.  I hope that this has not been done haphazardly: my aim has been to include proofs of any results which have not appeared explicitly in prior literature, while being more concise regarding those results which can easily be looked up in extant papers.  An exception to this heuristic is Section \ref{Chapter4}, where the proofs of correspondences between Wilson loops and certain correlation functions are particularly illustrative.

Additionally, it should be possible to read many of the sections in a self-contained manner.  Section \ref{Chapter2} should provide background for all the gauge theory considerations covered in this review, and all the new machinery needed for gravity is covered in the beginning of Section \ref{Chapter5}.  So a reader who is only interested in null limits of correlation functions can skip to Section \ref{Chapter4} without missing anything essential in Section \ref{Chapter3}.  The two appendices on $\cN=4$ superconnections (Appendix \ref{Appendix1}) and the Coulomb branch of $\cN=4$ Yang-Mills (Appendix \ref{Appendix2}) are also largely stand-alone, and composed of mostly unpublished material.  Throughout, I have also tried to assemble a relatively comprehensive list of references, which the reader should find helpful in filling the many gaps which are sure to be found in this review.

Finally, I have appropriated terms such as `lemma' or `proposition' in order to highlight concrete, important results.  Some proofs are obviously more rigorous than others, and we often take for granted such constructs as manipulation inside a path integral, or working with a loop integrand.  I have attempted to foreground any such assumptions, and to be honest about the degree to which they are essential in any given proof.

\subsection{Summary of Results}

For the reader's convenience, we list here the main results presented in this review:

\begin{itemize}
\item (Proposition \ref{MHVpropn}) Derivation of the MHV formalism from the twistor action.

\item (Section \ref{TScatAmps}) The full tree-level S-matrix of $\cN=4$ super-Yang-Mills theory as scattering amplitudes of the twistor action.

\item (Proposition \ref{corrWL}) Proof of the supersymmetric correlation function / Wilson loop correspondence.

\item (Proposition \ref{locP1}, \ref{locP2}) Proofs of conjectures relating mixed Wilson loop / local operator correlators to null limits of correlation functions. 

\item (Proposition \ref{recurpropn}) BCFW-like recursion relations for mixed Wilson loop / local operator correlators.

\item (Proposition \ref{CGDS}) Equivalence between the MHV generating functionals of conformal and Einstein gravity on de Sitter space.

\item (Section \ref{CGPerT}) Perturbation theory for the conformal gravity twistor action reduced to Einstein degrees of freedom.

\item (Section \ref{MHVLambda}) MHV amplitude with cosmological constant on twistor space.

\item (Proposition \ref{CCBCFW}) BCFW formula for the gravitational MHV amplitude with cosmological constant in twistor space.

\item (Proposition \ref{spropn}) Derivation of $\cN=4$ super-Yang-Mills superconnections from integrability conditions.

\item (Proposition \ref{CBProp}) Twistor action for the Coulomb branch of $\cN=4$ super-Yang-Mills.

\item (Appendix \ref{MMHVForm}) Derivation/proof of the massive MHV formalism from the Coulomb branch twistor action.

\end{itemize}


\section{Background Material}
\label{Chapter2}

This section reviews what will be considered as background material for the remainder of this paper.  We begin with an overview of the basics of twistor theory, which is the primary geometric framework for all our studies, establishing notational conventions and listing some important facts.  Since it was first described by Penrose in 1967 \cite{Penrose:1967wn}, twistor theory has had a long and varied history.  It is not the purpose of this section to serve as an extensive review of twistor theory and its many facets; the interested reader need only consult one of the many books or papers reviewing the subject (e.g., \cite{Penrose:1972ia, Huggett:1985, Penrose:1986ca, Ward:1990}) for a more detailed exposition.  We then introduce a calculus of distributional forms which will prove very useful for representing physical and geometric data on twistor space.  Finally, we provide a brief overview on $\cN=4$ super-symmetric Yang-Mills theory, and some of the surprising properties of its scattering amplitudes.  

The reader who is already familiar with twistor theory may wish to simply skim this section to familiarize themselves with notation, before moving on to the more interesting later sections.


\subsection{Twistor Theory}


\subsubsection{Basic formalism}

\subsubsection*{\textit{Spinor-helicity formalism}}

We begin with complexified 4-dimensional Minkowski space-time $\M_{b}\cong\C^{4}$ in Lorentzian signature $(+,-,-,-)$, with coordinates $x^{\mu}$ (for $\mu=0,\ldots,3$).  The complexified spin group is $\SO(4,\C)$, which is isomorphic to $\SL(2,\C)\times\SL(2,\C)/\Z_{2}$. Two-component Weyl spinors on $\M_{b}$ are in the $(\mathbf{\frac{1}{2}},0)$ and $(0,\mathbf{\frac{1}{2}})$ representations of $\SL(2,\C)\times\SL(2,\C)$, and we denote spinor indices with a capital Roman letter $A$, $A'$ respectively (we work in Penrose's abstract index notation \cite{Penrose:1984}).  We will refer to the spinor representations $(\mathbf{\frac{1}{2}},0)$ and $(0,\mathbf{\frac{1}{2}})$ as the `negative chirality' and `positive chirality' spinors, respectively.  Since vectors on $\M_{b}$ are in the $(\mathbf{\frac{1}{2}},\mathbf{\frac{1}{2}})$ representation, this allows us to associate a vector index $\mu$ with a pair of spinor indices $AA'$.  For instance, given a vector $v=v^{\mu}\partial_{\mu}\in T\M_{b}$, we have
\be{spinordecomp}
v^{\mu} \leftrightarrow v^{AA'}=\frac{1}{\sqrt{2}}\left(
\begin{array}{cc}
v^{0}+v^{1} & v^{2}+iv^{3} \\
v^{2}-iv^{3} & v^{0}-v^{1}
\end{array}
\right).
\ee
  
We can raise and lower spinor indices using the $\epsilon$-spinors:
\begin{equation*}
\epsilon_{AB}=\left(
\begin{array}{cc}
0 & 1 \\
-1 & 0
\end{array}\right) = \epsilon_{A'B'}
\end{equation*}
according to the usual conventions: 
\begin{equation*}
v_{A}=v^{B}\epsilon_{BA}, \qquad v^{A'}=\epsilon^{A'B'}v_{B'}.
\end{equation*}
It is easy to see that the Minkowski metric is then given in terms of these $\epsilon$-spinors: 
\begin{equation*}
\eta(v,w)=\eta_{\mu\nu}v^{\mu}w^{\nu}=\epsilon_{AB}\epsilon_{A'B'}v^{AA'}w^{BB'}.
\end{equation*}
Using these rules for lowering spinor indices, we can also see that the spinor decomposition \eqref{spinordecomp} of any vector has a nice formulation in terms of Pauli matrices, with $v_{AA'}=\sigma^{\mu}_{AA'}v_{\mu}$.  

An important point about the spinor-helicity formalism is that it is particularly well-adapted to studying null vectors.  Suppose $v^{AA'}$ corresponds to a null vector in $\M_{b}$; then $v^{AA'}v_{AA'}=\det(v)=0$. Since $v^{AA'}$ is a $2\times 2$ matrix, and the rank of a $2\times 2$ matrix is less than two if and only if it's determinant vanishes, this implies that we can write $v^{AA'}_{\mathrm{null}}=\lambda^{A}\tilde{\lambda}^{A'}$ for a pair of spinors (one of each chirality). Additionally, the $\epsilon$-spinors provide us with a $\SL(2,\C)$-invariant inner product between pairs of spinors of the same chirality:
\be{sip}
\la v w \ra = \epsilon_{AB}v^{A}w^{B}, \qquad [v w]=\epsilon_{A'B'}v^{A'}w^{B'}.
\ee  

For much of this review, we will be concerned with $\cN=4$ super-Yang-Mills theory (this will be introduced properly later in this chapter).  The natural setting for this theory is chiral Minkowski super-space; we denote its complexification as $\M\cong\C^{4|8}$, and chart it with coordinates $(x^{AA'}, \theta^{A a})$, where $a=1,\ldots,4$ indexes the $\SU(4)$ R-symmetry of the theory, and the $\theta$s are anti-commuting/Grassmann/fermionic coordinates.  Everything we have said about the spinor-helicity formalism goes through precisely the same in this setting for the bosonic coordinates; the only extension necessary is the notion of a `null vector' in $\M$.  For this, we simply extend the observation we just made, and state that the fermionic component of a vector is null if and only if it can be decomposed as $v^{Aa}=\lambda^{A}\eta^{a}$, for some spinor $\lambda^{A}$ and some Grassmann parameter $\eta^{a}$.

\subsubsection*{\textit{Twistor space}}  

Rather than define twistor space for both $\M_{b}$ and it's super-extension $\M$, we will simply give all definitions in the super-symmetric language \cite{Ferber:1977qx}.  The analogous statements for the bosonic category should be perfectly clear from this exposition.  For our purposes, twistor space, $\PT$, will be a suitable open subset of the complex projective super-space $\P^{3|4}$; its bosonic truncation $\PT_{b}$ is then just a suitable open subset of $\P^{3}$.  Na\"{i}vely, $\P^{3|4}$ is just the complex projective space $\P^{3}$ with four anti-commuting `dimensions' added to it.  More formally, it can be realized as a `super-scheme' \cite{Manin:1997}: that is, as the topological space $\P^{3}$ with a modified structure sheaf:
\begin{equation*}
\cO_{\P^{3|4}}=\cO\left(\bigoplus_{k=0}^{4}\wedge^{k}\cO_{\P^{3}}(-1)^{\oplus 4}\right).
\end{equation*}
Readers interested in a more formal treatment of super-schemes in the context of twistor theory may consult \cite{Wolf:2006me, Adamo:2012cd}; for this review though, the na\"{i}ve perspective on super-geometry will suffice.

\begin{figure}[t]
\centering
\includegraphics[width=80mm]{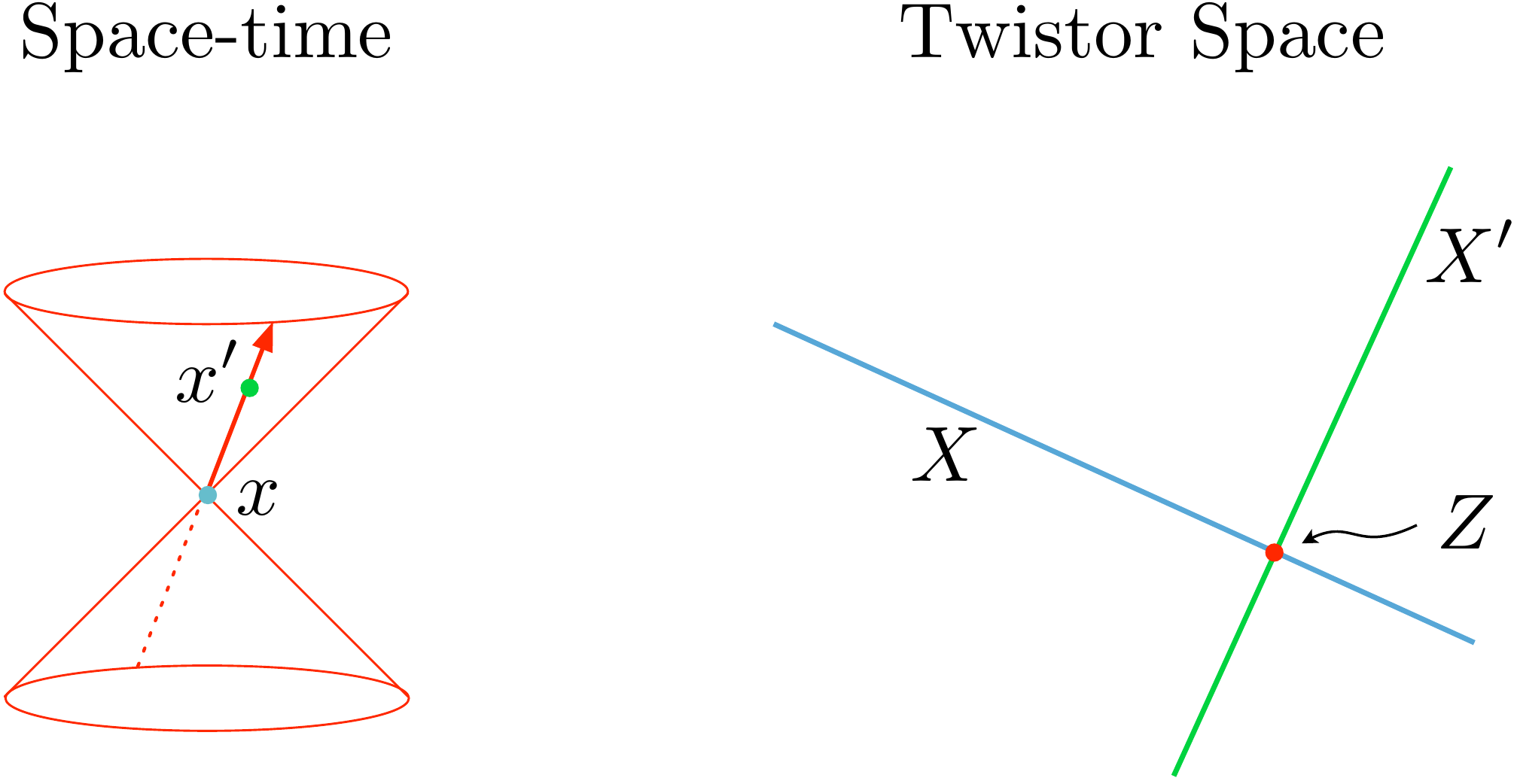}
\caption{\emph{Points in space-time correspond to complex lines in twistor space. Two space-time points are null separated if and only if their corresponding twistor lines intersect.}}
\label{tcorr}
\end{figure}

In this spirit, $\PT$ can be charted by homogeneous coordinates
\be{tsc} 
Z^{I}=(Z^{\alpha},\chi^{a})=(\lambda_{A},\mu^{A'},\chi^{a}), 
\ee 
where $\lambda_{A}$ and $\mu^{A'}$ are 2-component complex Weyl spinors of opposite chirality, and $\chi^a$ is an anti-commuting Grassmann coordinate\footnote{These conventions, first adopted in \cite{Witten:2003nn}, are essentially \emph{dual} to the original Penrose conventions \cite{Penrose:1986ca}.}, with $a=1,\ldots,4$ indexing the $\cN=4$ R-symmetry as before. Being homogeneous coordinates, the $Z^I$s are defined only up to the re-scalings $Z^I\sim r Z^I$ for any $r\in\C^{*}$.  The space $\P^{3|4}$ is a Calabi-Yau super-manifold, in the sense that it has trivial first super-Chern class and its Berezinian sheaf has a canonical global section: $\Ber_{\P^{3|4}}\cong\cO_{\P^{3|4}}$ \cite{Sethi:1994ch, Schwarz:1995ak, Manin:1997}.  This means that $\PT$ is equipped with a global holomorphic measure (the canonical section of $\Ber_{\P^{3|4}}$), which we write as:
\be{volm}
\D^{3|4}Z=\epsilon_{\alpha\beta\gamma\delta}Z^{\alpha}\d Z^{\beta}\wedge\d Z^{\gamma}\wedge\d Z^{\delta}\wedge\d^{4}\chi .
\ee

The most basic relation in twistor theory is the geometric correspondence between a point $(x,\theta)\in\M$ and a complex line\footnote{That is, a linearly embedded Riemann sphere.} $X\subset\PT$.  This complex line is the representation in twistor space of the sphere of null directions uniquely associated to the point in space-time, so if two points in space-time are null separated they share a common null geodesic and hence their associated twistor lines intersect, as illustrated in Figure \ref{tcorr}. Since the conformal structure of $\M$ is given by specifying light cones, and this is equivalent to specifying complex lines in twistor space, we see that giving a complex structure on $\PT$ is the same a giving a conformal structure on $\M$.

This correspondence is captured by the \emph{incidence relations}, which are just algebraic equations relating the coordinates of $\PT$ to $\M$:
\be{inc}
\mu^{A'} =ix^{AA'}\lambda_{A}\, , \qquad \chi^a =\theta^{Aa}\lambda_{A},
\ee 
where we can interpret $\lambda_A$ as homogeneous coordinates on the Riemann sphere $X\cong\P^{1}$, and $(x,\theta)$ are the parameters of the linear embedding.  These also encode the space-time interpretation of a point in twistor space: let $X$ and $X'$ be lines in twistor space which intersect at the point $Z=(\lambda,\mu,\chi)$.  Subtracting the incidence relations for the two points gives $(x-x')^{AA'}\lambda_A=0$ and $(\theta-\theta')^{Aa}\lambda_A=0$, so $(x-x')^{AA'}=\tilde\lambda^{A'}\lambda^A$ and $(\theta-\theta')^{Aa} = \eta^a\lambda^A$ for some Weyl spinor $\tilde\lambda$ and some Grassmann parameter $\eta$. If we vary the possible choices of $(\tilde\lambda,\eta)$, the vectors $(\tilde\lambda\lambda,\eta\lambda)$ span a totally null complex $2|4$-dimensional plane in $\M$, so every point $Z\in\PT$ is assigned such a complex null plane by the incidence relations.  These totally null planes are also known as (super) $\alpha$-planes (e.g., \cite{Huggett:1985, Penrose:1986ca}).  Further, since any two points $Z_{1},Z_{2}\in\PT$ define a line, a point $(x,\theta)\in\M$ can be represented on twistor space by a skew bi-twistor $X^{IJ}=Z_{1}^{[I}Z_{2}^{J]}$.
   
A canonical way of encapsulating the relationship between twistor space and $\M$ is via the `double fibration':
\begin{equation*}
\xymatrix{
 & \PS \ar[ld]_{p} \ar[rd]^{q} & \\
 \PT & & \M }
\end{equation*}
Here, $\PS\cong\M\times\P^{1}$ is the projective un-primed spinor bundle over $\M$ with coordinates $(x,\theta,\lambda)$.  The map $q:\PS\rightarrow\M$ is just the trivial projection, while $p:\PS\rightarrow\PT$ is specified by the incidence relations \eqref{inc}.  The double fibration provides a heuristic picture for how geometrical data on twistor space can be pulled back to $\PS$ and then pushed down to physical data on $\M$; later we will explore a few striking examples of this relationship.

A basic fact about twistor space that is that it carries a natural action of the super-conformal group.  The (complexified) superconformal algebra of $\M$ is $\mathfrak{psl}(4|4,\C)$, and its generators can be written in twistor space as \cite{Drummond:2009fd}:
\be{scongen}
J^{I}_{J}=Z^{I}\frac{\partial}{\partial Z^{J}},
\ee
excluding the Euler homogeneity operator $\Upsilon=Z^{I}\partial_{I}$ and the fermionic homogeneity operator $\chi^{a}\frac{\partial}{\partial \chi^{a}}$.  As we will see, this makes twistor theory an ideal tool for studying physical theories which have conformal symmetry.  Conformal invariance is broken by specifying an `infinity twistor' $I_{IJ}$ which obeys $X^{IJ}I_{JK}=0$ when $X$ corresponds to a point at infinity in $\M$.  In terms of the spinor decomposition of a twistor, the bosonic components of $I$ are given by:
\be{infty}
I_{\alpha\beta}=\left(
\begin{array}{cc}
\epsilon^{AB} & 0 \\
0 & 0
\end{array} \right), \qquad 
I^{\alpha\beta}=\left(
\begin{array}{cc}
0 & 0 \\
0 & \epsilon^{A'B'}
\end{array}\right).
\ee  
A contraction of the form $I_{IJ}Z_{1}^{I}Z_{2}^{J}$ thus breaks $\mathrm{PSL}(4,\C)$ conformal invariance, but maintains invariance under space-time translations (which do not `shift' the location of infinity).  Conformally invariant contractions between bosonic twistors take the form:
\begin{equation*}
(Z_{1},Z_{2},Z_{3},Z_{4})=\epsilon_{\alpha\beta\gamma\delta}Z_{1}^{\alpha}Z_{2}^{\beta}Z_{3}^{\gamma}Z_{4}^{\delta},
\end{equation*}
since this quantity is invariant under $\mathrm{PSL}(4,\C)$ transformations.


\subsubsection{Reality structures and space-time signature}

Thus far, we have consider space-time to be a complex 4-manifold $\M$; for many calculations which take place purely twistorially, this is fine since we can work holomorphically on $\PT$ and perform our computations in the framework of complex analysis and Dolbeault cohomology.  However, on space-time our computations should be taking place on a real slice $\M_{\R}\subset\M$; while real physics happens on the Lorentzian-real slice of signature $(1,3)$, there is no (mathematical) obstruction to choosing other signatures for $\M_{\R}$.  These different choices of space-time signature correspond to different reality structures on twistor space.  From time-to-time we will need to make an explicit choice for $\M_{\R}$, so we provide a brief overview of three choices of space time signature and their consequences on twistor space.  Since the distinctions between reality structures are captured entirely at the bosonic level, we leave the fermionic degrees of freedom out of this discussion.

\subsubsection*{\textit{Lorenztian signature}} 

If we choose $\M_{\R}$ to be real Minkowski space with its metric of signature $(1,3)$, then the natural conjugation on Weyl spinors is the usual complex conjugation of special relativity.  This maps the $(\mathbf{\frac{1}{2}},0)$ and $(0,\mathbf{\frac{1}{2}})$ spinor representations to one another:
\begin{equation*}
v^{A}=(a,b)\mapsto\bar{v}^{A'}=(\bar{a},\bar{b}), \qquad w^{A'}=(c,d)\mapsto \bar{w}^{A}=(\bar{c},\bar{d}).
\end{equation*}
In the spinor-helicity formalism, this means that a vector $v^{\mu}$ is real valued if and only if its $2\times 2$ spinor decomposition is Hermitian: $v^{AB'}=\bar{v}^{BA'}$.  In other words, if $x^{\mu}$ is a position vector in $\M$, then it corresponds to a real point in $\R^{1,3}$ if and only if $x^{AA'}$ is Hermitian.

On twistor space, this induces corresponding reality conditions.  Since complex conjugation exchanges spinor chiralities, it acts as an anti-holomorphic map from twistor space to \emph{dual} twistor space, $\PT^{\vee}$:
\begin{equation*}
Z^{\alpha}=(\lambda_{A},\mu^{A'})\mapsto \bar{Z}_{\alpha}=(\bar{\mu}^{A},\bar{\lambda}_{A'})\in\PT^{\vee}.
\end{equation*}
This defines a pseudo-Hermitian metric of signature $(2,2)$ on $\PT$ which preserves the Lorentzian real form of the conformal group, $\SU(2,2)$:
\begin{equation*}
Z\cdot \bar{Z}\equiv g_{\alpha\bar{\beta}}Z^{\alpha}\bar{Z}^{\bar{\beta}}=Z^{\alpha}\bar{Z}_{\alpha}=\la \lambda \bar{\mu}\ra + [\bar{\lambda}\mu].
\end{equation*}

Since $\PT$ is a projective space, the exact value of this $\SU(2,2)$-inner product is meaningless; its sign is an invariant notion, though.  Hence, we can accordingly partition $\PT$ into three sectors \cite{Huggett:1985, Penrose:1986ca}:
\be{tsets}
\PT^{\pm}=\left\{Z\in\PT | \pm Z\cdot\bar{Z}\geq 0\right\}, \qquad \PN = \left\{Z\in\PT | Z\cdot\bar{Z}=0\right\}.
\ee
The sets $\PT^{\pm}$ correspond to the future and past tubes of $\M$ respectively; that is, the the sets on which the imaginary part of $x^{AA'}$ is past or future pointing time-like respectively.  This follows from the fact that if we take $x=u+iv$, then $Z\cdot\bar{Z}= -v^{AA'}\bar{\lambda}_{A'}\lambda_{A}$ via \eqref{inc}.  This has a definite sign, depending on whether $v$ is time-like and future- or past-pointing, as claimed.  This indicates that $\PT^{\pm}$ is the natural choice for the regions of $\PT$ corresponding to positive/negative frequency fields on $\M$.  For instance, a field of positive frequency, whose Fourier transform is supported on the future lightcone in momentum space, automatically extends over the future tube because $\e^{ip\cdot x}$ is rapidly decreasing there, bounded by its values on the real slice.

If a line $X$ lies entirely in $\PN$, then \eqref{inc} tells us that
\be{Minkincidence}
0= i(x-x^\dagger)^{AA'}\lambda_A\bar\lambda_{A'}\qquad\hbox{for all }\lambda\ ,
\ee
which is possible if and only if the matrix $x^{AA'}$ is Hermitian, so $x\in\M_{\R}=\R^{1,3}$. Conversely, a point $Z\in\PN$ corresponds to a unique real null ray (the intersection of the complex $\alpha$-plane with $\M_{\R}$).  Hence, the portion of twistor space corresponding to the real slice $\R^{1,3}$ is the set $\PN$.

\subsubsection*{\textit{Euclidean signature}}

Now suppose we choose our real slice to have Euclidean signature $(+,+,+,+)$,\footnote{Actually, the procedure described here results in a \emph{negative} definite metric.  In practice this simply requires a change of sign at the end of calculations to obtain actual Euclidean results, so we will ignore the distinction from now on.} so that $\M_{\R}$ is the real Euclidean 4-space $\E\cong\R^{4}$ or, in the conformal compactification picture, $S^{4}$.  The Euclidean real form of the spin group is locally isomorphic to $\SO(4,\R)\cong \SU(2)\times\SU(2)/\Z_{2}$.  The Euclidean conjugation of Weyl spinors no longer interchanges spinor representations, and is given by \cite{Atiyah:1978wi, Woodhouse:1985id}:
\begin{equation*}
v^{A}=(a,b)\mapsto\hat{v}^{A}=(\bar{b},-\bar{a}), \qquad w^{A'}=(c,d)\mapsto\hat{w}^{A'}=(-\bar{d},\bar{c}).
\end{equation*}
This means that a position vector $x^{\mu}$ corresponds to a real point in $\E$ if and only if $x^{AA'}=\hat{x}^{AA'}$.  Note that $\hat{\hat{v}}^{A}=-v^{A}$, leading to the nomenclature `quaternionic conjugation' for this reality structure.

On twistor space, this induces an anti-holomorphic involution $\sigma:\PT\rightarrow\PT$ which has no fixed points and obeys $\sigma^{2}=-\mathrm{id}$.  This means that there are no points in twistor space preserved by the reality structure, which is just another way of saying that there are no real null vectors in $\E$ (i.e., the $\alpha$-plane corresponding to $Z\in\PT$ does not intersect the real slice in a null ray).  However, it is clear that $\sigma$ acts as the antipodal map on $\P^{1}$, so although it has no fixed points, it does have \emph{fixed lines}, given by $X^{\alpha\beta}=Z^{[\alpha}\hat{Z}^{\beta]}$.  Hence, each point in $\PT$ corresponds to a unique point in $\E$ by this construction, which can be written explicitly using \eqref{inc}:
\be{eucmap}
\rho: \PT\rightarrow\E, \qquad Z^{\alpha}=(\lambda_{A},\mu^{A'})\mapsto x^{AA'}=-i\frac{\mu^{A'}\hat\lambda^A-\hat\mu^{A'}\lambda^A}{\la\lambda\hat\lambda\ra} .
\ee
So in Euclidean signature, twistor space is just a $\P^{1}$ fibration $\PT\rightarrow\E$, and there is no need for the double fibration picture.  The conformally compactified version of this picture, with the $\P^{1}$ fibration $\PT\rightarrow S^{4}$ plays an important role in the ADHM construction of Yang-Mills instantons \cite{Atiyah:1978ri}.

This fibration allows us to define a complex structure for $\PT$ in terms of coordinates on $\E$.  To do this, we first specify a basis of $(0,1)$-forms on twistor space \cite{Mason:2005zm, Jiang:2008}:
\be{bforms}
\hat{e}^{0}=\frac{\la\hat{\lambda}\d\hat{\lambda}\ra}{\la\lambda\hat{\lambda}\ra^2}, \qquad \hat{e}^{A'}=\frac{\hat{\lambda}_{A}\d x^{AA'}}{\la\lambda\hat{\lambda}\ra},
\ee
and a dual basis for $T^{0,1}\PT_{b}$:
\be{bvects}
\dbar_{0}=\la\lambda\hat{\lambda}\ra\lambda_{A}\frac{\partial}{\partial\hat{\lambda}_{A}}, \qquad \dbar_{A'}=\lambda^{A}\partial_{AA'}.
\ee
Then the complex structure is given by the anti-holomorphic Dolbeault operator
\begin{equation*}
\dbar=\d\hat{Z}^{\alpha}\frac{\partial}{\partial \hat{Z}^{\alpha}}=\hat{e}^{0}\dbar_{0}+\hat{e}^{A'}\dbar_{A'}.
\end{equation*}
Additionally, we have Woodhouse's operator \cite{Woodhouse:1985id}
\be{WHO}
\dhat=\d\hat{Z}^{\alpha}\frac{\partial}{\partial Z^{\alpha}},
\ee
which acts as a holomorphic derivative in the anti-holomorphic directions and obeys $\dhat^{2}=0$, $\dbar\dhat=-\dhat\dbar$.

Note that although Euclidean signature is less realistic than Lorentzian, it also allows us to be very explicit in relating twistorial quantities to their space-time counterparts.  We will take advantage of this fact at several points later on.

\subsubsection*{\textit{Split signature}}

Finally, we discuss the consequences of choosing the real slice $\M_{\R}$ to have split signature $(+,+,-,-)$.  \emph{A priori}, this is the least physical choice of real slice: Euclidean signature is also non-physical, but there is the long-standing Wick-rotation prescription for moving between the Lorentzian and Euclidean regimes -- in the split signature case, there is neither a notion of `time' or of `space.'  However, in split signature, the spin group is locally isomorphic to $\SO(2,2,\R)\cong \SL(2,\R)\times\SL(2,\R)$, so the spinor representations are real and we get a substantial simplification.  The real form of the conformal group in this case is $\mathrm{PSL}(4,\R)$, so the reality condition on twistor space is simply to take all twistors to be real-valued; that is, $\PT_{\R}\subset\RP^{3|4}$.  

Hence, the benefit of choosing this un-physical signature is in the ability to work entirely with real variables (both with the spinor-helicity formalism and on twistor space).  This allows for a high degree of explicit calculation, for instance, constructing Yang-Mills instantons using twistor data \cite{Mason:2005qu}.  Split signature was highly utilized in the early developments of twistor-string theory, as well.  In Witten's original formulation \cite{Witten:2003nn}, it allowed functions of null momenta (e.g., scattering amplitudes) to be written as twistor functions on $\RP^{3|4}$ using a simple integral transform that would come to be known as the `half-Fourier' transform (c.f., \cite{Mason:2009sa}).  Berkovits' subsequent re-formulation of twistor string theory as an open string theory with world-sheet boundaries ending on $\RP^{3|4}$ also relied on this choice of signature \cite{Berkovits:2004hg}.

In this review, we avoid the split signature perspective, preferring to compute in the complexified setting.  When a choice of space-time signature is necessary, we only ever use Lorentzian or Euclidean signatures.  While this means that we lose access to the real-analytic methods of the split signature setting, we still have considerable computational power thanks to holomorphic tools at our disposal.  As we shall see, the result is a more general methodology for performing twistorial calculations and a minimized reliance on an explicit choice of space-time signature.


\subsubsection{Some important facts}

We conclude our lightning review of twistor theory by stating some of the fundamental theorems that have emerged from twistor theory over the past forty years.  These results serve as the primary tools in the remainder of our studies, and are examples of the general twistor philosophy: physical data on space-time is translated into pure geometry on twistor space.

\subsubsection*{\textit{The Penrose transform}}

One of the earliest results in twistor theory is a statement which allows us to represent zero-rest-mass (z.r.m.) fields on space-time in terms of cohomological data on twistor space.  If $\phi_{A_{1}\cdots A_{n}}(x)$ is a spinor field on $\M_{b}$ (with $n$ symmetric negative chirality spinor indices) which satisfies the linear partial differential equation
\begin{equation*}
\partial^{A_{1}A'}\phi_{A_{1}\cdots A_{n}}(x)=0,
\end{equation*}
then we say that it is a \emph{z.r.m. field of helicity} $-\frac{n}{2}$.  Similarly, we define z.r.m. fields of helicity $\frac{n}{2}$ and zero (i.e., scalars) by solutions to the equations:
\begin{equation*}
\partial^{AA'_{1}}\phi_{A'_{1}\cdots A'_{n}}(x)=0, \qquad \Box \phi(x)=0.
\end{equation*}

The Penrose transform manifests itself in the following manner \cite{Penrose:1969ae}:
\begin{thm}[Penrose transform]\label{PenTran}
Let $\PT_{b}\subset \P^{3}$ be a suitably chosen open subset, and $U' \subset \M_{b}$ be the corresponding open subset under the twistor double-fibration: $U'= q \circ p^{-1}(\PT_{b})$.  Then we have the following isomorphism:
\begin{equation*}
H^{1}(\PT_{b},\cO(2h-2))\cong \left\{\mbox{On-shell z.r.m. fields on}\;U'\;\mbox{of helicity}\;h\right\},
\end{equation*}
where $H^{1}$ denotes analytic cohomology and $\cO(k)$ is the sheaf of holomorphic functions which are homogeneous of degree $k$.
\end{thm}

Proving this isomorphism in detail is actually rather involved (c.f., \cite{Eastwood:1981jy, Ward:1990}), so we will ignore the details of the proof.  For our purposes, the choice of open subset $\PT_{b}$ can usually be made to coincide with one of \eqref{tsets}, or with the exclusion of the `point at infinity' in $\M$.  From now on, we cease to mention this explicitly to avoid cumbersome notation; the choice of twistor space should always be obvious from the context.

In this paper, we will work with the Dolbeault representation for the cohomology of twistor space; this should be contrasted against much of the older twistor literature, where a \v{C}ech representation is utilized \cite{Penrose:1969ae, Eastwood:1981jy, Ward:1990}.  To distinguish between the two representations, Dolbeault cohomology will be denoted $H^{0,k}$, and in this framework cohomology classes are given by $\dbar$-closed $(0,k)$-forms modulo $\dbar$-exact ones.  Choosing the Dolbeault representation is in keeping with our general complex/holomorphic philosophy and will lead to considerably simpler computations in many cases.

The Penrose transform can be realized quite explicitly thanks to beautiful integral formulae.  First consider a free massless scalar $\Phi(x)$ which is a solution to the wave equation $\Box\Phi(x)=0$.  Theorem \ref{PenTran} indicates that this field can be constructed from a cohomology class on (bosonic) twistor space $\phi(Z)\in H^{0,1}(\PT_{b},\cO(-2))$.  Take
\be{scalar}
\Phi(x) =  \frac{1}{2\pi i}\int_{X} \D\lambda \wedge \phi(Z)|_{X} \, ,
\ee
where the restriction of $\phi$ to $X\cong\P^{1}$ is given by the incidence relations \eqref{inc}, and $\D\lambda=\la \lambda \d \lambda \ra$ is the weight $+2$ holomorphic measure on $X$.  The fact that this yields a solution to the wave equation follows by differentiating under the integral and using the incidence relations.

If instead we had worked in a \v{C}ech representation, the transform would have been given by a representative $\phi\in\check{H}^{1}(\PT_{b},\cO(-2))$ and an integral
\begin{equation*}
\Phi(x)=\frac{1}{2\pi i}\oint\limits_\Gamma \D\lambda \, \phi(Z)|_{X} \,,
\end{equation*}
where the expression is really a contour integral on $X$ with contour $\Gamma$ specified by the cohomology class.  This demonstrates the advantage working with Dolbeault cohomology: no choice of contour is needed and we avoid the combinatorics of Leray covers required in the \v{C}ech setup. 

For other fields, the transform is realized by the natural extensions:
\be{nhel}
\phi_{A_{1}\cdots A_{k}}(x)=\frac{1}{2\pi i}\int_{X}\lambda_{A_{1}}\cdots\lambda_{A_{k}}\;\D\lambda\wedge\phi(Z)|_{X}, 
\ee
\be{phel}
\phi_{A'_{1}\cdots A'_{k}}(x)=\frac{1}{2\pi i}\int_{X}\D\lambda\wedge\;\frac{\partial}{\partial\mu^{A'_{1}}}\cdots\frac{\partial}{\partial\mu^{A'_{k}}}\;\phi(Z)|_{X},
\ee
for negative and positive helicity respectively.  Once again, the fact that the z.r.m. equations are obeyed follows by differentiating under the integral sign and noting that 
\begin{equation*}
\partial_{AA'}=i\lambda_{A}\frac{\partial}{\partial \mu^{A'}},
\end{equation*}
thanks to the incidence relations.

The Penrose transform extends naturally to deal with $\cN=4$ supersymmetry, as we will discuss shortly.

\subsubsection*{\textit{Momentum eigenstates}}

The Penrose transform lets us define on-shell physical momentum eigenstates in terms of twistor cohomology; this will be particularly useful when we want to compare twistorial calculations to known space-time results.  To begin, recall that on the complex plane $\C$ the delta function supported at the origin is naturally interpreted as a $(0,1)$-form\footnote{Actually, this is really a $(0,1)$-\emph{current}, but we will not make this distinction here.}:
\begin{equation*}
\bar{\delta}^{1}(z)\equiv\delta(x)\delta(y)\, \d \bar z =\frac1{2\pi i}\,\d \bar z\frac{\del}{\del\bar z}\, \frac1z\, , \qquad z=x+iy.
\end{equation*} 
as a result of the Cauchy kernel for the $\dbar$-operator \cite{Witten:2004cp}.  

If $\lambda_{A}$ is chosen to be a homogeneous coordinate on $\P^1$, then this extends naturally to the Riemann sphere by taking into account a scaling integral.  This allows us to define a projective version of $\bar{\delta}^{1}(z)$:
\be{meig1}
\bar\delta^1_m(\lambda, \lambda ')\equiv \int_{\C} \frac{\rd s}{ s^{1+m}}\,\bar\delta^2(s\lambda_{A}+\lambda'_{A})\, ,
\ee
which is supported only when $\lambda$ and $\lambda'$ coincide projectively.  One can check that $\bar\delta^1_m(\lambda,\lambda')$ has homogeneity $m$ in $\lambda$ and $-m-2$ in $\lambda'$, so that
\begin{equation*}
f(\lambda') = \int_{\P^1} f(\lambda)\;\bar\delta^1_m(\lambda,\lambda')\wedge\D\lambda
\end{equation*}
for any function $f$ of homogeneity $-m-2$ on $\P^1$.

Consider a particle of helicity $h$ with on-shell momentum $p_{AA'}=p_{A}\tilde{p}_{A'}$; Theorem \ref{PenTran} tells us that this will be represented by a twistor cohomology class taking values in the sheaf $\cO(2h-2)$.    Using \eqref{meig1}, we define the twistor momentum eigenstate
\be{meig2}
f_{2h-2}(\mu,\lambda)=\int_{\C}\frac{\d s}{s^{2h-1}}\bar{\delta}^{2}(s\lambda_{A}+p_{A})\;e^{s[\mu\;\tilde{p}]}.
\ee
Using the integral formulae \eqref{nhel}-\eqref{phel}, we see that this evaluates to give the appropriate z.r.m. fields on space-time
\begin{equation*}
p_{A_{1}}\cdots p_{A_{|2h|}}e^{ip\cdot x}, \qquad \tilde{p}_{A'_{1}}\cdots\tilde{p}_{A'_{|2h|}}e^{ip\cdot x},
\end{equation*}
depending on whether $h$ is negative or positive.  When we work with $\cN=4$ supersymmetry, \eqref{meig2} is easily modified by taking into account the full on-shell supermomentum $(p_{A}\tilde{p}_{A'},p_{A}\eta_{a})$:
\be{meig3}
f(\mu, \lambda ,\chi)=\int_{\C} \frac{\rd s}{ s}\bar\delta^2(s\lambda_{A} + p_{A})\,e^{s[[\mu\;\tilde{p}]]} ,  
\ee
where $[[\mu\tilde{p}]]=[\mu\tilde{p}]+\chi^{a}\eta_{a}$.

\subsubsection*{\textit{Woodhouse representatives}}

We have seen that the Penrose transform allows us to construct integral formulae for z.r.m. fields of arbitrary integer or half-integer helicity on $\M$ from cohomology classes on twistor space.  But if we are given the z.r.m. field on space-time, can we construct a twistorial cohomology class which manifests the space-time degrees of freedom?  We saw previously that when the real slice of $\M$ is chosen to be the Euclidean space $\E$, twistor space is just a $\P^1$ bundle $\PT\rightarrow\E$ equipped with a quaternionic conjugation.  It turns out that this additional structure is very useful, and lets us write down explicit cohomological representatives for negative helicity fields.

\begin{thm}[Woodhouse \cite{Woodhouse:1985id}]\label{Wrep1}
Let $\phi_{A\cdots B}(x)$ be a field on $\E$ with $2h$ symmetric spinor indices, $\dhat:\Omega^{0,k}\rightarrow\Omega^{0,k+1}$ as in \eqref{WHO}, and
\begin{equation*}
F_{\phi}\equiv \frac{\phi_{A\cdots B}\hat{\lambda}^{A}\cdots\hat{\lambda}^{B}}{\la\lambda\hat{\lambda}\ra^{2h+1}}.
\end{equation*}
Then $\dhat F_{\phi}\in H^{0,1}(\PT, \cO(-2h-2))$ if and only if $\phi_{A\cdots B}$ is a z.r.m. field of helicity $-h$.  Furthermore, $\dhat F_{\phi}$ is holomorphic upon restriction to the $\P^{1}$ fibers of $\PT$, and every class in $H^{0,1}(\PT, \cO(-2h-2))$ has a unique representative which is $\dhat$-exact.
\end{thm}
This result shows that by choosing Euclidean signature, we can build representatives for negative helicity fields which are holomorphic upon restriction to the $\P^{1}$ fibers of twistor space.  Constructing such representatives for fields with a primed spinor index is a bit more complicated, since there is no longer a potential definition:
\begin{thm}[Woodhouse \cite{Woodhouse:1985id}]\label{Wrep2}
Let $\phi_{A'A\cdots B}(x)$ be a field on $\E$ with $n$ symmetric un-primed spinor indices and define
\begin{equation*}
\alpha_{\phi}=-\frac{\phi_{A'A\cdots B}\lambda^{A}\cdots\lambda^{B}}{i^{n}\la\lambda\hat{\lambda}\ra}\;\hat{\lambda}_{C}\d x^{CA'}.
\end{equation*}
Then $\alpha_{\phi}\in H^{0,1}(\PT,\cO(n-1))$ if and only if $\partial_{A'(C}\phi^{A'}_{A)\cdots B}=0$.  Furthermore, $\alpha_{\phi}$ is holomorphic upon restriction to the $\P^{1}$ fibers of $\PT$.
\end{thm}

For the remainder of this paper, we will refer to the representatives $\dhat F_{\phi}$, $\alpha_{\phi}$ as \emph{Woodhouse representatives}.  They will be particularly useful in our study of the twistor action Feynman rules in the following section.

\subsubsection*{\textit{The Ward correspondence}}

The final classic result of twistor theory that will be used extensively in our discussion of gauge theory is the \emph{Ward correspondence}.  Heuristically, this can be thought of as a non-linear version of the Penrose transform dealing with Yang-Mills instantons (c.f., \cite{Ward:1990,Jost:2008} for a more detailed review of gauge theory).

Let $E\rightarrow\M$ be a principal $G$-bundle over space-time, with $G$ some Lie group.  The bundle $E$ encodes the information of the gauge group in the sense that $\End(E)\cong\mathfrak{g}^{\C}$, where $\mathfrak{g}^{\C}$ is the complexified Lie algebra of $G$.  On $E$ we can define a connection $\nabla$ which can be written locally as 
\begin{equation*}
\nabla=\d +A, \qquad A\in\Omega^{1}(\M, \End(E)).
\end{equation*}
Since $\M$ is a 4-manifold, the curvature of this connection $F$ can be decomposed into its self-dual (SD) and anti-self-dual (ASD) parts using the Hodge star:
\begin{equation*}
F=\d A+A\wedge A=F^{+}+F^{-}, \qquad *F^{\pm}=\pm F^{\pm}.
\end{equation*}
The connection $\nabla$ is said to be a \emph{Yang-Mills connection} if it is an extremum of the Yang-Mills action functional and satisfies the Yang-Mills equation:
\be{YMF}
S[A]=\int_{\M}\tr\left(F\wedge *F\right), \qquad \nabla *F=0,
\ee
where the trace is over $\End(E)$.  We call the connection $\nabla$ a \emph{Yang-Mills instanton} if its curvature is purely SD: $F=F^{+}$ (such a connection is automatically Yang-Mills by the Bianchi identity).  

The Ward correspondence tells us that there is a duality between Yang-Mills instantons and holomorphic vector bundles over twistor space satisfying certain conditions, and is true for a wide variety of gauge groups $G$ (e.g., \cite{Atiyah:1979}), although it was first formulated by Ward for $\GL(N,\C)$ \cite{Ward:1977ta}.  In this paper, we almost always work with $G=\SU(N)$ or $\U(N)$, so we state that version of the theorem here:
\begin{thm}[Ward Correspondence]\label{WardCorr}
Let $\PT$ be a suitable open subset of $\P^{3}$ and $\M$ the corresponding open subset in space-time.  There is a one-to-one correspondence between:
\begin{enumerate}
\item $\SU(N)$ Yang-Mills instantons on $\M$, and
\item holomorphic rank-$N$ vector bundles $V\rightarrow\PT$ such that \emph{(a.)} $V|_{X}$ is topologically trivial for $X\cong\P^{1}$ corresponding to $x\in\M$; \emph{(b.)} $\det V$ is trivial; and \emph{(c.)} $V$ admits a positive real form.
\end{enumerate}
\end{thm}
  
Here, the conditions (b.) and (c.) are not terribly important.  The condition that $\det V$ be trivial amounts to requiring that $V$ has a nowhere-vanishing holomorphic section, and a positive real form can be built from a reality structure on $\PT$ and the Killing form (this can even be done uniquely \cite{Atiyah:1978wi}). The important `moral' of the Ward correspondence is the equivalence between Yang-Mills instantons and holomorphic vector bundles on twistor space.  The proof of this theorem relies heavily upon the integrability of the SD Yang-Mills equations and illustrates a key trend: that twistors are powerful tools for describing integrable systems.

The Ward correspondence can be used to build explicit examples of Yang-Mills instantons.  This was first explored for $G=\SU(2)$ by Atiyah and Ward \cite{Atiyah:1977pw} and later generalized to the ADHM construction \cite{Atiyah:1978ri}.  For our purposes, it will be important to note that the correspondence continues to hold for \emph{supersymmetric} Yang-Mills theories \cite{Manin:1997}.


\subsection{A Calculus of Distributional Forms}

We have just seen that the Penrose transform allows us to write down momentum eigenstates as $(0,1)$-form cohomology classes in twistor space.  As we will see in the next chapter, it is often easier to work with twistor states which are dual to these eigenstates in a particular way.  In this section, we introduce a calculus of distributional forms on twistor space which will greatly facilitate our later discussions.  Here, our presentation will focus on the $\cN=4$ supersymmetric setting, trusting the reader to understand the generalization to other amounts of supersymmetry ($\cN=0,8$ will be the most important other cases).  This builds off earlier work in the real setting \cite{Mason:2009sa, Mason:2009qx} and was first set out in the complex setting by \cite{Adamo:2011cb}.

\subsubsection*{\textit{Distributional forms}}

Building from our earlier discussions, we begin by defining a Dolbeault delta-function of $\C^{4|4}$ by
\be{DF1}
\bar{\delta}^{4|4}(Z)=\prod_{\alpha=0}^{3}\bar{\delta}(Z^{\alpha})\prod_{a=1}^{4}\chi ^{a}=\bigwedge_{\alpha=0}^{3}\dbar\left(\frac{1}{Z^{\alpha}}\right)\prod_{a=1}^{4}\chi^{a}.
\ee
This is a $(0,4)$-form on $\C^{4|4}$ of weight zero and obeys the delta-function property in the fermionic coordinates thanks to the usual Berezinian integration rule: $\int\chi \d\chi=1$ \cite{Manin:1997}.  From this, we can define a projective delta-function by:
\be{DF2}
\bar{\delta}^{3|4}(Z_{1},Z_{2})=\int_{\P^{1}}\frac{\D c}{c_{1}c_{2}}\bar{\delta}^{4|4}(c_{1}Z_{1}+c_{2}Z_{2})=\int_{\C}\frac{\d s}{s}\bar{\delta}^{4|4}(Z_{1}+s Z_{2}),
\ee
where $\D c=c_{1}\d c_{2}+c_{2}\d c_{1}$.  This is a homogeneous $(0,3)$-form on $\P^{3|4}\cong\PT$ which enforces the projective coincidence of its arguments, is antisymmetric under their interchange, and obeys the natural identity
\begin{equation*}
f(Z')=\int_{\PT}f(Z)\bar{\delta}^{3|4}(Z,Z')\wedge\D^{3|4}Z.
\end{equation*}  
In other words, $\bar{\delta}^{3|4}$ acts as the anti-holomorphic Dirac current on $\PT$.

By integrating against a further parameter, we can obtain the $\delta$-function
\be{DF3}
\begin{aligned}
	\bar{\delta}^{2|4}(Z_{1},Z_{2},Z_3)&=\int_{\P^{2}}\frac{\D^{2} c}{c_{1}c_{2}c_3}\,\bar{\delta}^{4|4}(c_1Z_{1}+c_2Z_{2}+c_3Z_3)\\
	&=\int_{\C^2}\frac{\d s}{s}\frac{\d t}{t}\,\bar{\delta}^{4|4}(Z_3+sZ_{1}+tZ_{2}) \\
	&= \int_\C \frac{\d s}{s}\, \bar{\delta}^{3|4}(Z_{1},Z_{2}+s Z_3)\ ,
\end{aligned}
\ee
where $\D^{2}c=c_{1}\d c_{2}\wedge\d c_{3}+$ cyclic permutations.  This has support when $Z_1$, $Z_2$ and $Z_3$ are projectively collinear, and is manifestly superconformally invariant, weightless in each $Z_{i}$, and antisymmetric under exchange of any two.  Further, this is a homogeneous $(0,2)$-form on $\PT$ which has simple poles when any two of its arguments coincide.  

Following this pattern, we can similarly define
\be{DF4}
\begin{aligned}
	\bar{\delta}^{1|4}(Z_{1},Z_{2},Z_3, Z_4)&=\int_{\P^{3}}\frac{\D^{3} c}{c_{1}c_{2}c_3c_4}\,\bar{\delta}^{4|4}(c_1Z_{1}+c_2Z_{2}+c_3Z_3+c_4Z_4) \\
	&=\int_{\C^3}\frac{\d s}{s}\frac{\d t}{t}\frac{\d u}{u}\,\bar{\delta}^{4|4}(Z_4+sZ_3+tZ_{2}+uZ_{1})\\
	&= \int_\C \frac{\d s}{s} \bar{\delta}^{2|4}(Z_{1},Z_{2},Z_3+s Z_4).
\end{aligned}
\ee
This $(0,1)$-form is supported where its arguments lie on the same plane $\P^{2}\subset\PT$, and is singular when any three are collinear.  Finally, we have the rational object
\be{DF5}
\begin{aligned}
	\bar{\delta}^{0|4}(Z_{1},Z_{2},Z_3,Z_4,Z_5) \equiv[1,2,3,4,5]&=\int_{\P^4}\frac{\D^4 c}{c_1c_2c_3c_4c_5}\bar{\delta}^{4|4}\left(\sum_{i=1}^5c_iZ_i\right)\\
	&=\frac{\left( (1234)\chi_5 + \mbox{cyclic}\right)^4}{(1234)(2345)(3451)(4512)(5123)}\ ,
\end{aligned}
\ee
where the second line is obtained by integration against the delta functions and $(1234)\equiv \epsilon_{\alpha\beta\gamma\delta}Z_1^\alpha Z_2^\beta Z_3^\gamma Z_4^\delta$ \cite{Mason:2009qx}.  As there are no remaining bosonic delta-functions, this forces its five arguments to inhabit the same bosonic `body' $\P^3\subset\PT$ of twistor space.  The notation $\bar\delta^{0|4}(Z_1,Z_2,Z_3,Z_4,Z_5)=[1,2,3,4,5]$ indicates that if we were working with \emph{momentum} twistors, this would be the standard dual superconformal invariant: the R-invariant of \cite{Drummond:2008vq}.  On twistor space, this is the standard invariant of the regular superconformal group $\mathrm{PSL}(4|4,\C)$.

\subsubsection*{\textit{Some identities}}

We now briefly state a few properties of these distributional forms.  These will prove very useful later on.

\begin{lemma}\label{dfprop1}
Let $\dbar$ be the anti-holomorphic Dolbeault operator with respect to the twistor coordinates $Z^{I}$.  Then
\begin{equation*}
\dbar\bar{\delta}^{r|4}(Z_{1},\ldots,Z_{5-r})=2\pi i \sum_{i=1}^{5-r}(-1)^{i+1}\bar{\delta}^{r+1|4}(Z_{1},\ldots,\widehat{Z}_{i},\ldots,Z_{5-r}),
\end{equation*}
where $\widehat{Z}_{i}$ is omitted, for $r=0,\ldots,3$. Further, $\dbar\bar{\delta}^{3|4}(Z,Z')=0$.  
\end{lemma}
\proof  The fact that $\dbar\bar{\delta}^{3|4}(Z,Z')=0$ follows immediately from the fact that the $Z^{I}$s are homogeneous coordinates if the general relation holds.  Since $\bar{\delta}^{4|4}$ is a top-degree form on $\C^{4|4}$, it is $\dbar$-closed.  Thus,
\begin{equation*}
\dbar_{\mathrm{T}}\bar{\delta}^{4|4}\left(\sum_{i=1}^{5-r}c_{i}Z_{i}\right)=0,
\end{equation*}
where $\dbar_{\mathrm{T}}=\dbar+\dbar_{c}$ is the total $\dbar$-operator on the space of parameters $\{c_{i}\}$ together with the twistors $Z_{i}$, and $\dbar_{c}$ is the that on the $c_{i}$s alone.  Therefore, we have
\begin{multline*}
\dbar\bar{\delta}^{r|4}(Z_{1},\ldots,Z_{5-r})=\int_{\P^{4-r}}\frac{\D^{4-r}c}{c_{1}\cdots c_{5-r}}\dbar\bar{\delta}^{4|4}\left(\sum_{i=1}^{5-r}c_{i}Z_{i}\right) \\
=-\int_{\P^{4-r}}\frac{\D^{4-r}c}{c_{1}\cdots c_{5-r}}\dbar_{c}\bar{\delta}^{4|4}\left(\sum_{i=1}^{5-r}c_{i}Z_{i}\right) \\
=\int_{\P^{4-r}}\dbar_{c}\left(\frac{\D^{4-r}c}{c_{1}\cdots c_{5-r}}\right)\bar{\delta}^{4|4}\left(\sum_{i=1}^{5-r}c_{i}Z_{i}\right)\\
=\int_{\P^{4-r}}\D^{4-r}c \left(\sum_{i=1}^{5-r}\frac{1}{c_{1}\cdots\widehat{c}_{i}\cdots c_{5-r}}\dbar_{c}\frac{1}{c_{i}}\right)\bar{\delta}^{4|4}\left(\sum_{i=1}^{5-r}c_{i}Z_{i}\right),
\end{multline*}
where the third line is obtained by integrating by parts.

Now, we use the fact that $\dbar_{c}c_{i}^{-1}=2\pi i\bar{\delta}^{1}(c_{i})$ to obtain:
\begin{multline*}
\dbar\bar{\delta}^{r|4}(Z_{1},\ldots,Z_{5-r}) \\
=2\pi i \int_{\C^{3-r}}\frac{\d s_{1}}{s_{1}}\cdots\frac{\d s_{3-r}}{s_{3-r}}\left(\bar{\delta}^{4|4}(Z_{2}+s_{1}Z_{3}+\cdots +s_{3-r}Z_{5-r})+\mathrm{cyclic}\right) \\
=2\pi i \left(\bar{\delta}^{r+1|4}(Z_{2},\ldots, Z_{5-r})+\mathrm{cyclic}\right),
\end{multline*}
as required.     $\Box$

\medskip

Additionally, using these distributional forms, many integrals can be performed algebraically:
\begin{lemma}\label{dfprop2}
\begin{eqnarray*}
\bar{\delta}^{1|4}(Z_{1},Z_{2},Z_{3},Z_{4}) & = & \int_{\PT}\bar{\delta}^{2|4}(Z_{1},Z_{2},Z)\bar{\delta}^{2|4}(Z,Z_{3},Z_{4})\;\D^{3|4}Z, \\
\left[1,2,3,4,5\right] & = & \int_{\PT}\bar{\delta}^{2|4}(Z_{1},Z_{2},Z)\bar{\delta}^{1|4}(Z,Z_{3},Z_{4},Z_{5})\;\D^{3|4}Z.
\end{eqnarray*}
\end{lemma}
\proof By the definitions \eqref{DF3}, \eqref{DF4} we have
\begin{multline*}
\int_{\PT}\bar{\delta}^{2|4}(Z_{1},Z_{2},Z)\bar{\delta}^{2|4}(Z,Z_{3},Z_{4})\;\D^{3|4}Z \\
=\int_{\PT\times \C}\frac{\d s}{s}\bar{\delta}^{3|4}(Z_{1}+sZ_{2},Z)\bar{\delta}^{2|4}(Z,Z_{3},Z_{4})\;\D^{3|4}Z \\
=\int_{\C}\frac{\d s}{s}\bar{\delta}^{2|4}(Z_{1}+s Z_{2},Z_{3},Z_{4})=\bar{\delta}^{1|4}(Z_{1},Z_{2},Z_{3},Z_{4}).
\end{multline*}
The other identity follows in a similar fashion.     $\Box$


\subsection{$\cN=4$ Super-Yang-Mills Theory}

The gauge theory portion of this review focuses on maximally supersymmetric (i.e., $\cN=4$) super-Yang-Mills (SYM) theory in four space-time dimensions.  This theory is special for a wide variety of reasons which make it the simplest four dimensional gauge theory.  It was originally obtained by dimensional reduction from $\cN=1$ Yang-Mills theory in ten dimensions \cite{Brink:1976bc}, is UV finite and superconformal, and its space-time Lagrangian has only two tunable parameters: the gauge group and the coupling.  The AdS/CFT correspondence has indicated that it has a gravitational dual in the form of a Type IIB string theory on $AdS_{5}\times S^{5}$ \cite{Maldacena:1997re, Witten:1998qj, Gubser:1998bc, Aharony:1999ti}.  This provides a method for performing strong coupling calculations, and the widely believed integrability of the theory in the planar limit has enabled remarkable computational advances for physical observables (see \cite{Beisert:2010jr} for a review).

While $\cN=4$ SYM is a highly idealized version of the theories we believe to actually describe the interactions of the electromagnetic, strong, and weak forces (indeed, $\cN=4$ SYM contains no bound states, and therefore cannot describe these forces), it nevertheless captures many of the essential qualities that underlie actual physical theories such as QED or QCD.  For instance, the computation of 1-loop gluon scattering amplitudes in QCD can be facilitated by considering the 1-loop $\cN=4$ amplitude, along with $\cN=1$ chiral and scalar corrections (c.f., \cite{Ita:2011hi}).  In this section, we provide a brief review of the space-time formulation of this theory and some of the interesting properties that its scattering amplitudes exhibit.


\subsubsection{Space-time Lagrangian formulation}

Let the gauge group be $G=\SU(N)$.  The field content of $\cN=4$ SYM is encoded in a single vector multiplet which includes: gluons of helicity $\pm 1$ ($g^{\pm}$); 4 fermions of each $\pm\frac{1}{2}$ helicity ($\tilde{\Psi}_{a\;A'}$ and $\Psi^{a}_{A}$ respectively); and 6 complex scalars (which we can write in the $\mathbf{6}$ vector representation of $\SU(4)_{R}$ as $\Phi_{ab}$).  This theory is naturally chiral, since encoding the multiplet into an on-shell superfield results in
\be{ossf}
X(\eta)=g^{+}+\eta^{a}\tilde{\Psi}_{a}+\cdots+\frac{1}{4!}\eta^{4}g^{-},
\ee
so it is naturally expressed in the chiral Minkowski super-space $\M$.

This motivates adopting a manifestly chiral expression of the theory, known as the Chalmers-Siegel formulation \cite{Chalmers:1996rq, Chalmers:1997sg}.  This entails introducing an auxiliary ASD 2-form:
\begin{equation*}
G\in\Omega^{2\;-}(\M, \mathfrak{sl}_{N}), \qquad G=G_{AB}\d x_{AA'}\wedge \d x^{A'}_{B}.
\end{equation*}
The space-time Lagrangian is then written as:
\be{CS1}
\cL=\frac{N}{8\pi^2}\left(\cL_{1}+\frac{\lambda}{2}\cL_{2}\right),
\ee
where $\lambda$ is the 't Hooft coupling
\begin{equation*}
\lambda=\frac{g^{2}_{\mathrm{YM}}N}{8\pi^2},
\end{equation*}
and the two Lagrangian terms are
\be{CS-SD}
\cL_{1}=\tr\left(G^{AB}F_{AB}+\Psi^{a}_{A}D^{AA'}\tilde{\Psi}_{a\;A'}-\frac{1}{4}D_{AA'}\Phi_{ab}D^{AA'}\bar{\Phi}^{ab}+\tilde{\Psi}_{a\;A'}\tilde{\Psi}_{b}^{A'}\bar{\Phi}^{ab}\right),
\ee
\be{CS-int}
\cL_{2}=\tr\left(\frac{1}{16}\left[\bar{\Phi}^{ab},\bar{\Phi}^{cd}\right]\left[\Phi_{ab},\Phi_{cd}\right]+2\Psi^{a}_{A}\Psi^{b\;A}\Phi_{ab}-G_{AB}G^{AB}\right).
\ee
Here, $D_{AA'}=\partial_{AA'}+A_{AA'}$ is the gauge-covariant derivative, $F_{AB}$ is the ASD portion of the curvature via the decomposition
\begin{equation*}
F_{AA'BB'}=[D_{AA'},D_{BB'}]=\epsilon_{AB}F_{A'B'}+\epsilon_{A'B'}F_{AB},
\end{equation*}
and $\bar{\Phi}^{ab}=\frac{1}{2}\epsilon^{abcd}\Phi_{cd}$.

To see why such a Lagrangian should be equivalent to the usual $\cN=4$ SYM Lagrangian, it suffices to investigate the pure gauge theory sector, where the Chalmers-Siegel action functional looks like:
\begin{equation*}
S[A,G]=\int_{\M}\tr\left(F\wedge G\right)-\frac{\lambda}{2}\int_{\M}\tr\left(G\wedge G\right).
\end{equation*}
This gives the field equations
\begin{equation*}
F^{-}=\lambda G, \qquad \nabla G=0,
\end{equation*}
which can be seen to imply the Yang-Mills equations thanks to the Bianchi identity:
\begin{equation*}
\nabla *F=\nabla(F^{+}-F^{-})=\nabla(F-2F^{-})=0.
\end{equation*}
This means that the Chalmers-Siegel formulation agrees with the classical Yang-Mills theory up to a topological term (which is irrelevant for perturbation theory).

The field equations for $\cN=4$ SYM in Chalmers-Siegel form are:
\begin{eqnarray}
D^{B}_{A'}G_{AB} & = & \left\{\tilde{\Psi}_{a\;A'},\Psi^{a}_{A}\right\}-\frac{1}{2}\left[\Phi_{ab},D_{AA'}\bar{\Phi}^{ab}\right], \label{yFE1} \\
D^{AA'}\tilde{\Psi}_{a\;A'} & = & \lambda \left[\Psi^{b}_{A},\Phi_{ab}\right], \label{yFE2} \\
\Box \Phi_{ab} & = & \left\{\tilde{\Psi}_{[b}^{A'},\tilde{\Psi}_{a]\;A'}\right\}+\lambda\epsilon_{abcd}\left\{\Psi^{c}_{A},\Psi_{d\;A}\right\}+\lambda\left[\Phi_{c[a},[\bar{\Phi}^{cd},\Phi_{b]d}]\right], \label{yFE3} \\
D^{AA'}\Psi^{a}_{A} & = & \left[\tilde{\Psi}_{b\;A'},\bar{\Phi}^{ab}\right], \label{FE4} \\
F_{AB} & = & \lambda G_{AB} \label{FE5}.
\end{eqnarray}
Supersymmetry acts on the multiplet of fields $\{A,\tilde{\Psi},\Phi,\Psi, G\}$ via the generators $\delta_{\varepsilon}$, $\delta_{\tilde{\varepsilon}}$, where $\varepsilon_{A}^{a}$ and $\tilde{\varepsilon}_{a\;A'}$ are spinors.  Explicitly, this action is given by:
\be{eqn: susyg1}
\delta_{\varepsilon}\left(
\begin{array}{c}
A_{AA'} \\
\tilde{\Psi}_{a\;A'} \\
\Phi_{ab} \\
\Psi^{a}_{A} \\
G_{AB}
\end{array}
\right) = \left(
\begin{array}{c}
\varepsilon^{a}_{A}\tilde{\Psi}_{a\;A'} \\
\varepsilon^{b\;A}D_{AA'}\Phi_{ab} \\
\frac{1}{2}\varepsilon^{A\;[c}\Psi^{d]}_{A}\epsilon_{abcd} \\
\frac{1}{2}\varepsilon^{b}_{A}[\Phi_{cb},\bar{\Phi}^{ca}]-\varepsilon^{a\;B}G_{AB} \\
\varepsilon^{a}_{(A}[\Psi_{B)}^{b},\Phi_{ab}]
\end{array}
\right),
\ee
\be{eqn: susg2}
\delta_{\tilde{\varepsilon}}\left(
\begin{array}{c}
A_{AA'} \\
\tilde{\Psi}_{a\;A'} \\
\Phi_{ab} \\
\Psi^{a}_{A} \\
G_{AB}
\end{array}
\right) = \left(
\begin{array}{c}
\lambda\tilde{\varepsilon}_{a\;A'}\Psi^{a}_{A} \\
\tilde{\varepsilon}_{a}^{B'}F_{A'B'}+\frac{\lambda}{2}\tilde{\varepsilon}_{b\;A'}[\bar{\Phi}^{bc},\Phi_{ca}] \\
\tilde{\varepsilon}^{A'}_{[a}\tilde{\Psi}_{b]\;A'} \\
\tilde{\varepsilon}_{b}^{A'}D_{AA'}\bar{\Phi}^{ab} \\
\tilde{\varepsilon}_{a}^{A'}D_{A'(A}\Psi_{B)}^{a}
\end{array}
\right),
\ee
One can verify that $\{\delta_{\varepsilon},\delta_{\tilde{\varepsilon}}\}=0$ up to the field equations.

An interesting fact about $\cN=4$ SYM is that the field equations can be encoded by a system of constraints for a superconnection on a gauge bundle over $\M$ \cite{Witten:1978xx, Witten:1985nt, Harnad:1985bc}; this construction has deep connections with twistor theory (see \cite{Harnad:1988rs} for a review).  Additionally, the field content of $\cN=4$ SYM is easily encoded using a supersymmetric extension of the Penrose transform.  For an abelian theory, we can take the $(0,1)$-form on twistor space:
\begin{multline}\label{superfield1}
	\cA(Z,\bar Z,\chi)=
	a(Z,\bar Z)+\chi^{a}\tilde{\psi}_{a}(Z,\bar Z)+\frac{\chi^{a}\chi^{b}}{2}\phi_{ab}(Z,\bar Z) \\
	+\frac{\epsilon_{abcd}}{3!}\chi^{a}\chi^{b}\chi^{c}\,\psi^{d}(Z,\bar Z)+\frac{\epsilon_{abcd}}{4!}\chi^a\chi^b\chi^c\chi^{d}\, g(Z,\bar Z)   
\end{multline} 
where $a$, $\tilde{\psi}$, $\phi$, $\psi$, and $g$ have homogeneity $0$ $-1$, $-2$, $-3$ and $-4$ respectively, corresponding on-shell (i.e., $\dbar\cA=0$) to z.r.m. fields $\{F_{A'B'},\tilde{\Psi}_{a;A'}, \Phi_{ab}, \Psi_{A}^a, G_{AB}\}$ on space-time by theorem \ref{PenTran}. The integral formulae \eqref{nhel}-\eqref{phel} can now be used to build on-shell superfields on space-time encoding the $\cN=4$ multiplet:
\be{superfield2}
\begin{aligned}
	\cF_{A'B'} &=
	\int _{X } \frac{\del^2}{\del\mu ^{A'}\del\mu^{B'}} \,
	\cA(ix^{AA'}\lambda_{A}, \lambda_{A},\theta^{Aa}\lambda_{A} )\wedge \D\lambda\\
 	&=F_{A'B'}+\theta^{Aa}\p_{AA'}\left[\widetilde\Psi_{aB'}+ \theta^{Bb}\p_{BB'}\left(\frac{\Phi_{ab}}{2}+
    \theta^{Cc}\varepsilon_{abcd}\left(\frac{\Psi_{C}^d}{3!}+
    \theta^{Dd}\frac{G_{CD}}{4!}  \right) \right)\right] \nonumber\\
\end{aligned}
\ee
and
\be{Susy-Pint2}
\begin{aligned}
	\cF_{ab} &=
	\int _{X }  \frac{\del^2}{\del\chi^a\del\chi^b}\,
	\cA (ix^{AA'}\lambda_{A}, \lambda_{A},\theta^{Aa}\lambda_{A} )\wedge \D\lambda\\
	&=\Phi_{ab} +\theta^{Cc}\varepsilon_{abcd}(\Psi_{C}^d+\theta^{Dd}\frac{G_{CD}}2)\ ,
\end{aligned}
\ee
and another component $\cF_{aA'}$ (which has a formula as above with a mixed $\mu$ and $\chi$ derivative).  These can be interpreted as the non-zero parts of the curvature $2$-form
\be{susy-curv}
\cF=\cF_{A'B'}\varepsilon_{AB}\rd x^{AA'}\wedge \rd
x^{BB'}+\cF_{aA'} \varepsilon_{AB}\rd x^{AA'}\wedge\rd\theta ^{Ba}+ \cF_{ab}\varepsilon_{AB}\rd \theta^{Aa}\wedge \rd 
\theta^{Bb}
\ee
of the on-shell space-time superconnection
\begin{equation*}
\CA=\Gamma_{AA'}(x,\theta)\d x^{AA'}+\Gamma_{a\;A}\d\theta^{Aa}.
\end{equation*}
In appendix \ref{Appendix1}, we demonstrate how this superconnection can be obtain explicitly for abelian and $\SU(N)$ gauge groups, but it can be understood geometrically by a supersymmetric extension of the Ward correspondence \cite{Manin:1997}. 


\subsubsection{Special properties of scattering amplitudes}
\label{SAP}

One of the fundamental observables that can be calculated in any quantum field theory are scattering amplitudes.  Not only are these realistic observables in the sense that they are related to quantities measured in experimental particle physics, but they also tell us a great deal about the underlying mathematical structure of the theory.  The data for a scattering amplitude is usually specified in terms of incoming on-shell states composed of a momentum and polarization vector; in four-dimensions it is convenient to replace the polarization information with the helicity data.  Hence, a $n$-particle scattering amplitude $\cA_{n}$ is a function of on-shell momenta $p_{AA'}=p_{A}\tilde{p}_{A'}$ and helicity data (for gluon scattering, this is simply a $\pm 1$ label for each particle).

In $\cN=4$ SYM any classical, or \emph{tree-level}, scattering amplitude can be written in the form:
\be{colorstrip1}
\cA^{0}_{n}=g_{\mathrm{YM}}^{n-2}\sum_{\sigma\in S_{n}/\Z_{n}}\tr\left(\mathsf{T}^{a_{\sigma(1)}}\cdots \mathsf{T}^{a_{\sigma(n)}}\right)A^{0}_{n}(\sigma(1),\ldots, \sigma(n)),
\ee
where the $\mathsf{T}^{a}$s are the generators of the fundamental representation of the gauge group (which we take to be $\SU(N)$), and the sum runs over all non-cyclic permutations of the $n$ particles.  The $A^{0}_{n}$ is called a \emph{color-stripped} amplitude, and clearly enjoys a cyclic symmetry in its arguments by definition.  At higher-order in perturbation theory, amplitudes cannot be color-stripped so easily; a general $l$-loop amplitude will contain $l+1$ color traces over the gauge group.  However, if we consider the planar limit of the gauge theory (i.e., $N\rightarrow\infty$) then a single-trace term dominates (c.f., \cite{Dixon:2011xs}):
\be{colorstrip2}
\cA^{l}_{n}\xrightarrow{\mathrm{planar}\:\mathrm{limit}} (8\pi^2)^{l}g_{\mathrm{YM}}^{n-2}\lambda^{l} \sum_{\sigma\in S_{n}/\Z_{n}}\tr\left(\mathsf{T}^{a_{\sigma(1)}}\cdots \mathsf{T}^{a_{\sigma(n)}}\right)A^{l}_{n}(\sigma(1),\ldots, \sigma(n)).
\ee
Hence, in the planar limit, all amplitudes are uniquely determined by their color-stripped subamplitudes.  For the remainder of this review, we will exclusively consider the color-stripped amplitudes $A^{l}_{n}$ in gauge theory.

Even calculating tree-level amplitudes using a space-time Lagrangian such as \eqref{CS1} can be an involved process.  Using traditional methods, amplitudes are computed using the space-time Lagrangian's Feynman rules, and the number of diagrams required grows roughly factorially with particle number!  However, by organizing amplitudes according to helicity information, remarkable simplifications occur.  In pure gauge theory (i.e., $\cN=0$), one can show that for $n$ gluon configurations where all the particles have the same helicity or only one particle has a different helicity, the scattering amplitudes vanish.  This result follows (in a sense) because of the integrability of the Yang-Mills instanton equations.  

The truly remarkable result appears when we consider the first non-vanishing tree amplitude: this occurs when two gluons have a different helicity than the rest, and is referred to as the Maximal-Helicity-Violating (MHV) case. To standardize conventions, we consider MHV amplitudes to involve 2 gluons of negative helicity, and the rest of positive helicity.  In this case, the tree-level scattering amplitude takes the famous Parke-Taylor form \cite{Parke:1986gb, Berends:1987me}:
\be{ParkeTaylor0}
\delta^{4}\left(\sum_{i=1}^{n}p_{i}\right)\frac{\la l\;m\ra^{4}}{\la 1\;2\ra\cdots \la n\;1\ra},
\ee
where gluons $l$ and $m$ have negative helicity, and $\la i\;j\ra=\epsilon_{AB}p_{i}^{A}p_{j}^{B}$, etc.  This can be generalized to $\cN=4$ SYM by considering scattering amplitudes as functionals of the on-shell superfield \eqref{ossf}, and extracting the portion which is homogeneous of degree $8$ in the Grassmann variables $\eta_{i}$ \cite{Nair:1988bq}.  More generally, an amplitude $A^{l}_{n}$ can be expanded as:
\begin{equation*}
A^{l}_{n}=A^{l}_{n,0}+A^{l}_{n,1}+\cdots +A^{l}_{n, n-4},
\end{equation*}
where $A^{l}_{n,k}$ has homogeneity $4(k+2)$ in $\eta_{i}$ and is referred to as a N$^k$MHV amplitude.  This is the natural generalization from the $\cN=0$ setting, where a N$^k$MHV amplitude has $k+2$ gluons of negative helicity and $n-k-2$ of positive helicity.

This leads to the $\cN=4$ version of the Parke-Taylor formula:  
\be{ParkeTaylor}
A^{0}_{n,\mathrm{MHV}}=\frac{\delta^{4|8}\left(\sum_{i=1}^{n}p_{i}\right)}{\la 1\;2\ra\cdots \la n\;1\ra} ,
\ee
where the super-momentum-conserving delta function $\delta^{4|8}$ is given by
\begin{eqnarray*}
\delta^{4|8}\left(\sum_{i=1}^{n}p_{i}\right) & = & \delta^{4}\left(\sum_{i=1}^{n}p_{i\;A}\tilde{p}_{i\;A'}\right)\delta^{0|8}\left(\sum_{i=1}^{n}\eta^{a}_{i}p_{i\;A}\right), \\
\delta^{0|8}\left(\sum_{i=1}^{n}\eta^{a}_{i}p_{i\;A}\right) & = & \prod_{a,A}\left(\sum_{i=1}^{n}\eta^{a}_{i}p_{i\;A}\right).
\end{eqnarray*}
It is easy to check that \eqref{ParkeTaylor} is superconformally invariant and is homogeneous of degree $8$ in each of the $\eta_{i}$s as required.  Performing a fermionic integral to extract the appropriate $\cN=0$ component of this expression produces the factor of $\la i\;j\ra^{4}$ appearing in \eqref{ParkeTaylor0}.

The fact that the MHV tree amplitude has such an elegant and simple expression is completely obscured by the traditional Lagrangian or Feynman diagram formulation of the gauge theory.  Indeed, for $n=6$, there are over 200 traditional Feynman diagrams which would contribute to \eqref{ParkeTaylor}.  This is a strong indicator that the theory is in fact simpler than the space-time formulation appears, and this simplification takes the form of a hidden \emph{dual} superconformal symmetry \cite{Drummond:2008vq}.

It is widely believed that $\cN=4$ SYM is integrable in the planar limit; this means that it possesses an infinite-dimensional symmetry algebra (a Yangian algebra) $\mathcal{Y}[\mathfrak{psl}(4|4, \C)]$ \cite{Dolan:2004ps, Drummond:2009fd, Bargheer:2011mm}.  In a loose sense, this Yangian algebra is generated by the standard superconformal algebra $\mathfrak{psl}(4|4,\C)$ (which acts on space-time) and another copy of this algebra which acts on a dual space-time: the affine space parametrizing particle momenta.  Invariance of physical observables such as scattering amplitudes under this dual conformal symmetry has proven an immensely powerful tool.

One well-known example of this is the Bern-Dixon-Smirnov (BDS) ansatz for the all-loop structure of MHV amplitudes in $\cN=4$ SYM \cite{Bern:2005iz}, which takes the form:
\be{BDSansatz}
A_{n,0}=A^{0}_{n,0}\exp\left[D_{n}(\Gamma_{\mathrm{cusp}}, G_{\mathrm{collinear}}) + F_{n,0}(p_{1},\ldots,p_{n}\;\lambda)\right],
\ee
where $D_{n}$ captures the IR divergences of the amplitude and is a function of the cusp anomalous dimension $\Gamma_{\mathrm{cusp}}$ and collinear anomalous dimension $G_{\mathrm{collinear}}$, while $F_{n,0}$ is a finite contribution depending on the kinematics and coupling in a specific way.  Since $\Gamma_{\mathrm{cusp}}$ can be fixed completely with integrability \cite{Beisert:2006ez} and $G_{\mathrm{collinear}}$ has been calculated up to 4-loops \cite{Cachazo:2007ad}, the most interesting part of \eqref{BDSansatz} is the finite contribution $F_{n,0}$, which is explicitly specified by the details of the BDS ansatz.  Although this ansatz turns out to fail at two-loops and $n=6$ in perturbation theory \cite{Bern:2008ap, Drummond:2008aq} and for large $n$ in the strong coupling regime \cite{Alday:2007he}, it does so in a way that is constrained by dual superconformal invariance.

Twistors have proven a valuable tool for analysing dual superconformal invariance, thanks to Hodges' momentum twistors, which assign a twistor space to the dual affine space of null momenta \cite{Hodges:2009hk}.  In this setting the dual superconformal generators take the form displayed in \eqref{scongen}, and can also be expressed in ordinary twistors as \cite{Drummond:2009fd, Mason:2009qx}:
\be{dscongen}
J^{(1)\;I}_{J}=\sum_{i<j}(-1)^{K}\left[Z^{I}_{i}\frac{\partial}{\partial Z_{i}^{K}}Z^{K}_{j}\frac{\partial}{\partial Z^{J}_{j}}-(i\leftrightarrow j)\right].
\ee    
In this fashion, the integrability of $\cN=4$ SYM becomes a powerful method for constraining physical observables such as scattering amplitudes.

In this review, we will focus primarily on two other simplifying structures for which emerge as a result of the hidden simplicity of $\cN=4$ SYM: the Britto-Cachazo-Feng-Witten (BCFW) recursion relations, and the Maximal-Helicity-Violating (MHV) formalism of Cachazo, Svrcek, and Witten.  These and other properties of scattering amplitudes in various theories are discussed at great length in the comprehensive review of Elvang and Huang \cite{Elvang:2013cua}.

\subsubsection*{\textit{BCFW recursion}}

First conjectured in \cite{Britto:2004ap} and later proven in \cite{Britto:2005fq}, the BCFW recursion relations give a recursive procedure for obtaining any gluon tree amplitude in gauge theory, and are easily extended to $\cN=4$ SYM \cite{Brandhuber:2008pf, ArkaniHamed:2008gz}.  This can be derived by picking two external momenta for a scattering amplitude and analytically continuing them with a complex variable $z$ while keeping them on-shell and maintaining overall momentum conservation.  The amplitude then becomes a complex function $A^{0}_{n,k}(z)$: it has simple poles wherever internal propagators go on-shell, and $A^{0}_{n,k}(0)$ is the original amplitude.  These simple poles correspond to the terms arising in the BCFW recursion, so provided $A^{0}_{n,k}(z\rightarrow\infty)$ vanishes, Cauchy's theorem implies that
\be{BCFR1}
0=\frac{1}{2\pi i}\int \frac{\d z}{z}A^{0}_{n,k}(z)=A^{0}_{n,k}(0)+\mbox{BCFW terms}.
\ee

More specifically, take the incoming particles $1$ and $n$ with on-shell supermomenta $(p_{i\;A}\tilde{p}_{i\;A'},\eta_{i\;a}p_{i\;A})$, and perform the shift:
\begin{equation*}
\tilde{p}_{n}\rightarrow \hat{\tilde{p}}_{n}=\tilde{p}_{n}+z\tilde{p}_{1}, \qquad \eta_{n}\rightarrow \hat{\eta}_{n}=\eta_{n}+z\eta_{1}, \qquad p_{1}\rightarrow \hat{p}_{1}=p_{1}-zp_{n}.
\end{equation*}
At certain values $z=z_{i}$, internal propagators in the Feynman diagram expansion of $A^{0}_{n,k}$ will go on-shell.  Furthermore, one can show that as $z\rightarrow\infty$, $A^{0}_{n,k}(z)\sim z^{-1}$ \cite{Britto:2005fq, ArkaniHamed:2008yf}, so by \eqref{BCFR1} this leads to an expansion of the amplitude
\be{BCFR2}
A^{0}_{n,k}=\sum \int \d^{4}\eta A^{0}_{i+1\;L}(\hat{1},\ldots, i, \{-\hat{p},\eta\})\frac{1}{p^{2}}A^{0}_{n-i+1\;R}(\{\hat{p},\eta\},i+1,\ldots,\hat{n}).
\ee
Here the fermionic integration selects the correct sub-amplitudes $A_{L},A_{R}$ which are compatible with the overall N$^k$MHV degree, the sum is over the the possible partitions of external states between the sub-amplitudes, and $p=\sum_{j\in L}p_{j}$.

Since the reduced sub-amplitudes can themselves be recursively calculated in a similar fashion, this gives a simplified way of computing the tree-level S-matrix of $\cN=4$ SYM.  Indeed, a recursive formula for all tree-amplitudes has been obtained from BCFW recursion \cite{Drummond:2008cr}.  A particularly simple example is the MHV tree amplitudes of \eqref{ParkeTaylor}; in this case the entire recursion is composed of `homogeneous' terms where $A_{R}$ is a 3-point anti-MHV subamplitude.  Beyond tree-level, BCFW recursion can be used to compute the loop integrand of $\cN=4$ SYM to all orders in the planar limit via the all-loop recursion relations of \cite{ArkaniHamed:2010kv}.  

Furthermore, the BCFW shift becomes extremely simple when expressed on twistor space: $Z_{n}\rightarrow Z_{n}-zZ_{1}$, and the recursion relation \eqref{BCFR2} can be obtained via half-Fourier transform \cite{Mason:2009sa}
\begin{multline}\label{BCFR3}
A^{0}_{n,k}(Z_{1},\ldots, Z_{n})= \\
\sum \int_{\PT\times\C}\D^{3|4}Z\frac{\d z}{z}A^{0}_{i+1\;L}(Z_{1},\ldots,Z_{i},Z) A^{0}_{n-i+1\;R}(Z,Z_{i+1},\ldots,Z_{n}-zZ_{1}).
\end{multline}
The twistorial form of the BCFW recursion will be useful in our later discussion of Wilson loops and local operators in twistor space, as well as gravity.

\subsubsection*{\textit{The MHV formalism}}

While BCFW gives a recursive procedure for computing scattering amplitudes, the MHV rules of \cite{Cachazo:2004kj, Cachazo:2004zb, Cachazo:2004by} provide a Feynman diagram formalism for $\cN=4$ SYM which is dramatically more efficient than standard space-time Lagrangian techniques.  This formalism arose by considering the geometry of the instanton moduli space of twistor-string theory near the boundary \cite{Gukov:2004ei}.  In the twistor-string picture, a $n$-point N$^k$MHV tree amplitude is given by an integral over the moduli space of $n$-pointed, degree $d=k+1$ curves in $\PT$; on the boundary of this moduli space, such a curve can degenerate into $k+1$ intersecting lines, each of which corresponds to a MHV vertex.  The MHV formalism asserts that N$^k$MHV tree amplitudes of $\cN=4$ SYM can be constructed entirely from such disconnected configurations: MHV vertices joined by massless scalar propagators, $1/p^2$ \cite{Cachazo:2004kj}.

The MHV formalism has now been proven to be correct at tree-level (for all Yang-Mills theories) via a complex analysis argument which uses a BCFW momentum shift extended to \emph{all} the external states \cite{Risager:2005vk, Elvang:2008na, Elvang:2008vz}.  It can also be extended to loop level, albeit with some caveats: it can be shown to give the correct 1-loop MHV amplitude in $\cN=4$ SYM \cite{Brandhuber:2004yw} and can be expressed in momentum twistor space \cite{Bullimore:2010pj}, where it was shown to produce the correct planar integrand to all loops for supersymmetric theories which are cut-constructible \cite{Bullimore:2010dz}.\footnote{Such loop integrands are divergent upon integrating over loop momenta or region variables, and require regularization.  We will discuss this in more detail later.}  In this case, a $l$-loop N$^k$MHV amplitude will involve diagrams containing $k+l+1$ MHV vertices.

A key point of the MHV formalism is that the scalar propagators connecting MHV vertices in a diagram are off-shell.  This means that we cannot decompose $p_{AA'}$ into a tensor product of two Weyl spinors of opposite chirality.  But given the Parke-Taylor formula \eqref{ParkeTaylor}, we need a spinor $p_{A}$ for the MHV vertices to be well-defined.  This is accomplished by choosing an arbitrary reference spinor $\hat{\iota}^{A'}$ (which we call the \emph{CSW reference spinor}), and defining
\be{CSWspinor}
p_{A}=p_{AA'}\hat{\iota}^{A'},
\ee
for the off-shell propagators.  It can be shown that dependence on the choice of CSW spinor drops out of the final amplitude after all MHV diagrams have been summed over.

Despite its simplicity and utility, the origins of the MHV formalism have remained mysterious.  There are significant gaps in any explanation coming from twistor-string theory, and the formalism is non-obvious from the gauge theory's space-time Lagrangian (although a transformation which does produce the MHV formalism from the Lagrangian can be engineered \cite{Mansfield:2005yd}).  A major goal of Section \ref{Chapter3} will be to show that the MHV formalism follows naturally as the gauge-fixed Feynman rules of the twistor action for $\cN=4$ SYM.  Although prior efforts had indicated that this might be true using momentum eigenstates \cite{Boels:2007qn}, our derivation will be based entirely in twistor space and will manifest superconformal invariance.  

\medskip

The second half of this review will investigate how the themes explored in the context of $\cN=4$ SYM can be extended to gravity.  Unlike BCFW recursion, which extends to Einstein (super)gravity \cite{Bedford:2005yy, Cachazo:2005ca, Benincasa:2007qj}, the na\"{i}ve MHV vertex expansion defined by the Risager all-line shift fails \cite{BjerrumBohr:2005jr, Bianchi:2008pu} due to the `non-holomorphicity' of graviton scattering amplitudes.


\section{Twistor Action for $\cN=4$ Super-Yang-Mills}
\label{Chapter3}

In this section, we will apply our background knowledge of twistor theory to formulate maximally supersymmetric gauge theory in four dimensions (i.e., $\cN=4$ SYM) as a gauge theory on twistor space.  Our primary tool will be the twistor action for $\cN=4$ SYM \cite{Mason:2005zm, Boels:2006ir}, which can be thought of as an effective action for twistor-string theory which captures only the gauge theory contributions, eliminating the conformal gravity modes.  After recalling the basic definition and properties of the twistor action, we set out a rigorous derivation of its Feynman rules in a particular axial gauge.  The main result is a proof that these Feynman rules reproduce the MHV formalism of \cite{Cachazo:2004kj}; this provides a proof of the MHV formalism (at tree level), indicates its twistorial nature, and allows us to easily compute IR finite scattering amplitudes on twistor space.  We demonstrate how all tree-level scattering amplitudes can be calculated in this manner, and also discuss the status of loop-level calculations in perturbation theory.  The main advantage of this formalism (besides being dramatically more efficient than space-time techniques) is that it manifest superconformal invariance--up to the choice of reference spinor in the MHV formalism.


\subsection{Definition and Basic Properties}

The setting for this chapter will be $\cN=4$ supersymmetric twistor space $\PT\subset\P^{3|4}$; we have a gauge group $G=\SU(N)$, bundle $E\rightarrow\PT$ (which can be thought of as the space-time gauge bundle pulled back to twistor space) with $\End(E)\cong\mathfrak{sl}_{N}$.  For simplicity, we will assume that this bundle is topologically trivial: $c_{1}(E)=0$, although this assumption can be relaxed to $c_{1}(E|_{X})=0$ without serious consequences to any of our results.  As we noted earlier, the field content of $\cN=4$ SYM can be encoded in a homogeneous $(0,1)$-form $\cA\in\Omega^{0,1}(\PT,\cO\otimes\End(E))$,
\be{tsconn}
\cA=a+\chi^{a}\tilde{\psi}_{a}+\frac{\chi^{a}\chi^{b}}{2!}\phi_{ab}+\frac{\epsilon_{abcd}}{3!}\chi^{a}\chi^{b}\chi^{c}\psi^{d}+\frac{\chi^4}{4!}g,
\ee
which has no components in the anti-holomorphic directions, and each bosonic component corresponds to a space-time field via the Penrose transform.\footnote{Technically, for $G\neq\U(1)$, this requires a non-abelian generalization of the Penrose transform.  This can be defined by finding a holomorphic trivialization of the bundle $E|_{X}$ on the $\P^{1}$ fibers of twistor space; we will discuss this explicitly in the next chapter.}  This acts as a $(0,1)$-connection (or, equivalently, an endomorphism-valued complex structure) on $E$.

We want an action functional on twistor space which mimics the structure of the Chalmers-Siegel action \eqref{CS1}; this requires an instanton term plus an ASD interaction term.  By theorem \ref{WardCorr}, the bundle $E$ with connection $\dbar+\cA$ is equivalent to a $\cN=4$ Yang-Mills instanton on $\M$ provided $F^{0,2}=\dbar\cA+\cA\wedge\cA=0$.  Since $F^{0,2}=0$ are the Euler-Lagrange equations of the holomorphic Chern-Simons functional, we can account for the SD portion of the theory on twistor space with:
\be{TASD}
S_{1}[\cA]=\frac{i}{2\pi}\int_{\PT}\D^{3|4}Z\wedge\tr\left(\cA\wedge\dbar\cA+\frac{2}{3}\cA\wedge\cA\wedge\cA\right).
\ee
Beyond the Ward Correspondence, there is substantial motivation for this action capturing the instanton sector.  An early form of this action was derived by Sokatchev for self-dual $\cN=4$ SYM \cite{Sokatchev:1995nj}, and the holomorphic Chern-Simons functional can be seen as an artifact of twistor-string theory, which in Witten's original formulation is a topological B model with target $\PT$ \cite{Witten:2003nn}.  Open strings are stretched between $D5$-branes wrapped on $\P^{3|4}$; hence the theory of open $D5-D5$ strings is described by precisely the holomorphic Chern-Simons functional \cite{Witten:1992fb}.

Accounting for the ASD interactions of the gauge theory is a more subtle problem, though.  In twistor-string theory, these interactions take the form of $D1-D5$ instantons in Witten's model, or world-sheet instantons in the heterotic model \cite{Mason:2007zv, ReidEdwards:2012tq}.  At the level of a generating functional, such a contribution looks like
\begin{equation*}
\int_{\M_{\R}}\d^{4|8}X\,\det\left(\dbar+\cA\right)|_{X},
\end{equation*}
where $\d^{4|8}X$ is the measure over the space of $X\cong\P^{1}$ in $\PT$ corresponding to points in the chosen real slice $\M_{\R}\subset\M$, and $(\dbar+\cA)|_{X}$ is the complex structure induced by $\cA$ restricted to the line $X$.  However, under gauge transformations this determinant picks up exponential anomalies which lead to the conformal gravity modes of the twistor-string, so such a generating functional is not suitable for a twistor action which contains only the gauge theory.

The correct non-local term is given by taking the logarithm of the twistor-string generating functional:
\be{TAInt}
S_{2}[\cA]=\int_{\M_{\R}}\d^{4|8}X\,\log\det\left(\dbar+\cA\right)|_{X},
\ee
which essentially amounts to a WZW action, as first noted in \cite{Abe:2004ep}.  Of course, we are glossing over some subtleties here, because $\det(\dbar+\cA)|_{X}$ is not a function, but rather a section of a determinant line bundle over the space of connections:
\begin{equation*}
\det(\dbar+\cA)|_{X}\in\Gamma(\cL), \qquad \cL\rightarrow\mathrm{Conn}(E\rightarrow\PT)|_{X}\cong\mathrm{Conn}(E\rightarrow\P^{1}).
\end{equation*}
Hence, $\det(\dbar+\cA)|_{X}$ should be understood as a $\zeta$-regularized determinant.  The determinant line bundle comes equipped with a natural connection (the Quillen connection) \cite{Quillen:1985}, whose curvature can be computed using the Bismut-Freed index theorem \cite{Bismut:1986, Freed:1986zx}.  In our case, the data for this is given by the diagram:
\begin{equation*}
\xymatrix{
 & \cL \ar[d] \\
\mathrm{Conn}(E\rightarrow \P^{1})\times\M \ar[d] \ar[r]^{e} &  \mathrm{Conn}(E\rightarrow \P^{1}) \\
\M & }
\end{equation*}
where $e$ is the natural evaluation map.  The Bismut-Freed index theorem then states that
\begin{equation*}
F^{(\cL)}=\int \mathrm{Td}(\M)\ch(T\M\oplus E|_{X}) = 0,
\end{equation*}
since $E$ and $\M$ are both topologically trivial.  In other words, $\log\det(\dbar+\cA)|_{X}$ can safely be treated as a function on $\M$ (at least locally), so \eqref{TAInt} is well-defined.

We now have the full $\cN=4$ SYM twistor action as originally derived in \cite{Boels:2006ir}:
\be{TwistorAction}
S[\cA]=S_{1}[\cA]+\lambda S_{2}[\cA],
\ee
which is invariant under gauge transformations on twistor space:\footnote{In the real category, the normalization of a Chern-Simons action is fixed by requiring that the partition function is gauge invariant under `large' gauge transformations; this results in the familiar $\frac{k}{4\pi}$ normalization.  A holomorphic Chern-Simons theory on a Calabi-Yau 3-fold $M$ is defined up to periods of $H_{3}(M,\Z)$, which are generically dense in $\C$ \cite{Thomas:1997}.  However, since $H_{3}(\P^{3},\Z)=0$ we avoid this ambiguity and our arbitrary normalization$\frac{i}{2\pi}$ is fine.}
\be{gaugefreedom}
(\dbar+\cA)\rightarrow\gamma(\dbar+\cA)\gamma^{-1}, \qquad \gamma\in\Gamma(E,\End(E)),
\ee   
with $\gamma$ homotopic to the identity, and $\gamma\rightarrow\mathbb{I}_{N}$ asymptotically.  This follows because the exponential anomalies caused by gauge transformations in the determinant of \eqref{TAInt} become additional terms thanks to the logarithm; these terms vanish upon performing the fermionic integrations in $\d^{4|8}X$ (c.f., \cite{Boels:2006ir}).

Notice that bosonically, a gauge transformation $\gamma$ is a function of three complex, or six real, variables.  This means that the twistor action has substantially more gauge freedom than the space-time $\cN=4$ SYM Lagrangian.  This freedom can be fixed or reduced by imposing gauge conditions; two of these will be particularly important for our purposes.

\subsubsection*{\textit{Woodhouse Gauge}}

The \emph{Woodhouse gauge} condition \cite{Woodhouse:1985id}:
\be{WGauge}
\dbar^{*}|_{X}\cA|_X = 0
\ee
imposes the condition that $\cA$ is holomorphic upon restriction to all fibers $X\cong\P^1$ of twistor space. There are residual gauge transformations preserving this gauge condition
\begin{equation*}
\dbar^{*}|_{X}\dbar|_{X}\gamma =\Delta_{\P^{1}}\gamma =0,
\end{equation*}
for each $\P^{1}$ fiber of twistor space.  But this is just the homogeneous harmonicity condition, so $\gamma$ cannot depend on the fiber coordinate. The remaining gauge freedom is reduced to precisely that of space-time gauge transformations: $\gamma=\gamma(x)$.  In addition, recall that with Euclidean reality conditions explicit cohomological representatives can be constructed in Woodhouse gauge using theorems \ref{Wrep1}-\ref{Wrep2}.  These facts are crucial in the following theorem, which establishes that the twistor action indeed provides a full perturbative description of $\cN=4$ SYM:
\begin{thm}[Boels, Mason, \& Skinner \cite{Boels:2006ir}]\label{BMSthm}
The twistor action $S[\cA]$ is classically equivalent to the Chalmers-Siegel action \eqref{CS1} in the sense that solutions to its Euler-Lagrange equations are in one-to-one correspondence with solutions to the field equations \eqref{yFE1}-\eqref{FE5} up to space-time gauge transformations.  Additionally, upon fixing Woodhouse gauge and Euclidean reality conditions, $S[\cA]$ is equal to the Chalmers-Siegel action.
\end{thm}

This theorem confirms that the twistor action describes $\cN=4$ SYM at the Lagrangian level, and also indicates that any results which we prove using the twistor action will also be true for the space-time theory (at least perturbatively).  

\subsubsection*{\textit{Axial/CSW Gauge}}
  
The gauge freedom of the twistor action can also be reduced by choosing an axial gauge on twistor space.  This corresponds to a choice of holomorphic 1-dimensional distribution $D\subset T^{1,0}\PT$ with the requirement that $\cA|_{\overline{D}}=0$.  More concretely, if we take $D$ to be the span of some holomorphic vector field $V$, then the axial gauge is the condition that
\begin{equation*}
\overline{V}\lrcorner\cA=0.
\end{equation*}
The simplest axial gauge available on twistor space is when $V$ corresponds to a null translation in space-time.  This is known as the \emph{CSW gauge} after \cite{Cachazo:2004kj}, and corresponds to the choice of a reference twistor $Z_{*}$ which induces a foliation of $\PT$ by those lines which pass through $Z_{*}$.  The CSW gauge is the condition that $\cA$ vanish when restricted to the leaves of this foliation:
\be{CSWgauge}
\overline{Z_{*} \cdot \frac{\partial}{\partial Z}}\lrcorner \cA =0.
\ee
It was initially argued using momentum eigenstates that the Feynman rules for the twistor action in the CSW gauge corresponded to the MHV formalism \cite{Boels:2007qn} on momentum space.  However, this argument was far from rigorous, and was not self-contained on twistor space.  We now present the rigorous derivation of the CSW gauge-fixed Feynman rules for the twistor action, and a purely twistorial derivation of the MHV formalism \cite{Adamo:2011cb}.


\subsection{Feynman Rules}  

We begin by fixing CSW gauge, making the choice of fixed reference twistor to correspond to the `point at infinity' in $\M$:
\begin{equation*}
Z_{*}=(0,\iota^{A'},0)\in\P^{3|4}.
\end{equation*}
The gauge condition \eqref{CSWgauge} reduces the number of independent components of the $(0,1)$-connection $\cA$ from three to two; this eliminates the cubic Chern-Simons vertex in $S_{1}[\cA]$.  Since this cubic vertex corresponds to the anti-MHV three-point amplitude, the choice of CSW gauge eliminates this vertex; anti-MHV amplitudes will of course still exist, but are now constructed from the remaining vertices of the theory.  The gauged-fixed twistor action is therefore reduced to:
\be{CSWgf}
S[\cA]=\frac{i}{2\pi}\int_{\PT}\D^{3|4}Z\wedge\tr\left(\cA\wedge\dbar\cA\right)+\lambda\int_{\M_{\R}}\d^{4|8}X\;\log\det(\dbar+\cA)|_{X}.
\ee
As usual, the propagator is determined by the quadratic portion of the action. However, there are two such contributions in \eqref{CSWgf}: one from the kinetic Chern-Simons portion and another from the perturbative expansion of the $\log\det$ (see below).  Since the latter occurs as part of a generating functional of vertices, we choose to treat it perturbatively at the expense of including a two-point vertex in our formalism (as we discuss below). 


\subsubsection{Vertices}

In the CSW gauge, all vertices of the twistor action come from the $\log\det$, and can be made explicit by perturbatively expanding \cite{Boels:2006ir, Boels:2007qn}:
\be{detexp}
\log\det(\dbar+\cA)|_{X}=\tr\left(\log \dbar|_{X}\right)+\sum_{n=2}^{\infty}\frac{1}{n}\int_{X^n}\tr\left(\dbar^{-1}|_{X}\cA_{1}\dbar^{-1}|_{X}\cA_{2}\cdots\dbar^{-1}|_{X}\cA_{n}\right),
\ee
where $\cA_{i}$ is the field inserted at a point $Z_{i}\in X$, and $\dbar^{-1}|_{X}$ is the Green's function for the $\dbar$-operator restricted to $X$.  Since the line $X$ can be written as a skew bi-twistor $X^{IJ}=Z_{A}^{[I}Z_{B}^{J]}$, we can introduce a coordinate $\sigma^{A}=(\sigma^{0},\sigma^{1})$ on $X$ and write
\begin{equation*}
Z(\sigma)=Z_{A}\sigma^{0}+Z_{B}\sigma^{1}.
\end{equation*}
This allows us to express $\dbar^{-1}|_{X}$ in terms of the Cauchy kernel in these coordinates:
\begin{equation*}
(\dbar^{-1}|_{X}\cA)(\sigma_{i-1})=\frac{1}{2\pi i}\int_{X}\frac{\cA(Z(\sigma_{i}))\wedge\D\sigma_{i}}{(i-1\;i)},
\end{equation*}
with $(i-1\;i)=\epsilon_{AB}\sigma^{A}_{i-1}\sigma^{B}_{i}$ the $\SL(2,\C)$-invariant inner product on the $\P^{1}$ coordinates.

Therefore, the $n^{\mathrm{th}}$ term in the perturbative expansion of the $\log\det$ gives:
\be{detexp2}
\frac{1}{n}\left(\frac{1}{2\pi i}\right)^{n}\int_{\M_{\R}}\d^{4|8}X \int_{X^n}\tr\left(\prod_{i=1}^{n}\frac{\cA(Z(\sigma_{i}))\wedge\D\sigma_{i}}{(i-1\;i)}\right).
\ee
Note that we consider the index $i$ modulo $n$ (i.e., $i=i+n$); this corresponds to the cyclic particle ordering of a color-stripped amplitude.  In order to obtain a formula for the vertex which manifests conformal invariance, we represent the measure $\d^{4|8}X$ as a volume form on the moduli space of degree one maps $Z:\P^{1}\rightarrow\PT$:
\be{vmeasure}
\d^{4|8}X=\frac{\d^{4|4}Z_{A}\wedge\d^{4|4}Z_{B}}{\vol\;\GL(2,\C)}.
\ee
The quotient by the volume of $\GL(2,\C)$ transformations accounts for the redundancy in $\sigma$ and $Z_{A,B}$; this is the $\SL(2,\C)$ automorphism group of $\P^{1}$ and the $\C^{*}$ scaling freedom.  

This choice allows us to write down the superconformally invariant formula
\be{TAvertex}
V(1,\ldots,n)=\int_{\CM_{n,1}}\frac{\d^{4|4}Z_{A}\wedge\d^{4|4}Z_{B}}{\vol\;\GL(2,\C)}\int_{X^n}\tr\left(\prod_{i=1}^{n}\frac{\cA(Z(\sigma_{i}))\wedge\D\sigma_{i}}{(i-1\;i)}\right),
\ee
with $\CM_{n,d}$ the space of maps $Z:\P^{1}\rightarrow\PT$ of degree-$d$ and $n$ marked points.  This is easily recognizable as the twistor-string formulation of the MHV amplitude as an integral over the space of lines in $\PT$ \cite{Roiban:2004yf}, and is a Dolbeault analogue of Nair's original twistor formulation \cite{Nair:1988bq}.  Indeed, the Parke-Taylor amplitude can be recovered explicitly by inserting the on-shell momentum eigenstates:
\be{YMeig}
\cA_{i}=\int_{\C}\frac{\d s}{s}\bar{\delta}^{2}(s\lambda_{i\;A}-p_{i\;A})e^{s[[\mu_{i}\tilde{p}_{i}]]}.
\ee
We fix the $\GL(2,\C)$ freedom by setting $\sigma_{i}=\lambda_{i}$ and quotienting out by the scale of $Z_{A,B}$.  This gives
\begin{multline}
\int_{\CM_{n,1}}\frac{\d^{4|4}Z_{A}\wedge\d^{4|4}Z_{B}}{\vol\;\GL(2,\C)}\int_{X^n}\,\prod_{i=1}^{n}\frac{\cA(Z(\sigma_{i}))\wedge\D\sigma_{i}}{(i-1\;i)} \\
=\int_{\M_{\R}}\d^{4|8}x \int\; \prod_{i=1}^{n}\frac{\d s_{i}}{s_{i}}\bar{\delta}^{2}(s\lambda_{i\;A}-p_{i\;A})e^{s[[\mu_{i}\tilde{p}_{i}]]}\frac{\D\lambda_{i}}{\la i-1\;i\ra} \\
=\frac{1}{\prod_{i=1}^{n}\la i\;i+1\ra}\int_{\M_{\R}}\d^{4|8}x\;\exp\left(i\sum_{i=1}^{n}p_{i}\cdot x+\eta_{a\;i}p_{A\;i}\theta^{Aa}\right) \\
=\frac{\delta^{4|8}\left(\sum_{i=1}^{n}p_{i}\right)}{\la 1\;2\ra\cdots \la n\;1\ra}=A^{0}_{n,0},
\end{multline}
as expected.  In the final step, we used Nair's lemma \cite{Nair:1988bq} to express the delta-function as an integral over the real space-time.  

Hence, we see that on-shell the vertices of the twistor action are the MHV amplitudes of $\cN=4$ SYM.  Determining the form of the twistor propagator in CSW gauge is a bit more involved, though.


\subsubsection{Propagator}

The propagator is fixed by the kinetic part of the holomorphic Chern-Simons action
\begin{equation*}
\int_{\PT}\D^{3|4}Z\wedge\tr\left(\cA\wedge\dbar\cA\right),
\end{equation*}  
to be the inverse of the $\dbar$-operator on $\PT$ acting on $(0,1)$-forms in the CSW gauge:
\begin{equation*}
\dbar \Delta(Z_{1},Z_{2})=\bar{\delta}^{3|4}(Z_{1},Z_{2}), \qquad \overline {Z_{*}\cdot\frac{\partial}{\partial Z_{1}}}\, \lrcorner
\,\Delta=\overline {Z_{*}\cdot\frac{\partial}{\partial Z_{2}}}\, \lrcorner \,\Delta=0.
\end{equation*}   
In the end, we will see that the correct form of the propagator is given simply by one of our distributional forms: $\Delta(Z_{1},Z_{2})=\bar{\delta}^{2|4}(Z_{1},*,Z_{2})$.  However, the steps necessary for a careful derivation of this using cohomological representatives are rather involved.

We reduce the problem to one on bosonic twistor space $\PT_{b}$ by performing the $\d^{4}\chi$ fermionic integrals in the kinetic portion of the action to obtain: 
\be{kin}
 \int_{\PT_{b}}\D^{3}Z\wedge\tr\left(g\wedge\dbar a+\psi^{a}\wedge\dbar\tilde{\psi}_{a}+\frac{\epsilon^{abcd}}{4}\phi_{ab}\wedge\dbar\phi_{cd}\right).
\ee
From this, we see that the propagator must be a sum of terms, each of which is a kernel for $\dbar$ on $\PT_{b}$ taking values in the proper homogeneity configurations.  More formally, we have:
\be{prop1}
\Delta=(\chi_{2})^{4}\Delta_{0,-4}+\chi_{1}(\chi_{2})^{3}\Delta_{-1,-3}+(\chi_{1})^{2}(\chi_{2})^{2}\Delta_{-2,-2},
\ee
where each bosonic propagator obeys:
\begin{equation*}
\Delta_{k,l}\in H^{0,2}((\PT_{b}\times\PT_{b})\setminus\Delta, \cO(k,l)), \qquad \dbar \Delta_{k,l}=(\dbar_{1}+\dbar_{2})\Delta_{k,l}=\bar{\delta}_{\Delta},
\end{equation*}
for $\Delta\subset\PT_{b}\times\PT_{b}$ the diagonal and $\bar{\delta}_{\Delta}$ the anti-holomorphic Dirac current.

The inverse of the $\dbar$-operator on non-projective complex manifolds is given locally by the Bochner-Martinelli kernel \cite{Griffiths:1978}.  Most attempts at building kernels for $\dbar$ on $\P^{n}$ are rooted in complex analysis (e.g., \cite{Polyakov:1987, Berndtsson:1988}), and geometric efforts work with a positive definite (i.e., Fubini-Study) metric \cite{Gotmark:2008}.  While these results are impressive in their generality, they are unwieldy for physical calculations.  By using the natural machinery of twistor theory reviewed in Section \ref{Chapter2}, we can obtain a simple answer in CSW gauge.  

The basic roadmap is to begin with a space-time representative for the propagator in Feynman gauge, transform it to twistor space using Woodhouse representatives, and then make a gauge transformation to arrive at CSW gauge.  These calculations are rather involved, so we only outline them here; the interested reader need only consult appendix D of \cite{Adamo:2011cb}.  Let us consider the propagator component $\Delta_{-2,-2}$ as an example, since this is when the computations are easiest.

In order to utilize Woodhouse representatives, we need to impose Euclidean reality conditions, for which the CSW gauge condition reads:
\be{EuclCSW}
\hat{\iota}^{A'}\lambda^{A}\partial_{AA'}\lrcorner\Delta_{-i,j}=N^{\alpha}\frac{\partial}{\partial\hat{Z}^{\alpha}}\lrcorner\Delta_{i,j}=0, \qquad N^{\alpha}=(0,\hat{\iota}^{A'}).
\ee
Now, on space-time, $\Delta_{-2,-2}$ is just the scalar propagator
\begin{equation*}
 \Delta_{-2,-2}(x_{1},x_{2})=\frac{1}{(x_1-x_2)^2},
\end{equation*}
which is a z.r.m. field on $\E\times\E$ away from the diagonal.  Hence, we can apply theorem \ref{Wrep1} to construct a Woodhouse representative for the propagator:
\begin{multline}\label{eqn: wg22}
\Delta^{\mathrm{W}}_{-2,-2}(Z_{1},Z_{2})=\dhat_{1}\dhat_{2}\left(\frac{1}{(1,\hat{1},2,\hat{2})}\right) \\
=2\frac{(\d\hat{Z}_{1},\hat{1},2,\hat{2})\wedge(1,\hat{1},\d\hat{Z}_{2},\hat{2})}{(1,\hat{1},2,\hat{2})^3}-\frac{(\d\hat{Z}_{1},\hat{1},\d\hat{Z}_{2},\hat{2})}{(1,\hat{1},2,\hat{2})^2},
\end{multline}
using the fact that
\begin{equation*}
(x_{1}-x_{2})^{2}=\frac{(1,\hat{1},2,\hat{2})}{\la \lambda_{1}\hat{\lambda}_{1}\ra \la\lambda_{2}\hat{\lambda}_{2}\ra}.
\end{equation*}
The twistor propagator \eqref{eqn: wg22} is a $(0,2)$-form on $\PT_{b}\times\PT_{b}$, is $\dbar$-closed away from the diagonal, and is in Woodhouse gauge on each factor.  

Now, we want to exploit the freedom of adding a $\dbar$-exact $(0,1)$-form on $\PT_{b}\times\PT_{b}$ in order to transform \eqref{eqn: wg22} into CSW gauge:
\begin{equation*}
 N\cdot\dhat_{i}\lrcorner(\Delta^{\mathrm{W}}_{-2,-2}+\dbar f)=0,
\end{equation*}
with $i=1,2$ labelling the factor of twistor space.  Such a $f$ can indeed be found, and after accounting for potential gauge anomalies resulting from delta-functions, we find
\be{eqn: csw22*}
\Delta_{-2,-2}=\frac{(N,\hat{1},2,\hat{2})(1,\hat{1},N,\hat{2})}{(1,\hat{1},2,\hat{2})}\dbar_{1}\left(\frac{1}{(1,\widehat{N},2,\hat{2})}\right)\wedge\dbar_{2}\left(\frac{1}{(1,\hat{1}, 2,\widehat{N})}\right),
\ee
which is obviously in CSW gauge since the form component is skewed with $N$.

\begin{figure}
\centering
\includegraphics[width=4 in, height=3 in]{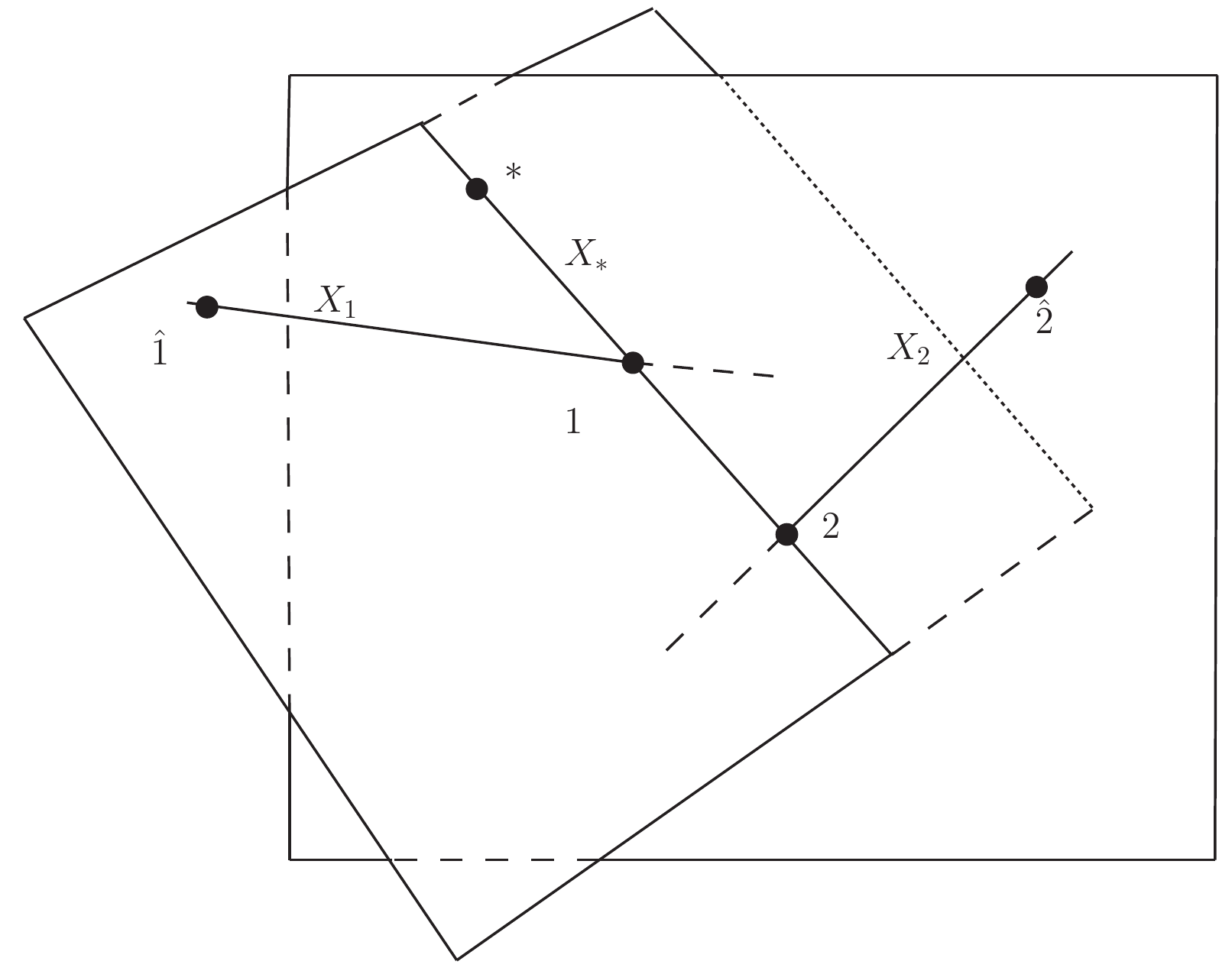}\caption{\textit{Twistor space support of the propagator}}\label{PropGeo}
\end{figure}

This procedure can be carried out for $\Delta_{-1,-3}$ and $\Delta_{0,-4}$ (with a few additional subtleties) by again applying theorems \ref{Wrep1}-\ref{Wrep2} and finding the appropriate gauge transformation, resulting in \cite{Adamo:2011cb}:
\be{eqn: csw13*}
 \Delta_{-1,-3}=i\frac{[(1,\hat{1},N,\hat{2})]^{2}}{(1,\hat{1},2,\hat{2})}\dbar_{1}\left(\frac{1}{(1,\widehat{N},2,\hat{2})}\right)\wedge\dbar_{2}\left(\frac{1}{(1,\hat{1},2,\widehat{N})}\right),
\ee
\be{eqn: csw04*}
 \Delta_{0,-4}=2\frac{\la\hat{\lambda}_{2}\lambda_{1}\ra[(1,\hat{1},N,\hat{2})]^2}{(1,\hat{1},2,\hat{2}) \la\lambda_{2}\hat{\lambda}_{2}\ra}\dbar_{1}\left(\frac{1}{(1,\widehat{N},2,\hat{2})}\right)\wedge\dbar_{2}\left(\frac{1}{(1,\hat{1},2,\widehat{N})}\right).
\ee
These bosonic components define the full supersymmetric propagator via \eqref{prop1}, where the homogeneity factors appearing at the front of each of \eqref{eqn: csw22*}-\eqref{eqn: csw04*} are balanced by the fermionic coordinates of twistor space.  Furthermore, the reference twistor $Z_{*}=-\widehat{N}$, so the full propagator in CSW gauge contains an overall factor of
\begin{equation*}
\dbar_{1}\left(\frac{1}{(1,\rf,2,\hat{2})}\right)\wedge\dbar_{2}\left(\frac{1}{(1,\hat{1},2,\rf)}\right),
\end{equation*}
which is supported only on the set $(1,\rf,2,\hat{2})=(1,\hat{1},2,\rf)=0$.  This restricts $Z_1$ to lie in the plan spanned by $\{Z_{*},Z_{2},\hat{Z}_2\}$, and $Z_2$ to lie in the plane spanned by $\{Z_{*},Z_1, \hat{Z}_1\}$.  In $\PT$, these two planes must intersect in a line containing the reference twistor: $(\rf ,2,\hat{2})\cap(\rf, 1,\hat{1})=X_{*}$.  But this is only possible if $Z_{1}$ and $Z_{2}$ are also contained in $X_{*}$; in other words: $Z_{1}$, $Z_{2}$ and $Z_{*}$ are collinear in twistor space (see Figure \ref{PropGeo}).

Our distributional forms then allow us to represent the twistor space propagator for $\cN=4$ SYM in CSW gauge by
\be{TAprop}
\Delta(Z_{1},Z_{2})=\bar{\delta}^{2|4}(Z_{1},\rf, Z_{2}).
\ee
Note that although our derivation here began in Euclidean signature, the result for the full propagator is signature-independent and superconformally invariant up to choice of $Z_{\rf}$.  The fact that $\Delta$ is in CSW gauge follows from our derivation, and it is the propagator for the kinetic operator $\dbar$ thanks to lemma \ref{dfprop1}:
\begin{equation*}
\dbar\Delta=2\pi i\left( \bar{\delta}^{3|4}(Z_{1},Z_{2})+\bar{\delta}^{3|4}(Z_{1},\rf)+ \bar{\delta}^{3|4}(Z_{2},\rf)\right) .
\end{equation*} 
The first term is the anti-holomorphic Dirac current we want, and the other two terms do not contribute to the physical portion of the propagator.  This is because $\Delta(Z_{1},Z_{2})$ should be a $(0,1)$-form in each variable; the second and third terms in the above expression have $(0,2)$-form components in $Z_{1}$ and $Z_{2}$, however.  In any case, these contributions correspond to unphysical poles in momentum space, which are an expected feature of the axial gauge we are working in.  They can be removed entirely by restricting ourselves to the twistor space $\PT\subset\P^{3|4}$ that excludes the `point at infinity' (i.e., $\PT=\{Z | \lambda\neq 0\}$), where the error terms do not have support.

Finally, the tensor structure of the propagator is included by accounting for the gauge group, and writing these gauge indices explicitly gives:
\be{TAcolorprop}
\Delta(Z_{1},Z_{2})^{i k}_{j l}=\bar{\delta}^{2|4}(Z_{1},\rf, Z_{2})\left(\delta^{i}_{l}\delta^{k}_{j}-\frac{1}{N}\delta^{i}_{j}\delta^{k}_{l}\right).
\ee
We often suppress the color structure of the propagator in our following discussions.


\subsection{The MHV Formalism}

Having derived the CSW gauge-fixed Feynman rules of the twistor action, it is natural to ask what they correspond to on momentum space.  Given that we have shown the twistor vertices to be equivalent to the MHV amplitudes of $\cN=4$ SYM on-shell, a natural guess would be the MHV formalism of \cite{Cachazo:2004kj}.  In this subsection, we will demonstrate that this is, in fact, true.  We choose $Z_{*}=(0,\iota^{A'},0)$, and assume (without loss of generality) that $\iota$ is normalised with respect to Euclidean reality conditions: $[\hat{\iota}\iota]=1$.

To begin, we pull back the twistor propagator $\Delta(Z,Z')$ to the primed spinor bundle using the twistor double fibration:
\begin{equation*}
p:\PS\rightarrow\PT, \qquad (x^{AA'},\theta^{aA},\lambda_{A})\mapsto (\lambda_{A}, ix^{AA'}\lambda_{A},\theta^{aA}\lambda_{A}).
\end{equation*}
This can be accomplished using the definition of $\bar{\delta}^{2|4}$ and the incidence relations which define the map $p$:
\begin{multline}\label{mMHV1}
p^{*}\Delta=\int_{\C^2}\frac{\d s}{s}\frac{\d t}{t}\bar{\delta}^{2}(s\iota^{A'}-ix^{AA'}\lambda_{A}-itx^{\prime AA'}\lambda_{A}^{\prime})\bar{\delta}^{2|4}(\lambda_{A}+t\lambda_{A}^{\prime}) \\
=\int_{\C^2}\frac{\d s}{s}\frac{\d t}{t}\bar{\delta}^{2}(s\iota^{A'}-iy^{AA'}\lambda_{A})\bar{\delta}^{2|4}(\lambda_{A}+t\lambda_{A}^{\prime}).
\end{multline}
where the second line follows from the support of the delta functions and we abuse notation by writing
\begin{equation*}
\bar{\delta}^{2|4}(\lambda_{A}+t\lambda'_{A})\equiv \bar{\delta}^{2}(\lambda_{A}+t\lambda'_{A})\bar{\delta}^{0|4}(\theta^{aA}\lambda_{A}+t\theta^{\prime aA}\lambda'_{A}).
\end{equation*}
Since this expression is independent of $x+x'$, we are free to perform a Fourier transform in $y=x-x'$ to obtain the momentum space version of the propagator:
\be{mMHV2}
\widetilde{\Delta}=\int\d^{4|8}y\; e^{ip\cdot y}\frac{\d s}{s}\frac{\d t}{t}\bar{\delta}^{2}(s\iota^{A'}-iy^{AA'}\lambda_{A})\bar{\delta}^{2|4}(\lambda_{A}+t\lambda_{A}^{\prime}).
\ee
Note that this Fourier transform takes into account the \emph{super}-momentum via
\begin{equation*}
p\cdot y=p_{AA'}(x-x')^{AA'}+p_{A}\eta_{a}(\theta-\theta')^{aA}.
\end{equation*}

Now, $\bar{\delta}^{2}(s\iota^{A'}-iy^{AA'}\lambda_{A})$ is a $(0,2)$-form multiplied by four real delta-functions, which uniquely restrict $y$ to be
\begin{equation*}
y^{AA'}=i\frac{s\iota^{A'}\hat{\lambda}^{A}-\bar{s}\hat{\iota}^{A'}\lambda^{A}}{\la\lambda\hat{\lambda}\ra}.
\end{equation*}
Taking into account the $(0,2)$-form components as well as the resulting Jacobian factor leaves us with:
\begin{equation*}
\widetilde{\Delta}=\int\;\exp\left[-p_{AA'}\frac{s\iota^{A'}\hat{\lambda}^{A}-\bar{s}\hat{\iota}^{A'}\lambda^{A}}{\la\lambda\hat{\lambda}\ra} \right]\frac{\d s}{s}\frac{\d t}{t}\frac{s\d\bar{s}\;\D\hat{\lambda}}{\la\lambda\hat{\lambda}\ra^2}\bar{\delta}^{2}(s\iota^{A'}-iy^{AA'}\lambda_{A})\bar{\delta}^{2|4}(\lambda_{A}+t\lambda_{A}^{\prime}).
\end{equation*}
The $s$-integrals can now be performed to give a holomorphic delta-function:
\begin{equation*}
\widetilde{\Delta}=\int \frac{\d t}{t}\frac{\bar{\delta}^{1}_{0}(p|\hat{\iota}],\lambda)}{[\iota |p|\lambda\ra^2}\bar{\delta}^{2|4}(\lambda_{A}+t\lambda_{A}^{\prime})=\frac{\bar{\delta}^{1}_{0}(p|\hat{\iota}],\lambda)\wedge\bar{\delta}^{1}_{0}(p|\hat{\iota}],\lambda')}{p^2}.
\end{equation*}
The fermionic dependence is easily re-introduced using the delta function:
\be{mMHV4}
\tilde{\Delta}(p,\lambda,\lambda')=\frac{\bar{\delta}^{1}([\hat{\iota}|p|\lambda\ra)\wedge\bar{\delta}^{1}([\hat{\iota}|p|\lambda'\ra)}{p^2}\delta^{0|4}\left(p_{AA'}\hat{\iota}^{A'}(\theta-\theta')^{aA}\right).
\ee
But this is precisely the propagator for the MHV formalism on momentum space: the scalar $p^{-2}$ propagator and the prescription that off-shell momentum spinors are defined using the CSW reference spinor and the rule $p_{A}=p_{AA'}\hat{\iota}^{A'}$.\footnote{In \eqref{mMHV4}, the prescription is actually that $\lambda_{A}$ for a propagator leg is defined by $\lambda_{A}=\hat{\iota}^{A'}p_{AA'}$.  After inserting momentum eigenstates, this is easily seen to reduce to the CSW prescription.}  Additionally, since the proof required Euclidean reality conditions (i.e., the Fourier transform to momentum space is performed on the Euclidean-real slice), this definition of the propagator automatically incorporates the Feynman $i\epsilon$-prescription.

To complete a proof that the twistor Feynman rules are equivalent to the momentum space MHV formalism, we must still show that the vertices can be extended off-shell and that the 2-point vertex does not enter the formalism.  For the first point, it suffices to demonstrate that off-shell momentum eigenstates reduce to the standard states \eqref{YMeig} on-shell.  For simplicity, we will prove this for the abelian case, but the proof can be extended with only notational complications to $\SU(N)$.  

We begin with the abelian superconnection for $\cN=4$ SYM (derived in appendix \ref{Appendix1}), $\CA=\Gamma_{AA'}\d x^{AA'}+\Gamma_{a\;A}\d\theta^{aA}$.  With Euclidean reality conditions and Woodhouse gauge, the multiplet of the theory takes the form:
\begin{eqnarray*}
A_{AA'}=e^{ip\cdot x}\varepsilon_{AA'}, \qquad \tilde{\Psi}_{a\;A'}=e^{ip\cdot x}\xi_{A'}\eta_{a}, \qquad \Phi_{ab}=\frac{e^{ip\cdot x}}{2}\eta_{a}\eta_{b}, \\
\Psi^{a}_{A}=\frac{e^{ip\cdot x}}{3!}p_{a}\epsilon^{abcd}\eta_{b}\eta_{c}\eta_{d}, \qquad G_{AB}=\frac{e^{ip\cdot x}}{4!}p_{A}p_{B}\eta^{4},
\end{eqnarray*}
where the polarization spinors are defined in relation to the CSW reference spinor:
\begin{equation*}
p_{A}=p_{AA'}\hat{\iota}^{A'}, \qquad \hat{\iota}^{A'}p^{A}\varepsilon_{AA'}=1, \qquad \hat{\iota}^{A'}\xi_{A'}=1.
\end{equation*}
The superconnection components can be written in terms of these Woodhouse representatives, and then pulled back to $\PS$ to give an off-shell momentum eigenstate $\cA$ in the Woodhouse gauge.

The transformation to CSW gauge requires finding a function $\gamma$ such that 
\begin{equation*}
\hat{\iota}^{A'}\lambda^{A}\partial_{AA'}\lrcorner(\cA+\d\gamma)=0.  
\end{equation*}
A short calculation shows that the required gauge transformation is:
\be{osmr}
\gamma=i\frac{e^{ip\cdot x}}{\la p\lambda\ra}\left[[\hat{\iota}|\varepsilon|\lambda\ra+(\eta\cdot\chi)\left(1+i\frac{(\eta\cdot\tilde{\chi})}{2}-\frac{(\eta\cdot\tilde{\chi})^2}{3!}-i\frac{(\eta\cdot\tilde{\chi})^3}{4!}\right)\right],
\ee
where $\chi^{a}=\theta^{aA}\lambda_{A}$ and $\tilde{\chi}^{a}=\theta^{aA}p_{A}$.  It is then easy to see that the off-shell momentum eigenstate in CSW gauge takes the form
\begin{multline}\label{osmr2}
\cA^{\mathrm{off-shell}}=\bar{\delta}^{1}(\la p\lambda\ra)e^{ip\cdot x}\left[[\hat{\iota}|\varepsilon|\lambda\ra+(\eta\cdot\chi)\left(1+i\frac{(\eta\cdot\tilde{\chi})}{2}-\frac{(\eta\cdot\tilde{\chi})^2}{3!}-i\frac{(\eta\cdot\tilde{\chi})^3}{4!}\right)\right] \\
+\cA_{AA'}\d x^{AA'}+\cA_{a\;A}\d\theta^{aA}.
\end{multline}
The precise form of the remaining components of the eigenstate are not important because on-shell, they vanish.  Additionally, it is easy to see that the first component of \eqref{osmr2} reduces to \eqref{YMeig} on-shell, and it descends from $\PS$ to $\PT$ as required.

Finally, the 2-point vertex vanishes in momentum space for trivial reasons of momentum conservation.  The most nontrivial case is when the vertex is in the middle of a Feynman diagram with propagators attached to each leg with supermomenta $(p,\eta)$ and $(p',\eta')$.  The fermionic part of the momentum conserving delta function then reduces to $\la p\,p'\ra^4\delta^{0|4}(\eta)\delta^{0|4}(\eta')$ and so the spinor products
cancel those in the Parke-Taylor denominator, yielding an overall $\la p\, p'\ra^2$ in the numerator.  The bosonic delta function then forces $p+p'=0$ and the numerator factor forces the vertex to vanish.

In summary, we have proven the following fact:
\begin{propn}\label{MHVpropn}
After the choice of Euclidean reality conditions, the Feynman rules of the twistor action in CSW gauge are equivalent to the MHV formalism on momentum space.
\end{propn}
Note the importance of Euclidean reality conditions in this proposition.  Although the twistor space vertices and propagator are independent of the choice of space-time signature, their translation to momentum space is not.  This can be seen as a consequence of the Feynman $i\epsilon$-prescription for the propagator, as well as the need to write down explicit representatives when pulling back to the spinor bundle.  We will now see that calculating amplitudes on twistor space (where signature choices need not be made) avoids many of the technical issues encountered in this subsection while working with momentum space representatives.


\subsection{Scattering Amplitudes in Twistor Space}
\label{TScatAmps}


\subsubsection{Amplitudes and cohomology}

Scattering amplitudes are functionals of asymptotic states; via the Penrose transform, we can represent these using momentum eigenstates which take values in $H^{0,1}(\PT,\cO)$ as given in \eqref{YMeig}.  As we will see, the twistor space MHV formalism provides a natural way to calculate the \emph{kernel} for scattering amplitudes.  For a $n$-particle scattering amplitude, such a kernel will live in the $n$-fold product of the dual of $H^{0,1}(\PT,\cO)$; in other words, the amplitude itself is obtained by pairing the kernel with momentum eigenstates.  At first, a natural choice for this pairing seems to be a Hilbert space structure on $H^{1}(\PT,\cO(k))$; however, this requires a choice of space-time signature and is actually non-local on twistor space \cite{Eastwood:1981}.

A much more natural pairing is given by the duality between $(0,1)$-forms and distributional $(0,2)$-forms with compact support on twistor space.  This is given simply by:
\begin{equation*}
\Omega^{0,1}(\PT,\cO)\times\Omega^{0,2}_{c}(\PT,\cO)\rightarrow\C, \qquad (\phi,\alpha)\mapsto\int_{\PT}\D^{3|4}Z\wedge\phi\wedge\alpha.
\end{equation*}
Hence, on twistor space, we will represent a $n$-particle amplitude as an element of
\begin{equation*}
A(1,\ldots, n)\in\Omega^{0,2}_{c}\left(\bigoplus_{i=1}^{n}\PT_{i},\cO\right).
\end{equation*}
The region of compact support is determined by ensuring that amplitudes manifest crossing symmetry: scattering states can be chosen to have both positive and negative frequency.  For instance, if we work in Lorentzian signature, then crossing symmetry dictates that the compact support of the twistor space kernel be contained in $\PN=\PT^{+}\cap\PT^{-}$.  Furthermore, the amplitude should be independent of the choice of momentum eigenstates within the same cohomology class of $H^{0,1}(\PT,\cO)$.  By taking $\phi=\dbar f$ in the above pairing, this requires the compactly supported $(0,2)$-form to be $\dbar$-closed.  Hence, we should find that the twistor space amplitude takes values in $H^{0,2}_{c}(\oplus_{i=1}^{n}\PT_{i},\cO)$.

Now, recall the form of the twistor space MHV vertex from \eqref{TAvertex}:
\begin{equation*}
V(1,\ldots,n)=\int_{\CM_{n,1}}\frac{\d^{4|4}Z_{A}\wedge\d^{4|4}Z_{B}}{\vol\;\GL(2,\C)}\int_{X^n}\tr\left(\prod_{i=1}^{n}\frac{\cA(Z(\sigma_{i}))\wedge\D\sigma_{i}}{(i-1\;i)}\right)
\end{equation*}
To obtain the kernel amplitude on twistor space, we want to insert a $(0,2)$-form representative for $\cA$ rather than a $(0,1)$-form momentum eigenstate.  This is accomplished by using the elemental state:
\be{elemental}
\cA_{i}=\bar{\delta}^{3|4}(Z_{i},Z(\sigma_{i})), \qquad Z(\sigma_{i})=Z_{A}\sigma_{i}^{0}+Z_{B}\sigma^{1}_{i}.
\ee
This forces the twistor for the $i^{\mathrm{th}}$ external state to lie on the line parametrized by $\sigma_{i}$, and after integrating with respect to $\D\sigma_{i}$, reduces to a $(0,2)$-form as desired.  The external twistors $Z_{i}$ are then integrated out against the $(0,1)$-form wavefunctions to obtain the final amplitude.  Hence, the twistor space vertex can be written as:
\be{Tvertex}
V(1,\ldots,n)=\int \frac{\d^{4|4}Z_{A}\wedge\d^{4|4}Z_{B}}{\vol\;\GL(2,\C)}\int_{X^n}\prod_{i=1}^{n}\frac{\bar{\delta}^{3|4}(Z_{i},Z(\sigma_{i}))\wedge\D\sigma_{i}}{(i-1\;i)},
\ee
ignoring the color trace.\footnote{For the remainder of this section, we take the color trace to be implicit; this is fine as long as we continue to impose the cyclic symmetry of color-stripped amplitudes.}

An unfortunate consequence of this choice is that the MHV vertex can no longer be interpreted as taking values in cohomology, since
\begin{equation*}
\dbar V(1,\ldots, n)=2\pi i \sum_{i=1}^{n}\bar{\delta}^{3|4}(i,i+1)V(1,\ldots,\widehat{i},\ldots n),
\end{equation*}
meaning that MHV amplitude will not take values in $H^{0,2}_{c}(\oplus_{i=1}^{n}\PT_{i},\cO)$.  However, these anomalies are supported at the collinear limits and are expressing the standard IR singularity structure of the amplitude.  This indicates that the collinear IR divergences of an amplitude lead to anomalies in gauge invariance, since our pairing with external wavefunctions is no longer independent of the choice of cohomological representative.  Since we have already fixed the CSW gauge, this means that a different choice of gauge would lead to different expressions for the amplitudes.  But this is an expected phenomenon in quantum field theory; for instance, a similar mechanism gives rise to anomalies in superconformal invariance of $\cN=4$ SYM \cite{Beisert:2010gn} and can be dealt with by performing suitable deformations of the super-conformal algebra generators.  Hence, we treat these IR divergences as a relic of our choice of gauge, and choose generic external twistors in the same way that one usually chooses generic (i.e., non-collinear) external momenta for a scattering amplitude.

With \eqref{Tvertex}, we can readily verify that the MHV vertices of the twistor action obey the inverse soft limit \cite{ArkaniHamed:2009si}:
\begin{lemma}\label{isl}
The MHV vertex \eqref{Tvertex} obeys the \emph{inverse soft limit}:
\begin{equation*}
V(1,\ldots,n+1)=V(1,\ldots,n)\bar{\delta}^{2|4}(n,n+1,1).
\end{equation*}
\end{lemma}

\proof  Define a non-projective coordinate $s$ on the $n+1^{\mathrm{st}}$ copy of $X$ by
\begin{equation*}
s=\frac{(n+1\;1)}{(n\;n+1)}.
\end{equation*}
Under this change of variables, we have
\begin{equation*}
Z(\sigma_{n+1})=Z(\sigma_{n})+sZ(\sigma_{1}), \qquad \frac{\d s}{s}=\frac{\D\sigma_{n+1}}{(n\;n+1)(n+1\;1)},
\end{equation*}
and the vertex becomes:
\begin{multline*}
V(1,\ldots,n+1)=\int \frac{\d^{4|4}Z_{A}\wedge\d^{4|4}Z_{B}}{\vol\;\GL(2,\C)}\prod_{i=1}^{n}\frac{\bar{\delta}^{3|4}(Z_{i},Z(\sigma_{i}))\wedge\D\sigma_{i}}{(i-1\;i)} \\
\times \frac{\d s}{s}\bar{\delta}^{3|4}(Z_{n+1},Z(\sigma_{n})+sZ(\sigma_{1})).
\end{multline*}
On the support of the delta functions involved, we can set $Z(\sigma_{n})=Z_{n}$ and $Z(\sigma_{1})=Z_{1}$ in the last factor and the proof is complete.     $\Box$

\medskip

This property was first noted in twistor space using split signature reality conditions \cite{Mason:2009sa}, and can be applied repeatedly to the vertex.  If every such application of the inverse soft limit is taken with respect to $Z_{1}$, then one obtains the formula
\be{Tvertex2}
V(1,\ldots,n)=V(1,2)\prod_{i=2}^{n}\bar{\delta}^{2|4}(1,i-1,i),
\ee
where $V(1,2)$ is the two-point vertex of the theory.  While this minimizes the number of remaining integrals and manifests superconformal invariance, it no longer exhibits the explicit cyclic symmetry of the twistor-string version of the vertex.  Indeed, there are many possible reductions of the $n$-point vertex to formulae like \eqref{Tvertex2}, depending on how the inverse soft limits are taken.  One can move between these (equivalent) representations by repeated application of the cyclic identity for the four-point vertex:
\be{4ptcyclic}
V(1,2,3)\bar{\delta}^{2|4}(1,3,4)=V(2,3,4)\bar{\delta}^{2|4}(2,4,1).
\ee

The two-point vertex is an essential part of the reduced form of the MHV vertex on twistor space.  We have seen that it cannot contribute to the Feynman diagram calculus of the twistor action due to momentum conservation; however, it would be nice to have a purely twistorial argument for this.  

\subsubsection*{\textit{The two-point vertex}}

The fermionic integrals in the twistor expression for $V(1,2)$ can be performed algebraically, resulting in a Jacobian factor of $(1\; 2)^4$ and leaving
\be{2pt1} 
V(1,2)=\int\limits_{\M_{\R}\times(\P^1)^2}\frac{\d^{4}Z_{A}\wedge\d^{4}Z_{B}}{\vol\;\GL(2,\C)}\,\D\sigma_{1}\D\sigma_{2}(1\;2)^{2}\bar{\delta}^{3}_{0,-4}(Z_{1},Z(\sigma_{1}))\bar{\delta}^{3}_{0,-4}(Z_{2},Z(\sigma_{2})). 
\ee 
Here, we define
\begin{equation*}
\bar{\delta}^{3}_{p,-p-4}(Z_1,Z_2)= \int_{\C}\frac{\d s}{s^{1+p}} \, \bar \delta^4(sZ_1+Z_2)
\end{equation*}
where the subscripts denote the homogeneity in the first and second entry respectively.

The $\vol\;\GL(2,\C)$ quotient can be fixed by setting $\sigma_{1}=(1,0)$ and $\sigma_{2}=(0,1)$ on $\P^{1}$, and then reducing $\d^{4}Z_{A}\d^{4}Z_{B}$ to projective integrals. Removing the appropriate Jacobian factor we obtain 
\be{2pt2} 
V(1,2)=\oint\limits_{\M_{\R}\times(\P^1)^2}\D^{3}Z_{A}\wedge\D^{3}Z_{B}\,\bar{\delta}^{3}_{0,-4}(Z_{1},Z_{A})\bar{\delta}^{3}_{0,-4}(Z_{2},Z_{B}), 
\ee 
where the contour is now understood as arising from integrating $Z_A$ and $Z_B$ over the $\P^1$ corresponding to $x\in\M$ and then integrating over the real slice $\M_{\R}$.  This is an integral of a $12$-form over an $8$-dimensional contour so that we are left with a $(0,2)$-form in each factor of $Z_{1}$ and $Z_{2}$, as expected for a twistor space vertex.  Now, using a simple bosonic extension of lemma \ref{dfprop1} we can see that:
\begin{equation*}
\dbar \bar{\delta}^{2}_{0,0,-4}(Z_{1},Z_{2},Z_{3})=2\pi i\left( \bar{\delta}^{3}_{0,-4}(Z_{1},Z_{3})+\bar{\delta}^{3}_{0,-4}(Z_{2},Z_{3})\right).
\end{equation*}
Note that there are only two terms here because in
\begin{equation*}
\bar{\delta}^{2}_{0,0,-4}(Z_{1},Z_{2},Z_{3})=\int_{\P^2}\frac{c_{3}^{3}\;\D^{2}c}{c_{1}c_{2}} \bar{\delta}^{4}(c_{1}Z_{1}+c_{2}Z_{2}+c_{3}Z_{3}),
\end{equation*}
there is no pole in $c_{3}$.  This means that we can write the two-point vertex as a $\dbar$-exact form:
\be{2pt3}
V(1,2)=\dbar\left(\frac{1}{2\pi i}\oint\limits_{\M_{\R}\times(\P^1)^2}\D^{3}Z_{A}\wedge\D^{3}Z_{B}\,\bar{\delta}^{2}_{0,0,-4}(Z_{1},Z_{B},Z_{A})\bar{\delta}^{3}_{0,-4}(Z_{2},Z_{B})\right),
\ee
since $\D^{3}Z_{A}\wedge\D^{3}Z_{B}=0$ on the support of $\bar{\delta}^{3}_{0,-4}(Z_{B}, Z_{A})$.  

This immediately indicates that the two-point vertex cannot enter the Feynman diagram calculus when one of its legs is an external particle, since in this case we could integrate by parts to get a $\dbar$-operator acting on an external wavefunction, which lives in cohomology and therefore gives zero.  Unfortunately, it is not so obvious that its contribution vanishes when inserted on an internal leg, since in that case integration by parts moves the $\dbar$-operator onto a propagator rather than a cohomology class.  More explicitly, if we consider the two-point vertex inserted between two twistor propagators connecting lines spanned by $(i,j)$ and $(k,l)$, this gives:
\begin{equation*}
\int\limits_{\M_{\R}\times(\P^{1})^{2}}\D^{3}Z_{A}\D^{3}Z_{B}\,\bar{\delta}^{1}_{0,0,0,-4}(i,j,*,Z_{A})\;\bar{\delta}^{1}_{0,0,0,-4}(k,l,*,Z_{B}).
\end{equation*}
We can re-write these delta functions as
\begin{equation*}
\bar{\delta}^{1}_{0,0,0,-4}(i,j,*,Z_{A})=\int_{\C^{2}}\frac{\d^{2}t}{t_{i}t_{j}}\;\bar{\delta}^{1}_{-4,0}(\lambda_{A},t_{i}\lambda_{i}+t_{j}\lambda_{j})\;\bar{\delta}^{2}(\mu_{A}+\iota+ t_{i}\mu_{i}+t_{j}\mu_{j}),
\end{equation*}
and break explicit conformal invariance by setting $\D^{3}Z_{A}\D^{3}Z_{B}=\d^{4}x\D\lambda_{A}\D\lambda_{B}\la A\;B\ra^{2}$.  Hence, the internal two-point contribution can be re-written as:
\begin{multline*}
\int \d^{4}x\;\D\lambda_{A}\;\D\lambda_{B}\la A\;B\ra^{2}\frac{\d^{4}t}{t_{i}t_{j}t_{k}t_{l}}\bar{\delta}^{1}_{-4,0}(\lambda_{A},t_{i}\lambda_{i}+t_{j}\lambda_{j}) \bar{\delta}^{1}_{-4,0}(\lambda_{B},t_{k}\lambda_{k}+t_{l}\lambda_{l})\\
\bar{\delta}^{2}(\mu_{A}+\iota+ t_{i}\mu_{i}+t_{j}\mu_{j})\bar{\delta}^{2}(\mu_{B}+\iota+ t_{k}\mu_{k}+t_{l}\mu_{l}).
\end{multline*}

A lengthy calculation allows us to perform all the parameter and $\P^{1}$ integrations against the delta functions, and several applications of the Schouten identity leaves the result:
\be{Kermit}
\int_{\M_{\R}}\frac{[\iota|(y-x)|(z-x)|\iota]^{2}\;\la i\;j\ra\la k\;l\ra }{[\iota|(y-x)|i\ra [\iota|(y-x)|j\ra [\iota|(z-x)|k\ra [\iota|(z-x)|l\ra}\frac{\d^{4}x}{(y-x)^{2}(z-x)^{2}},
\ee
where $y$ corresponds to the line $(i,j)\subset\PT$ and $z$ corresponds to $(k,l)$.  The integrand of \eqref{Kermit} is exactly the same as that arising in the computation of the so-called `Kermit' diagrams in the momentum twistor MHV formalism \cite{Bullimore:2010pj, Lipstein:2012vs, Lipstein:2013}, where it plays a non-vanishing role in the 1-loop MHV integrand.  If such contributions are to vanish (as we know they must from our momentum space arguments), the crucial difference must manifest itself at the level of the real contour $\M_{\R}$ which is chosen.  In other words, we demand that $\M_{\R}$ be chosen such that \eqref{Kermit} vanishes, while in the momentum twistor formalism, a different real contour must be chosen.

Finally, we present a few additional expressions of the twistor two-point vertex which are simpler than \eqref{2pt3}, but require an explicit choice of space-time signature.  If we choose Euclidean signature, then twistor space becomes a $\P^{1}$-bundle over $\M$, and we can perform the $\D^{3}Z_{A}$ integral over the whole of twistor space, leaving
\be{2pt4}
V(1,2)=\int_{X_{1}}\D^{3}Z_{B}\;\bar{\delta}^{3}_{0,-4}(Z_{2},Z_{B}),
\ee
where $X_{1}\cong\P^{1}$ is the Euclidean line in $\PT$ containing $Z_{1}$ (i.e., $X_{1}=Z_{1}\wedge\hat{Z}_{1}$).  This means that we can parametrize $Z_{B}=\hat{Z}_{1}+tZ_{1}$, and integrate
\begin{multline}\label{2pt5}
V(1,2)=(1,\hat{1},\d\hat{Z}_{1},\d\hat{Z}_{1})\int_{\C^2}t^{2}\;\d s\;\d t\;\bar{\delta}^{4}(Z_{2}+sZ_{1}+t\hat{Z}_{1}) \\
=(1,\hat{1},\d\hat{Z}_{1},\d\hat{Z}_{1})\bar{\delta}^{2}_{0,-1,-3}(2,1,\hat{1}).
\end{multline}
Although we will not use this form of the two-point vertex in our scattering amplitude calculations, it demonstrates that all residual integrations in the MHV vertices of the theory can be performed explicitly if one is willing to make a choice of space-time signature.


\subsubsection{Tree-level amplitudes}

Having determined the CSW gauge-fixed Feynman rules of the twistor action, we will now compute the full tree-level S-matrix of $\cN=4$ SYM in twistor space.  Using the twistor kernel formulation, we will see that generic diagram contributions to a $n$-point N$^k$MHV tree amplitude can be computed algebraically (where generic means $n>>k$).  Non-generic diagrams will fall into two classes: boundary diagrams and boundary-boundary diagrams.  In all cases, we will see that the twistor MHV formalism provides an efficient calculation at tree-level.

The classification of Feynman diagrams into generic, boundary, and boundary-boundary is essentially geometric from the twistor point of view.  In twistor space a MHV vertex corresponds to a line with the legs of the vertex given by points on this linearly embedded $\P^1$.  The twistor propagator forces a marked point on one line to be collinear with the reference twistor $Z_{*}$ and a marked point on another line.  For any two lines in general position, there is a unique line connecting these two marked points which passes through $Z_{*}$; hence any diagram contributing to a N$^k$MHV tree amplitude will contain $k+1$ MHV vertices/lines and $k$ propagators. 

\emph{Generic} diagrams will be those with no adjacent propagator insertions on any of the vertices. \emph{Boundary} diagrams are those which have adjacent propagator insertions on at least one vertex, but have at least two external particles attached to all vertices.  This last condition means that after propagators are integrated out, a line in twistor space can still be associated to each vertex using the external legs.  \emph{Boundary-boundary} diagrams have adjacent propagator insertions on a vertex with fewer than two external legs; in this case not all integrals can be performed algebraically without making some choices.

To illustrate how calculations proceed in the twistor framework, we begin by considering the NMHV tree amplitude, since there are only generic diagrams which contribute here.

\subsubsection*{\textit{Example: NMHV tree}}

\begin{figure}
\centering
\includegraphics[width=5.5 in, height=1.5 in]{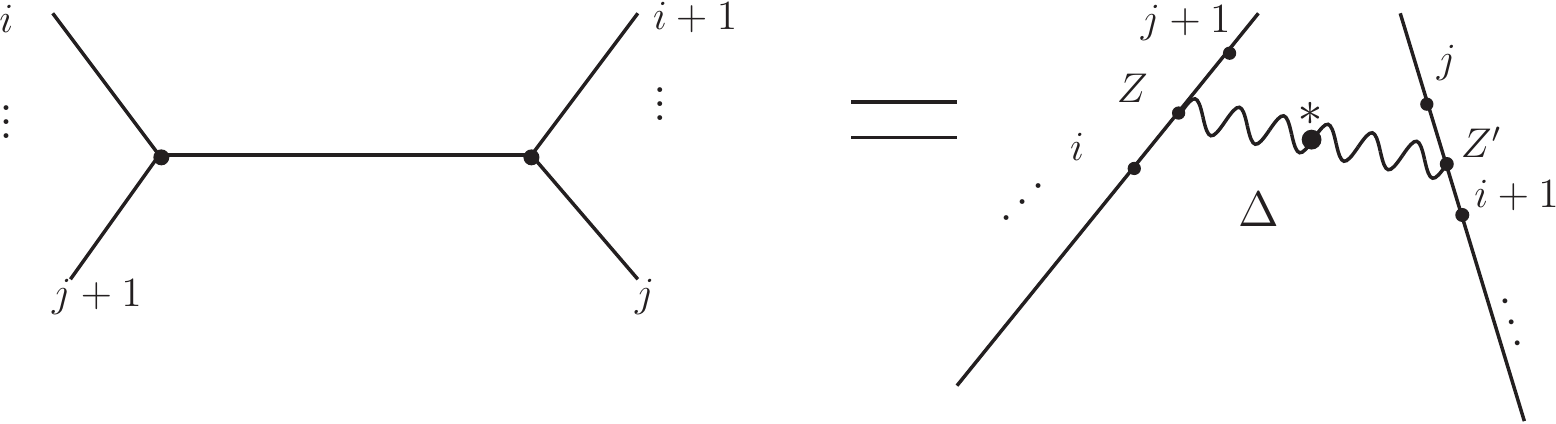}\caption{\textit{Twistor support of a NMHV tree diagram}}\label{TNMHV}
\end{figure}  

For a $n$-point tree-level NMHV amplitude, the only diagrams which contribute are those with two vertices joined by a single propagator, as illustrated in Figure \ref{TNMHV}.  Our discussion of the two-point vertex indicates that every possible diagram in this situation will be generic (i.e., at least two external particles on each vertex).  The twistor space Feynman rules therefore indicate that the contribution from such a diagram is given by:
\be{NMHV1}
\int_{\PT^{2}}\D^{3|4}Z\;\D^{3|4}Z'\;V(j+1,\ldots,i,Z)\;\bar{\delta}^{2|4}(Z,*,Z')\;V(Z',i+1,\ldots,j).
\ee
We can use the inverse soft limit to write this as
\begin{multline}\label{NMHV2}
V(j+1,\ldots,i)\;V(i+1,\ldots,j) \times \\
\int_{\PT^{2}}\D^{3|4}Z\;\D^{3|4}Z'\;\bar{\delta}^{2|4}(i,j+1,Z)\;\bar{\delta}^{2|4}(Z,*,Z')\;\bar{\delta}^{2|4}(j, i+1,Z') \\
=V(j+1,\ldots,i)\;V(i+1,\ldots,j)\;[i,j+1,*,i+1,j],
\end{multline}
with the last line following by lemma \ref{dfprop2}.  Hence, these generic diagrams can be evaluated algebraically against the delta functions on twistor space, and correspond to the two lines (each remaining vertex factor) together with their unique transversal through the reference twistor (the R-invariant).

The NMHV tree amplitude on twistor space is then given by a sum over the contributing diagrams, which is equivalent to
\be{NMHV3}
A^{0}_{n,1}=\sum_{i<j}V(j+1,\ldots,i)\;V(i+1,\ldots,j)\;[i,j+1,*,i+1,j].
\ee
Since each vertex is conformally invariant, and the R-invariant is the standard invariant of the superconformal group, this twistorial form of the amplitude manifests the superconformal symmetry of $\cN=4$ SYM, up to the choice of reference twistor $Z_{*}$.

\subsubsection*{\textit{Generic diagrams}}

\begin{figure}
\centering
\includegraphics[width=2 in, height= 1 in]{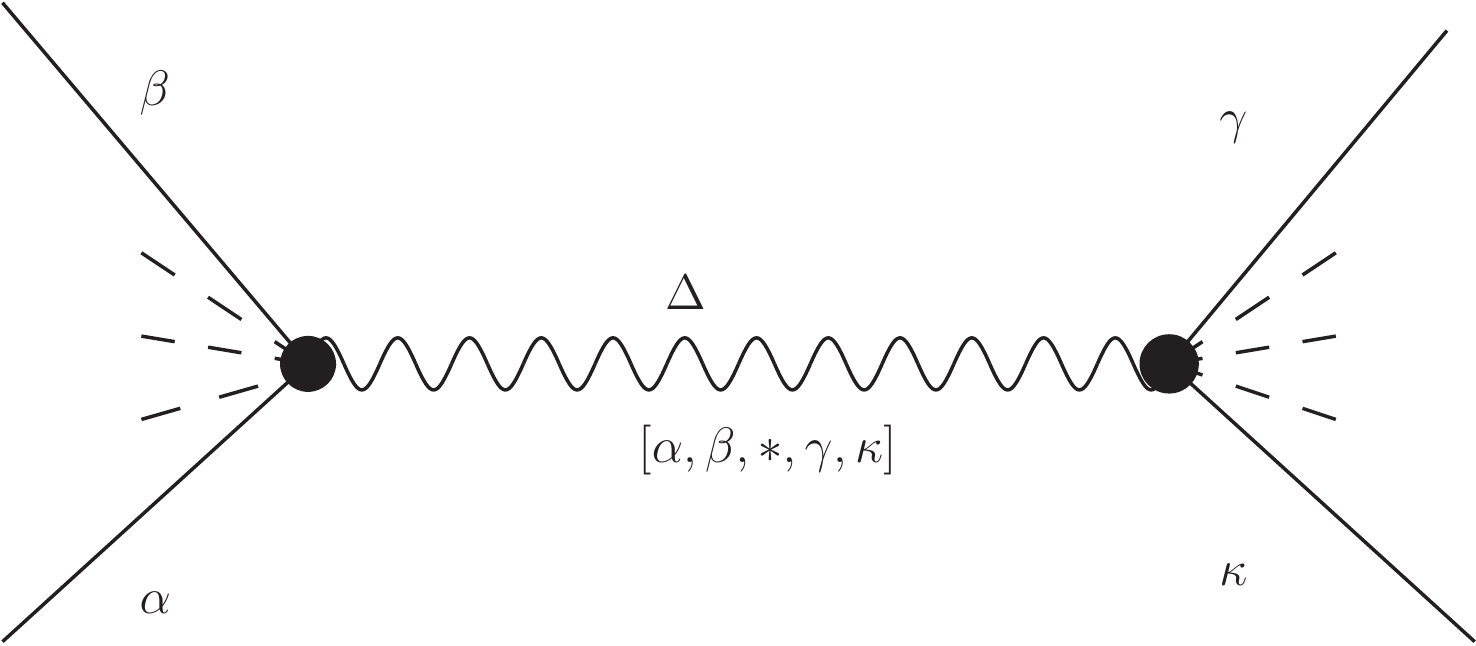}\caption{\textit{Propagator contributions for generic diagrams}}\label{PropR} 
\end{figure}

The NMHV calculation above extends directly to each propagator of a generic N$^k$MHV diagram, where there are no adjacent propagator insertions, as depicted in Figure \ref{PropR}.  Using the inverse soft limit, we can strip off a $\bar{\delta}^{2|4}$ from each vertex leaving MHV vertices which no longer depend on the propagator insertion points $Z_{1}$, $Z_{2}$.  Hence, the propagator leads to a R-invariant factor:
\be{generic}
\int\D^{3|4}Z_{1}\D^{3|4}Z_{2}\;\bar{\delta}^{2|4}(\alpha,\beta,Z_{1})\;\bar{\delta}^{2|4}(Z_{1},*,Z_{2})\;\bar{\delta}^{2|4}(\gamma,\kappa,Z_{2})=[\alpha,\beta,*,\gamma,\kappa].
\ee
Here, $\alpha$ and $\beta$ are the two nearest particles on the left-hand side of the propagator insertion, and $\gamma$ and $\kappa$ are the nearest particles on the right-hand side of the insertion.
When $n>>k$, we can see that these sorts of diagrams will dominate the contributions to the tree-level N$^k$MHV amplitude.  Proceeding inductively, a tree-level generic N$^k$MHV diagram gives a product of $k$ R-invariants, one for each propagator depending on $Z_{*}$ and each adjacent external twistor. These are multiplied by the $k+1$ MHV vertices depending only on the external particles.

\subsubsection*{\textit{Boundary diagrams}}

For boundary terms, there are some vertices which have adjacent propagator insertions.  The resulting formulae are similar to the generic case: we obtain a product of $k+1$ MHV vertices (one for each vertex containing only the external particles at that vertex) and $k$ R-invariants (one for each propagator).  However, because of adjacent propagator insertions, some of the entries in the R-invariants are now shifted.

\begin{figure}
\centering
\includegraphics[width=3.5 in, height=2 in]{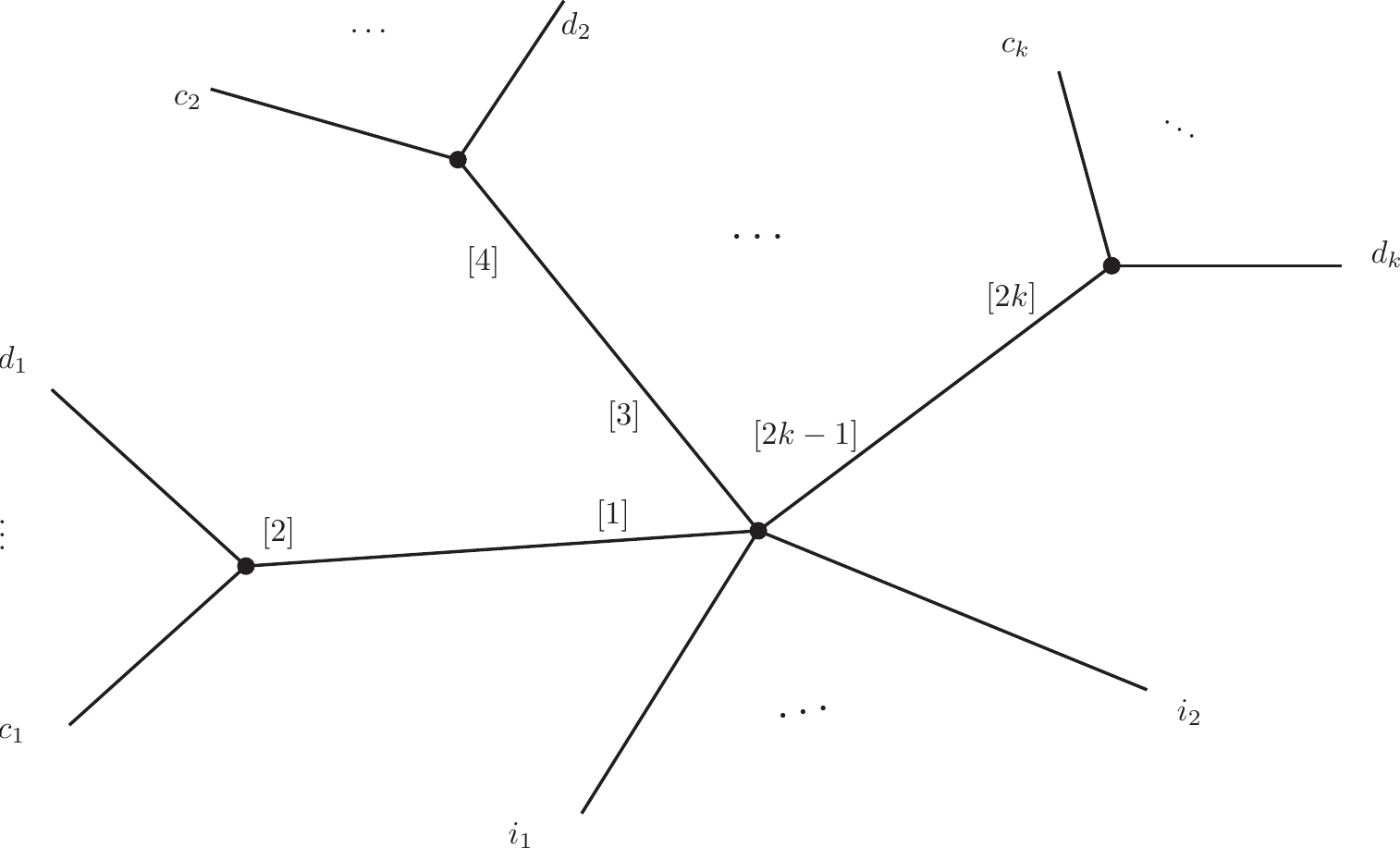}\caption{\textit{N$^k$MHV boundary term with $k$ adjacent propagators}}\label{kboundary}
\end{figure}

The rule for the shifts can be obtained by studying each end of the propagator separately; to give the most general case, we compute the shifts at a vertex with $k$ adjacent propagators, as in Figure \ref{kboundary}.  As in the generic case, we can decompose the central vertex into
\begin{equation*}
V(i_{1},\ldots,i_{2})\;\bar{\delta}^{2|4}(i_{2},[2k-1],i_{1})\;\prod_{j=1}^{k-1} \bar{\delta}^{2|4}(i_{2},[2j-1],[2j+1]).
\end{equation*} 
Clearly, we have made a choice by taking this form of the decomposition, both with respect to the overall orientation of the diagram and to each inverse soft limit.  Other choices will yield equivalent final answers upon utilizing cyclical identities.   

The factor of $V(i_{1},\ldots, i_{2})$ will be left as part of our final answer, but we want to use the delta functions to integrate out the $Z_{[2j-1]}$ corresponding to propagator insertions.  The relevant integrals are:
\be{adjacent-props}
\int \prod_{j=1}^{k}\D^{3|4}Z_{[2j-1]} \; \bar{\delta}^{2|4}([2j],*,[2j-1])\;\bar{\delta}^{2|4}(i_{2},[2j-1],[2j+1]),
\ee
where $Z_{[2k+1]}=Z_{i_1}$.  We can proceed inductively using the fact that $Z_{[2j]}$ must lie on the line $(c_{j},d_{j})$ in twistor space.  Indeed, performing the $\D^{3|4}Z_{[2k-1]}$ integral leaves $\bar{\delta}^{1|4}([2k],*,i_{2},i_{1})$, which forces its three arguments to be co-planar.  But since $Z_{[2k-1]}\in (i_{1},i_{2})$ already, this indicates that we must have $Z_{[2k-1]}=(i_{1},i_{2})\cap(*,c_{k},d_{k})$.  In this fashion, we easily deduce that \eqref{adjacent-props} is equal to:
\be{adjp2}
\prod_{j=1}^{k}\bar{\delta}^{1|4}([2j],*,i_{2},[2j+1]), \qquad Z_{[2j-1]}=(i_{1},i_{2})\cap(*,c_{j},d_{j}).
\ee
Upon connecting with the propagator legs in the $(c_{j},d_{j})$ vertices, we obtain a product of R-invariants.  In other words, the total contribution for the diagram in Figure \ref{kboundary} reads:
\be{adjp3}
V(i_{1},\ldots,i_{2})\prod_{j=1}^{k}V(c_{j}\ldots,d_{j})\;[[2j-1],i_{2},*,c_{j},d_{j}].
\ee

This immediately leads to a general rule for computing R-invariant contributions in both generic and boundary diagrams.
\begin{itemize}
\item Each MHV vertex in a diagram contributes a factor of the MHV tree amplitude that depends only on the external legs at that vertex.

\item Each propagator contributes a factor $[\widehat{i}_{1},i_{2},*,\widehat{j}_{1},j_{2}]$, where $i_{1}$, $i_{2}$ are the external particles nearest to one side of the propagator insertion and $j_{1}$, $j_{2}$ are nearest to the opposite side, with $i_{1}<i_{2}$ and $j_{1}<j_{2}$ in the cyclic ordering.  Let $p$ be the propagator insertion point on a vertex; then
\be{shiftrule1}
Z_{\widehat{i}_1}=\left\{
\begin{array}{cl}
Z_{i_1} & \mbox{if } p\;\mbox{is adjacent to } i_{1} \\
(i_{1},i_{2})\cap(*,c,d) & \mbox{if } p\;\mbox{is adjacent to the propagator} \\
 & \mbox{connecting to the line } (c,d).
\end{array} \right.
\ee
The rule for $\widehat{j}_{1}$ follows by taking $i\leftrightarrow j$.
\end{itemize}

\subsubsection*{\textit{Boundary-boundary diagrams}}

\begin{figure}
\centering
\includegraphics[width=5 in, height=1.5 in]{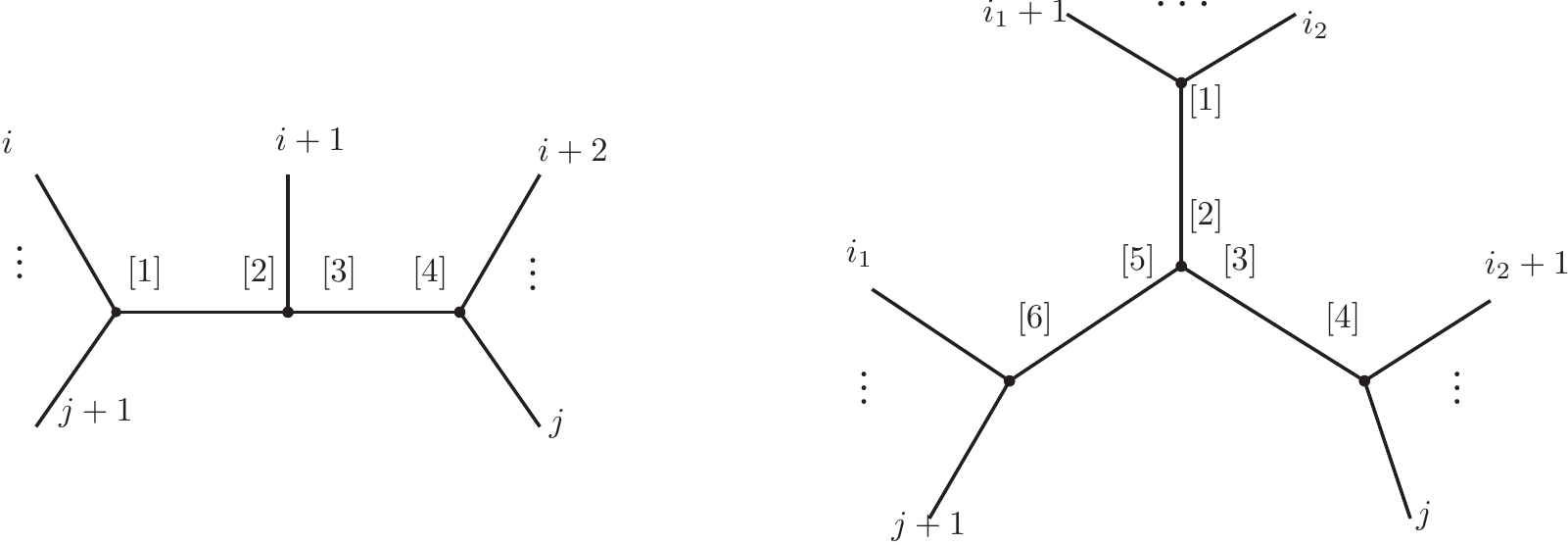}\caption{\textit{Boundary-boundary terms at N$^2$MHV and N$^3$MHV}}\label{boundary} 
\end{figure}

The final class of potential MHV diagrams in twistor space are those in which some vertices have fewer than two external legs.  In this case, the prescription of \eqref{shiftrule1} breaks down, as there is no line $(i_{1},i_{2})$ to use in the definition of the shifts.  See Figure \ref{boundary} for simple examples of such diagrams; the worst case is when there are \emph{no} external legs on the vertex in question and the simplest example of such a diagram is the N$^3$MHV `cartwheel' diagram. 

Using our standard techniques, the cartwheel can be reduced to:
\begin{multline*}
V(j+1,\ldots ,i_{1})V(i_{1}+1,\ldots ,i_{2})V(i_{2}+1,\ldots ,j) \times \\
\int\D^{3|4}Z_{[3]}\D^{3|4}Z_{[5]}\bar{\delta}^{1|4}(j+1,i_{1},\rf, [5])\;\bar{\delta}^{1|4}(i_{2}+1,j,\rf, [3])\;V([3],[5])\;\left[i_{1}+1,i_{2},\rf, [3],[5]\right],
\end{multline*}
where $V(\cdot, \cdot)$ is the two-point MHV amplitude given by \eqref{2pt1}.  Clearly, we cannot perform the two remaining twistor integrals without specifying additional constraints.  The case where there is a single external particle leaves one un-resolved integral.

Note that although we cannot reduce boundary-boundary terms to a simple expression in terms of shifted twistors, they are still fully described by the twistorial MHV formalism.  It is possible to reduce these further using the remaining delta functions, but it seems to be impossible to obtain an expression built only out of R-invariants and MHV vertices.  However, with a choice of real contour these remaining integrals could be performed (and do not introduce divergences); this would simply entail the introduction of new signature-dependent machinery such as the form of the two-point vertex presented in \eqref{2pt5}

A full N$^k$MHV tree amplitude is computed by summing the contributions for all generic, boundary, and boundary-boundary diagrams for the given specification of external particles and MHV degree.  We conclude our discussion of the tree-level amplitudes with a more non-trivial example.

\subsubsection*{\textit{Example: N$^2$MHV tree}}

\begin{figure}
\centering
\includegraphics[width=6 in, height=1.5 in]{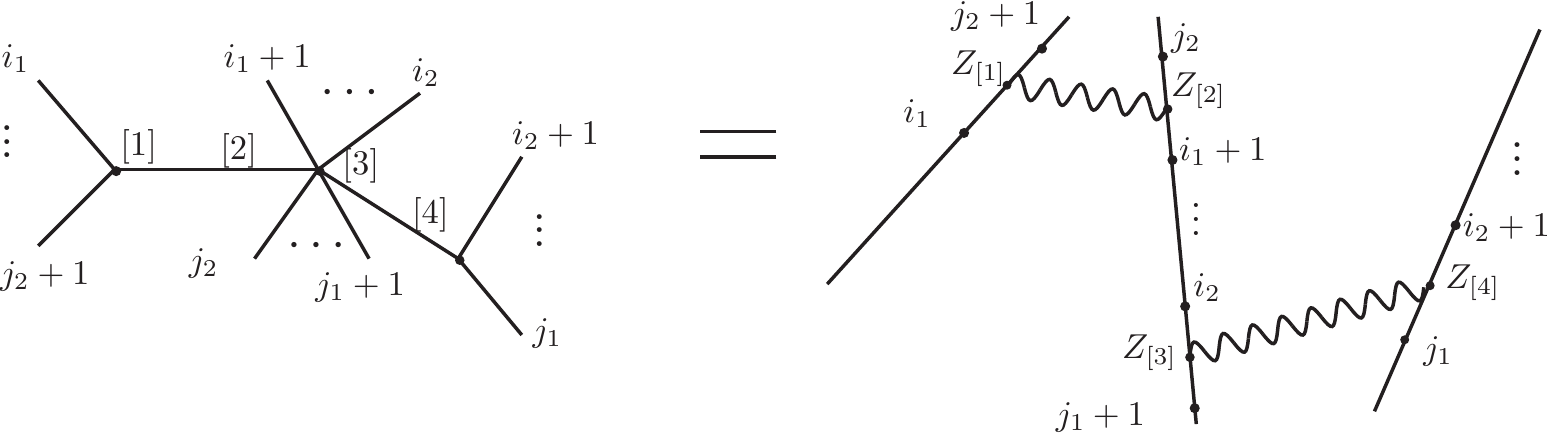}\caption{\textit{Twistor support of a generic N$^{2}$MHV tree diagram}}\label{TN2MHV}
\end{figure}

N$^2$MHV tree amplitudes provide the simplest example where all three classes of diagram contribute. The twistor space support of the generic diagrams is illustrated in Figure \ref{TN2MHV}.  Applying our usual rules gives a contribution from all generic diagrams of the form:
\begin{multline}\label{N2Gen}
A^{\mathrm{gen}}_{n,2}=\sum_{i_{1}+1<i_{2}<j_{1}-1<j_{2}-2}[i_{1},j_{2}+1,*,i_{1}+1,j_{2}]\;[i_{2},j_{1}+1,*,i_{2}+1,j_{1}] \\
\times V(j_{2}+1,\ldots,i_{1})\;V(i_{1}+1,\ldots,i_{2},j_{1}+1,\ldots,j_{2})\;V(j_{1},\ldots,i_{2}+1).
\end{multline}
The R-invariants are obtained by integrating out the internal twistors in the usual algebraic fashion.  

\begin{figure}
\centering
\includegraphics[width=6 in, height=1.5 in]{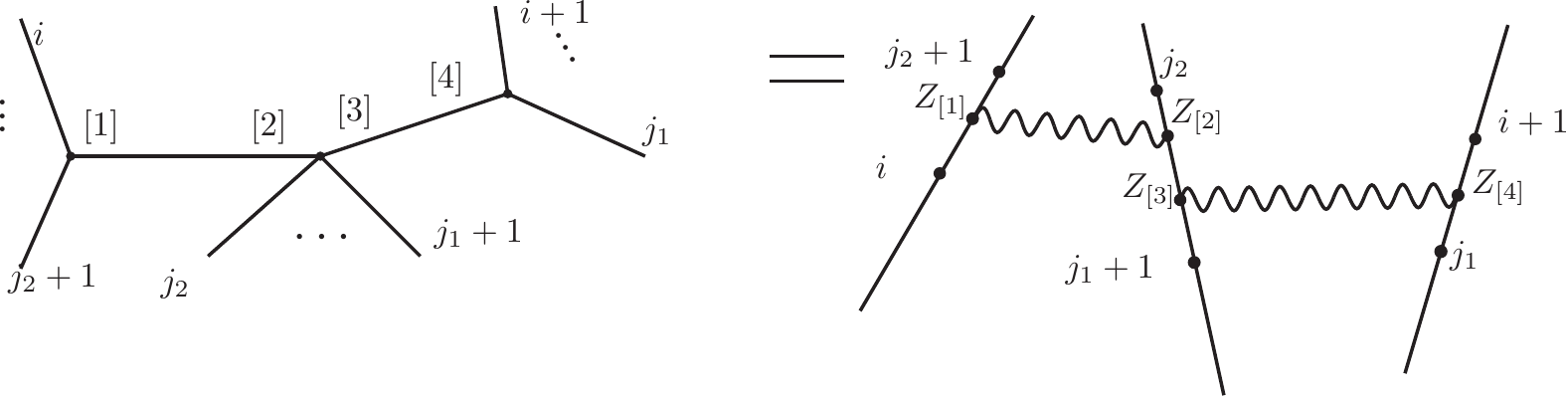}\caption{\textit{Twistor support of a boundary N$^{2}$MHV tree diagram}}\label{TN2C}
\end{figure}

A boundary diagram, on the other hand, is one in which the propagator insertions are adjacent on the middle vertex (see Figure \ref{TN2C}).  In this case we must apply the rules given above for assigning R-invariants to the propagators, since some of the entries must be shifted.  Performing the inverse soft limit decomposition with respect to $Z_{[3]}$ and then $Z_{[2]}$, we obtain:
\begin{multline}\label{N2B}
 A^{\mathrm{bound}}_{n,2}=\sum_{i+1<j_{1}<j_{2}-1}[i,j_{2}+1,*,j_1+1,j_{2}]\; [\widehat{j_2},j_1+1,\rf,i+1,j_1] \\
\times V(j_{2}+1,\ldots, i)\; V(j_1+1,\ldots,j_2)\; V(i+1,\ldots,j_1), 
\end{multline}
where 
\begin{equation*}
Z_{\widehat{j_2}}= (j_2,j_1+1)\cap (*,i,j_2+1)\, .
\end{equation*}
An equivalent shifted contribution could be defined by taking the inverse soft limit with respect to different internal twistors.

Finally, we must account for the boundary-boundary contributions.  Our discussion of the two point vertex narrows such diagrams down to those with a single external particle on the middle vertex (an example is given by the first diagram of Figure \ref{boundary}).  For such a diagrams, we obtain the contribution
\begin{multline}\label{N2BB}
A^{\mathrm{bb}}_{n,2}=\sum_{j+1<i<j-2} V(j+1,\ldots , i)\;V(i+2,\ldots ,j) \\
\times \int \D^{3|4}Z_{[2]} \;  V([2],i+1)\; [i+2,j,* ,i+1,[2] ]\;\bar{\delta}^{1|4}(j+1,i,*,[2])\, . 
\end{multline}
As discussed before, the remaining integral could be performed in various ways, for instance by introducing Euclidean reality conditions.

The full N$^{2}$MHV amplitude is a sum over generic, boundary and boundary-boundary diagrams using \eqref{N2Gen}, \eqref{N2B} and \eqref{N2BB}:
\begin{equation*}
A^{0}_{n,2}=A^{\mathrm{gen}}_{n,2}+A^{\mathrm{bound}}_{n,2}+A^{\mathrm{bb}}_{n,2}.
\end{equation*}


\subsubsection{Loop-level amplitudes}   

Clearly, the twistor action provides an efficient mechanism for calculating tree-level scattering amplitudes in $\cN=4$ SYM via the MHV formalism.  If we truly want to think of the twistor action as defining a quantum field theory on twistor space, then we must be able to describe computations at all loop orders in perturbation theory, though.  While $\cN=4$ SYM is UV finite, its generic scattering amplitudes have IR divergences at loop-level, which require regularization.

We begin by illustrating that our twistorial formalism extends directly from the tree-level setting in the case of those loop diagrams which are IR finite.  Then, we will discuss how IR divergences appear for a generic amplitude and potential regularization strategies on twistor space.

\subsubsection*{\textit{Finite examples}}

\begin{figure}
\centering
\includegraphics[width=4 in, height=2.25 in]{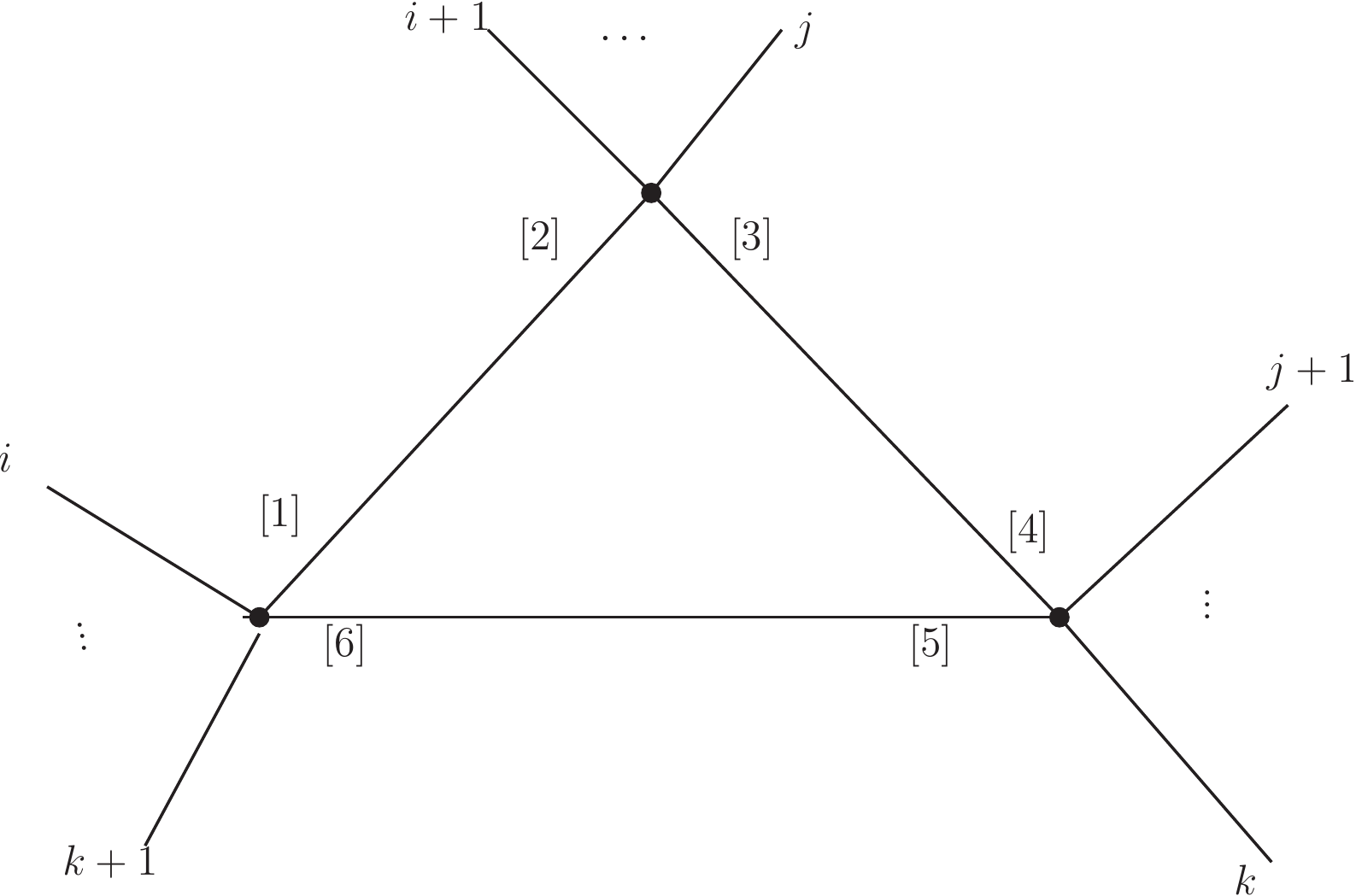}\caption{\textit{Triangular 1-loop NMHV diagram}}\label{LNMHV}
\end{figure}

Using the MHV formalism for $\cN=4$ SYM, one can easily identify loop level diagrams which are finite.  Although such diagrams are certainly not generic, they nevertheless provide an interesting example of how our twistor methods enable simple calculations.  For instance, in the planar sector at $1$-loop NMHV, the triangle diagrams of the form illustrated in Figure \ref{LNMHV} are finite.  On twistor space, this corresponds to being able to perform all integrals in a well-defined and algebraic fashion.

The contribution coming from a diagram of this form can be computed by first performing the integrals in $Z_{[2]}$, $Z_{[4]}$, and $Z_{[6]}$ trivially, and then using the boundary diagram rule to do the remaining integrals in terms of shifted twistors.  The result is
\begin{multline}\label{LNMHV1}
V(i+1,\ldots, j)\;V(j+1,\ldots, k)\;V(k+1,\ldots, i) \\
\times [k+1,i,*,\widehat{i},i+1]\;[i+1,j,*,\widehat{j},j+1]\;[j+1,k,*,\widehat{k},k+1],
\end{multline}
where the shifted twistors are:
\begin{equation*}
Z_{\widehat{i}}=(i+1,j)\cap(j+1,k,*),\quad Z_{\widehat{j}}=(j+1,k)\cap(k+1,i,*), \quad Z_{\widehat{k}}=(k+1,i)\cap(i+1,j,*).
\end{equation*}
Unpacking the definition of the R-invariants, it is easy to see that \eqref{LNMHV1} is indeed finite.

\begin{figure}
\centering
\includegraphics[width=3 in, height=2.25 in]{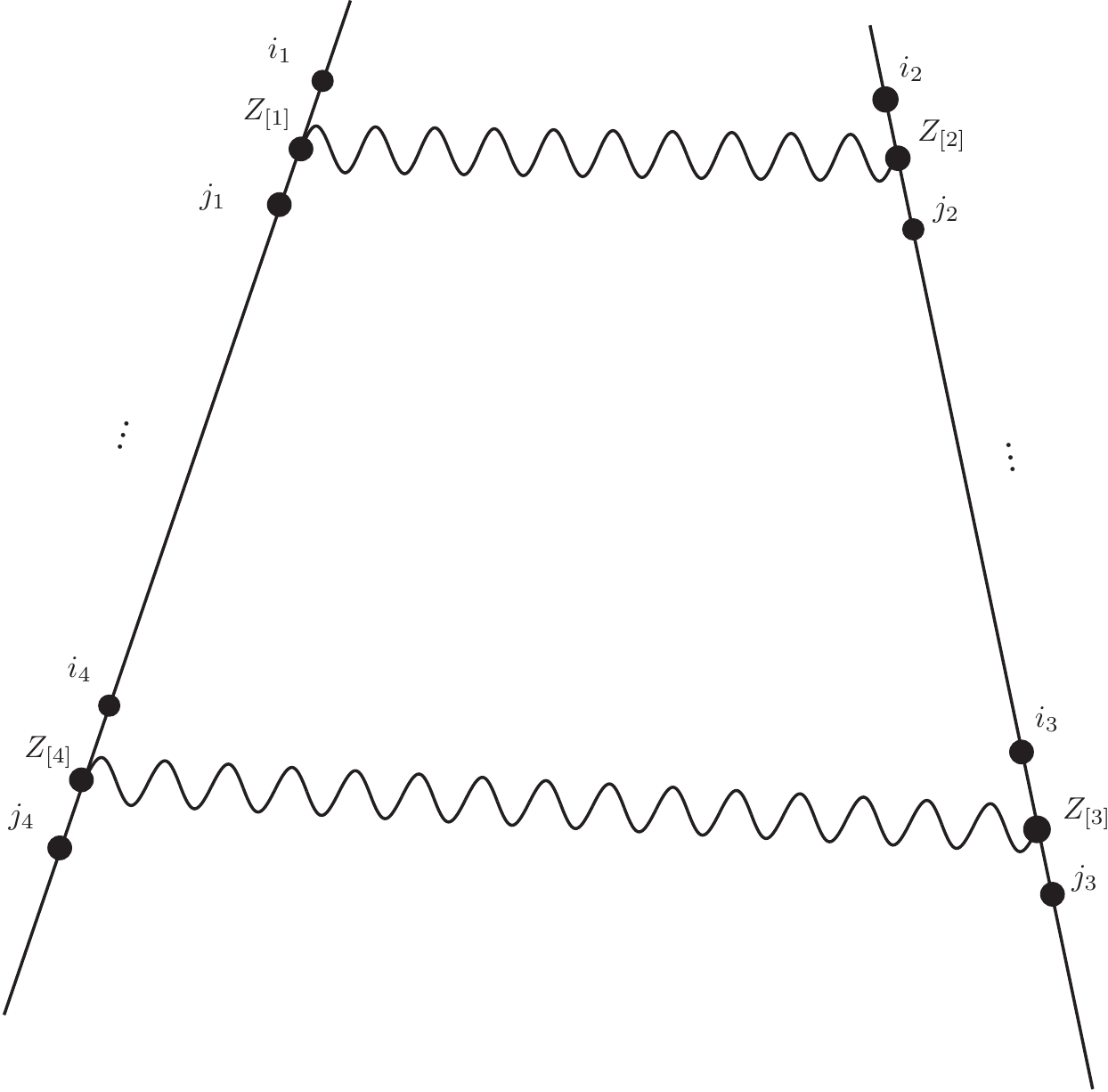}\caption{\textit{Strictly non-planar 1-loop MHV diagram in twistor space}}\label{NPMHV}
\end{figure}

An even simpler example is available if we allow ourselves to consider the non-planar sector: a (strictly) non-planar 1-loop MHV diagram is not only finite, but its twistor space support remains planar, as illustrated in Figure \ref{NPMHV}.  In this case, all the integrals can be performed as in a generic diagram, and we are left with a contribution of the form:
\be{NPMHV1}
V(i_{1},j_{1},\ldots, i_{4},j_{4},\ldots)\;V(i_{2},j_{2},\ldots, i_{3},j_{3},\ldots )\;[i_{1},j_{1},*,i_{2},j_{2}]\;[i_{3},j_{3},*,i_{4},j_{4}].
\ee
Although the answer is simple, the geometry of this situation is still quite important.  The twistor propagator forces insertion points on each vertex to lie on the transversal through the reference point $Z_{*}$.  Since the transversal between two lines and a point in $\PT$ is unique, this forces the two insertion points on each vertex to be equal.  In the strictly non-planar setting this is finite because the external states separate the propagator insertions in the color ordering; however, when propagator insertions are adjacent a pole arises from the Parke-Taylor denominator.  As we shall see, in the planar 1-loop MHV this leads to a double divergence.

\subsubsection*{\textit{Generic loop diagrams}}
 
\begin{figure}
\centering
\includegraphics[width=5.5 in, height=2 in]{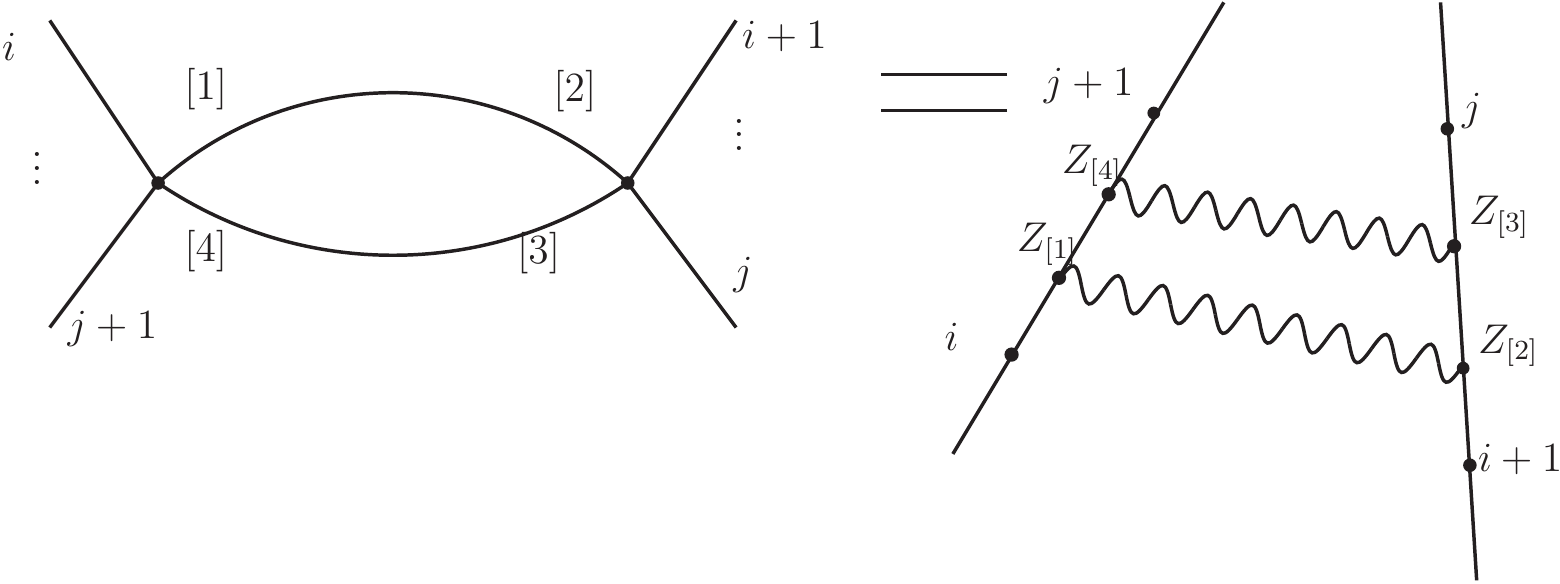}\caption{\textit{Twistor support of the 1-loop MHV amplitude}}\label{TLoop}
\end{figure}

Generic loop diagrams in $\cN=4$ SYM will contain IR divergences coming from when internal momenta are collinear with the external states.  The simplest example of this is captured at 1-loop by the planar MHV amplitude; see Figure \ref{TLoop} for the twistor support of a generic diagram contributing to this amplitude.  For such a diagram, the geometry in twistor space forces the propagator insertions on each line to coincide: $Z_{[1]}=Z_{[4]}$ and $Z_{[2]}=Z_{[3]}$.

We can easily evaluate all the integrals contributing to this diagram; this results in some shifted twistors due to the adjacent propagator insertions:
\be{pMHV1-loop}
V(i+1, \ldots, j)\;V(j+1,\ldots, i)\;[i, j+1, *,\widehat{j},i+1]\;[j,i+1,*,\widehat{i},j+1],
\ee
with shifts given by the usual rule
\begin{equation*}
\widehat{i}=(i,j+1)\cap(j,i+1,*), \qquad \widehat{j}=(j,i+1)\cap(i,j+1,*).
\end{equation*}
Recalling the definition of the R-invariant \eqref{DF5}, we can see a single divergence coming from each propagator factor since
\begin{equation*}
[j,i+1,*,\widehat{i},j+1]=\frac{\delta^{0|4}\left((j,i+1,*,\widehat{i})\chi_{j+1}+\mathrm{cyclic}\right)}{(j,i+1,*,\widehat{i})\cdots(j+1,j,i+1,*)}.
\end{equation*}
The shifts indicate that $Z_{\widehat{i}}$ is coplanar with $Z_{j}$, $Z_{i+1}$, and $Z_{*}$, so $(j,i+1,*,\widehat{i})=0$.  Similarly, we get a divergence in the other R-invariant from $(i,j+1,*,\widehat{j})=0$.

However, we also get a numerator factor of zero in \eqref{pMHV1-loop} from the fermionic portions of the R-invariants.  It is easy to see that the two $\delta^{0|4}$s are proportional, so their product must vanish.  This leaves us with a `$0/0$' type expression for the diagram in Figure \ref{TLoop}, which clearly requires a regularization scheme.  Indeed, at 1-loop it is known that the only true IR divergences should come from those diagrams in which one of the vertices is a 3-point vertex (i.e., only two external particles) \cite{Brandhuber:2004yw, Bena:2004xu, Lipstein:2013}.  Hence, the required regularization on twistor space must first treat the fermionic contributions carefully and then implement a correct IR regularization mechanism.

It is easy to see that at higher loops and MHV degree, the divergence structure we have observed here will persist.  To say that the twistor action for $\cN=4$ SYM is well-defined as a \emph{quantum} field theory, we must be able to give (at least in principle) a regularization scheme on twistor space.  Of course, the most widely used regularization scheme is dimensional regularization; this is particularly well-adapted to traditional Feynman integrals which appear in space-time calculations, although it is manifestly unphysical, breaks dual superconformal symmetry, and obscures the underlying integrability of the theory.  Furthermore, dimensional regularization seems impossible to implement on twistor space, which is not well-defined for a $4-2\varepsilon$-dimensional space-time.  

Motivated by the AdS/CFT correspondence, Alday, Henn, Plefka and Schuster proposed a `mass regularization' scheme for scattering amplitudes that preserves dual superconformal symmetry \cite{Alday:2009zm}.  In this scheme, one gives some of the scalars in $\cN=4$ SYM a vacuum expectation value (VEV) by moving out onto the Coulomb branch in a particular direction; at loop level this keeps external particles as well as totally internal loop lines massless, while particles running around the external edges of the loops acquire masses (see \cite{Henn:2011xk} for a review).  These masses then act as a regulator for the theory.  

Building off the earlier findings of \cite{Craig:2011ws}, Kiermaier proposed a \emph{massive MHV formalism} which demonstrates that the MHV formalism at the origin of the moduli space extends to the Coulomb branch of $\cN=4$ SYM \cite{Kiermaier:2011cr}.  In accordance with the scalar VEV structure associated to the Coulomb branch, there are three types of vertices which serve the purpose of the single MHV vertex in the original formalism, and the massless scalar propagator is replaced by a massive one. This massive MHV formalism has been shown to be correct using recursive arguments \cite{Elvang:2011ub}.

In Appendix \ref{Appendix2}, we show that the Coulomb branch of $\cN=4$ SYM can be accessed on twistor space, leading to a twistor action derivation of Kiermaier's massive MHV formalism.  Combined with the mass regularization scheme, this provides a mechanism for regularizing IR divergences on twistor space.  Unfortunately, the Coulomb branch MHV rules on twistor space are not as elegant as those at the origin of the moduli space.\footnote{In particular, the massive propagator on twistor space takes the form of an infinite series. Upon translating this to space-time, we see that it can be re-summed to $(p^{2}-m^{2})^{-1}$, but there does not appear to be an elegant resummation procedure that is self-contained in twistor space.}  So rather than study this formalism here, we take the following facts:
\begin{itemize}
\item There is a twistor action for the Coulomb branch of $\cN=4$ SYM.

\item Its Feynman rules in CSW gauge are equivalent to the massive MHV formalism of \cite{Kiermaier:2011cr}.

\item This allows us (in principle) to implement the mass regularization scheme on twistor space.
\end{itemize}      

The interested reader need only consult Appendix \ref{Appendix2} for proofs of these facts.  An alternative option could be provided by the work of Lipstein and Mason \cite{Lipstein:2013}, which provides a mechanism for correctly regulating 1-loop Kermit integrals in \emph{momentum} twistor space.  This formalism incorporates the Feynman $i\epsilon$-prescription and correctly regulates the integral, albeit in a non-trivial way.  If one could adopt this methodology to the twistor space integrals here, it could provide another regularization mechanism (which is related to both dimensional and mass regularization) for computing loop amplitudes twistorially.   

Clearly, the incomplete picture of regulation for loop amplitudes is a shortcoming of the twistor action approach as presented here.  Rather than dwell on this issue, let us instead consider how the twistor action can be used to study other interesting gauge theoretic observables.


\section{Wilson Loops, Local Operators, and Correlation Functions}
\label{Chapter4}

In the study of any gauge theory, interesting physical observables include correlation functions of local operators and Wilson loops, and gauge theory on twistor space is no exception.  In the previous section, we demonstrated that scattering amplitudes at tree-level and beyond could be computed efficiently using the twistor action for $\cN=4$ SYM; now we further explore the utility of the twistor action by studying gauge invariant local operators, null polygonal Wilson loops, and their expectation values in $\cN=4$ SYM on twistor space.

Recall that in a (bosonic) gauge theory, a Wilson loop is given by computing the trace of the holonomy of a connection 1-form $A$ around some closed path $\gamma$:
\be{bWilsonloop}
W_{R}[\gamma]=\tr_{R}\mathrm{Hol}[A,\gamma]=\tr_{R}\cP \exp\left(-\oint_{\gamma}A\right),
\ee
where $R$ is the representation of the gauge group in which the trace is taken, and $\cP$ is the `path-ordering' symbol.  Besides forming a natural class of gauge invariant observables, Wilson loops arise in a wide variety of applications in both pure mathematics and physics.  In $\cN=4$ SYM, these operators can be extended to compute the holonomy of the full $\cN=4$ superconnection:
\be{sWilsonloop}
W_{R}[\gamma]=\tr_{R}\cP\exp\left(-\oint_{\gamma}\CA\right)=\tr_{R}\cP\exp\left(-\oint_{\gamma}\Gamma_{AA'}\d x^{AA'}+\Gamma_{aA}\d\theta^{aA}\right),
\ee
where $\gamma\subset\M$ is now understood to be a curve in the full chiral superspace.

Null polygonal Wilson loops (i.e., when the curve $\gamma$ is a null polygon $C\subset\M$) are of particular interest beyond belonging to this class of important operators.  Motivated by the AdS/CFT correspondence, Alday and Maldacena first conjectured the duality between the expectation value of a $n$-cusp null polygonal Wilson loop in the fundamental representation of the gauge group and $n$-particle gluon scattering amplitudes by studying these objects in the strong coupling regime (i.e., using string theory in the $AdS_{5}\times S^{5}$ geometry near the boundary) \cite{Alday:2007hr}.  In this picture, the gluon null momenta of the scattering process become the edges of the null polygon.  

This amplitude / Wilson loop duality can be understood as arising from a non-compact T-duality which maps the string scattering worldsheet on the AdS-boundary to a minimal surface with the null polygon as its boundary, and interchanges the superconformal and dual superconformal groups \cite{Berkovits:2008ic}.  From a purely gauge-theoretic point of view, this means that the Wilson loop lives in a dual affine Minkowski space, on which the dual superconformal group acts.  Differences between points in this space correspond to momenta, and the momentum conservation condition is automatically encoded by the fact that the null polygon $C$ is closed.

Since the original conjecture of Alday and Maldacena, a wide variety of studies have been performed at both strong and weak coupling which indicate that the duality should be true (c.f., \cite{Drummond:2007aua, Brandhuber:2007yx, Drummond:2007cf, Drummond:2007bm, Drummond:2008aq, Gorsky:2009dr}).  For the fully supersymmetric Wilson loop of $\cN=4$ SYM \eqref{sWilsonloop}, performing explicit computations can be rather complicated due to the form of the superconnection $\CA$ (see Appendix \ref{Appendix1}), so proving general statements was difficult.  

Translating the supersymmetric Wilson loop to twistor space has provided an efficient means of checking the amplitude / Wilson loop duality for arbitrary MHV degree and loop order at the level of the integrand (for both the Wilson loop and scattering amplitudes) \cite{Mason:2010yk}.  Furthermore, it has been shown that the twistor Wilson loop has the same singularity structure as scattering amplitudes; this essentially constitutes a twistor-theoretic proof of the original conjecture at the level of the integrand \cite{Bullimore:2011ni}.

However, the duality between Wilson loops and other gauge-theoretic objects does not stop at scattering amplitudes.  In this section, we study some conjectured correspondences between null polygonal Wilson loops and correlation functions of local operators.  By working with these objects on twistor space, we not only obtain analytic proofs, but can also derive efficient calculational mechanisms, just as in the case of scattering amplitudes.


\subsection{Local Operators and Wilson Loops in Twistor Space}

Gauge invariant local operators in $\cN=4$ SYM include Konishi, dilaton, or indeed any chiral primary operators.  In this review, we restrict our attention to the `1/2-BPS' operators; these have a non-anomalous conformal dimension and do not require renormalization \cite{Alday:2010zy}.  Later, we will be working at the level of the loop integrand; although the integrand of a correlation function is simply a rational function for \emph{any} choice of local operators, protected operators such as the 1/2-BPS operators allow us to more plausibly extend our claims to the full loop \emph{integral}.

These 1/2-BPS operators are built from pairs of scalars:
\be{BPS}
\cO(x)=\cO_{abcd}(x)=\tr(\Phi_{ab}(x)\Phi_{cd}(x))-\frac{\epsilon_{abcd}}{12}\tr(\Phi^{2}(x)).
\ee
For an abelian gauge group, it is easy to see how to express $\cO$ in twistor space using the Penrose transform:
\begin{multline*}
\cO^{\U(1)}(x)=\int_{X\times X}\D\lambda\wedge\D\lambda'\wedge\phi_{ab}(\lambda)\wedge\phi_{cd}(\lambda') \\
-\frac{\epsilon_{abcd}}{12}\int_{X\times X}\D\lambda\wedge\D\lambda'\wedge\phi^{ef}(\lambda)\wedge\phi_{ef}(\lambda'),
\end{multline*}
where $\phi_{ab}(\lambda)$ denotes the pullback of $\phi_{ab}$ to the line $X$ charted by $\lambda$.  This can be naturally generalized to $\cN=4$ supersymmetry by using $\frac{\partial^{2}\cA}{\partial\chi^{2}}$ instead of $\phi_{ab}$:
\begin{multline}\label{abelian}
\cO^{\U(1)}(x,\theta)=\int_{X\times X}\D\lambda\wedge\D\lambda'\wedge\frac{\partial^{2}\cA}{\partial\chi^{a}\partial\chi^{b}}(\lambda)\wedge\frac{\partial^{2}\cA}{\partial\chi^{c}\partial\chi^{d}}(\lambda') \\
-\frac{\epsilon_{abcd}}{12}\int_{X\times X}\D\lambda\wedge\D\lambda'\wedge\frac{\partial^{2}\cA}{\partial\chi_{e}\partial\chi_{f}}(\lambda)\wedge\frac{\partial^{2}\cA}{\partial\chi^{e}\partial\chi^{f}}(\lambda'),
\end{multline}
where
\begin{equation*}
\frac{\partial^{2}\cA}{\partial\chi^{a}\partial\chi^{b}}=\phi_{ab}+\epsilon_{abcd}\chi^{c}\psi^{d}+\frac{1}{2!}\epsilon_{abcd}\chi^{c}\chi^{d}g.
\end{equation*}

Of course, for a non-abelian gauge group, the twistorial operator \eqref{abelian} is not well-defined: we cannot integrate $\phi_{ab}$ over $X$ because it takes values in the the Lie algebra of the gauge group.  What we need is a frame for $E\rightarrow\PT$ which provides a holomorphic trivialization of $E|_{X}$.  Now, $E|_{X}$ is holomorphic (because $X$ has only one complex dimension) and topologically trivial by assumption (i.e., $c_{1}(E)=0$), so all that is required is a gauge transformation $\gamma$ which obeys:
\begin{equation*}
\gamma(\dbar+\cA)|_{X}\gamma^{-1}=\dbar|_{X}.
\end{equation*}
As it turns out, such a $\gamma$ can be found generically.  Since $X$ is rational and $E|_{X}$ is topologically trivial and holomorphic, the Birkhoff-Grothendieck theorem tells us that
\begin{equation*}
E|_{X}\cong\bigoplus^{r}_{i=1}\cO(a_{i}), \qquad \sum_{i=1}^{r}a_{i}=0,
\end{equation*} 
where $r=\mathrm{rank} E$.  When $\cA=0$, all of the $a_{i}=0$ and we are just working on the trivial bundle $\cO^{\oplus r}$.  As we work perturbatively around this trivial background, the holomorphic trivialization will continue to hold generically provided $\cA$ is sufficiently small.  Since $X$ is linearly embedded, the holomorphic trivialization given by $\gamma$ is unique.

\begin{figure}
\centering
\includegraphics[width=1.7 in, height=1 in]{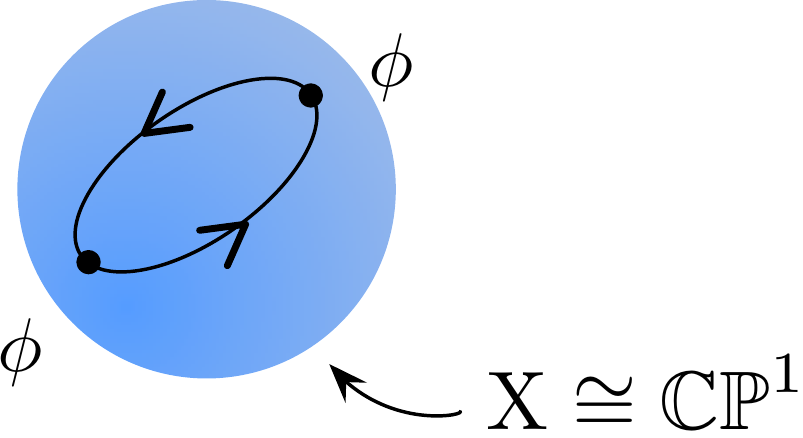}\caption{\textit{The twistor space form of the local space-time operator} $\tr\Phi^2(x)$, \textit{involving holomorphic Wilson lines; arrows indicate the flow of the color trace.}}\label{operator}
\end{figure}

If we define
\be{frame}
U_{X}(\lambda,\lambda')=\gamma(x,\lambda)\gamma^{-1}(x,\lambda'),
\ee
then $U_{X}$ is formally a Green's function for $(\dbar+\cA)|_{X}$, and acts as
\begin{equation*}
U_{X}(\lambda,\lambda)=\mathbb{I}, \qquad U_{X}(\lambda,\lambda'): E|_{\lambda'}\rightarrow E|_{\lambda}.
\end{equation*}
Thus, it is natural to interpret $U_{X}$ as the twistor space parallel propagator for the gauge bundle $E$ along $X$.  This allows us to write down an immediate non-abelian generalization of \eqref{abelian} for our 1/2-BPS operators:
\begin{multline}\label{nabelian}
\cO(x,\theta)=\int_{X\times X}\D\lambda\; \D\lambda'\; \tr\left[U_{X}(\lambda,\lambda')\frac{\partial^{2}\cA(\lambda')}{\partial\chi^{a}\partial\chi^{b}}U_{X}(\lambda',\lambda)\frac{\partial^{2}\cA(\lambda)}{\partial\chi^{c}\partial{\chi}^{d}}\right] \\
-\frac{\epsilon_{abcd}}{12}\int_{X\times X}\D\lambda\; \D\lambda'\; \tr\left[U_{X}(\lambda,\lambda')\frac{\partial^{2}\cA(\lambda')}{\partial\chi_{e}\partial{\chi}_{f}}U_{X}(\lambda',\lambda)\frac{\partial^{2}\cA(\lambda)}{\partial\chi^{e}\partial{\chi}^{f}}\right].
\end{multline}
The bosonic portion of this operator is depicted in Figure \ref{operator}.  The following lemma ensures us that this supersymmetric operator is well-defined.

\begin{lemma}
$\cO(x,\theta)$ is a well-defined, gauge invariant operator on $\PT$, and corresponds to the chiral half of the 1/2-BPS multiplet of $\cN=4$ SYM.
\end{lemma}
\proof  By \eqref{frame}, $U_{X}$ is the unique solution of
\begin{equation}\label{gfunct}
(\dbar+\cA)|_{X}U_{X}=0.
\end{equation}
Since $\cA$ depends on $\theta$ only through $\chi^{a}=\theta^{aA}\lambda_{A}$, we can differentiate with respect to $\theta$ to get:
\begin{equation*}
(\dbar+\cA)|_{X}(\lambda^{A}\partial_{aA}U_{X})=0.
\end{equation*}
This implies that $U^{-1}_{X}(\lambda^{A}\partial_{aA}U_{X})$ is globally holomorphic on $X\cong\P^{1}$; by Liouville's theorem, we then have
\begin{equation}\label{gfunct2}
U^{-1}_{X}(\lambda^{A}\partial_{aA}U_{X})=\lambda^{A}\Gamma_{aA}(x,\theta).
\end{equation}
One can show that $\nabla_{aA}=\partial_{aA}+\Gamma_{aA}$ transforms as a connection and obeys the condition
\begin{equation*}
\left\{\nabla_{a(A},\nabla_{B)b}\right\}=0.
\end{equation*}
By lemma \ref{lemma: odd}, this means that $\Gamma_{aA}$ is the odd superconnection of $\cN=4$ SYM, with curvature given by $\cF_{ab}$.  Using the fact that $U_{X}(\lambda,\lambda)=\mathbb{I}$, it follows that $\cF_{ab}=\partial^{A}_{[a}\Gamma_{b]A}$.

By \eqref{gfunct}, we have that
\begin{equation*}
\int_{X}\frac{\la\lambda'' \lambda'\ra\;\D\lambda}{\la\lambda'' \lambda\ra \la\lambda \lambda' \ra}U_{X}(\lambda'',\lambda)(\dbar+\cA)|_{X}U_{X}(\lambda,\lambda')=0.
\end{equation*}
Noting that
\begin{multline*}
\int_{X}\frac{\la\lambda'' \lambda'\ra\;\D\lambda}{\la\lambda'' \lambda\ra \la\lambda \lambda' \ra}U_{X}(\lambda'',\lambda)\dbar U_{X}(\lambda,\lambda') \\
=-\int_{X}\dbar\left(\frac{\la\lambda'' \lambda'\ra\;\D\lambda}{\la\lambda'' \lambda\ra \la\lambda \lambda' \ra}\right)U_{X}(\lambda'',\lambda)U_{X}(\lambda,\lambda')=-U_{X}(\lambda'',\lambda'),
\end{multline*}
we can differentiate with respect to $\theta$ to obtain
\begin{equation*}
\frac{\partial U_{X}(\lambda'',\lambda')}{\partial\theta^{Aa}}=\int_{X}\frac{\la\lambda'' \lambda'\ra\;\D\lambda}{\la\lambda'' \lambda\ra \la\lambda \lambda' \ra}U_{X}(\lambda'',\lambda)\lambda_{A}\frac{\partial\cA(\lambda)}{\partial\chi^{a}}U_{X}(\lambda,\lambda').
\end{equation*}

From \eqref{gfunct2}, we have 
\begin{equation*}
\Gamma_{aA}=\frac{1}{\la\lambda'' \lambda'\ra}U_{X}^{-1}(\lambda'',\lambda')\lambda^{''A}\frac{\partial U_{X}(\lambda'',\lambda')}{\partial\theta^{Aa}} =\int_{X}\frac{\D\lambda}{\la\lambda \lambda' \ra}U_{X}(\lambda',\lambda)\frac{\partial\cA(\lambda)}{\partial\chi^{a}}U_{X}(\lambda,\lambda').
\end{equation*}
This indicates that the fermionic curvature is given by
\begin{equation*}
\cF_{ab}=\partial^{A}_{[a}\Gamma_{b]A}=-\int_{X}\D\lambda\;U_{X}(\lambda',\lambda)\frac{\partial^{2}\cA(\lambda)}{\partial\chi^{a}\partial\chi^{b}}U_{X}(\lambda,\lambda'),
\end{equation*}
and hence that
\begin{equation*}
\cO(x,\theta)=\tr\left(\cF_{ab}\cF_{cd}\right)-\frac{\epsilon_{abcd}}{12}\tr\left(\cF^{2}\right).
\end{equation*}
Since $\cF_{ab}$ is a curvature of the $\cN=4$ superconnection, the operator $\cO(x,\theta)$ is manifestly gauge invariant on $\PT$.

Finally, we can use the results of Appendix \ref{Appendix1} to expand the operator in powers of $\theta$:
\begin{equation*}
\cO(x,\theta)=\cO(x)+3\theta^{eA}\tr\left(\Phi_{ab}\Psi_{cdeA}\right)-\theta^{gA}\frac{\epsilon_{abcd}}{4}\tr\left(\bar{\Phi}^{ef}\Psi_{efgA}\right)+O(\theta^{2}),
\end{equation*}
which is equivalent to the 1/2-BPS supermultiplet with $\bar{\theta}=0$, as desired.     $\Box$.

\medskip

This procedure can be duplicated for practically any choice of local operator in $\cN=4$ SYM, including those which are not protected (see \cite{Adamo:2011dq} for the Konishi operator).  

It turns out that $U_{X}$ also provides the definition of a null polygonal Wilson loop in twistor space \cite{Mason:2010yk, Bullimore:2011ni}.  Recall that for $\cN=4$ SYM, the fully supersymmetric Wilson loop $W[C]$ is given by \eqref{sWilsonloop}: the trace of the holonomy of $\CA$ around a null polygon $C$ in $\M$ with $n$ cusps labelled by $(x_{i},\theta_{i})$ (fixing $R$ to be the fundamental representation for now).  The basic twistor correspondence tells us that each cusp $x_{i}$ is equivalent to a line $X_{i}\cong\P^{1}$ in $\PT$; since these cusps are pairwise null-separated, $X_{i}$ intersects $X_{i-1}$ and $X_{i+1}$ in points $Z_{i-1}$ and $Z_{i}$ respectively.  This translates the space-time null polygon into a nodal elliptic curve in twistor space.  Likewise, the space-time superconnection $\CA$ is translated into the $(0,1)$-connection $\cA$ on $E\rightarrow\PT$.  To compute the Wilson loop, we simply need to parallel transport $\cA$ around the nodal elliptic curve using $U_{X}$.  

This leads us to the definition:
\be{WL}
W[C]=\tr\;\mathrm{Hol}_{Z_{n}}[C]=\tr\left[ U_{X_n}(Z_{n},Z_{n-1})U_{X_{n-1}}(Z_{n-1},Z_{n-2})\cdots U_{X_1}(Z_{1},Z_{n})\right],
\ee
where we abuse notation by writing $C$ for both the space-time nully polygon and the twistor nodal elliptic curve.  We will also abbreviate coordinates in $\M$ by their bosonic part: $(x,\theta)$ will be written $x$.  It has now been confirmed that \eqref{WL} coincides with the supersymmetric space-time Wilson loop of Caron-Huot \cite{CaronHuot:2010ek} up to terms proportional to the equations of motion \cite{Belitsky:2011zm}.

While this is clearly a well-defined and gauge invariant object on twistor space, it can also be seen to be equivalent to the na\"ive holomorphic generalization of a \emph{real} Wilson loop given by \eqref{sWilsonloop}.  Indeed, since the holomorphic frame is a solution to \eqref{gfunct}, we can formally expand it as a Born series:
\begin{multline}
U_{X}(Z,Z')=\frac{1}{1+\dbar^{-1}|_{X}\cA}=\mathbb{I}+\sum_{k=1}^{\infty}(-1)^{k}\int_{X^{k}}\bigwedge_{i=1}^{k}\omega_{Z,Z'}(Z_{i})\wedge\cA(Z_{i}) \\
\equiv \cP\exp\left(-\int_{X}\omega_{Z,Z'}\wedge\cA\right),
\end{multline}
where $\omega_{Z,Z'}$ is a meromorphic 1-form on $X$ with simple poles at $Z$, $Z'$ and
\begin{equation*}
\bigwedge_{i=1}^{k}\omega_{Z,Z'}(Z_{i})=\frac{\la\lambda \lambda'\ra}{\la\lambda\lambda_{1}\ra\la\lambda_{1}\lambda_{2}\ra\cdots\la\lambda_{k}\lambda'\ra}\frac{\D\lambda_{1}}{2\pi i}\wedge\cdots\wedge\frac{\D\lambda_{k}}{2\pi i}.
\end{equation*}
The concatenation of these frames about the nodal curve $C$ then gives the rather aesthetically appealing identification:
\be{WL2}
W[C]=\tr\;\mathrm{Hol}_{Z_{n}}[C]=\tr\;\cP\exp\left(-\int_{C}\omega\wedge\cA\right),
\ee
where $\omega$ is now the meromorphic 1-form on $C$ with simple poles at each node $Z_{i}$.


\subsection{Correlation Function / Wilson Loop Correspondence}

The duality between scattering amplitudes and supersymmetric null polygonal Wilson loops received its first analytic proof using the twistor Wilson loop (at the level of the loop integrand) \cite{Bullimore:2011ni}, and this picture allows the computation of the all-loop integrand for scattering amplitudes in a remarkably efficient fashion \cite{Mason:2010yk} (see \cite{Adamo:2011pv} for a review emphasizing the role of the twistor action in these developments).  The correspondence between null polygonal Wilson loops and other physically interesting observables doesn't stop here though.

In \cite{Eden:2010zz, Alday:2010zy, Eden:2010ce} it was conjectured that, in the limit where the insertion points become pairwise null separated, the ratio of certain $n$-point correlation functions in $\cN=4$ SYM is equal to the expectation value of a null polygonal Wilson loop in the adjoint representation.  More formally, if $\{\cO(x_{i})\}_{i=1,\ldots,n}$ are gauge invariant local operators in $\cN=4$ SYM, then this conjecture takes the form:
\be{corrW}
\lim_{x_{i,i+1}^{2}\rightarrow 0}\frac{\la \cO(x_{1})\ldots\cO(x_{n})\ra}{\la\cO(x_{1})\ldots\cO(x_{n})\ra^{\mathrm{tree}}}=\la W_{\mathrm{adj}}[C]\ra \xrightarrow{\mathrm{planar}\:\mathrm{limit}} \la W[C]\ra^{2},
\ee
where $x^{2}_{i,j}=(x_{i}-x_{j})^{2}$, $C$ is the resulting null polygon with $n$ cusps, $W_{\mathrm{adj}}$ is the Wilson loop in the adjoint representation, and $W$ the Wilson loop in the fundamental representation.  Difficult calculations in perturbation theory on space-time have confirmed this conjecture through examples \cite{Eden:2011yp, Eden:2011ku}, and it is was also expected to hold at the level of the loop integrand \cite{Eden:2010ce}.\footnote{The loop integrand indicates allowing all possible Feynman diagrams at the given order in perturbation theory, but we do not perform the integrals over the locations of Lagrangian insertions corresponding to the loop variables.  For the Wilson loop, the integrand is always well-defined, whereas for scattering amplitudes it is well-defined only in the planar limit \cite{ArkaniHamed:2010kv}.}  

Beyond its intrinsic interest as a conjecture about two interesting classes of observables in gauge theory, \eqref{corrW} also has practical implications.  In \cite{Belitsky:2011zm}, it was conjectured that the correlation function / Wilson loop correspondence should actually be \emph{more} robust than the amplitudes / Wilson loop duality.  Indeed, carefully considering the null limit on the left-hand side of \eqref{corrW} can be thought of as providing a regularization mechanism for the Wilson loop on the right-hand side \cite{Alday:2010zy}.  In some sense, this means that the proper strong-coupling tool for studying scattering amplitudes is actually the null limit of correlation functions, since these objects are well-behaved (i.e., lack the cusp divergences of the Wilson loop that is approached in the limit).

We want to evaluate
\be{CW1}
\lim_{x_{i,i+1}^{2}\rightarrow 0}\frac{\la \cO(x_{1})\ldots\cO(x_{n})\ra}{\la\cO(x_{1})\ldots\cO(x_{n})\ra^{\mathrm{tree}}}
\ee
using our twistorial local operators \eqref{nabelian} with respect to the $\cN=4$ SYM twistor action.  It is well known that the tree level contribution in the denominator goes as
\begin{equation*}
\la \cO(x_{1})\cdots\cO(x_{n})\ra^{\mathrm{tree}} \sim \frac{1}{x_{12}^{2}x_{23}^{2}\cdots x_{n1}^{2}},
\end{equation*}
so we can neglect any contribution from the numerator which does not have a counterbalancing divergence in the null limit.

\begin{figure}[t]
\centering
\includegraphics[width=100mm]{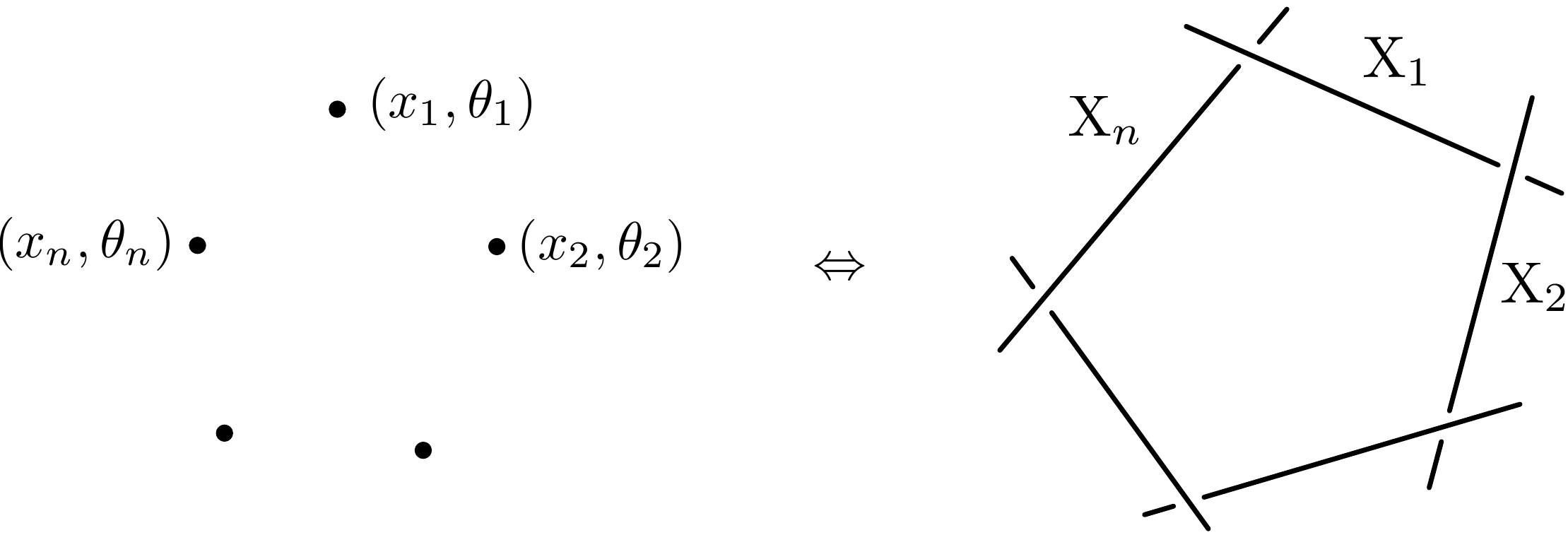}\caption{\textit{As the $n$ generic points $(x,\theta)$ become null separated in $\M$, the corresponding $n$ $\P^1$s in $\PT$ intersect to form the nodal elliptic curve $C$.}}\label{polygons}
\end{figure}
In twistor space, the geometry of the null limit is elegantly manifested (see Figure \ref{polygons}).  As the null polygon is approached in space-time, the $n$ lines $\{X_{i}\}$ intersect sequentially to form the nodal elliptic curve $C\subset\PT$.  To evaluate the numerator of \eqref{CW1}, we apply Wick's theorem to the twistorial path integral
\begin{equation*}
\int [\cD\cA] \cO(x_{1})\cdots\cO(x_{n})\;e^{-S[\cA]},
\end{equation*}
under the assumptions of \emph{normal ordering} and \emph{genericity}.  The normal ordering assumption means that we can exclude any contractions between fields or frames inserted on the same lines in twistor space; the genericity assumption means that the MHV vertices generated by the second term $S_{2}[\cA]$ in the twistor action \eqref{TAInt} are not null separated from any of the operator insertions.  The latter condition is simply that the lines appearing in the perturbative expansion \eqref{detexp} do not intersect any of the lines where an operator insertion lives.

Contractions will occur between insertions of the twistor $(0,1)$-connection $\cA$.  These appear in operator insertions $\frac{\partial^{2}\cA}{\partial\chi^{2}}$, the perturbative expansion of the holomorphic frames $U_{X}$, and in the MHV vertex insertions from $S_{2}[\cA]$.  We will thus have three classes of contractions to consider:
\begin{itemize}
\item Contractions involving a MHV vertex.

\item Contractions between non-adjacent $X_{i}$s.

\item Contractions between adjacent $X_{i}$s (i.e., between fields on $X_{i}$ and $X_{i\pm 1}$).
\end{itemize}

We are free to choose a gauge on twistor space in which to perform these calculations; following the lessons of Section \ref{Chapter3}, let us fix CSW gauge.  Then the twistor space propagator is given by \eqref{TAprop}.  Lines in $\PT$ can be parametrized by $Z(s_{i})=Z_{A_i}+s_{i}Z_{B_i}$, with $s_{i}$ acting as an inhomogeneous coordinate on $X_{i}$.  The measure in homogeneous coordinates $\D\lambda_{i}$ is then written as $\la A_{i} B_{i}\ra \d s_{i}$.  Without loss of generality, we can choose $Z_{A_i}$ and $Z_{B_i}$ to be the intersection points that $X_{i}$ develops with $X_{i-1}$ and $X_{i+1}$ respectively in the null limit (i.e., as $x_{i,i+1}^{2}\rightarrow 0$, $Z_{B_{i}}\rightarrow Z_{A_{i+1}}$).  Similarly, we parametrize a line $X$ corresponding to an arbitrary MHV vertex from the twistor action as $Z(t)=Z_{A}+t Z_{B}$.  Without loss of generality, we assume that the fixed CSW reference twistor $Z_{*}$ has no fermionic part (i.e., $\chi_{*}=0$).

\subsubsection*{\textit{Contractions involving an MHV vertex}}

Consider an arbitrary MHV vertex supported on $X$ and the operators and frames supported on any of the $X_{i}$.  The contraction between a field $\cA$ in a holomorphic frame $U_{X_i}$ and a field $\cA$ on $X$ is given by 
\be{C1}
\left\la \overbrace{\cA|_{X_{i}}\cA|_{X}} \right\ra=\int_{\C^{2}} \frac{\d s_{i}}{s_{i}}\frac{\d t}{t} \Delta(Z(s_{i}), Z(t))=[A_{i},B_{i},*,A,B].
\ee
Genericity means that (even in the null limit) $X_{i}\cap X=\emptyset$, so contractions of the type \eqref{C1} are always finite by the definition of the R-invariant.  Additionally, we can have a contraction between an insertion of $\frac{\partial^{2}\cA}{\partial\chi^{2}}$ on $X_{i}$ and a field $\cA$ on $X$, which leads to
\begin{multline}\label{C2}
\left\la \overbrace{\frac{\partial^{2}\cA}{\partial\chi^{a}\partial\chi^{b}}|_{X_{i}}\cA|_{X}} \right\ra = \frac{\partial^{2}}{\partial\chi_{A_i}^{a}\partial\chi_{B_{i}}^{b}}[A_{i},B_{i},*,A,B] \\
=\frac{\delta^{0|2}_{ab}\left(\chi_{A_i}(B_{i}*A B) +\chi_{B_{i}}(*ABA_{i})+\chi_{A}(BA_{i}B_{i}*)+\chi_{B}(A_{i}B_{i}*A)\right)}{(A_{i}B_{i}AB)(BA_{i}B_{i}*)(A_{i}B_{i}*A)}.
\end{multline}
This is also finite thanks to genericity, so we can neglect all contributions to \eqref{CW1} due to contractions between an operator and MHV vertices.

\subsubsection*{\textit{Contractions between non-adjacent $X_{i}$s}}

Contractions between non-adjacent operator insertions on $X_{i}$ and $X_{j}$ (for $j\neq i+1,\; i-1$) follow in a similar fashion to \eqref{C1} and \eqref{C2}.  In this class, we have potential contributions from contractions between frames, between a frame and an insertion of $\frac{\partial^{2}\cA}{\partial\chi^{2}}$, or between two insertions of $\frac{\partial^{2}\cA}{\partial\chi^{2}}$.  Short calculations show that these are given by:
\be{C3}
\left\la \overbrace{\cA|_{X_{i}}\cA|_{X_{j}}} \right\ra=\int_{\C^{2}} \frac{\d s_{i}}{s_{i}}\frac{\d s_{j}}{s_j} \Delta(Z(s_{i}), Z(s_j))=[A_{i},B_{i},*,A_{j},B_{j}],
\ee
\begin{multline}\label{C4}
\left\la \overbrace{\frac{\partial^{2}\cA}{\partial\chi^{a}\partial\chi^{b}}|_{X_{i}}\cA|_{X_{j}}} \right\ra = \frac{\partial^{2}}{\partial\chi_{A_i}^{a}\partial\chi_{B_{i}}^{b}}[A_{i},B_{i},*,A_{j},B_{j}] \\
=\frac{\delta^{0|2}_{ab}\left(\chi_{A_i}(B_{i}*A_{j}B_{j}) +\chi_{B_{i}}(*A_{j}B_{j}A_{i})+\chi_{A_{j}}(B_{j}A_{i}B_{i}*)+\chi_{B_{j}}(A_{i}B_{i}*A_{j})\right)}{(A_{i}B_{i}A_{j}B_{j})(B_{j}A_{i}B_{i}*)(A_{i}B_{i}*A_{j})},
\end{multline}
\begin{multline}\label{C5}
\left\la \overbrace{\frac{\partial^{2}\cA}{\partial\chi^{a}\partial\chi^{b}}|_{X_{i}}\frac{\partial^{2}\cA}{\partial\chi^{c}\partial\chi^{d}}|_{X_{j}}} \right\ra =\frac{\partial^{4}}{\partial\chi^{a}_{A_{i}}\partial\chi^{b}_{B_{i}}\partial\chi^{c}_{A_{j}}\partial\chi^{d}_{B_{j}}}[A_{i},B_{i},*,A_{j},B_{j}] \\
=\frac{\epsilon_{abcd}}{(A_{i}B_{i}A_{j}B_{j})},
\end{multline}
respectively.  Because the $X_{i}$ only become \emph{pairwise} null separated in the limit, $X_{i}\cap X_{j}=\emptyset$ and all three of \eqref{C3}-\eqref{C5} are finite.  So no contractions from this class contribute to the ratio \eqref{CW1} in the null limit.

\subsubsection*{\textit{Contractions between adjacent $X_{i}$s}}

Finally, we must consider contractions between operator insertions on $X_{i}$ and $X_{i+1}$.  In this case, we need to be careful because the two lines will intersect in the null limit, and a regularization procedure is needed to isolate the behavior of the Wick contractions as the limit is approached.  The simplest mechanism is given by a framing procedure: take two copies of the singular configuration in twistor space separated by a small parameter, and then consider the limit as this parameter is taken to zero.  While this is not gauge invariant on space-time, we work at the level of the loop integrand in twistor space and the framing regulator is perfectly well-defined at the level of this rational function.

More precisely, as $x^{2}_{i,i+1}\rightarrow 0$, we assume that $Z_{A_{i+1}}=Z_{B_i}+\varepsilon Z$ for some twistor $Z$ and $\varepsilon$ our small parameter.  In this scheme, the numerator of the R-invariant becomes:
\begin{multline*}
\delta^{0|4}\left(\varepsilon\chi_{A_{i}}(B_{i}* Z B_{i+1})+\;\mathrm{cyclic}\right)=\prod_{a=1}^{4}\left[\varepsilon\chi^{a}_{A_{i}}(B_{i}* Z B_{i+1})+ \chi^{a}_{B_i}(Z B_{i}B_{i+1}A_{i}) \right.\\
 \left. +\varepsilon\chi^{a}_{B_i}(* Z B_{i+1}A_{i})+\chi^{a}_{B_i}(B_{i+1}A_{i}B_{i}*)+\varepsilon\chi^{a}_{B_{i+1}}(A_{i}B_{i}* Z)\right] \sim O(\varepsilon^{4}),
\end{multline*}
while the denominator behaves as:
\begin{multline*}
\varepsilon^{4}(B_{i+1}A_{i}B_{i}* )(A_{i}B_{i}* Z)(B_{i}* Z B_{i+1})(* Z B_{i+1} A_{i})(Z B_{i+1} A_{i} B_{i}) \\
+ \varepsilon^{3}(B_{i+1}A_{i}B_{i}*)(A_{i}B_{i}* Z)(B_{i}* Z B_{i+1})(* B_{i}B_{i+1}A_{i})(Z B_{i+1}A_{i}B_{i}).
\end{multline*}
Let us apply this to contractions between the frames $U_{X_i}$ and $U_{X_{i+1}}$:
\begin{equation*}
\left\la \overbrace{\cA|_{X_{i}}\cA|_{X_{i+1}}} \right\ra=\int_{\C^{2}} \frac{\d s_{i}}{s_{i}}\frac{\d s_{i+1}}{s_{i+1}} \Delta(Z(s_{i}), Z(s_{i+1}))=[A_{i},B_{i},*,A_{i+1},B_{i+1}].
\end{equation*}
The null limit gives
\be{C6}
\lim_{x^{2}_{i,i+1}\rightarrow 0}\left\la \overbrace{\cA|_{X_{i}}\cA|_{X_{i+1}}} \right\ra= \lim_{\varepsilon\rightarrow 0}[A_{i},B_{i},*, (B_{i}+\varepsilon Z), B_{i+1}] \sim \lim_{\varepsilon\rightarrow 0} \frac{\varepsilon^{4}}{\varepsilon^{4}+\varepsilon^{3}} =0,
\ee
as a consequence of $\cN=4$ supersymmetry in the numerator of the R-invariant.  Similarly, the contraction between an insertion of $\frac{\partial^{2}\cA}{\partial\chi^{2}}$ on $X_{i}$ and a field $\cA$ in $U_{X_{i+1}}$ gives
\begin{multline*}
\left\la \overbrace{\frac{\partial^{2}\cA}{\partial\chi^{a}\partial\chi^{b}}|_{X_{i}}\cA|_{X_{i+1}}} \right\ra = \frac{\partial^{2}}{\partial\chi_{A_i}^{a}\partial\chi_{B_{i}}^{b}}[A_{i},B_{i},*,A_{i+1},B_{i+1}] = \\
\frac{\delta^{0|2}_{ab}\left(\chi_{A_i}(B_{i}*A_{i+1}B_{i+1}) +\chi_{B_{i}}(*A_{i+1}B_{i+1}A_{i})+\chi_{A_{i+1}}(B_{i+1}A_{i}B_{i}*)+\chi_{B_{i+1}}(A_{i}B_{i}*A_{i+1})\right)}{(A_{i}B_{i}A_{i+1}B_{i+1})(B_{i+1}A_{i}B_{i}*)(A_{i}B_{i}*A_{i+1})},
\end{multline*}
which is finite upon passing to the null limit:
\be{C7}
\lim_{x^{2}_{i,i+1}\rightarrow 0} \left\la \overbrace{\frac{\partial^{2}\cA}{\partial\chi^{a}\partial\chi^{b}}|_{X_{i}}\cA|_{X_{i+1}}} \right\ra \sim \lim_{\varepsilon\rightarrow 0}\frac{\varepsilon^{2}}{\varepsilon^{2}} =1.
\ee
Hence, these contributions also contribute nothing to the overall ratio \eqref{CW1} in the null limit.

Finally, we must consider the contraction between insertions of $\frac{\partial^{2}\cA}{\partial\chi^{2}}$ on each of $X_{i}$ and $X_{i+1}$.  From \eqref{C5}, it is easy to see that
\begin{multline*}
\left\la \overbrace{\frac{\partial^{2}\cA}{\partial\chi^{a}\partial\chi^{b}}|_{X_{i}}\frac{\partial^{2}\cA}{\partial\chi^{c}\partial\chi^{d}}|_{X_{i+1}}} \right\ra =\frac{\partial^{4}}{\partial\chi^{a}_{A_{i}}\partial\chi^{b}_{B_{i}}\partial\chi^{c}_{A_{i+1}}\partial\chi^{d}_{B_{i+1}}}[A_{i},B_{i},*,A_{i+1},B_{i+1}] \\
=\frac{\epsilon_{abcd}}{(A_{i}B_{i}A_{i+1}B_{i+1})}.
\end{multline*}
Rather than regulate this contraction, note that its behavior is evident after integrating the contraction over the respective operator insertion sites:
\begin{multline}\label{C8}
\int\limits_{X_{i}\times X_{i+1}}\D\lambda_{i}\;\D\lambda_{i+1} \left\la \overbrace{\frac{\partial^{2}\cA}{\partial\chi^{a}\partial\chi^{b}}|_{X_{i}}\frac{\partial^{2}\cA}{\partial\chi^{c}\partial\chi^{d}}|_{X_{i+1}}} \right\ra\\
=\la A_{i}B_{i}\ra \la A_{i+1}B_{i+1}\ra \frac{\partial^{4}}{\partial\chi^{a}_{A_{i}}\partial\chi^{b}_{B_{i}}\partial\chi^{c}_{A_{i+1}}\partial\chi^{d}_{B_{i+1}}} \int_{\C^{2}}\frac{\d s_{i}}{s_{i}}\frac{\d s_{i+1}}{s_{i+1}} \Delta(Z(s_{i}),Z(s_{i+1})) \\
=\epsilon_{abcd}\frac{\la A_{i}B_{i}\ra \la A_{i+1}B_{i+1}\ra}{(A_{i}B_{i}A_{i+1}B_{i+1})}=\frac{\epsilon_{abcd}}{(x_{i}-x_{i+1})^{2}}.
\end{multline}
Such a contraction thus diverges as $x^{2}_{i,i+1}\rightarrow 0$ in the null limit, precisely the singular behavior needed to counterbalance the tree-level denominator of \eqref{CW1}.  Hence, \eqref{C8} are the \emph{only} contractions which survive in the null limit. 

\medskip

\begin{figure}[tp]
\centering   
\includegraphics[width=150mm]{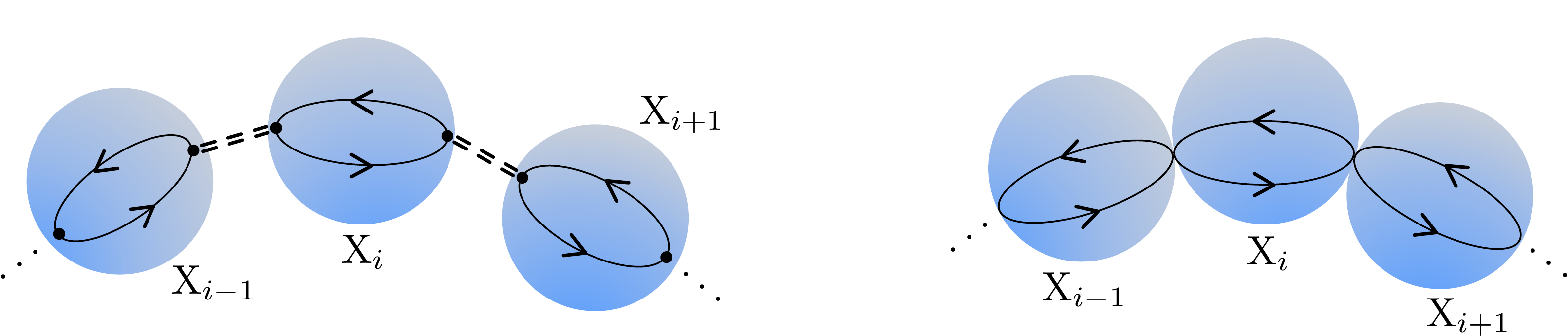}\caption{\textit{The surviving contributions in the null limit form a doubled trace of the holonomy around $C$ corresponding to a Wilson loop in the adjoint representation.}}\label{limit}
\end{figure}

So in the numerator of \eqref{CW1}, only those contributions which contract all adjacent insertions of $\frac{\partial^{2}\cA}{\partial\chi^{2}}$ survive.  These cancel the tree-level denominator and leave two holomorphic frames $U_{X}$ on each of the $X_{i}$.  In the null limit, the trace around these intersecting lines yields the integrand of the twistor Wilson loop in the adjoint representation; see Figure \ref{limit}.  Hence, we have shown
\be{CW2}
\lim_{x_{i,i+1}^{2}\rightarrow 0}\frac{\la \cO(x_{1})\ldots\cO(x_{n})\ra}{\la\cO(x_{1})\ldots\cO(x_{n})\ra^{\mathrm{tree}}}=\left\la W_{\mathrm{adj}}[C]\right\ra,
\ee
at the level of the integrand and for gauge group $\SU(N)$ or $\U(N)$.

In the planar limit (i.e., $N\rightarrow\infty$), the twistor propagator suppresses any mixing between fundamental and anti-fundamental representations thanks to the color structure \eqref{TAcolorprop}:
\begin{equation*}
\Delta(Z_{1},Z_{2})^{i k}_{j l}=\bar{\delta}^{2|4}(Z_{1},\rf, Z_{2})\left(\delta^{i}_{l}\delta^{k}_{j}-\frac{1}{N}\delta^{i}_{j}\delta^{k}_{l}\right).
\end{equation*}
Hence, we can decompose the adjoint representation into the product of fundamental and anti-fundamental representations, and at the level of the Wilson loop, we have:
\begin{equation*}
\left\la W_{\mathrm{adj}}[C]\right\ra = \left\la W[C]\;\widetilde{W}[C]\right\ra = \left\la W[C]\right\ra^{2}.
\end{equation*}

This means we have proven the supersymmetric correlation function / Wilson loop correspondence \eqref{corrW}, as first reported in \cite{Adamo:2011dq}.  More formally, our result is:
\begin{propn}\label{corrWL}
Let $\{\cO(x_{i})\}_{i=1,\ldots, n}$ be gauge invariant local operators in $\cN=4$ SYM and $C$ be the null polygon resulting from the limit where these operators become pairwise null separated (i.e., $x_{i,i+1}^{2}=0$).  Then at the level of the integrand,
\be{corrW2}
\lim_{x_{i,i+1}^{2}\rightarrow 0}\frac{\la \cO(x_{1})\ldots\cO(x_{n})\ra}{\la\cO(x_{1})\ldots\cO(x_{n})\ra^{\mathrm{tree}}}=\la W_{\mathrm{adj}}[C]\ra \xrightarrow{\mathrm{planar}\:\mathrm{limit}} \la W[C]\ra^{2},
\ee
where all expectation values are assumed to be generic and normal ordered, and $W[C]$ is the Wilson loop in the fundamental representation.
\end{propn}

From our basic knowledge of the twistor action, twistor Wilson loop, and the scattering amplitudes/Wilson loop duality, there are several immediate corollaries of this fact.  The resulting null polygon $C\subset\M$ defines a set of $n$ null (super-)momenta, which satisfy momentum conservation and hence define data for a scattering amplitude.  This allows us to relate the supersymmetric correlation function / Wilson loop correspondence to scattering amplitudes in the planar limit: 
\begin{corol}\label{CcorrWL}
Fix the planar limit of $\cN=4$ SYM.  The following statements hold at the level of the loop integrand: 
\be{corrW3}
\lim_{x_{i,i+1}^{2}\rightarrow 0}\frac{\la \cO_{b}(x_{1})\ldots\cO_{b}(x_{n})\ra}{\la\cO(x_{1})\ldots\cO(x_{n})\ra^{\mathrm{tree}}}= \left(\sum_{l=0}^{\infty}\frac{A^{l}_{n,0}}{A^{0}_{n,0}}\right)^{2},
\ee
\be{corrW4}
\lim_{x_{i,i+1}^{2}\rightarrow 0}\frac{\la \cO(x_{1})\ldots\cO(x_{n})\ra^{\mathrm{SD}}}{\la\cO(x_{1})\ldots\cO(x_{n})\ra^{\mathrm{tree}}}=\left(1+\frac{A^{0}_{n,1}}{A^{0}_{n,0}}+\cdots +\frac{A^{0}_{n,n-4}}{A^{0}_{n,0}}\right)^{2},
\ee
where all expectation values are assumed to be generic and normal ordered, $\{\cO_{b}(x_{i})\}_{i=1,\ldots,n}$ are the bosonic version of the local operators, and $\la \cO(x_{1})\ldots\cO(x_{n})\ra^{\mathrm{SD}}$ denotes the expectation value with respect to the self-dual portion of the theory.
\end{corol}
\proof  \eqref{corrW3} follows from the fact that for the bosonic operators in the planar limit, we will recover the square of the bosonic Wilson loop \eqref{bWilsonloop} about the contour $C\subset\M_{b}$ \cite{Mason:2010yk}.  The most basic form of the scattering amplitude / Wilson loop duality indicates that this is the ratio of the all loop MHV amplitude divided by the tree-level MHV amplitude, all squared.  For \eqref{corrW4}, the self-dual truncation indicates taking the expectation value with respect to $S_{1}[\cA]$ in twistor space.  This eliminates the MHV vertex insertions from $S_{2}[\cA]$, which constitute the loop corrections.  Hence, the fundamental Wilson loop in the self-dual theory gives all the $n$-point tree-level amplitudes, normalized by the tree-level MHV amplitude.     $\Box$ 

\medskip

There are several facts worth noting before we move on.  Firstly, proposition \ref{corrWL} establishes the correspondence between correlation functions and Wilson loops for finite-rank gauge group.  This immediately confirms that this correspondence is more robust than the scattering amplitudes / Wilson loop duality, which only holds in the planar limit.  Of course, this is to be expected: scattering amplitudes and Wilson loops are defined in \emph{dual} spaces related by a sort of T-duality, whereas our correlation functions are defined on the same space as the Wilson loops.  If one wanted to extend this correspondence to account for full loop \emph{integrals}, then this indicates that the same regularization procedure can be used for both the Wilson loop and the correlation function.  Upon passing to the planar limit, corollary \ref{CcorrWL} tells us that this will define the (square of the) regularized scattering amplitudes.


\subsection{Mixed Wilson Loop / Local Operator Correlators}

The supersymmetric correlation function / Wilson loop correspondence can naturally be generalized by considering null limits of local operator insertions in which some local operators remain in general position (i.e., not null separated).  In the planar limit, Alday, Buchbinder, and Tseytlin conjectured that such a process would lead to mixed Wilson loop / local operator correlators \cite{Alday:2011ga}
\be{locW1}
\lim_{x_{i,i+1}^{2}\rightarrow 0}\frac{\la \cO(x_{1})\ldots\cO(x_{n})\cO(y)\ra}{\la\cO(x_{1})\ldots\cO(x_{n})\ra}\sim \frac{\la W^{n}[C]\cO(y)\ra}{\la W^{n}[C]\ra}\equiv \cC^{n}_{1}(W^{n}, y).
\ee
The intuition for this is based upon the case when $\cO(y)=\cO_{\mathrm{dil}}(y)$, the dilaton operator.  Since $\cO_{\mathrm{dil}}$ is (up to a re-scaling) the $\cN=4$ SYM Lagrangian (c.f., \cite{Liu:1999kg}), one can use proposition \ref{corrWL} in conjunction with integration-by-parts inside the path integral to arrive at the right-hand side of \eqref{locW1}--albeit with the position $y$ of the dilaton operator integrated over.

While this proposal has been confirmed at weak coupling for twist-2 local operators using dimensional regularization \cite{Engelund:2011fg}, it is hardly obvious that the integral over position can be omitted or that \eqref{locW1} holds for arbitrary local operators.  It turns out that the twistorial point of view once again allows us to prove these claims with relative ease (at the level of the integrand). 

Note that there are many reasons to be interested in the mixed correlators which are conjectured to appear on the right-hand side of \eqref{locW1}.  These mixed correlators are a natural candidate for interpolating between Wilson loops and generic correlation functions; their structure is highly constrained by conformal invariance; and studying $\cC^{n}_{1}$ provides information about the Wilson loop OPE \cite{Berenstein:1998ij}.  Indeed, for $n=4$ (where the strong coupling solution for the Wilson loop is explicitly known \cite{Alday:2007hr}) one can show that $\cC^{4}_{1}$ is a function of a single conformal cross-ratio, and hence explicit strong coupling calculations are possible.  In this setting, the functional dependence can be determined precisely in the strong coupling regime using a semi-classical string theory approximation in the AdS-geometry \cite{Zarembo:2002ph, Roiban:2010fe, Zarembo:2010rr, Costa:2010rz, Buchbinder:2010ek, Alday:2011ga, Hernandez:2012zj}. This method has also been used to study a similar correlator involving a circular Wilson loop (i.e., $n\rightarrow\infty$) \cite{Alday:2011pf}, and in this case some progress can also be made for the inclusion of two local operators in general position \cite{Buchbinder:2012vr}. 

Furthermore, while null polygonal Wilson loops have UV divergences coming from their cusps, the mixed correlators appear to be UV finite since these divergence should cancel between the numerator and denominator.  This has been checked explicitly to two loops for the $n=4$ Wilson loop and the dilaton operator \cite{Alday:2012hy, Alday:2013ip}, and is expected to hold to all orders in perturbation theory.  This indicates that studying mixed correlators at the level of the loop integrand is in fact a mathematically safer endeavour than for scattering amplitudes or null polygonal Wilson loops on their own.


\subsubsection{Null limits in twistor space}

We now seek to confirm the conjecture of \cite{Alday:2011ga}: that mixed correlators are equivalent to null limits of ratios of correlation functions.  Without loss of generality, let all operators in question be 1/2-BPS operators given on space-time by \eqref{BPS} and twistor space by \eqref{nabelian}.  Recall that we could easily modify our discussion to account for any local operators (such as Konishi or dilaton); as in the proof of proposition \ref{corrWL}, the only thing that matters is the null limit.  

Provided all limits exist (as we will show), we can separate the limit of interest as
\be{eqn: l2}
\lim_{x^{2}_{i,i+1}\rightarrow 0}\frac{\la \cO(x_{1})\cdots\cO(x_{n})\cO(y)\ra}{\la \cO(x_{1})\cdots\cO(x_{n})\ra^{\mathrm{tree}}}\times \lim_{x^{2}_{i,i+1}\rightarrow 0}\frac{\la \cO(x_{1})\cdots\cO(x_{n})\ra^{\mathrm{tree}}}{\la \cO(x_{1})\cdots\cO(x_{n})\ra},
\ee
where the insertion $y$ is in general position (i.e., not null separated from any of the $x_{i}$). However, using proposition \ref{corrWL}, it is easy to see that the limit we are actually interested in calculating is computed by:
\be{newlimit}
\lim_{x^{2}_{i,i+1}\rightarrow 0}\frac{\la \cO(x_{1})\cdots\cO(x_{n})\cO(y)\ra}{\la \cO(x_{1})\cdots\cO(x_{n})\ra^{\mathrm{tree}}}\times \frac{1}{\la W^{n}_{\mathrm{adj}}[C]\ra}.
\ee
Once again, the tree level contribution in the denominator goes as
\begin{equation*}
\la \cO(x_{1})\cdots\cO(x_{n})\ra^{\mathrm{tree}} \sim \frac{1}{x_{12}^{2}x_{23}^{2}\cdots x_{n1}^{2}},
\end{equation*}
so we are interested in extracting those contributions from the numerator which counterbalance this classical factor in the null limit.  

\begin{figure}
\centering
\includegraphics[width=3 in, height=1.5 in]{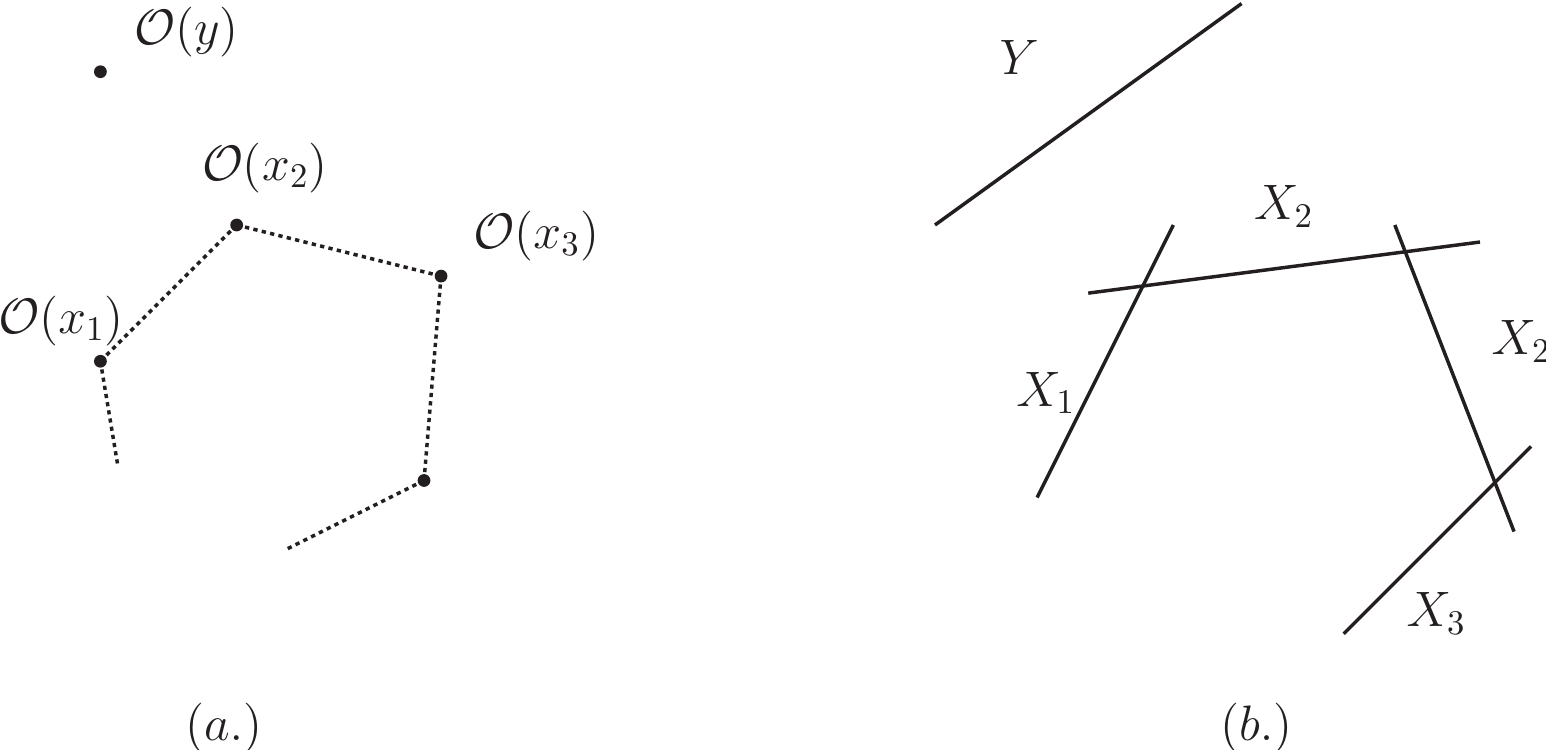}\caption{\textit{The geometry of the null limit in} (a.) \textit{space-time, and} (b.) \textit{twistor space.}}\label{locops1}
\end{figure}
As before, this situation has a nice formulation in twistor space: we begin with $n+1$ lines $X_{1},\ldots,X_{n},Y\subset\PT$ for each local operator.  In the limit, the first $n$ of these intersect each other sequentially to form the nodal curve corresponding to the resulting null polygon in $\M$; the final operator in general position, $\cO(y)$, lies on a line $Y$ which does not intersect \emph{any} of the others.  This configuration is illustrated in Figure \ref{locops1}.  

We can now obtain the desired result, first reported in \cite{Adamo:2011cd}:
\begin{propn}\label{locP1}
Let $\{\cO(x_{i}), \cO(y)\}_{i=1,\ldots,n}$ be gauge invariant local operators in $\cN=4$ SYM, and $C$ be the null polygon resulting from the limit where the first $n$ of these operators become pairwise null separated (i.e., $x_{i,i+1}^{2}=0$).  Then at the level of the integrand,
\be{locW}
\lim_{x^{2}_{i,i+1}\rightarrow 0}\frac{\la \cO(x_{1})\cdots\cO(x_{n})\cO(y)\ra}{\la \cO(x_{1})\cdots\cO(x_{n})\ra} = \frac{\la W^{n}_{\mathrm{adj}}[C]\cO(y)\ra}{\la W^{n}_{\mathrm{adj}}[C]\ra} \xrightarrow{\mathrm{planar}\:\mathrm{limit}} 2\frac{\la W^{n}[C]\cO(y)\ra}{\la W^{n}[C]\ra},
\ee
where all expectation values are assumed to be generic and normal ordered, and $W^{n}[C]$ is the Wilson loop in the fundamental representation.
\end{propn}
\proof  In this setting, we have the same classes of contractions as in proposition \ref{corrWL}, along with an additional class involving frames and operator insertions along the line $Y$.  By the genericity assumption, any contractions involving $Y$ and a MHV vertex will produce a R-invariant, or a second derivative of a R-invariant, which will be finite in the null limit.  Additionally, since $Y$ does not intersect any of the $\{X_{i}\}$, all other possible contributions from the local operator in general position will be finite in the null limit (this follows using methods identical to those for the proof of proposition \ref{corrWL}).  This leaves two holomorphic frames $U_{X}$ on each of the $X_{i}$ and the local operator $\cO(y)$ in general position after the null limit, so we have
\begin{equation*}
\lim_{x^{2}_{i,i+1}\rightarrow 0}\frac{\la \cO(x_{1})\cdots\cO(x_{n})\cO(y)\ra}{\la \cO(x_{1})\cdots\cO(x_{n})\ra^{\mathrm{tree}}}=\la W^{n}_{\mathrm{adj}}[C]\cO(y) \ra.
\end{equation*}

\begin{figure}
\centering
\includegraphics[width=4 in, height=1.5 in]{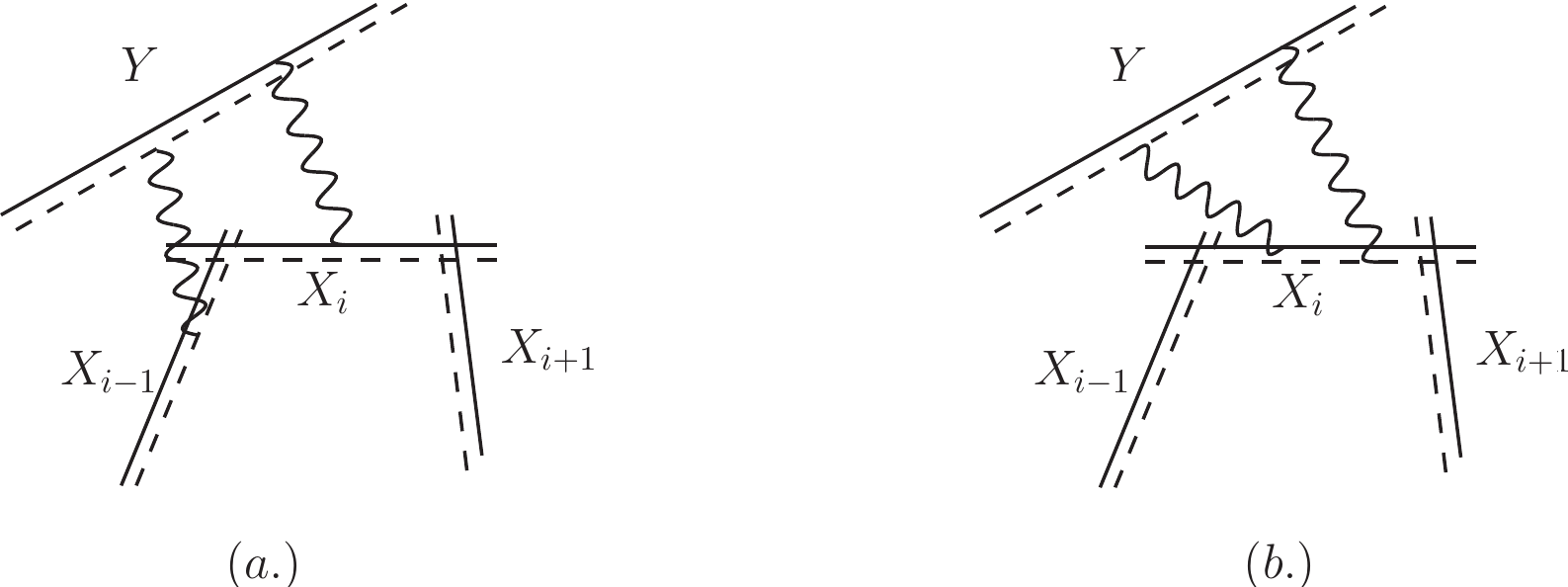}\caption{\textit{Contractions which are} (a.) \textit{leading, and} (b.) \textit{suppressed in the planar limit.  The solid and dashed lines are meant to distinguish the fundamental and anti-fundamental representations.}}\label{locops2}
\end{figure}

Now, passing to the planar limit of the gauge theory (i.e., $N\rightarrow\infty$), we can decompose the adjoint representation into the product of fundamental and anti-fundamental representations to write:
\begin{equation*}
\la W^{n}_{\mathrm{adj}}[C]\cO(y) \ra = \la W^{n}[C] \widetilde{W}^{n}[C]\cO(y) \ra.
\end{equation*}
The tensor structure of the propagator \eqref{TAcolorprop} suppresses contractions between operators and frames on $Y$ with the Wilson loops which mix fundamental and anti-fundamental representations in the planar limit, as depicted in Figure \ref{locops2}.  This means that in the large $N$ limit, we have
\begin{equation*}
\la W^{n}[C] \widetilde{W}^{n}[C]\cO(y) \ra = 2 \la W^{n}[C] \ra \la W^{n}[C] \cO(y)\ra,
\end{equation*}
as required.     $\Box$

Note that the last step in this proof is a rather explicit manifestation of the following general heuristic for a planar gauge theory: given three operators $\cO_{1}$, $\cO_{2}$, and $\cO_{3}$, their expectation value should obey
\begin{equation*}
\la \cO_{1} \cO_{2} \cO_{3} \ra = \la \cO_{1}\ra \la \cO_{2} \cO_{3}\ra + \la \cO_{2}\ra \la \cO_{3} \cO_{1}\ra + \la \cO_{3}\ra \la \cO_{1} \cO_{2}\ra.
\end{equation*}
In the case of interest, two of these terms are equal while the third vanishes due to normal ordering.

\subsubsection*{\textit{Additional operators and null limits}}

There are natural generalizations of the null limit we have considered here; in particular, we could leave an arbitrary number of local operators in general position.  This extension was first proposed in \cite{Alday:2011ga} and has been investigated at weak \cite{Engelund:2011fg} and strong coupling \cite{Adamo:2011cd, Buchbinder:2012vr}.  For instance, consider
\begin{equation*}
\lim_{x^{2}_{i,i+1}\rightarrow 0}\frac{\la \cO(x_{1})\cdots\cO(x_{n})\cO(y_{1})\cdots\cO(y_{k})\ra}{\la \cO(x_1)\cdots\cO(x_n)\ra},
\end{equation*}
where the $k$ operators $\cO(y_{1}),\ldots,\cO(y_k)$ remain in general position relative to the $x_{i}$ and each other.  The proof of proposition \ref{locP1} can easily be adapted to this situation, giving 
\begin{equation*}
\frac{1}{\la W_{\mathrm{adj}}[C]\ra} \sum_{j=0}^{k-2} \sum_{\{i_{1},\ldots, i_{j}\}\subset \{1,\ldots, k\}} \la W_{\mathrm{adj}}[C]\cO(y_{i_{1}})\cdots\cO(y_{i_{j}})\ra\: \la\cO(y_{i_{j+1}})\cdots\cO(y_{i_{k}})\ra,
\end{equation*} 
with the range of the sum dictated by normal ordering.  Taking the planar limit splits the first factor into two correlators with fundamental Wilson loops as before, and introduces another sum over partitions of the remaining operators.

Of course, we can generalize this further by also allowing the $k$ additional operators to become pairwise null separated, forming a second null polygon $D$.  This results in new divergences which must be balanced by the appropriate denominator; the natural choice is:
\be{locW2*}
\lim_{x^{2}_{i,i+1},y^{2}_{j,j+1}\rightarrow 0} \frac{\la \cO(x_{1})\cdots\cO(x_{n})\cO(y_{1})\cdots\cO(y_{k})\ra}{\la \cO(x_{1})\cdots\cO(x_{n})\ra \la\cO(y_{1})\cdots\cO(y_{k})\ra}.
\ee
By proposition \ref{corrWL}, this is:
\begin{equation*}
\lim_{x^{2}_{i,i+1},y^{2}_{j,j+1}\rightarrow 0} \frac{\la \cO(x_{1})\cdots\cO(x_{n})\cO(y_{1})\cdots\cO(y_{k})\ra}{\la \cO(x_{1})\cdots\cO(x_{n})\ra^{\mathrm{tree}} \la\cO(y_{1})\cdots\cO(y_{k})\ra^{\mathrm{tree}}}\frac{1}{\la W_{\mathrm{adj}}[C]\ra \la W_{\mathrm{adj}}[D]\ra}, 
\end{equation*}
where the tree-level denominator has the expected singularity structure:
\be{diverge}
\la \cO(x_{1})\cdots\cO(x_{n})\ra^{\mathrm{tree}} \la \cO(y_{1})\cdots\cO(y_{k})\ra^{\mathrm{tree}} \sim \frac{1}{x_{12}^{2}x_{23}^{2}\cdots x_{n1}^{2}} \times \frac{1}{y_{12}^{2}y_{23}^{2}\cdots y_{k1}^{2}}.
\ee
So once again, we need to extract compensating divergences from the numerator of \eqref{locW2*}.

We can break the numerator into a sum of connected and disconnected components:
\begin{multline*}
\frac{1}{\la W_{\mathrm{adj}}[C]\ra \la W_{\mathrm{adj}}[D]\ra} \lim_{x^{2}_{i,i+1},y^{2}_{j,j+1}\rightarrow 0}\left( \frac{\la \cO(x_{1})\cdots\cO(x_{n})\cO(y_{1})\cdots\cO(y_{k})\ra^{\mathrm{conn}}}{\la \cO(x_{1})\cdots\cO(x_{n})\ra^{\mathrm{tree}} \la\cO(y_{1})\cdots\cO(y_{k})\ra^{\mathrm{tree}}} \right. \\
+ \left. \frac{\la \cO(x_{1})\cdots\cO(x_{n})\ra \la\cO(y_{1})\cdots\cO(y_{k})\ra}{\la \cO(x_{1})\cdots\cO(x_{n})\ra^{\mathrm{tree}} \la\cO(y_{1})\cdots\cO(y_{k})\ra^{\mathrm{tree}}} + \frac{\{\mbox{all other disconnected}\}}{\la \cO(x_{1})\cdots\cO(x_{n})\ra^{\mathrm{tree}} \la\cO(y_{1})\cdots\cO(y_{k})\ra^{\mathrm{tree}}} \right),
\end{multline*}
and analyse each term by performing all contractions in twistor space and looking at their degree of divergence.  Because none of the $X_{i}$ and $Y_{j}$ ever intersect in twistor space (we assume that the two sets of operators become pairwise null separated independently), the proof of proposition \ref{corrWL} indicates that the only contractions which produce the correct degree of divergence in the first term are those between $\frac{\partial^{2}\cA}{\partial \chi^{2}}$ on adjacent $X$s and adjacent $Y$s.  Hence,
\begin{equation*}
\lim_{x^{2}_{i,i+1},y^{2}_{j,j+1}\rightarrow 0}\frac{\la \cO(x_{1})\cdots\cO(x_{n})\cO(y_{1})\cdots\cO(y_{k})\ra^{\mathrm{conn}}}{\la \cO(x_{1})\cdots\cO(x_{n})\ra^{\mathrm{tree}} \la\cO(y_{1})\cdots\cO(y_{k})\ra^{\mathrm{tree}}}=\la W_{\mathrm{adj}}[C] W_{\mathrm{adj}}[D]\ra^{\mathrm{conn}}.
\end{equation*}

The second term is easily evaluated by applying proposition \ref{corrWL}:
\begin{equation*}
\lim_{x^{2}_{i,i+1},y^{2}_{j,j+1}\rightarrow 0}\frac{\la \cO(x_{1})\cdots\cO(x_{n})\ra \la\cO(y_{1})\cdots\cO(y_{k})\ra}{\la \cO(x_{1})\cdots\cO(x_{n})\ra^{\mathrm{tree}} \la\cO(y_{1})\cdots\cO(y_{k})\ra^{\mathrm{tree}}}= \la W_{\mathrm{adj}}[C]\ra \la W_{\mathrm{adj}}[D]\ra .
\end{equation*}
The remaining terms (composed of all other disconnected components from the correlation function) involve all the usual contractions which give a vanishing contribution (e.g., contractions between non-adjacent lines, contractions between operators and fields on any line with a MHV vertex), and additionally contain no connected component with enough lines in twistor space to form a full Wilson loop in the null limit.  So any term in this sum of disconnected components will contain some divergences of the form $x_{i,i+1}^{-2} y_{j,j+1}^{-2}$, but never the full array appearing in \eqref{diverge}.  Thus, all remaining disconnected terms vanish in the null limit.

In both of the generalizations discussed here, we can pass to the planar limit in twistor space by invoking the twistor propagator with its color structure given by \eqref{TAcolorprop}.  More formally, we have: 
\begin{propn}\label{locP2}
Let $\{\cO(x_{i}), \cO(y_{j})\}^{i=1,\ldots,n}_{j=1,\ldots,k}$ be gauge invariant local operators in $\cN=4$ SYM, $C$ be the null polygon resulting from the limit where the $\{\cO(x_{i})\}$ become pairwise null separated (i.e., $x_{i,i+1}^{2}=0$), and $D$ be the null polygon when the $\{\cO(y_{j})\}$ become null separated ($y^{2}_{j,j+1}=0$).  Then at the level of the integrand,
\begin{multline}\label{locW2}
\lim_{x^{2}_{i,i+1}\rightarrow 0}\frac{\la \cO(x_{1})\cdots\cO(x_{n})\cO(y_{1})\cdots\cO(y_{k})\ra}{\la \cO(x_1)\cdots\cO(x_n)\ra} \\
= \frac{1}{\la W_{\mathrm{adj}}[C]\ra} \sum_{j=0}^{k-2} \sum_{\{i_{1},\ldots, i_{j}\}\subset \{1,\ldots, k\}} \left\la W_{\mathrm{adj}}[C]\cO(y_{i_{1}})\cdots\cO(y_{i_{j}})\right\ra\: \la\cO(y_{i_{j+1}})\cdots\cO(y_{i_{k}})\ra \\
\xrightarrow{\mathrm{planar}\:\mathrm{limit}} \frac{1}{\la W[C]\ra^{2}} \sum_{\cP_{k}} \left\la W[C]\cO(y_{i_{1}})\cdots\cO(y_{i_{j}})\right\ra \;\la W[C]\cO(y_{i_{j+1}})\cdots\cO(y_{i_{l}})\ra \\
\times \la \cO(y_{i_{l+1}})\cdots\cO(y_{i_{k}})\ra,
\end{multline}
where all expectation values are assumed to be generic and normal ordered, $W[C]$ is the Wilson loop in the fundamental representation, and $\sum_{\cP_{k}}$ is the sum over relevant partitions of $\{1,\ldots,k\}$.  If we allow the remaining $k$ operators to also become null separated, then
\begin{multline}\label{locW3}
\lim_{x^{2}_{i,i+1},y^{2}_{j,j+1}\rightarrow 0} \frac{\la \cO(x_{1})\cdots\cO(x_{n})\cO(y_{1})\cdots\cO(y_{k})\ra}{\la \cO(x_{1})\cdots\cO(x_{n})\ra \la\cO(y_{1})\cdots\cO(y_{k})\ra} = 1+\frac{\la W_{\mathrm{adj}}[C] W_{\mathrm{adj}}[D]\ra^{\mathrm{conn}}}{\la W_{\mathrm{adj}}[C]\ra \la W_{\mathrm{adj}}[D]\ra} \\
\xrightarrow{\mathrm{planar}\:\mathrm{limit}} 1+2\frac{\la W[C]\; W[D]\ra^{\mathrm{conn}}}{\la W[C]\ra \la W[D]\ra}.
\end{multline}
\end{propn}  


\subsubsection{Recursion relations}

\begin{figure}
\centering
\includegraphics[width=3.5 in, height=2.5 in]{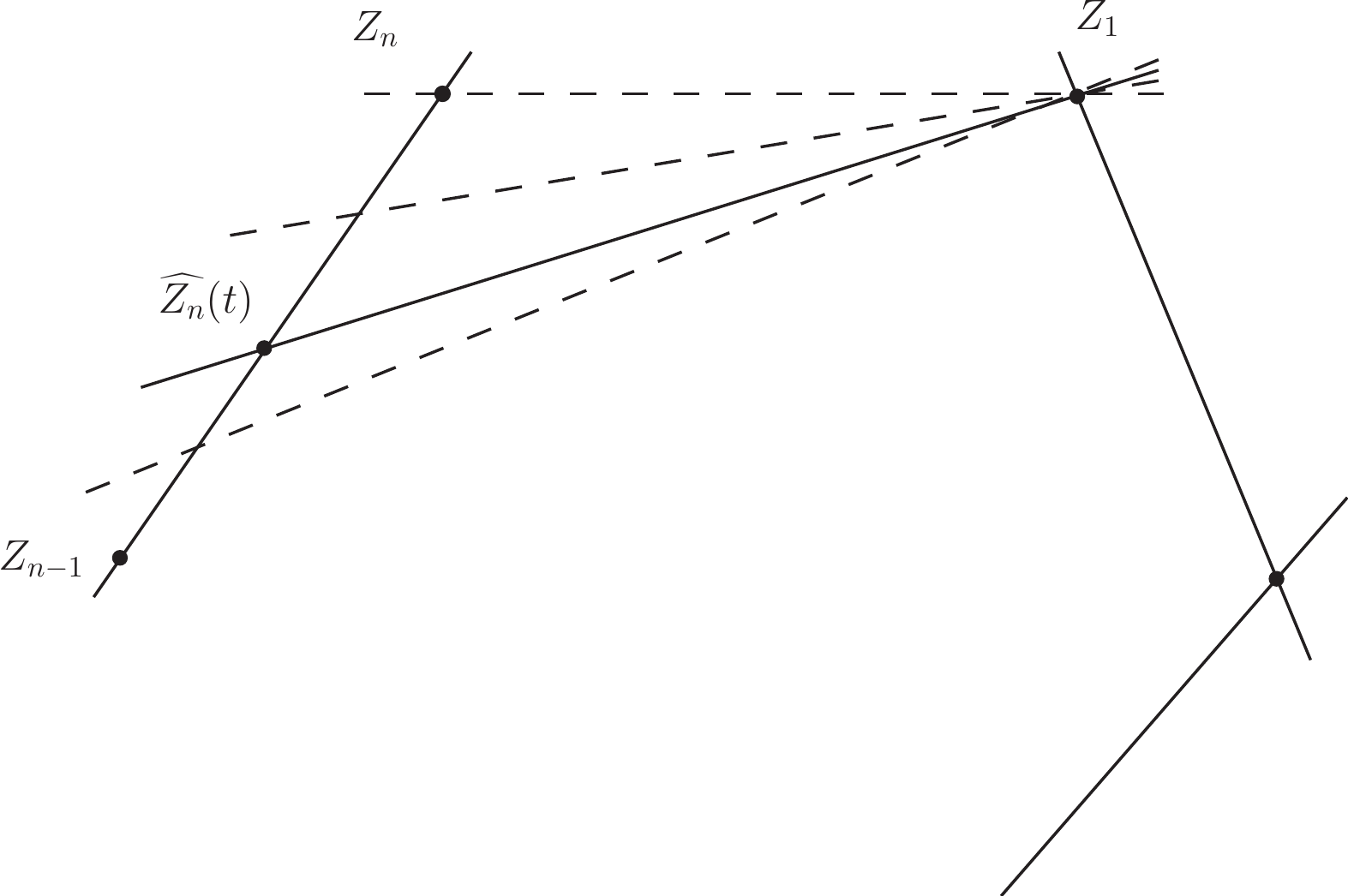}\caption{\textit{The BCFW-like deformation at the level of the twistor Wilson loop.}}\label{BCF1}
\end{figure}

The scattering amplitude / Wilson loop duality follows by showing that both objects obey the same recursive relations: the BCFW recursions (see Section \ref{SAP} for a quick reminder).  Proving this is particularly natural in twistor space, where the BCFW recursion is implemented by performing a one-parameter shift of one of the nodes.  Without loss of generality, this takes the form:
\be{BCFW1}
\widehat{Z_{n}}(t)=Z_{n}+tZ_{n-1}, \qquad t\in\C,
\ee
which shifts the $n^{\mathrm{th}}$ node along the line $(n-1, n)\cong\P^{1}\subset\PT$, as illustrated in Figure \ref{BCF1}.  It was shown in \cite{Bullimore:2011ni} that the variation of the expectation value of this deformed Wilson loop is supported only on self-intersections or intersections with Lagrangian insertions, precisely reproducing the all-loop BCFW recursion of \cite{ArkaniHamed:2010kv}.  This is a holomorphic analogue of the loop equations \cite{Makeenko:1979pb} and skein relations which arise in the study of real knot theory \cite{Witten:1988hf}.  

Beyond establishing the duality with amplitudes at the level of the loop integrand, these recursion relations also provide a means for actually computing the Wilson loop integrand.  We will see that a BCFW-like recursion relation for the correlator
\begin{equation*}
\la W^{n}[C]\cO(y)\ra,
\end{equation*}
can also be derived, enabling the computation of the integrand for such correlators.  Once again, the key to doing this will be studying the problem in twistor space.  

Recall the null polygonal Wilson loop in twistor space is given by:
\begin{multline}\label{WL*}
W[C]\equiv W[1,2,\ldots, n]=\tr\: \mathrm{Hol}_{Z_n}[C]= \\
\tr\left[ U(Z_{n},Z_{n-1})U(Z_{n-1},Z_{n-2})\cdots U(Z_{1},Z_{n})\right],
\end{multline}
where $\mathrm{Hol}_{Z}[C]$ denotes the holonomy about $C$ at base point $Z$ and the $Z_{i}$ are the nodes of the resulting curve in twistor space.  The BCFW shift \eqref{BCFW1} results in a one-parameter family of nodal curves in twistor space and their corresponding family of Wilson loops:
\begin{equation*}
C(t)=(1,2)\cup (2,3)\cup\cdots \cup (n-1, \hat{n}(t))\cup (\hat{n}(t), 1), \qquad W[C(t)]=W[1,\ldots n-1, \hat{n}(t)],
\end{equation*}
where we have adopted the shorthand $\hat{n}(t)$ for $\widehat{Z_{n}}(t)$.

Formally, we can think of $t\in \C$ as a coordinate on the moduli space of maps from $\Sigma\cong\P^{1}$ into $\PT$ with two fixed points (the nodes at $Z_{n}$ and $Z_{n-1}$).  We will be interested in the variation of our correlator with respect to \emph{anti-holomorphic} dependence on this coordinate; this requires a $\dbar$-operator on the moduli space $\overline{M}_{0,2}(\P^{3|4},1)$.\footnote{These moduli spaces are, strictly speaking, algebraic stacks.  However, for the case of a genus zero Riemann surface and target space $\PT$, they are unobstructed and have a versal family which can be treated as an algebraic space \cite{Adamo:2012cd}.}  Formally, this can be constructed by considering the diagram:
\begin{equation*}
\xymatrix{
\overline{M}_{0,3}(\P^{3|4},1)  \ar[d]^{\rho} \ar[rr]^{\Phi} & & \PT \\
\overline{M}_{0,2}(\P^{3|4},1) & & }
\end{equation*}
where $\rho$ is the forgetful functor which throws away an extra marked point, and $\Phi$ is the `universal instanton' \cite{Adamo:2012cd}.  Since the universal curve is just $\overline{M}_{0,3}(\P^{3|4},1)\cong \overline{M}_{0,2}(\P^{3|4},1)\times\Sigma$, this map simply takes $f\in \overline{M}_{0,2}(\P^{3|4},1)$ and $z\in\Sigma$ to $f(z)\in\PT$.  Hence, we can take the complex structure on $\PT$ given by $\dbar$, and define $\bar{\delta}$ on $\overline{M}_{0,2}(\P^{3|4},1)$ both formally and heuristically:
\begin{equation*}
\bar{\delta}=\rho_{*}\Phi^{*}\dbar, \qquad \bar{\delta}=\d\bar{t}\frac{\partial}{\partial \bar{t}}.
\end{equation*}

In \cite{Bullimore:2011ni}, the twistor action and Wilson loop were used to study $\bar{\delta}\la W[C(t)]\ra$; we will use the same methodology to study the correlator between a Wilson loop and single local operator.  The key relation is the following:
\begin{lemma}[Bullimore \& Skinner \cite{Bullimore:2011ni}]
The infinitesimal variation of $W[C]$ with respect to $\bar{t}$ is given by:
\be{WLvar}
\bar{\delta}\;W[C]=-\int_{C} \omega(Z)\wedge\d \bar{Z}^{\bar{\alpha}}\wedge\bar{\delta}\bar{Z}^{\bar{\beta}}\; \tr\left( F^{0,2}_{\bar{\alpha}\bar{\beta}}\mathrm{Hol}_{Z}[C]\right),
\ee 
where $\omega(Z)$ is a meromorphic 1-form on $C$ with simple poles at each node $Z=Z_{i}$, and $F^{0,2}=\dbar\cA+\cA\wedge\cA$ is the anti-holomorphic curvature of the gauge connection on twistor space.
\end{lemma}
By inserting this into the path integral for $\bar{\delta}\la W[C(t)]\ra$ with respect to the twistor action \eqref{TwistorAction}, a holomorphic analogue of the loop equations \cite{Makeenko:1979pb} was found which lead to the all-loop BCFW recursion relations of \cite{ArkaniHamed:2010kv}.  Since scattering amplitudes are also determined by BCFW recursion, this proves that the two observables are actually the same.

In our case, we want to consider $\bar{\delta}\la W[C(t)]\cO(y)\ra$ for any $\U(N)$ gauge group.  Since the BCFW-like deformation \eqref{BCFW1} only acts on the Wilson loop, we can use \eqref{WLvar} to consider:
\be{BCFW2}
\bar{\delta}\la W[C(t)]\cO(y)\ra =-\frac{1}{N}\int [\mathcal{D}\cA]\left[ \int_{C(t)} \omega(Z)\wedge\d \bar{Z}^{\bar{\alpha}}\wedge\bar{\delta}\bar{Z}^{\bar{\beta}} \tr\left( F^{(0,2)}_{\bar{\alpha}\bar{\beta}}\mathrm{Hol}_{Z}[C]\right) \cO(y)\right] e^{-S[\cA]},
\ee
where $\cO(y)$ is our 1/2-BPS operator \eqref{nabelian}, $S[\cA]$ is the twistor action \eqref{TwistorAction}, and we have included a normalization factor of $1/N$.  As noted earlier, the twistor action can be decomposed into a holomorphic Chern-Simons portion accounting for the SD sector of the theory (or tree-level for the Wilson loop) and a non-local contribution encoding the ASD interactions (or loop-level for the Wilson loop).  These are given by $S_{1}$ \eqref{TASD} and $S_{2}$ \eqref{TAInt} respectively. 

\subsubsection*{\textit{Holomorphic linking contribution}}

We begin by considering the classical piece of \eqref{BCFW2} corresponding to $S_{1}[\cA]$.  For an abelian gauge group, this will produce contributions corresponding to holomorphic linking between the irreducible components of $C$ \cite{Atiyah:1981, Penrose:1988, Khesin:2000ng, Frenkel:2005qk}.  For a general gauge group, this provides a formal path-integral definition for holomorphic linking.

Now, note that
\begin{equation*}
\frac{\delta S_{1}[\cA]}{\delta \cA(Z)}= N\;F^{(0,2)}(Z),
\end{equation*}
so that
\begin{equation*}
\bar{\delta}\la W[C(t)]\cO(y)\ra^{\mathrm{tree}}=\frac{1}{N^2}\int [\mathcal{D}\cA]\left[ \int_{C(t)}\omega(Z)\wedge\tr\left(\mathrm{Hol}_{Z}[C(t)]\frac{\delta}{\delta\cA(Z)}e^{-S_{1}[\cA]}\right)\cO(y)\right].
\end{equation*}
Integrating by parts within the path integral moves the variational derivative onto the holonomy, yielding \cite{Bullimore:2011ni}:
\begin{multline*}
\tr\left(\frac{\delta}{\delta\cA(Z)}\mathrm{Hol}_{Z}[C(t)]\right)= \\
\sum_{j=1}^{n}\int_{C_{j}(t)}\omega_{j-1,j}(Z')\wedge\bar{\delta}^{3|4}(Z,Z')\tr\left[U(Z,Z_{n})\cdots U(Z_{j},Z')\right]\tr\left[U(Z',Z_{j-1})\cdots U(Z_{1},Z)\right],
\end{multline*}
where $C_{j}(t)=(j-1,j)$ is the $j^{\mathrm{th}}$ component of the nodal curve $C(t)$, and $\omega_{j-1,j}(Z)$ is the meromorphic 1-form on $C_{j}(t)$ with poles at $Z_{j-1}$, $Z_{j}$.

\begin{figure}
\centering
\includegraphics[width=3.5 in, height=1.5 in]{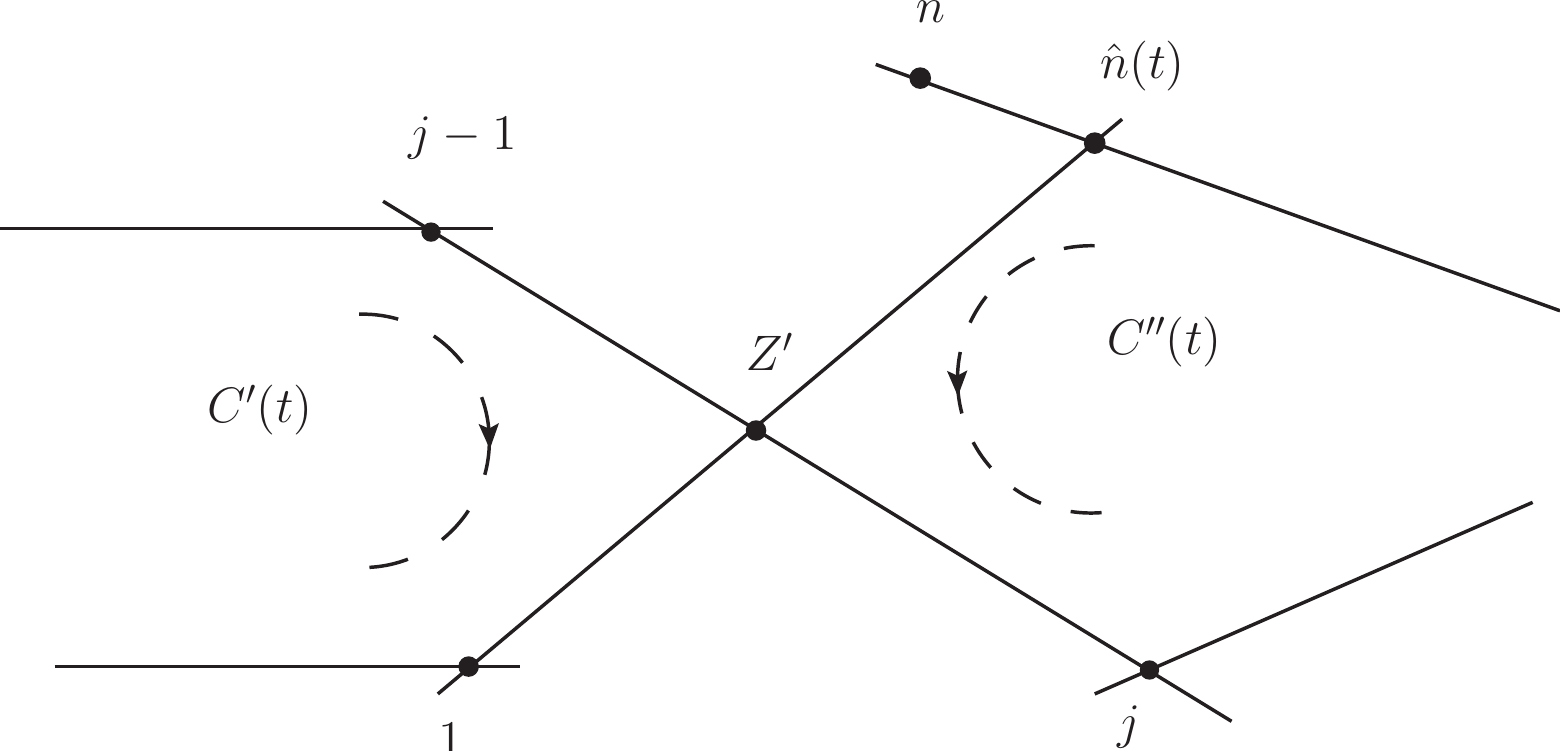}\caption{\textit{A holomorphic linking contribution when the curve $C(t)$ intersects itself.}}\label{BCF2}
\end{figure}
The $\bar{\delta}^{3|4}(Z,Z')$ only has support at $t\in\C$ where the deformed Wilson loop intersects itself, and the trace structure results in a factorization of the holonomy into two Wilson loops around the nodal curves $C'(t)$ and $C''(t)$, obtained by ungluing $C(t)$ at the intersection point $Z=Z'$.  This configuration is illustrated in Figure \ref{BCF2}.

This leaves us with:
\be{BCFWt}
\bar{\delta}\la W[C(t)]\cO(y)\ra^{\mathrm{tree}}=- \int\limits_{C(t)\times C(t)} \omega(Z)\wedge \omega(Z')\wedge\bar{\delta}^{3|4}(Z,Z') \left\la W[C'(t)]\;W[C''(t)]\;\cO(y)\right\ra,
\ee
where we have absorbed a normalization factor of $1/N$ into each Wilson loop.  This is the analogue of the holomorphic linking term of the loop equations derived in \cite{Bullimore:2011ni}, but now with a local operator in general position.  Note that in the planar limit of the gauge theory, the correlator can be re-written as
\begin{equation*}
\left\la W[C'(t)]\;W[C''(t)]\;\cO(y)\right\ra = \la W[C'(t)]\cO(y)\ra\; \la W[C''(t)]\ra + \la W[C'(t)]\ra\; \la W[C''(t)]\cO(y)\ra.
\end{equation*}

\subsubsection*{\textit{Contributions from MHV vertices and local operator}}

We still have to account for the contributions to $\bar{\delta}\la W[C(t)]\cO(y)\ra$ from $S_{2}[\cA]$ and the local operator $\cO(y)$.  Although the genericity assumption tells us that the nodal curve $C(t=0)$ never intersects any line $X$ corresponding to a MHV vertex or the line $Y$ corresponding to the local operator, as $t$ varies it sweeps out a plane $(n-1, n, 1)$ which \emph{all} lines in general position will intersect.  

The contribution from MHV vertices is given by
\be{PA1}
-\frac{\lambda}{N} \int[\mathcal{D}\cA]\int_{\Gamma} \d^{4|8}X\;\left[\int_{C(t)}\omega(Z)\wedge\tr\left( \delta\log\det(\dbar+\cA)|_{X}\mathrm{Hol}_{Z}[C(t)]\right)\cO(y)\right]e^{-S[\cA]},
\ee 
where the factor of $\lambda$ comes from $S_{2}[\cA]$, $1/N$ is for normalization, and $\Gamma=\M_{\R}\subset\M$ the real contour.  The variation of the logdet can be found by standard methods (c.f., \cite{Mason:2001vj}); if we assume that $X$ is given by the span of $Z_{A}$ and $Z_{B}$, then
\begin{equation*}
\delta\log\det(\dbar+\cA)|_{X}=\int\limits_{X\times S^{1}\times S^{1}} \omega_{A,B}(Z')\wedge \frac{\D\lambda_{A} \wedge\D\lambda_{B}}{\la AB\ra^{2}}\tr\left(U(Z_{B},Z')\delta\cA(Z')\right),
\end{equation*}
where $\omega_{A,B}(Z')$ is the meromorphic differential on $X$ with poles at $Z_{A}$ and $Z_{B}$, and $\lambda_{A},\lambda_{B}$ are the homogeneous coordinates of these points on $X$.  The integral over $S^{1}\times S^{1}$ is a contour integral surrounding the poles at $Z_{A}=Z_{B}=Z'$.

\begin{figure}
\centering
\includegraphics[width=4 in, height=1.5 in]{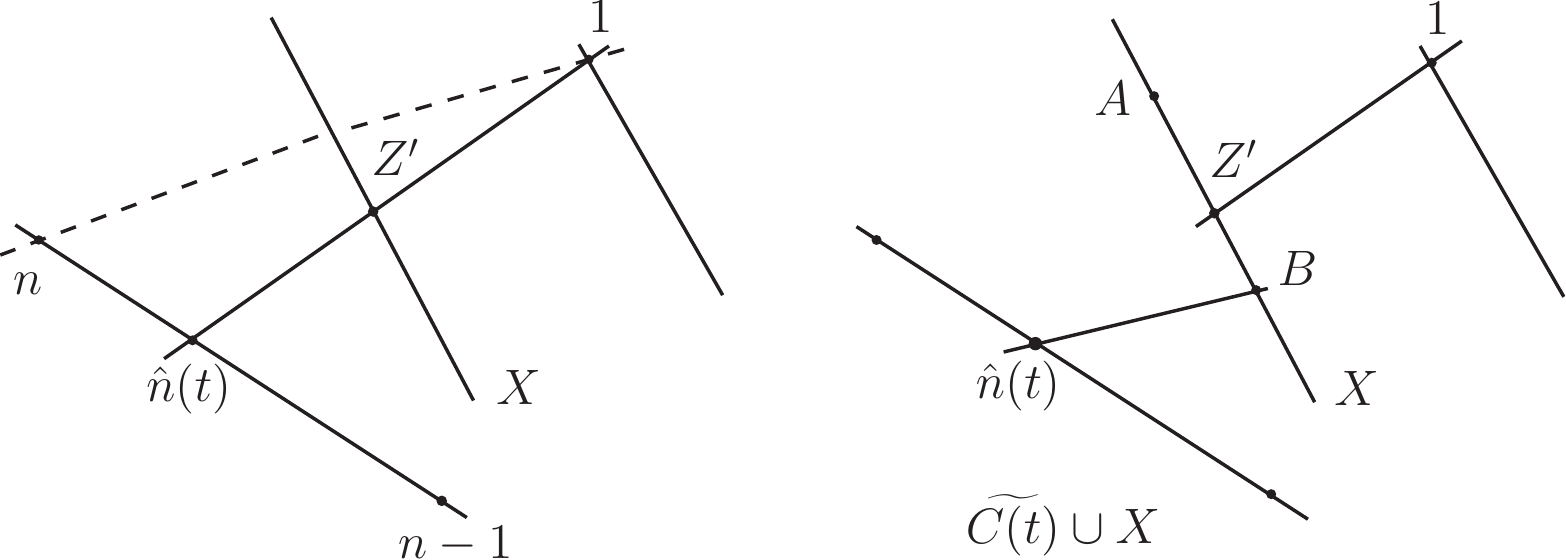}\caption{\textit{An intersection of the curve $C(t)$ with a MHV vertex $X$ (left) can be expressed as a forward limit of a new curve $\widetilde{C(t)}\cup X$ (right).}}\label{BCF3}
\end{figure}

The integral over the positions of $Z_{A}$ and $Z_{B}$ on $X$ can be combined with the measure $\d^{4|8}X$ to give a conformally invariant measure:
\begin{equation*}
\d^{4|8}X\wedge\frac{\D\lambda_{A} \wedge\D\lambda_{B}}{\la AB\ra^{2}}=\D^{3|4}Z_{A}\wedge\D^{3|4}Z_{B}.
\end{equation*}
Hence, the integrand of our path integral expression \eqref{PA1} is:
\begin{multline}\label{PA2}
-\frac{\lambda}{N}\oint\limits_{\Gamma\times S^{1}\times S^{1}}\D^{3|4}Z_{A}\wedge\D^{3|4}Z_{B}\:\int\limits_{C(t)\times X}\omega(Z)\wedge\omega_{A,B}(Z')\wedge\bar{\delta}^{3|4}(Z,Z')  \\
\times \tr\left(U(Z_{B},Z')\mathrm{Hol}_{Z}[C(t)]\right)\cO(y),
\end{multline}
with the $\bar{\delta}^{3|4}(Z,Z')$ ensuring that this is supported only when $C(t)$ intersects $X$ at $Z=Z'$.  As shown in \cite{Bullimore:2011ni}, this configuration can naturally be interpreted as a forward limit where the MHV vertex at $x\in\M$ becomes null separated from the point corresponding to the line $(\hat{n}(t),1)\subset\PT$.  More formally, we can replace $C(t)$ with a new curve $\widetilde{C(t)}$ which has an additional component such that:
\begin{equation*}
\widetilde{C(t)}\cap X=\{Z',Z_{B}\}, \qquad \lim_{Z_{B}\rightarrow Z'}\widetilde{C(t)}\rightarrow C(t),
\end{equation*}
\begin{equation*}
\widetilde{C(t)}\cup X=(1,2)\cup\cdots\cup (n-1,\hat{n}(t))\cup (Z',B)\cup (B,1).
\end{equation*}
This forward limit curve is pictured in Figure \ref{BCF3}.

Now the combined contours and delta-functions in \eqref{PA2} allow us to replace
\be{BCFW3}
\frac{1}{N}\tr\left(U(Z_{B},Z')\mathrm{Hol}_{Z}[C(t)]\right)=W[\widetilde{C(t)}\cup X],
\ee
and the contribution to $\bar{\delta}\la W[C(t)]\cO(y)\ra$ from the MHV vertices becomes:
\be{BCFWmhv}
-\lambda \oint\limits_{\Gamma\times S^{1}\times S^{1}}\D^{3|4}Z_{A}\wedge \D^{3|4}Z_{B} \left[ \int\limits_{C(t)\times X} \omega(Z)\wedge\omega_{A,B}(Z')\wedge\bar{\delta}^{3|4}(Z,Z')\left\la W[\widetilde{C(t)}\cup X]\;\cO(y)\right\ra \right].
\ee
Once again, this is the natural analogy of the second term in the holomorphic loop equations of \cite{Bullimore:2011ni}.

Finally, we must account for when $C(t)$ intersects $Y$, the line corresponding to the operator $\cO(y)$.  Clearly, the geometry of this configuration is identical to the intersection with a MHV vertex; the difference is that there are more fields and a more complicated R-symmetry on $Y$ due to the 1/2-BPS operator.  Nevertheless, we will see that this contribution can be treated similarly to the MHV vertices. 

For simplicity, take $\cO(y)=\cO_{abab}$, and suppose that $Y$ is given by the span of $Z_{C}$ and $Z_{D}$.  As before, we start with $\delta\log\det(\dbar+\cA)|_{Y}$, but now only integrated over a \emph{fermionic} contour in $\M$:
\begin{multline*}
- \oint\limits_{\tilde{\Gamma}\times S^{1}\times S^{1}}\d^{0|4}\theta^{abab} \wedge\frac{\D\lambda_{C}\wedge\D\lambda_{D}}{\la CD\ra^{2}}\\
\times\left[\int\limits_{C(t)\times Y}\omega(Z)\wedge\omega_{C,D}(Z')\wedge\bar{\delta}^{3|4}(Z,Z')\;\left\la W[\widehat{C(t)}\cup Y]\right\ra\right],
\end{multline*}
where $\d^{0|4}\theta^{abab}=\d\theta^{aA}\d\theta^{b}_{A}\d\theta^{aB}\d\theta^{b}_{B}$, $\tilde{\Gamma}$ is the corresponding fermionic contour (which, as usual, can be evaluated algebraically), and $\widehat{C(t)}$ is the forward limit curve associated with $Y$.  The R-symmetry of this measure extracts fermionic derivatives of the field $\cA$ from the holomorphic frames in $W[\widehat{C(t)}\cup Y]$, but supersymmetry dictates that they must be inserted at two different places on the line $Y$.  The remainder of the contour $\tilde{\Gamma}$ integrates these insertion points over $Y$.

Explicitly, in $W[\widehat{C(t)}\cup Y]$ we can use the properties of the holomorphic frame to write:
\begin{equation*}
U(Z_{D},Z')=U(Z_{D},Z_{C}) U(Z_{C},Z'),
\end{equation*}
and $Z'=Z_{D}$ on the support of the $S^{1}\times S^{1}$ contour and $\bar{\delta}^{3|4}(Z,Z')$.  The integral over $\d^{0|4}\theta^{abab}$ then pulls two derivatives from each of these frames on $Y$, and what is left is integrated over $Y$ to give:
\begin{equation*}
\int_{Y\times Y} \D\lambda_{C}\wedge\D\lambda_{D}\; U(Z_{D},Z_{C})\frac{\partial^{2}\cA(Z_{C})}{\partial\chi^{a}\partial\chi^{b}} U(Z_{C},Z_{D})\frac{\partial^{2}\cA(Z_{D})}{\partial\chi^{a}\partial\chi^{b}}.
\end{equation*}
Of course, this is our 1/2-BPS operator $\cO(y)=\cO_{abab}(y)$, and is inserted in the color trace running over the remaining holomorphic frames of the Wilson loop at the point $Z'$.  But this is precisely what we expect for the configuration where the deformed Wilson loop $C(t)$ intersects $Y$ at the point $Z=Z'$!  In other words, the 1/2-BPS operator is also captured by the variation of $S_{2}[\cA]$, but integrated over a partial fermionic contour corresponding to its R-symmetry structure.\footnote{Recall that $\log\det(\dbar+\cA)|_{X}$ is not locally gauge invariant; it must be integrated over some contour in order to kill the exponential anomalies associated with conformal gravity \cite{Boels:2006ir}.  The algebraic integral over $\tilde{\Gamma}$ gives the gauge invariance of the 1/2-BPS operators, as desired.}

Thus we obtain a third contribution to $\bar{\delta}\la W[C(t)]\cO(y)\ra$ of the form
\be{BCFWop}
-\oint\limits_{\tilde{\Gamma}\times S^{1}\times S^{1}}\d \mu^{abab} \left[\int\limits_{C(t)\times Y}\omega(Z)\wedge\omega_{C,D}(Z')\wedge\bar{\delta}^{3|4}(Z,Z') \left\la W[\widehat{C(t)}\cup Y]\right\ra \right],
\ee
where 
\begin{equation}\label{fmeasure}
\d\mu^{abab}=\d^{0|4}\theta^{abab} \wedge\frac{\D\lambda_{C}\wedge\D\lambda_{D}}{\la CD\ra^{2}}.
\end{equation}

\subsubsection*{\textit{The recursion relation}}

The all-loop BCFW-like recursion for our mixed correlator is given by combining \eqref{BCFWt}, \eqref{BCFWmhv}, \eqref{BCFWop} and then integrating over $t$:
\be{recur1}
-\int_{\C}\frac{\d t}{t}\wedge\bar{\delta}\la W[C(t)]\cO(y)\ra =\int_{\C} \frac{\d t}{t}\wedge \left( \Lambda^{\mathrm{tree}}+\Lambda^{\mathrm{MHV}}+\Lambda^{\mathrm{Op}}\right),
\ee
where the $\Lambda$s are given by the deformation contributions we just calculated.  Integration by parts immediately gives
\begin{equation*}
-\int_{\C}\frac{\d t}{t}\wedge\bar{\delta}\la W[C(t)]\cO(y)\ra = \left\la W[1,2,\ldots , n]\cO(y)\right\ra-\left\la W[1,2,\ldots ,n-1]\cO(y)\right\ra,
\end{equation*}
which is just the difference in the correlators at $t=0$ and $t=\infty$.  

Let us consider the contribution $\Lambda^{\mathrm{tree}}$ explicitly; the other two contributions can be treated in an identical manner.  Using \eqref{BCFWt}, we have
\begin{multline*}
\int_{\C}\frac{\d t}{t}\wedge \Lambda^{\mathrm{tree}}=  \int\limits_{\C\times C(t)\times C(t)} \frac{\d t}{t}\wedge\omega(Z)\wedge\omega(Z')\wedge\bar{\delta}^{3|4}(Z,Z') \\
\times \left\la W[C'(t)]W[C''(t)]\cO(y)\right\ra,
\end{multline*}
where $\bar{\delta}^{3|4}(Z,Z')$ has support only when the curve $C(t)$ intersects itself.  For every $j=3,\ldots n-1$ there will be some value of $t=t_{j}$ for which the line $(\hat{n}(t_{j}),1)$ intersects $(j-1,j)$.  If we label those intersection points as $I_{j}$, then clearly we have \cite{Bullimore:2011ni}
\begin{eqnarray*}
C'(t_{j})=(1,2)\cup(2,3)\cup\cdots\cup (j-1, I_{j}) \\
C''(t_{j})=(I_{j},j)\cup(j,j+1)\cup\cdots\cup(\hat{n}(t_{j}),1).
\end{eqnarray*}
For each such contribution at $t_{j}$ we can parametrize the positions of $Z$ and $Z'$ by
\begin{equation*}
Z=\widehat{Z_{n}}(t)+sZ_{1}=Z_{n}+tZ_{n-1}+sZ_{1}, \qquad Z'=Z_{j-1}+r Z_{j},
\end{equation*}
so the meromorphic differentials become
\begin{equation*}
\omega(Z)=\frac{\d s}{s}, \qquad \omega(Z')=\frac{\d r}{r}.
\end{equation*}

Thus, we have
\begin{multline}
\int_{\C}\frac{\d t}{t}\wedge \Lambda^{\mathrm{tree}}= \sum_{j=3}^{n-1}\int_{\C^{3}}\frac{\d t}{t}\frac{\d s}{s}\frac{\d r}{r}\wedge\bar{\delta}^{3|4}(Z_{n}+tZ_{n-1}+sZ_{1},Z_{j-1}+r Z_{j}) \\
\times \left\la W[1,\ldots, j-1, I_{j}]W[I_{j},j,\ldots, n-1,\hat{n}(t_{j})]\cO(y)\right\ra \\
= \sum_{j=3}^{n-1}[n-1,n,1,j-1,j]\left\la W[1,\ldots, j-1, I_{j}]W[I_{j},j,\ldots, n-1,\hat{n}_{j}]\cO(y)\right\ra,
\end{multline}
where $[A,B,C,D,E]$ is the standard R-invariant and we have abbreviated $\hat{n}(t_{j})=\hat{n}_{j}$.  We can perform similar parametrizations for $\Lambda^{\mathrm{MHV}}$ and $\Lambda^{\mathrm{Op}}$, leading to the full all-loop recursion relation:

\begin{propn}\label{recurpropn}
Let $W[C]=W[1,\ldots,n]$ be the Wilson loop in the fundamental representation around the n-cusp null polygon $C$, $\cO(y)$ be a local operator in general position, and $Y=\mathrm{span}\{Z_{C},Z_{D}\}$ be the $\P^{1}\subset\PT$ corresponding to this position.  Then
\begin{multline}\label{recur2}
\left\la W[1,\ldots, n]\cO(y)\right\ra = \left\la W[1,\ldots, n-1]\cO(y)\right\ra \\
+ \sum_{j=3}^{n-1}[n-1,n,1,j-1,j]\left\la W[1,\ldots, j-1, I_{j}]W[I_{j},j,\ldots, n-1,\hat{n}_{j}]\cO(y)\right\ra \\
+\lambda \oint\limits_{\Gamma\times S^{1}\times S^{1}}\D^{3|4}Z_{A}\wedge\D^{3|4}Z_{B} [n-1,n,1,A,B]\left\la W[1,\ldots, n-1,\hat{n}_{AB},Z',B]\cO(y)\right\ra \\
+ \oint\limits_{\tilde{\Gamma}\times S^{1}\times S^{1}} \d\mu^{abab}[n-1,n,1,C,D]\left\la W[1,\ldots, n-1, \hat{n}_{CD},Z',D]\right\ra,
\end{multline}
where the measure $\d\mu^{abab}$ is given by \eqref{fmeasure}; the contours $\Gamma$ and $\tilde{\Gamma}$ are over $(4|8)$- and $(0|4)$-dimensional real slices of the space of lines in $\PT$ respectively; and the contours $S^{1}\times S^{1}$ ensure $Z_{A,B}, Z_{C,D}\rightarrow Z'$ in their respective integrals.\footnote{As mentioned earlier, recall that in the planar limit $\la W[1,\ldots, j-1, I_{j}]W[I_{j},j,\ldots, n-1,\hat{n}_{j}]\cO(y)\ra = \la W[1,\ldots, j-1, I_{j}]\ra\; \la W[I_{j},j,\ldots, n-1,\hat{n}_{j}]\cO(y)\ra +\la W[1,\ldots, j-1, I_{j}]\cO(y)\ra\; \la W[I_{j},j,\ldots, n-1,\hat{n}_{j}]\ra$, so this indeed constitutes a recursion relation.}
\end{propn}

\subsubsection*{\textit{Some loop integrand computations}}

In the study of scattering amplitudes, the primary utility of the all-loop BCFW recursion relations has been their ability to enable simple computations of loop integrands (e.g., \cite{ArkaniHamed:2010kv}).  The recursion relation we have just defined holds at the level of the loop integrand of the correlator $\la W[1,\ldots, n]\cO(y)\ra$; this will be a rational function of the nodes $Z_{i}$, the line $Y$ as indexed by $Z_{C},Z_{D}$, and the loops as indexed by an internal region coordinate $X=\mathrm{span}\{Z_{A},Z_{B}\}$.  In computing the $l$-loop \emph{integral} of the correlator, the internal regions must be integrated over using
\begin{equation*}
\d^{4|8}X=\frac{\d^{4|4}Z_{A}\wedge\d^{4|4}Z_{B}}{\mathrm{vol}\;\GL(2,\C)},
\end{equation*}
in the usual fashion.  Denoting the $l$-loop integrand of the correlator by $G^{l}_{n}$, this means that
\be{cintegrand1}
G^{l}_{n}=G^{l}_{n}\left(Z_{1},\ldots, Z_{n},C,D; (A,B)_{1},\ldots, (A,B)_{l}\right),
\ee
with an implicit symmetrization over the loop variables.  This integrand can be further expanded in the fermionic twistor variables:
\be{cintegrand2}
G^{l}_{n}=G^{l}_{n,0}+G^{l}_{n,1}+G^{l}_{n,2}+\cdots +G^{l}_{n,n-4},
\ee
where $G^{l}_{n,k}$ is of order $4k$ in $\chi$, and can be thought of as the analogue of a N$^k$MHV integrand for our mixed correlators.

In this language, we can re-write our recursion relation in a slightly more appealing fashion:
\begin{multline}\label{recur3}
G^{l}_{n,k}=G^{l}_{n-1,k} \\
+\sum_{n_{1},k_{1},l_{1},j} [n-1,n,1,j-1,j] W^{l_{1}}_{n_{1},k_{1}}(1,\ldots, j-1,I_{j})\;G^{l_{2}}_{n_{2},k_{2}}(I_{j},j,\ldots, n-1,\hat{n}_{j}) \\
+\lambda \int \frac{\d^{4|4}Z_{A}\wedge\d^{4|4}Z_{B}}{\mathrm{vol}\;\GL(2,\C)} [n-1,n,1,A,B]\;G^{l-1}_{n+2,k+1}(1,\ldots, n-1,\hat{n}_{AB},\hat{A},B) \\
+\int \d^{0|4}\theta_{CD}\;  [n-1,n,1,C,D]\;W^{l}_{n+2,k}(1,\ldots, n-1,\hat{n}_{CD},\hat{C},D),
\end{multline}
where $W^{l}_{n,k}$ is the usual $l$-loop integrand of the Wilson loop and
\begin{eqnarray*}
n_{1}+n_{2}=n+2, \qquad k_{1}+k_{2}=k-1, \qquad l_{1}+l_{2}=l, \\
\hat{A}=(A,B)\cap (n-1,n,1), \qquad \hat{C}=(C,D)\cap (n-1,n,1).
\end{eqnarray*}
Since the $l$-loop integrand for the Wilson loop is known \cite{Mason:2010yk}, this makes it possible for us to compute the integrands $G^{l}_{n,k}$ recursively.

For instance, consider the tree-level analogue of the NMHV amplitude: $G^{0}_{n,1}$.  If we perform an implicit summation over the possible BCFW-type shifts, then \eqref{recur3} gives:
\begin{multline}\label{NMHVtree}
G^{0}_{n,1}=\sum_{i<j}[i-1,i,1,j-1,j] \mathbb{I}\times \mathbb{I} \\
+\int \d^{0|4}\theta_{CD}\;\sum_{i}[i-1,i,1,C,D]\left(\sum_{j<k} [j-1,j,1,k-1,k]\right),
\end{multline}
where the range of the second summation in the second term is over $1,\ldots,n-1,\hat{n}_{CD},\hat{C},D$.  Note that as predicted, $G_{n,1,0}=G_{n,1,0}(Z_{1},\ldots,Z_{n},C,D)$.  Usually we consider mixed correlators which are normalized by the expectation value of the Wilson loop $\la W[1,\ldots, n]\ra$; including this in the present calculation has the effect of eliminating the first term in \eqref{NMHVtree}, as it corresponds to the NMHV contribution from the Wilson loop itself.

If we wanted to compute the analogue of a 1-loop MHV integrand for our correlator, a quick inspection of \eqref{recur3} shows that we know all the required ingredients:
\begin{multline*}
G^{1}_{n,0}=\lambda \int \frac{\d^{4|4}Z_{A}\wedge\d^{4|4}Z_{B}}{\mathrm{vol}\;\GL(2,\C)} \sum_{i}[i-1,i,1,A,B]\;G^{0}_{n+2,1}(1,\ldots, n-1,\hat{n}_{AB},\hat{A},B)\\
+ \int \d^{0|4}\theta_{CD}\sum_{i}[i-1,i,1,C,D]\;W^{1}_{n+2,0}(1,\ldots, n-1,\hat{n}_{CD},\hat{C},D).
\end{multline*}
Using \eqref{NMHVtree} as well as the known contributions from the Wilson loop \cite{ArkaniHamed:2010kv, Mason:2010yk} gives
\begin{multline}\label{MHVloop}
G^{1}_{n,0}(Z_{1},\ldots,Z_{n},C,D;\;(A,B)_{1}) = \\
\lambda \int \frac{\d^{4|4}Z_{A}\wedge\d^{4|4}Z_{B}}{\mathrm{vol}\;\GL(2,\C)} \sum_{i}[i-1,i,1,A,B]\left(\sum_{j<k}[j-1,j,1,k-1,k]\right. \\
\left. +\int \d^{0|4}\theta_{CD}\sum_{j} [j-1,j,1,C,D]\sum_{k<l}[k-1,k,1,l-1,l]\right) \\
+\lambda \int \frac{\d^{4|4}Z_{A}\wedge\d^{4|4}Z_{B}}{\mathrm{vol}\;\GL(2,\C)}\d^{0|4}\theta_{CD}\sum_{i}[i-1,i,1,C,D]\sum_{j<k}[1,j-1,j,A,B'][1,k-1,k,A,B''],
\end{multline}
where the first sum in each term ranges from $i=1,\ldots, n$ and the remaining sums range over $1,\ldots,n-1,\hat{n}_{CD},\hat{C},D$.  In the second line, the shifted twistors $B'$, $B''$ correspond to the intersections between lines and planes given by:
\begin{equation*}
B'=(A,B)\cap(1,k-1,k), \qquad B''=(A,B)\cap(1,j-1,j).
\end{equation*}

\medskip

Of course, many of the terms here will actually vanish upon performing the fermionic integrals, and more will be subtracted from the first term if the quotient by the pure Wilson loop integrand is included.  It would be interesting to compare the results for the integrand generated by our recursion relation against other computations, such as the $\bar{Q}$-anomaly techniques of \cite{Bullimore:2011kg}.  

It is also worth mentioning that mixed Wilson loop / local operator correlators can be studied from a very different perspective when the Wilson loop under consideration is (topologically) circular, rather than a null polygon.\footnote{Although the Wilson loops considered in this setting only couple to three of the scalars of $\cN=4$ SYM \cite{Drukker:2007dw}, rather than the full superconnection $\CA$ as we considered here.}  In particular, if the Wilson loop is defined on a $S^{2}\subset\M$, then the configuration with an arbitrary number of scalar chiral primary operators also inserted on the sphere is $1/8$-BPS, and the computation can be localized to two-dimensional Yang-Mills theory on the sphere \cite{Pestun:2009nn}.  This in turn allows one to compute the correlator to all values of the coupling via a matrix model calculation (c.f., \cite{Giombi:2012ep} and the references therein).  It would be interesting to know if twistor theory has anything to add to this perspective, since it entails non-null data and non-perturbative results.      

Rather than pursue these issues further, we will now turn to the study of gravity, and attempt to apply the methods we have used in gauge theory to that new setting.


\section{Twistor Actions, Conformal and Einstein Gravity}
\label{Chapter5}

In the previous sections, we saw that by studying gauge theory on twistor space, we were able to learn many interesting things about the physical theory.  In particular, efficient calculational mechanisms like the MHV formalism were manifested explicitly on twistor space, and computations involving gauge invariant local operators and null polygonal Wilson loops were also streamlined.  It seems natural to ask if similar insights can be found in the study of gravity via twistor methods.

As one might expect, the story is much more complicated in this setting.  Dealing with generally curved space-times is a long-standing difficulty for twistor theory, referred to as the `googly problem' \cite{Penrose:1999cw}.  Twistor-string theory provides a perturbative solution to the googly problem for gauge theory, and there was hope that it would yield a similar mechanism for the study of gravity.  However, all twistor-string theories based on Witten's model contain conformal gravity degrees of freedom \cite{Berkovits:2004jj}; this theory has fourth-order equations of motion and is widely considered to be non-physical (see \cite{Fradkin:1985am} for a review).  Indeed, any attempt to remove these degrees of freedom by a gauging appears to result in a free theory which misses an entire self-duality sector \cite{AbouZeid:2006wu, Nair:2007md}.  The twistor-string of Skinner appears to correctly describe Einstein gravity (at least in the flat space limit) \cite{Skinner:2013xp}, but it is not clear in what way it connects to an action principle for gravity itself.  However, a twistor action for conformal gravity has been known for some time \cite{Mason:2005zm}.

There is also a mixture of similarities and differences between the basic structures of scattering amplitudes in gauge theory and gravity.  Graviton amplitudes posses the same `MHV-like' structure of gauge theory amplitudes, in the sense that $n$-graviton amplitudes involving $n$ or $n-1$ gravitons of the same helicity vanish.\footnote{The notion of helicity is well-defined in general relativity provided one restricts to positive-frequency fields \cite{Ashtekar:1986}.}  However, the functional form of gravity amplitudes is more complicated than their gauge theory counterparts.  This is due to the underlying permutation invariance of gravity, as there is no color trace to enforce a cyclic ordering on external particles.  Indeed, the analogue of a Parke-Taylor amplitude for gravity (Hodges' formula) was only recently discovered \cite{Hodges:2012ym}.

As it turns out, the ability for us to treat conformal gravity twistorially is actually an advantage rather than an obstruction, at least at tree-level.  This is due to an observation of Maldacena \cite{Maldacena:2011mk} that the tree-level S-matrices of these two theories are equivalent on a de Sitter background when Einstein scattering states are used.  In this section, we study the conformal gravity twistor action, with a view to extracting the Einstein gravity subsector.  After reviewing some basic facts about twistor theory for curved backgrounds, we give a brief summary of the Maldacena argument and apply it to amplitude generating functionals in Einstein and conformal gravity.  We then perform the reduction to Einstein gravity at the level of the twistor action, extracting a twistorial expression for the MHV amplitude generating functional.  This procedure also leads to a proposal for the twistor action of Einstein gravity itself.


\subsection{Background}

While our study of gauge theory took place on `flat' twistor space associated to Minkowski space-time, gravity requires twistor machinery adapted to curved backgrounds (possibly with cosmological constant).  We begin with a brief review of the necessary background material for this `curved' twistor theory, including some basic facts about de Sitter space, the non-linear graviton construction, and local twistor connection.  The reader need only consult the references for further details.

\subsubsection*{\textit{de Sitter geometry}}

The homogeneous Einstein geometries of Minkowski, de Sitter, and anti-de Sitter space are the simplest solutions to the field equations of general relativity: they are space-times with only scalar curvature (in the form of a cosmological constant), and are hence conformally flat (c.f., \cite{Hawking:1973}).  In four dimensions, each has a conformal compactification which is topologically $S^{1}\times S^{3}/\Z_{2}$ and can be realized as a quadric in $\RP^{5}$.  Although we will focus on de Sitter space (when the cosmological constant is positive) in much of what follows, many of the results in both this section and Section \ref{Chapter6} hold for \emph{anti-}de Sitter space as well.
\begin{figure}
\centering
\includegraphics[width=2 in, height=1.7 in]{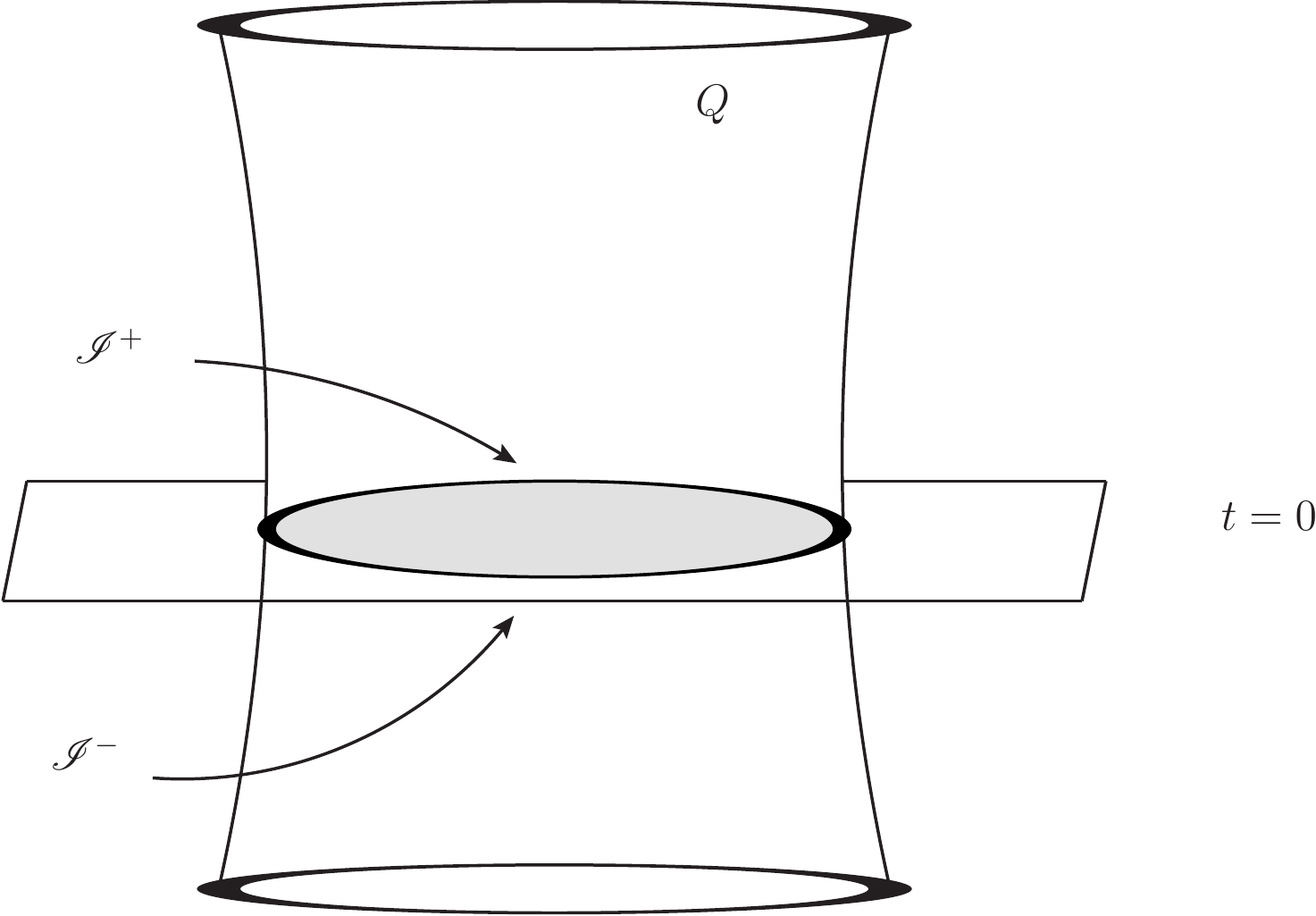}\caption{\textit{de Sitter space as the quadric} $Q\subset\RP^{5}$ \textit{and the identification of infinity.}}\label{dS1}
\end{figure}   

Before conformal compactification, de Sitter space is topologically $\R\times S^{3}$, and can be realized as the pseudosphere in $\R^{1,4}$ with coordinates $(w, x^{\mu})$, $\mu=0,\ldots, 3$ via the embedding relation:
\begin{equation*}
\eta_{\mu\nu}x^{\mu}x^{\nu}-w^{2}=x^{2}-w^{2}=-\frac{3}{\Lambda}, \qquad \eta_{\mu\nu}=\mathrm{diag}(1,-1,-1,-1),
\end{equation*}
where $\Lambda >0$ is the cosmological constant.  Writing de Sitter space in this fashion makes manifest its isometry group $\SO(1,4)$, which is the Lorentz group inherited from the embedding space.  We denote this space as $dS_{4}$.

The aforementioned conformal compactification embeds $dS_{4}$ into $\RP^{5}$ with homogeneous coordinates $(t,w,x^{\mu})$ as the $t\neq 0$ portion of the quadric:
\begin{equation*}
2Q\equiv t^{2}-w^{2}+x^{2}=0,
\end{equation*}
with scale-invariant metric
\be{dSmetric1}
\d s^{2}=\frac{3}{\Lambda}\frac{\d t^{2}-\d w^{2}+\eta_{\mu\nu}\d x^{\mu}\d x^{\nu}}{t^{2}}.
\ee
The intersection of $Q$ with the plane $t=0$ corresponds to the $S^{3}$ at infinity, and is the identification of the past ($\scri^{-}$) and future ($\scri^{+}$) infinities; see Figure \ref{dS1}.  Note that if we work on the patch where $t=\sqrt{3/\Lambda}$, then we recover the description of de Sitter space as the pseudosphere in $\R^{1,4}$. 

\begin{figure}
\centering
\includegraphics[width=3.25 in, height=1.5 in]{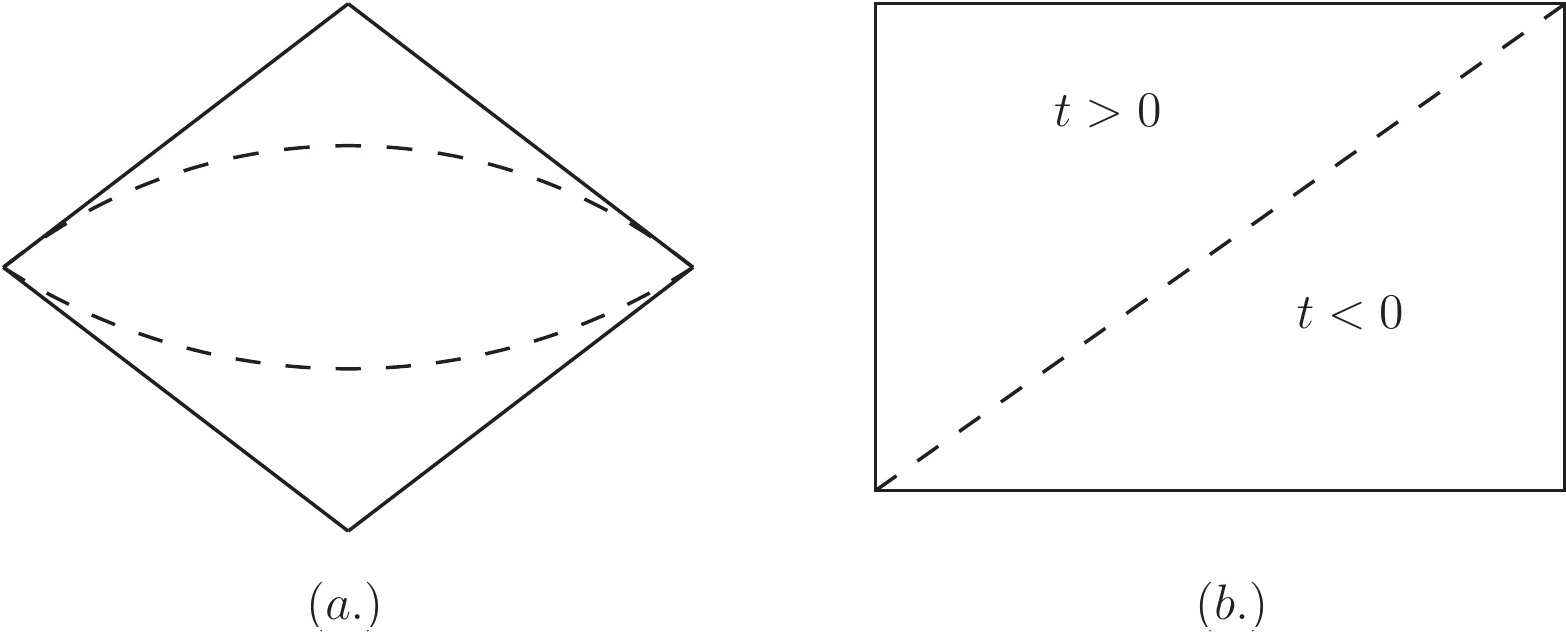}\caption{\textit{De Sitter space on the affine patch} (\emph{a}.), \textit{and the Poincar\'{e} patch} (\emph{b}.)}\label{dS2}
\end{figure}

Two particularly useful coordinate patches on de Sitter space are the affine and Poincar\'e patches.  The \emph{affine} patch is $t+w=1$; after a proper re-scaling of the affine coordinates $x^{\mu}$ the metric for this patch becomes
\be{dSmetric2}
\d s^{2}=\frac{\eta_{\mu\nu}\d x^{\mu} \d x^{\nu}}{(1-\Lambda x^{2})^2}.
\ee
In a sense, working with this slicing of global de Sitter space is rather awkward: de Sitter infinity is represented by finite points in the affine space where $x^{2}=\Lambda^{-1}$, and \emph{vice versa}.  Here, the null infinity of the affine space intersects the infinity of $dS_{4}$ in a $S^{2}$ at spatial infinity.  The main advantage of working with this slicing is that it is well-behaved in the $\Lambda\rightarrow 0$ limit: in this case \eqref{dSmetric2} simply becomes the usual Minkowski metric (see Figure \ref{dS2}, (\emph{a}.)).  

The more conventional \emph{Poincar\'{e}} patch of de Sitter space, where $x^{0}+w=1$, has metric:
\be{dSmetric3}
\d s^{2}=\frac{3}{\Lambda}\frac{\d t^{2}-\delta_{ij}\d x^{i}\d x^{j}}{t^{2}},
\ee
and $t=0$ is infinity minus a point.  The light cone of this point divides global de Sitter into two halves ($t>0$ and $t<0$) corresponding to what a physical observer situated at $\scri^{\pm}$ could actually see.  This slicing also manifests the three-dimensional rotation and translation symmetries of $dS_{4}$, but is certainly not well-behaved in the $\Lambda\rightarrow 0$ limit; see Figure \ref{dS2}, (\emph{b}.).

\subsubsection*{\textit{Non-linear graviton}}

It is natural to ask if the twistor formalism used in Sections \ref{Chapter2}-\ref{Chapter4} extends to the study of curved space-times.  The following result, known as the \emph{non-linear graviton} construction, establishes precisely how this can happen:
\begin{thm}[Penrose \cite{Penrose:1976js}, Ward \cite{Ward:1980am}]\label{NLG}
There is a one-to-one correspondence between: \emph{(a.)} self-dual space-times\footnote{A self-dual (SD) space-time is one whose anti-self-dual Weyl curvature and trace-free Ricci tensor vanish.} $M$, and \emph{(b.)} twistor spaces $\CPT$, a complex projective 3-manifold obtained as a complex deformation of $\PT$, containing a rational curve $X_{0}$ with normal bundle $\cN_{X_0}\cong\cO(1)\oplus \cO(1)$.  

There is a metric in this self-dual conformal class with scalar curvature $R=4\Lambda$ if and only if $\CPT$ is equipped with:
\begin{itemize}
\item a non-degenerate holomorphic contact structure specified by $\tau\in\Omega^{1,0}(\CPT, \cO(2))$, and
\item a holomorphic 3-form $\D^{3}Z\in\Omega^{3,0}(\CPT,\cO(4))$ obeying $\tau\wedge \d\tau=\frac{\Lambda}{3}\D^{3}Z$.
\end{itemize}
\end{thm}
Here, the line bundle $\cO(1)\rightarrow\CPT$ is defined to be the dual of the fourth-root of $\Omega^{3}(\CPT)\cong\cO(-4)$; this exists on a neighborhood of the rational curve $X_{0}$ by assumption on $\cN_{X_0}$.  The non-projective curved twistor space $\mathscr{T}$ is also defined as the total space of $\cO(-1)\rightarrow\CPT$.

The requirement that $\CPT$ arise as a (integrable) complex deformation of $\PT$ results in a four-parameter family of rational curves $\{X\}_{x\in\C^4}$ in a neighborhood of $X_0$, each with normal bundle $\cN_{X}\cong\cO(1)\oplus\cO(1)$.  This is a consequence of Kodaira-Spencer theory.  Thus, points $x\in M$ (for $M$ obeying the conditions of this theorem) correspond to rational, but not necessarily linearly embedded, curves $X\subset\CPT$.  The self-dual conformal structure on $M$ corresponds to requiring that if two of these curves $X,Y$ intersect in $\CPT$, then the points $x,y\in M$ are null separated.  

In the Einstein case, the contact 1-form $\tau$ serves as a holomorphic measure on these curves, while $\D^{3}Z$ provides a holomorphic measure on twistor space itself, as our notation suggests.  Furthermore, it is known that $\CPT$ is uniquely associated with $M$, in the sense that any two space-times which have the same twistor space will be isomorphic in a neighborhood of conformal infinity \cite{LeBrun:1982}.  The other important tools of twistor theory on $\PT$--namely the Penrose transform and Ward correspondence--still hold for $\CPT$ as well \cite{Hitchin:1980hp}.

As usual, $\CPT$ fits into the twistor double fibration:
\begin{equation*}
\xymatrix{
 & \PS \ar[ld]_{p} \ar[rd]^{q} & \\
 \CPT & & M }
\end{equation*}
Provided $M$ is not curved too severely, it follows that $\PS\cong M\times\P^{1}$, which can be charted with $(x^{\mu},\sigma_{A})$.  With this assumption, the map $q: \PS\rightarrow M$ is the trivial projection, while $p:\PS\rightarrow\CPT$ is specified by the generalized incidence relations:
\be{incidence}
Z^{\alpha}:M\times\P^{1}\rightarrow\CPT, \qquad Z^{\alpha}=Z^{\alpha}(x^{\mu},\sigma_{A})=\left(\lambda_{A}(x,\sigma), \mu^{A'}(x,\sigma)\right).
\ee
For consistency with the case $M=\M$, we demand that $Z^{\alpha}$ be homogeneous of degree one in $\sigma$.  In the Einstein case, when $\Lambda=0$ the contact structure $\tau$ becomes degenerate and we have a fibration $\CPT\rightarrow\P^{1}$; in this case `one-half' of the incidence relations become the identity map (i.e., $\lambda_{A}(x,\sigma)=\sigma_{A}$).

According to theorem \ref{NLG}, $\CPT$ arises as a complex deformation of $\PT$, and $Z^{\alpha}$ must be holomorphic with respect to the deformed complex structure.  In a coordinate-free language, this complex structure is specified by an endomorphism $J:T_{\CPT}\rightarrow T_{\CPT}$ which squares to $J^2=-1$.  This induces a splitting of complexified tangent bundle into holomorphic and anti-holomorphic parts, and defines Dolbeault operators $\partial^{J}$, $\dbar^{J}$.  Integrability of $J$ corresponds to the vanishing of its Nijenhuis tensor, $N_{J}\in\Omega^{0,2}(\CPT, T^{1,0}_{\CPT})$.  So by theorem \ref{NLG}, the field equations for $M$ to be self-dual correspond to $N_{J}=0$, while the requirement that the map $Z^{\alpha}$ be holomorphic is $\dbar^{J}Z^{\alpha}=0$.

It will often be convenient for us to work in coordinates, with a specific choice of background complex structure.  Taking as our background the flat complex structure of $\PT$, we can denote the complex structure on $\CPT$ as a (small but finite) deformation:    
\begin{equation*}
\dbar_{f}=\dbar +f= \d \bar{Z}^{\bar{\alpha}} \frac{\partial}{\partial \bar{Z}^{\bar{\alpha}}}+ f,
\end{equation*}
for $f\in\Omega^{0,1}(\PT, T^{1,0}_{\PT})$.  The corresponding coordinate basis for $T^{0,1}_{\CPT}$ and $\Omega^{1,0}(\CPT)$ is then:
\begin{eqnarray}
T^{0,1}_{\CPT}=\mathrm{span}\left\{\frac{\partial}{\partial\bar{Z}^{\bar{\alpha}}}+f^{\alpha}_{\bar{\alpha}}\frac{\partial}{\partial Z^{\alpha}}\right\}, \label{gbasis} \\
\Omega^{1,0}(\CPT)=\mathrm{span}\{\D Z^{\alpha}\}=\mathrm{span}\left\{\d Z^{\alpha}-f^{\alpha}\right\}, \label{gfbasis}
\end{eqnarray}
where we have denoted $f=f^{\alpha}\partial_{\alpha}=f^{\alpha}_{\bar{\alpha}}\d \bar{Z}^{\bar{\alpha}}\partial_{\alpha}$.  The requirement that the form $f^{\alpha}$ descend from $\mathscr{T}$ to $\CPT$ is satisfied so long as
\be{fgf}
\partial_{\alpha}f^{\alpha}=0, \qquad \bar{Z}^{\bar{\alpha}}f^{\beta}_{\bar{\alpha}}=0.
\ee 

With this choice of background, the integrability of the complex structure is equivalent to
\be{contact}
\dbar_{f}^{2}=\left(\dbar f^{\alpha}+\left[f,f\right]^{\alpha}\right)\partial_{\alpha}=0, \qquad \left[f,f\right]^{\alpha}=f^{\beta}\wedge\partial_{\beta}f^{\alpha},
\ee
and holomorphicity of the map $Z^{\alpha}$ is
\be{holomap}
\dbar|_{X} Z^{\alpha}-f^{\alpha}(Z)=0,
\ee
where $\dbar|_{X} =\d\bar{\sigma}\frac{\partial}{\partial\bar{\sigma}}$ is the $\dbar$-operator on $X\subset\CPT$ pulled back to $\PS$.  This equation has a four-complex parameter family of solutions regardless of whether \eqref{contact} is satisfied \cite{Penrose:1976js, Hansen:1978jz}, and when $\Lambda=0$ it can be thought of as the good cut equation for $M$ \cite{Eastwood:1982, Adamo:2010ey}.

While we have focused on the $\cN=0$ version of the non-linear graviton here, the construction has a natural generalization to $\cN>0$--just like the Ward Correspondence for Yang-Mills instantons \cite{Wolf:2007tx}.

\subsubsection*{\textit{Local twistor formalism}}

If we hope to implement the Penrose transform concretely on curved twistor spaces, we must have a means of defining things like twistor indices.  That is to say, our twistor coordinates $Z^{\alpha}(x,\sigma)$ are abstract on $\CPT$ until they are pulled back to the spinor bundle $\PS$.  To get a concrete coordinate basis on the curved twistor space, we must use the \emph{local twistor formalism}.  Note that this formalism will make sense for any (complex) space-time $M$, whether or not it satisfies the conditions of theorem \ref{NLG}.

Local twistors are defined at points $x\in M$, and so constitute a complex rank-four bundle over space-time:
\begin{equation*}
\xymatrix{
Z^{\underline{\alpha}}=(\lambda_{A},\mu^{A'}) \ar[r] & \mathbb{LT} \ar[d]\\
& M }
\end{equation*}
Let $\mathbf{t}\in T_{x}M$ be a vector at $x$; we can compute the infinitesimal variation of the local twistor bundle in the direction of $\mathbf{t}$ as \cite{Penrose:1986ca}
\be{LTT1}
\nabla_{\mathbf{t}} Z^{\underline{\alpha}}(x) =\left(t^{BB'}\nabla_{BB'}\lambda_{A}-it^{BB'}P_{ABA'B'}\mu^{A'},\;t^{BB'}\nabla_{BB'}\mu^{A'}-it^{BA'}\lambda_{B}\right),
\ee
where the tensor $P_{\mu\nu}$ is given by:
\begin{equation*}
P_{ABA'B'}=\Phi_{ABA'B'}-\Lambda\epsilon_{AB}\epsilon_{A'B'},
\end{equation*}
with $\Phi_{ABA'B'}$ the trace-free portion of the Ricci tensor. This local twistor transport along the vector $\mathbf{t}$ defines a \emph{local twistor connection}, which is equivalent to the Cartan conformal connection on $M$ \cite{Friedrich:1977}.    

The curvature of the local twistor connection is computed by considering
\be{ltc1}
i\left(\nabla_{\mathbf{t}}\nabla_{\mathbf{u}}-\nabla_{\mathbf{u}}\nabla_{\mathbf{t}}-\nabla_{[\mathbf{t},\mathbf{u}]}\right)Z^{\underline{\beta}}=Z^{\underline{\alpha}}F_{\underline{\alpha}}^{\underline{\beta}}(\mathbf{t},\mathbf{u}).
\ee
For a general complex space-time, we find \cite{Penrose:1986ca}:
\begin{equation*}
F^{\underline{\beta}}_{\underline{\alpha}}(\mathbf{t},\mathbf{u})=\left(
\begin{array}{cc}
it^{C}_{D'}u^{DD'}\Psi_{CDB}^{A} & t_{D}^{C'}u^{DD'}\nabla^{A}_{A'}\widetilde{\Psi}^{B'A'}_{C'D'}+t^{C}_{D'}u^{DD'}\nabla^{B'}_{B}\Psi^{BA}_{CD} \\
0 & -it^{C'}_{D}u^{DD'}\widetilde{\Psi}_{C'D'A'}^{B'}
\end{array}\right),
\end{equation*}
where $\Psi_{ABCD}$ and $\widetilde{\Psi}_{A'B'C'D'}$ are the ASD and SD Weyl spinors respectively.  So on a SD background $M$, the local twistor bundle $\LT$ is half-flat and the Ward transform applies \cite{Hitchin:1980hp}.  In other words, when $M$ satisfies the conditions of the non-linear graviton construction, we can obtain a rank-four bundle $\T^{\underline{\alpha}}\rightarrow\CPT$ by applying the Ward correspondence to $\LT\rightarrow M$.  More formally, it can be shown that $\T^{\underline{\alpha}}\cong (J^{1}\cO(-1))^{\vee}$, where $J^1$ is the first jet bundle \cite{LeBrun:1986}. Abusing terminology, we also refer to this bundle $\T^{\underline{\alpha}}\rightarrow\CPT$ as the `local twistor bundle.'

The bundle $\T^{\underline{\alpha}}$ lets us assign meaning to tensors on $\CPT$.  In particular, by choosing a holomorphic frame $H^{\underline{\alpha}}_{\alpha}$ for $\T^{\underline{\alpha}}$, we can translate twistor indices (in $\CPT$) into local twistor indices (in $\T^{\underline{\alpha}}$) \cite{Mason:1990}.  For instance, consider a tensor $f^{\alpha\cdots}_{\beta\cdots}\in H^{0,1}(\CPT,\cO(n-2))$ for $n<0$.  After contracting with the holomorphic frame, we get a $(0,1)$-form valued section of $\T^{\underline{\alpha}\cdots}_{\underline{\beta}\cdots}\otimes\cO(n-2)$, and can then apply the Penrose transform to obtain a field on $M$:
\begin{equation*}
\int_{X}\lambda_{A_{1}}\cdots\lambda_{A_{n}} f^{\underline{\alpha}\cdots}_{\underline{\beta}\cdots}\wedge\tau =\Gamma^{\underline{\alpha}\cdots}_{\underline{\beta}\cdots A_{1}\cdots A_{n}}, \qquad \nabla^{A_{1}A'}\Gamma^{\underline{\alpha}\cdots}_{\underline{\beta}\cdots A_{1}\cdots A_{n}}=0.
\end{equation*}
In the zero-rest-mass field equation, it is understood that the covariant derivative $\nabla^{AA'}$ acts via the local twistor connection on any local twistor indices of the object in question.  This is because the holomorphic frame $H^{\underline{\alpha}}_{\alpha}$ on $\T^{\underline{\alpha}}$ corresponds to a covariantly constant frame for $\LT\rightarrow M$.

From now on, we will drop the underline notation, and assume that the distinction between concrete and local twistor indices is clear from the context.  We can use \eqref{LTT1} to derive relevant expressions for how $\nabla$ acts on quantities with a single twistor index, say $\Gamma^{\beta}_{A\cdots}=(\Phi_{BA\cdots}, \Psi^{B'}_{A\cdots})$:
\be{ltc2}
\nabla^{AA'}\Gamma^{\beta}_{A\cdots}=\left(
\begin{array}{c}
\nabla^{AA'}\Phi_{BA\cdots} \\
\nabla^{AA'}\Psi^{B'}_{A\cdots}
\end{array}\right) + \left(
\begin{array}{cc}
0 & iP^{AA'}_{BB'} \\
i\epsilon^{AB}\epsilon^{A'B'} & 0 
\end{array}\right) \left(
\begin{array}{c}
\Phi_{BA\cdots} \\
\Psi^{B'}_{A\cdots}
\end{array}\right).
\ee
Similar rules for dual twistor indices as well as higher-rank tensors can be derived or looked up in \cite{Penrose:1986ca}.  The gauge freedom of such objects on space-time can be determined by computing the Penrose transform of $Z^{\gamma}f^{\alpha\cdots}_{\beta\cdots}$ and then imposing the condition $Z^{\beta}f^{\alpha\cdots}_{\beta\cdots}=0$ \cite{Mason:1990}.  


\subsection{Einstein Gravity from Conformal Gravity}

The main stumbling block for the twistor-string revolution was the presence of conformal gravity degrees of freedom \cite{Berkovits:2004jj}.  In the setting of twistor-strings, this appeared to correspond to non-minimal $\cN=4$ conformal supergravity (CSG) coupled to $\cN=4$ SYM.  As we saw in Sections \ref{Chapter3} and \ref{Chapter4}, this issue could be side-stepped in the study of gauge theory by working directly with a twistor action.  A first hope would be to attempt a similar procedure for the study of Einstein gravity; a particularly attractive route is presented by the embedding of Einstein gravity inside conformal gravity.  In this subsection, we review this embedding and derive a precise version of it at the level of MHV amplitudes.


\subsubsection{Conformal gravity}

Conformal gravity is obtained from the conformally invariant action
\be{CGA1}
S^{\mathrm{CG}}[g]=\frac{1}{\varepsilon^{2}}\int_{M}\d\mu\;C^{\mu\nu\rho\sigma}C_{\mu\nu\rho\sigma}=\frac{1}{\varepsilon^{2}}\int_{M}\d\mu\left(\Psi^{ABCD}\Psi_{ABCD}+\widetilde{\Psi}^{A'B'C'D'}\widetilde{\Psi}_{A'B'C'D'}\right),
\ee
where $\varepsilon^{2}$ is a dimensionless coupling constant, $\d\mu=\d^{4}x\sqrt{|g|}$ is the volume element, and $C_{\mu\nu\rho\sigma}$ is the Weyl curvature tensor.  The field equations of this action are the vanishing of the Bach tensor, $B_{\mu\nu}$, which can be written in a variety of different forms thanks to the Bianchi identities:
\begin{multline}\label{Bach}
B_{\mu\nu}=2\nabla^{\rho}\nabla^{\sigma}C_{\rho\mu\nu\sigma}+C_{\rho\mu\nu\sigma}R^{\rho\sigma} \\
=\left(2\nabla_{\rho}\nabla_{(\mu}R^{\rho}_{\nu)}-\Box R_{\mu\nu}-\frac{2}{3}\nabla_{\mu}\nabla_{\nu}R
-2R_{\rho\mu}R^{\rho}_{\nu}+\frac{2}{3}R_{\mu\nu}R
\right)_0\\
= 2(\nabla_{A'}^C\nabla_{B'}^D+\Phi^{CD}_{A'B'} )\Psi_{ABCD}=2(\nabla_{A}^{C'}\nabla_{B}^{D'}+\Phi^{C'D'}_{AB})\widetilde{\Psi}_{A'B'C'D'},
\end{multline}
where the subscript `0' denotes trace-free part. The last line implies that the field equations are satisfied whenever $M$ is Einstein, or when its Weyl curvature is either self-dual or anti-self-dual.  

In our study of Yang-Mills theory, the Chalmers-Siegel action \eqref{CS1} allowed us to expand around the SD sector.  We can perform a similar expansion for conformal gravity by first considering the action:\footnote{Note that the field equations of conformal gravity can be understood as the Yang-Mills equations of the local twistor connection \cite{Merkulov:1984nz}; hence, a Chalmers-Siegel-like expansion must exist.}
\be{CGA2}
S^{\mathrm{CG}}[g]=\frac{2}{\varepsilon^{2}}\int_{M}\d\mu\;\Psi^{ABCD}\Psi_{ABCD}.
\ee
This differs from \eqref{CGA1} by
\begin{equation*}
\frac{1}{\varepsilon^2}\int_{M}\d\mu\left(\Psi^{ABCD}\Psi_{ABCD}-\widetilde{\Psi}^{A'B'C'D'}\widetilde{\Psi}_{A'B'C'D'}\right),
\end{equation*}
which is equal to $\frac{12\pi^2}{\varepsilon^{2}}(\tau(M)-\eta(\partial M))$, where $\tau(M)$ is the signature of $M$ and $\eta(\partial M)$ is the $\eta$-invariant of the conformal boundary \cite{Hitchin:1997}.  Hence, \eqref{CGA2} is equal to the conformal gravity action up to a topological term which will be irrelevant in perturbation theory.

To expand around the SD sector, we introduce the totally symmetric spinor field $G_{ABCD}$ as a Lagrange multiplier, and write the action as \cite{Berkovits:2004jj}:
\be{CGA3}
S^{\mathrm{CG}}[g,G]=\int_{M}\d\mu \left(G^{ABCD}\Psi_{ABCD}-\varepsilon^{2}G^{ABCD}G_{ABCD}\right).
\ee 
This has field equations \cite{Mason:2005zm}
\be{CGFE}
\Psi^{ABCD}=\varepsilon^{2}G^{ABCD}, \qquad \left(\nabla^{C}_{A'}\nabla^{D}_{B'}+\Phi^{CD}_{A'B'}\right)G_{ABCD}=0,
\ee
so integrating out $G$ returns \eqref{CGA2}.  But now $\varepsilon^{2}$ becomes a parameter for expanding about the SD sector: when $\varepsilon=0$, we have a SD solution.  This means that $G_{ABCD}$ can be thought of as a linear ASD solution propagating on the SD background, and $\varepsilon^{2}$ plays the role of the 't Hooft coupling $\lambda$ as an expansion parameter around the SD sector.


\subsubsection{Embedding Einstein gravity in conformal gravity}

We now review a recent argument by Maldacena, which states that on-shell and after imposing certain boundary conditions, the conformal gravity and Einstein-Hilbert actions agree on de Sitter space \cite{Maldacena:2011mk}.  Note that many of the claims we will make were originally stated for asymptotically hyperbolic Riemannian four-manifolds; their extension to Lorentzian space-times which are asymptotically de Sitter follows by analytic continuation.

The Einstein-Hilbert action in the presence of a cosmological constant is
\begin{equation*}
S^{\mathrm{EH}}[g]=\frac{1}{\kappa^{2}}\int_{M}\d\mu\; (R-2\Lambda), \qquad \kappa^{2}=16\pi G_{N}.
\end{equation*}
On a de Sitter space, the field equations are $R_{\mu\nu}=\Lambda g_{\mu\nu}$, so the action reads
\begin{equation*}
S^{\mathrm{EH}}[dS_{4}]=\frac{2\Lambda}{\kappa^2}\int_{dS_{4}}\d\mu =\frac{2\Lambda}{\kappa^2}V(dS_{4}),
\end{equation*}
where $V(M)$ is the volume of $M$.  For any asymptotically de Sitter manifold, this volume will be infinite, so the action functional must be modified by the Gibbons-Hawking boundary term \cite{Gibbons:1976ue}.  Additionally, we must include the holographic renormalization counter-terms (which also live on the boundary) in order to render the volume finite \cite{Balasubramanian:1999re, Skenderis:2002wp}.  This leaves us with the so-called renormalized Einstein-Hilbert action \cite{Miskovic:2009bm}:
\begin{equation*}
S^{\mathrm{EH}}_{\mathrm{ren}}[g]=\frac{1}{\kappa^{2}}\left[\int_{M}\d\mu\left(R-2\Lambda\right)-2\int_{\partial M} \d\tilde{\mu}\; K -\int_{\partial M}\d\tilde{\mu}\;\mathcal{L}_{\mathrm{ct}}\right],
\end{equation*}
where $\d\tilde{\mu}$ is the volume element on the boundary, $K$ is the extrinsic curvature of $\partial M$, and $\mathcal{L}_{\mathrm{ct}}$ is the holographic renormalization Lagrangian of counter-terms.  For instance, on de Sitter space
\be{EHren2}
\mathcal{L}_{\mathrm{ct}}[dS_4]=\frac{2}{\ell_{dS}}+\frac{\ell_{dS}}{2}\tilde{R},
\ee
where $\ell_{dS}$ is the de Sitter curvature radius and $\tilde{R}$ is the intrinsic curvature tensor of the conformal boundary.

The important message is the fact that $S^{\mathrm{EH}}_{\mathrm{ren}}[M]$ is finite, and
\begin{equation*}
S^{\mathrm{EH}}_{\mathrm{ren}}[M]=\frac{2\Lambda}{\kappa^2}V_{\mathrm{ren}}(M),
\end{equation*}
where $V_{\mathrm{ren}}$ is the renormalized volume of the space-time \cite{Graham:1999jg}.  In other words, the on-shell renormalized Einstein-Hilbert action is equal (up to a constant proportional to $\Lambda$) to the renormalized volume of the asymptotically de Sitter space-time.  The next step is to relate this observation to conformal gravity.

Suppose that $M$ were an abstract Riemannian 4-manifold which was compact without boundary.  Then the Chern-Gauss-Bonnet formula states that
\begin{equation*}
\chi(M)=\frac{1}{8\pi^{2}}\int_{M}\d\mu \left(C^{\mu\nu\rho\sigma}C_{\mu\nu\rho\sigma}-\frac{1}{2}R_{\mu\nu}R^{\mu\nu}+\frac{1}{6}R^{2}\right).
\end{equation*}
If $M$ were additionally Einstein $(R_{\mu\nu}=\Lambda g_{\mu\nu})$, then this would immediately imply that
\be{CGB1}
S^{\mathrm{CG}}[M]=\frac{8\pi^{2}\chi(M)}{\varepsilon^{2}}-\frac{2\Lambda^{2}}{3\varepsilon^{2}}V(M).
\ee
Of course, when $M$ is (Lorentzian) asymptotically de Sitter, the Chern-Gauss-Bonnet formula requires a boundary term, and the volume needs renormalization.  However, the left-hand side of \eqref{CGB1} is canonically defined and independent of the conformal compactification of $M$, so all that is required is to properly renormalize the right-hand side.

A remarkable theorem of Anderson tells us that the relationship \eqref{CGB1} continues to hold when the boundary terms for the Euler characteristic and volume are taken into account \cite{Anderson:2001}.  In other words, we have
\begin{equation*}
S^{\mathrm{CG}}[M]=\frac{8\pi^{2}\widehat{\chi}(M)}{\varepsilon^{2}}-\frac{2\Lambda^{2}}{3\varepsilon^{2}}V_{\mathrm{ren}}(M),
\end{equation*}
where $\widehat{\chi}$ is the renormalized Euler characteristic.  Working on an asymptotically de Sitter background, we will always be perturbing around the topologically trivial flat case ($\chi=0$), so we have:
\be{CGB2}
S^{\mathrm{CG}}[dS_{4}]=-\frac{\Lambda\;\kappa^{2}}{3\varepsilon^{2}}S^{\mathrm{EH}}_{\mathrm{ren}}[dS_{4}].
\ee

What does this tell us about the scattering amplitudes of the two theories, though?  The answer is obvious using the perturbiner formalism \cite{Rosly:1996vr, Rosly:1997ap}.  Formally, the tree-level S-matrix of any theory with fields $\phi$ and action $S[\phi]$ is obtained by first taking asymptotic states $\{\phi_{1},\ldots,\phi_{n}\}$ which are positive frequency at $\scri^{-}$ if incoming, and negative frequency at $\scri^{+}$ if outgoing.  We then construct a classical solution $\phi_{\mathrm{cl}}$ (the scattering background) such that $\phi_{\mathrm{cl}}-\sum_{i}\epsilon_{i}\phi_{i}$ is positive frequency at $\scri^{+}$ and negative frequency at $\scri^{-}$.  Then the tree-level scattering amplitude on this classical background is given by:
\be{perturbiner}
\cA(\phi_{1},\ldots,\phi_{n})=\left.\frac{\partial^{n} S\left[\phi_{\mathrm{cl}}-\sum^{n}_{i=1}\epsilon_{i}\phi_{i}\right]}{\partial\epsilon_{1}\cdots\partial\epsilon_{n}}\right|_{\epsilon_{1}=\cdots =\epsilon_{n}=0}.
\ee
Hence, if two theories agree on a classical background then the tree-level S-matrix of one can be computed with the other, provided the asymptotic states can be singled out in a coherent way.

Equation \eqref{CGB2} confirms that conformal and Einstein gravity agree (up to constants) on a classical de Sitter background.  We also know that Einstein solutions sit inside the space of all solutions to the Bach equations of conformal gravity.  All that remains is to show that asymptotic Einstein scattering states can be consistently singled out within the conformally invariant theory.  Maldacena argues that this can be done by employing `Neumann' boundary conditions on the metric as follows \cite{Maldacena:2011mk}. 

For any asymptotically de Sitter space-time, we can expand the line element in Fefferman-Graham coordinates \cite{Fefferman:1985}.  On the Poincar\'{e} patch of \eqref{dSmetric3}, this looks like:
\be{FG}
\d s^{2}=\frac{-\d t^{2}+\d x^{i}\otimes \d x^{j}\left( g^{(0)}_{ij}(x)-t^{2}g^{(2)}_{ij}(x)-t^{3}g^{(3)}_{ij}(x)+\cdots\right)}{-t^{2}}.
\ee     
The important point is that this expansion has no $O(t)$ term in the numerator; since a conformal transformation can be made to eliminate the $t^{-2}$ factor, this means that asymptotically de Sitter space-times are conformal to metrics which obey $\partial_{t} g|_{t=0}=0$.  This is a Neumann boundary condition on the metric, and it can be made gauge invariant via: the $t=0$ slice of $dS_{4}$ is totally geodesic with respect to the ambient metric \cite{Maldacena:2011mk}.  Since conformal gravity has fourth-order equations of motion, one expects it to have four solutions given a single momentum in a Fourier transform picture.  Restricting our attention to positive frequency fields should eliminate two of these solutions, while the Neumann boundary condition gets rid of a third.  As asymptotically de Sitter spaces are conformal to solutions respecting these conditions, it follows that the remaining solution must be the Einstein one.

Hence, calculation of conformal gravity amplitudes at tree-level restricted to Einstein states will give $-\Lambda/3$ times the corresponding Einstein amplitudes. In particular, they will degenerate as $\Lambda \rightarrow 0$, but by construction we will find that the $n$-point conformal gravity amplitude will be a polynomial of degree $n-1$ in $\Lambda$, so it will be relatively straightforward in practice to divide by $\Lambda$ and take $\Lambda\rightarrow 0$.  


\subsubsection{Graviton scattering in de Sitter space}

We begin by showing how the relationship between conformal and Einstein gravity is manifested for generating functionals of scattering amplitudes involving two negative helicity gravitons.  To do this, we use the chiral formulation of general relativity.  For a general space-time $M$, the metric is given by a tetrad of 1-forms as $\d s^{2}=\epsilon_{AB}\epsilon_{A'B'}e^{AA'}\otimes e^{BB'}$.  This information can be packaged nicely into three ASD 2-forms:
\begin{equation*}
\Sigma^{AB}=e^{A'(A}\wedge e^{B)}_{A'},
\end{equation*}
and combined with the ASD spin connection $\Gamma_{AB}$ to provide the basic variables for Plebanski's chiral formulation of gravity \cite{Plebanski:1977zz}.  In the presence of a cosmological constant, this action takes the form:
\be{eqn: PA}
S[\Sigma, \Gamma]=\frac{1}{\kappa^2}\int_{M} \left(\Sigma^{AB}\wedge F_{AB}-\frac{\Lambda}{6}\Sigma^{AB}\wedge\Sigma_{AB}\right),
\ee
where 
\be{eqn: ASDcurv}
F_{AB}=\d\Gamma_{AB}+\Gamma^{C}_{A}\wedge \Gamma_{BC}
\ee
is the curvature of the ASD spin connection.  This action produces two field equations, to which we append a third (the condition that $\Sigma^{AB}$ be derived from a tetrad) \cite{Capovilla:1991qb}:
\begin{eqnarray}
\D \Sigma^{AB} & = & 0, \label{FE1} \\
F_{AB} & = & \Psi_{ABCD}\Sigma^{CD}+\frac{\Lambda}{3}\Sigma_{AB}, \label{FE2} \\
\Sigma^{(AB}\wedge\Sigma^{CD)} & = & 0 . \label{FE3}
\end{eqnarray}
Here, $\D$ is the covariant derivative with respect to the ASD spin connection, so explicitly,
\begin{equation*}
\D\Sigma^{AB}=\d\Sigma^{AB}+2\Gamma^{(A}_{C}\wedge\Sigma^{B)C}.
\end{equation*}

In the context of graviton scattering amplitudes, the MHV amplitude with two negative helicity gravitons can be pictured geometrically as the classical scattering of these two gravitons off a SD background. This SD background will be built perturbatively from the $n-2$ positive helicity gravitons in a $n$-particle graviton MHV amplitude \cite{Mason:2008jy}.  In such a background, $\Psi_{ABCD}=0$, which means that \eqref{FE2} can be solved for $\Sigma$ in terms of $F$ and then \eqref{FE1} and \eqref{FE3} may be combined to give a condition on the curvature of the ASD spin connection.  Hence, a SD solution $(\Sigma_{0}, \Gamma_{0})$ obeys \cite{Capovilla:1990qi}:
\begin{eqnarray}
\Sigma_{0}^{AB} & = & \frac{3}{\Lambda} F^{AB}_{0}, \label{SD1} \\
F_{0(AB}\wedge F_{0\; CD)} & = & 0. \label{SD2}
\end{eqnarray}
If we now consider small perturbations away from this SD background of the form $\Sigma= \Sigma_{0}+\sigma_{0}$, $\Gamma = \Gamma_{0}+\gamma$, then we obtain a set of linearized field equations:
\begin{eqnarray}
\D_{0}\sigma^{AB} & = & -2\gamma^{(A}_{C}\wedge\Sigma^{B)C}_{0}, \label{LFE1} \\
\D_{0}\gamma_{AB} & = & \psi_{ABCD}\Sigma^{CD}_{0}+\frac{\Lambda}{3}\sigma_{AB}, \label{LFE2} \\
\sigma^{(AB}\wedge\Sigma^{CD)}_{0} & = & 0, \label{LFE3}
\end{eqnarray}
where $\D_{0}$ is the covariant derivative with respect to the background ASD spin connection $\Gamma_{0}$.

\begin{lemma}\label{ZRM}
The linearized field $\psi_{ABCD}=\psi_{(ABCD)}$ may be interpreted as linearized ASD Weyl spinor propagating on the SD background $(\Sigma_{0},\Gamma_{0})$.
\end{lemma}
\proof It suffices to show that $\psi_{ABCD}$ obeys the zero-rest-mass equation for spin $-2$ fields: $\nabla^{AA'}\psi_{ABCD}=0$, where $\nabla$ is the background connection.  Act on both sides of \eqref{LFE2} with the background covariant derivative $\D_{0}$:
\begin{equation*}
\D_{0}^{2}\gamma_{AB}= 2F_{0\; C(A}\wedge\gamma_{B)}^{C}=(\D_{0}\psi_{ABCD})\Sigma^{CD}_{0}+\frac{\Lambda}{3}\D_{0}\sigma_{AB}.
\end{equation*}
Now use \eqref{LFE1} and \eqref{SD1} to obtain
\begin{equation*}
2F_{0\; C(A}\wedge\gamma_{B)}^{C}=(\D_{0}\psi_{ABCD})\Sigma^{CD}_{0}-2F_{0 (A}^{C}\wedge\gamma_{B)C} \qquad \Rightarrow \D_{0}\psi_{ABCD}=0,
\end{equation*}
as required.     $\Box$

\medskip

Geometrically, we can conceptualize the framework of linearized solutions on a SD background in the following way.  Let $\mathcal{S}$ be the space of solutions to the full field equations \eqref{FE1}-\eqref{FE3}; solutions to the linearized equations \eqref{LFE1}-\eqref{LFE3} form a vector space $V$.  We identify $V$ with the tangent space to $\mathcal{S}$ over the point $(\Sigma_{0},\Gamma_{0})$ representing an SD solution: $T_{(\Sigma_{0},\Gamma_{0})}\mathcal{S}= V$.  The vector space $V$ itself can be split into SD and ASD sectors by a short exact sequence resolution.  A linearized SD solution is completely characterized by the ASD spin connection, since the linearized SD field equations read
\be{eqn: SDLFE}
\sigma_{AB} =\frac{3}{\Lambda}\D_{0}\gamma_{AB}, \qquad \D_{0}\gamma^{(AB}\wedge F_{0}^{CD)} = 0.
\ee
Hence, we define the SD portion of $V$ by
\begin{equation*}
V^{+}=\left\{(\sigma,\gamma)\in V \: : \: \D_{0}\gamma^{(AB}\wedge F_{0}^{CD)} = 0\right\}.
\end{equation*}
We can therefore define  $V^{-}$ by the quotient map in the short exact sequence:
\begin{equation*}
0\longrightarrow V^{+} \hookrightarrow V \longrightarrow V^{-} \longrightarrow 0,
\end{equation*}
with
\begin{equation*}
V^{-}\equiv V/ V^{+}= \left\{(\sigma,\gamma)\in V\right\} / \left\{\gamma \: : \: \D_{0}\gamma^{(AB}\wedge F_{0}^{CD)} = 0\right\} .
\end{equation*}

The space of solutions to the field equations $\mathcal{S}$ comes equipped with a natural symplectic form $\omega$ given by the boundary term in the action \cite{Ashtekar:2008jw}:
\be{eqn: symp}
\omega = \frac{1}{\kappa^{2}}\int_{C}\delta\Sigma^{AB}\wedge\delta\Gamma_{AB},
\ee
where $C$ is a Cauchy surface in $M$ (when $\Lambda>0$, there is always a slicing where $C\cong S^3$ topologically) and $\delta$ is the exterior derivative on $\mathcal{S}$.  We have the following lemma:
\begin{lemma}
The form $\omega$ on $\mathcal{S}$ is independent of choice of Cauchy surface, and defines a symplectic form on $\mathcal{S}/\mathrm{Diff}^{+}_{0}(M)$.
\end{lemma}
\proof Let $C_{1}$ and $C_{2}$ be any two Cauchy surfaces in $M$ bounding some region $R$.  Then using $\delta^2=0$ and Stokes' theorem, it follows that
\begin{equation*}
\int_{C_{1}-C_{2}}\delta\Sigma^{AB}\wedge\delta\Gamma_{AB} = \delta\int_{\partial R}\Sigma^{AB}\wedge\delta\Gamma_{AB} = \delta \int_{R} \d\left(\Sigma^{AB}\wedge\delta\Gamma_{AB}\right).
\end{equation*}
Assuming the field equations hold in $R$, \eqref{FE1} implies that 
\begin{multline*}
\delta \int_{R} \d\left(\Sigma^{AB}\wedge\delta\Gamma_{AB}\right)= \delta\left( \int_{R} -2\Gamma^{(A}_{C}\wedge\Sigma^{B)C}\wedge\delta\Gamma_{AB}+\Sigma^{AB}\wedge\d\delta\Gamma_{AB} \right) \\
=\delta \left(\int_{R} \Sigma^{AB}\wedge\D\delta\Gamma_{AB}\right) \sim \delta (\delta S[\Sigma,\Gamma]) =0,
\end{multline*}
by the nilpotency of $\delta$ on $\mathcal{S}$.  Hence, $\omega$ is independent of choice of Cauchy surface, and is furthermore invariant under diffeomorphisms of $M$ as well as rotations of the spin frame (since all spinor indices are contracted).  By definition, it is easy to see that $\omega$ annihilates transformations of the form $\Sigma\rightarrow\Sigma+\delta\sigma$ or $\Gamma\rightarrow\Gamma+\delta\gamma$ and so descends to $\mathcal{S}/\mathrm{Diff}^{+}_{0}(M)$.  Finally, it is clear that $\delta\omega=0$, indicating that it is a symplectic form on this space.     $\Box$
\medskip

We can use this symplectic form to define an inner product between states in the fiber of the tangent space $V$.  Let $h_{i}, h_{j}\in V$ be two linearized solutions, and define their inner product to be:
\be{ip}
\la h_{i}|h_{j}\ra = -\frac{i}{\kappa^{2}}\int_{C}\sigma^{AB}_{j}\wedge\gamma_{i\;AB}.
\ee
An important fact about this inner product (which is obvious in the $\Lambda=0$ setting, c.f., \cite{Mason:2008jy}) is that it annihilates the SD sector:
\begin{lemma}\label{SDsec}
Let $h_{i},h_{j}\in V^{+}$ on the SD background with $(\Sigma_{0},\Gamma_{0})$.  Then $\la h_{i}|h_{j} \ra=0$, or equivalently: for all $h_{i}\in V^{+}$, $\la h_{i}|\cdot\ra|_{V^+}=0$.
\end{lemma}
\proof  The inner product must clearly be skew-symmetric under interchange of $h_{i}$ and $h_{j}$, so we have
\begin{equation*}
\la h_{i}|h_{j} \ra =-\frac{i}{2\kappa^{2}}\int_{C}\left(\sigma^{AB}_{j}\wedge\gamma_{i\;AB}-\sigma^{AB}_{i}\wedge\gamma_{j\;AB}\right).
\end{equation*}
Now, suppose $h_{j}\in V^{+}$; then from \eqref{LFE2} it follows that $\D_{0}\gamma_{j\;AB}=\frac{\Lambda}{3}\sigma_{j\;AB}$.  Furthermore, in the $\Lambda=0$ limit, we know that $\D_{0}\rightarrow\d$ so that any SD perturbation of the ASD spin connection must be pure gauge.  In other words, we know that $\gamma_{j}^{AB}|_{\Lambda=0}=0$, so we can write $\gamma_{j}^{AB}=\Lambda \nu_{j}^{AB}$ for some array of space-time 1-forms $\nu_{i}^{AB}$.  Then the linearized SD field equation gives $\sigma_{j\;AB}=3\D_{0}\nu_{j\;AB}$.  Feeding this into the inner product, gives:
\begin{multline*}
-\frac{i}{2\kappa^{2}}\int_{C}\left(3\d\nu^{AB}_{j}\wedge\gamma_{i\;AB}+6\Gamma_{0\;C}^{(A}\wedge\nu^{B)C}_{j}\wedge\gamma_{i\;AB}-\sigma^{AB}_{i} \wedge\gamma_{j\;AB}\right)\\
=\frac{i}{2\kappa^{2}}\int_{C}\left(3\nu^{AB}_{j}\wedge\D_{0}\gamma_{i\;AB}-\sigma^{AB}_{i}\wedge\gamma_{j\;AB}\right),
\end{multline*}  
where the second line follows by integration by parts and a re-arranging of index contractions.  Now, using $\gamma_{j\;AB}=\Lambda\nu_{j\;AB}$, we have:
\begin{equation*}
\la h_{i}|h_{j}\ra = \frac{i}{2\kappa^{2}}\int_{C}\nu_{j}^{AB}\wedge\left(3\D_{0}\gamma_{i\;AB}-\Lambda\sigma_{i\;AB}\right)=\frac{3i}{2\kappa^{2}}\int_{C}\nu_{j}^{AB}\wedge\psi_{i\;ABCD}\Sigma^{CD}_{0},
\end{equation*} 
using \eqref{LFE2} for $h_{i}$.  So, if $h_{i}\in V^{+}$ as well, $\psi_{i\;ABCD}=0$ and the inner product vanishes as desired.     $\Box$
\bigskip

Hence, if $h_{i}\in V^{+}$, it follows that the inner product annihilates all other states $h_{j}$ in $V^{+}$.  In other words, the inner product vanishes on linearized SD solutions.  To use this inner product to define ASD solutions at the boundary of our space-time, we simply take a one-parameter family of Cauchy hypersurfaces $C_{t}\rightarrow\scri^{\pm}$ as $t\rightarrow\pm\infty$.  Then we say that $h_{j}=(\sigma_{j},\gamma_{j})$ is ASD at $\scri^{\pm}$ if
\be{ASDlim}
\lim_{t\rightarrow\pm\infty} \int_{C_t}\sigma^{AB}_{j}\wedge\gamma_{i\;AB}=0 \qquad \mbox{for all}\:\: h_{i}=(\sigma_{i},\gamma_{i})\in V^{-}.
\ee

\subsubsection*{\textit{Graviton MHV amplitudes}}

\begin{figure}
\centering
\includegraphics[width=3.6 in, height=1.5 in]{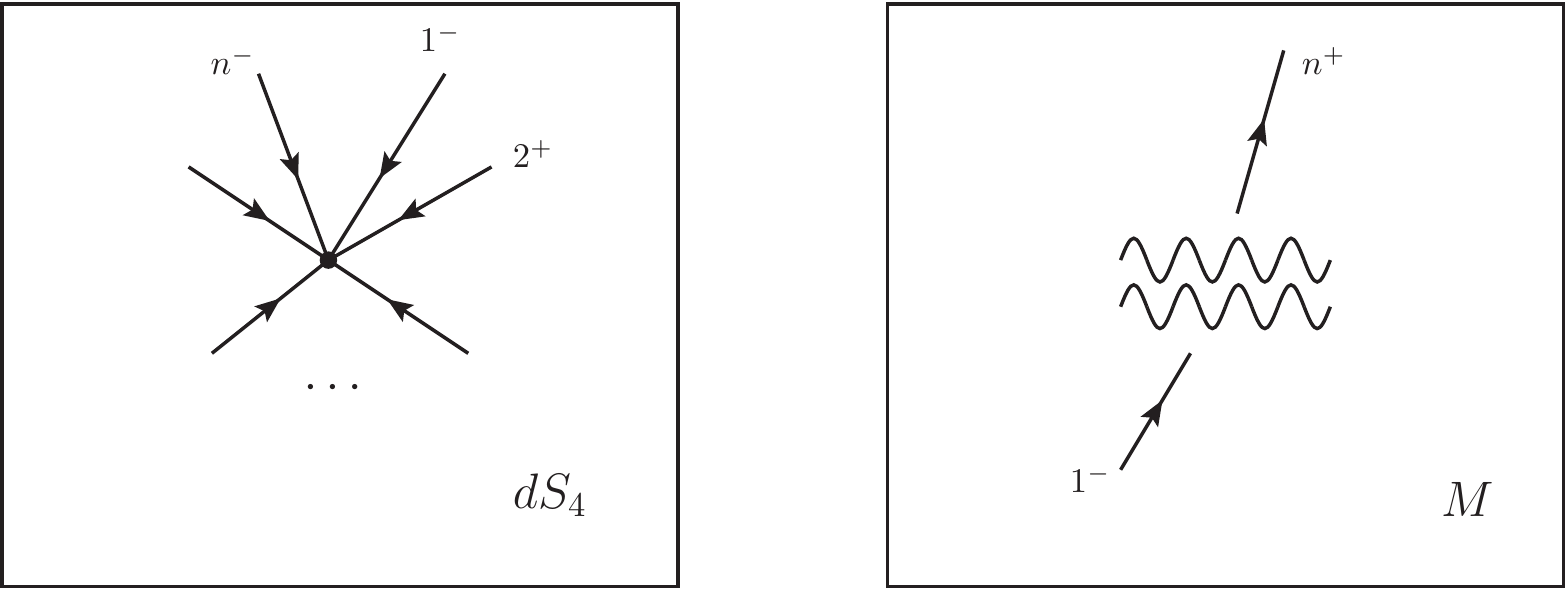}\caption{\textit{Geometric picture of MHV graviton scattering}}\label{dS3}
\end{figure}

A $n$-graviton MHV amplitude will consist of $n-2$ SD and $2$ ASD incoming gravitons.  Following \cite{Mason:2008jy}, we assume that the $n-2$ SD gravitons can be absorbed into a SD background space-time $M$, which can be perturbatively expanded to recover the individual particle content.  Reversing the momentum of one of the two remaining gravitons, the MHV amplitude is the probability for a pure ASD state at $\scri^{-}$ to propagate across $M$ and evolve into a SD state at $\scri^{+}$.  This is illustrated in Figure \ref{dS3}.

We can express this situation mathematically using our inner product \eqref{ip}.  For the incoming state, we take $h_{1}\in V^{-}$ at $\scri^{-}$; since the inner product annihilates the SD sector, the amplitude for it to evolve into something self-dual at $\scri^{+}$ is given by its contraction with a state $h_{2}\in V^{-}$ at $\scri^{+}$.  In other words, we need to compute the inner product between two states $h_{1}|_{\scri^{-}}\in V^{-}$, $h_{2}|_{\scri^{+}}\in V^{-}$ at the future conformal boundary $\scri^{+}$:\footnote{This form for the `scattering amplitude' does not actually constitute a \emph{physical} observable, since the measurement is performed by integrating over all of $\scri^{+}$.  This is a space-like hypersurface, so no physical observer can perform this measurement.  Hence, \eqref{ip*} is a `meta-observable' in the sense proposed by the dS/CFT correspondence \cite{Witten:2001kn, Strominger:2001pn}, but limits nicely to the asymptotically flat definition of a scattering amplitude as $\Lambda\rightarrow 0$.}
\be{ip*}
\la h_{2}|h_{1}\ra =-\frac{i}{\kappa^{2}}\int_{\scri^{+}}\sigma^{AB}_{1}\wedge\gamma_{2\;AB}.
\ee

Before proceeding, one might ask: how do we know that the all SD or one ASD graviton amplitudes vanish?  Even with a cosmological constant, the SD Einstein equations are integrable; this is captured in the chiral formalism by the fact that the SD sector is fully characterized by a single relation, \eqref{SD2}.  Furthermore,  lemma \ref{SDsec} tells us that the inner product on linearized spin-2 fields (i.e., gravitons) vanishes on the SD sector.  The first fact ensures the vanishing of the all SD graviton scattering amplitude, while the second fact tells us that any scattering amplitude involving only a single ASD graviton also vanishes.  Hence, Einstein gravity does indeed possess `MHV-like' behavior, as desired.

Now, we would like to get \eqref{ip*} into a form which is an integral over the SD background $M$; this would allow us to perturbatively expand the background to recover the $n-2$ SD gravitons of the scattering amplitude.  The following proposition allows us to do just that:

\begin{propn}
The amplitude $\la h_{2}|h_{1}\ra$ is given by the formula:
\be{MHV1}
\la h_{2}|h_{1}\ra =\frac{i}{\kappa^{2}}\int_{M}\left(\Sigma^{AB}_{0}\wedge\gamma_{1\;A}^{C}\wedge\gamma_{2\;CB}-\frac{\Lambda}{3}\sigma^{AB}_{1}\wedge\sigma_{2\;AB}\right),
\ee
where $M$ is a SD background space-time described by $(\Sigma_{0},\Gamma_{0})$.
\end{propn}
\proof  Recall that $\partial M=\scri^{+}-\scri^{-}$, so Stokes' theorem gives
\begin{equation*}
-\frac{i}{\kappa^{2}}\int_{\scri^{+}}\sigma^{AB}_{1}\wedge\gamma_{2\;AB}=-\frac{i}{\kappa^{2}}\int_{M}\left(\d\sigma_{1}^{AB}\wedge\gamma_{2\;AB}+\sigma^{AB}_{1}\wedge\d\gamma_{2\;AB}\right)-\frac{i}{\kappa^{2}} \int_{\scri^{-}}\sigma^{AB}_{1}\wedge\gamma_{2\;AB}.
\end{equation*}
Now, the second term on the right vanishes, since we have assumed that $h_{1}\in V^{-}$ at $\scri^{-}$.  Using the linearized field equations \eqref{LFE1}, \eqref{LFE2} it follows that
\begin{eqnarray*}
\d\sigma_{1}^{AB} & = & -2\gamma_{1\;C}^{(A}\wedge\Sigma^{B)C}_{0}-2\Gamma_{0\;C}^{(A}\wedge\sigma_{1}^{B)C},\\
\d\gamma_{2\;AB} & = & \psi_{2\;ABCD}\Sigma^{CD}_{0}+\frac{\Lambda}{3}\sigma_{2\;AB}-2\Gamma_{0\;C(A}\wedge\gamma_{2\;B)}^{C}.
\end{eqnarray*}
This means that we can re-write our amplitude as:
\begin{multline*}
\frac{i}{\kappa^{2}}\int_{M}\left(\Sigma_{0}^{AB}\wedge\gamma_{1\;A}^{C}\wedge\gamma_{2\;CB}+\sigma^{AB}_{1}\wedge\Gamma_{0\;A}^{C}\wedge\gamma_{2\;CB}+\sigma^{AB}_{1}\wedge\Gamma_{0\;CA}\wedge\gamma_{2\;B}^{C}\right. \\
\left. -\frac{\Lambda}{3}\sigma_{1}^{AB}\wedge\sigma_{2\;AB}-\sigma^{AB}_{1}\wedge\psi_{2\;ABCD}\Sigma_{0}^{CD}\right).
\end{multline*}
However, the final term vanishes due to the linearized field equation \eqref{LFE3} and the fact that $\psi_{ABCD}=\psi_{(ABCD)}$, while the second and third terms cancel after restructuring the spinor indices.  The resulting expression agrees with \eqref{MHV1}, but we must still verify that it has the correct gauge invariance: if one of the ASD states is pure gauge, the amplitude must vanish.  Without loss of generality, suppose that $h_{1}$ is pure gauge, so that $\psi_{1\;ABCD}=0$.  By \eqref{eqn: SDLFE}, we know that $\frac{\Lambda}{3}\sigma_{1}^{AB}=\D_{0}\gamma_{1}^{AB}$, and putting this into \eqref{MHV1} and integrating by parts gives us
\begin{equation*}
\la h_{2}| h_{1,\;\psi_{1}=0}\ra =\frac{i}{\kappa^{2}}\int_{M}\left(\Sigma_{0}^{AB}\wedge\gamma_{1\;A}^{C}\wedge\gamma_{2\;CB}+\gamma_{1}^{AB}\wedge\D_{0}\sigma_{2\;AB}\right)-\int_{\partial M}\gamma_{1}^{AB}\wedge\sigma_{2\;AB}.
\end{equation*}
The boundary term vanishes at $\scri^{+}$ since $h_{2}|_{\scri^{+}}\in V^{-}$, and also at $\scri^{-}$ since $h_{1}$ is pure gauge.  This leaves us with the bulk terms, which can be evaluated using the linearized field equation \eqref{LFE1} for $h_{2}$:
\begin{multline*}
\int_{M}\left(\Sigma_{0}^{AB}\wedge\gamma_{1\;A}^{C}\wedge\gamma_{2\;CB}+\gamma_{1}^{AB}\wedge\D_{0}\sigma_{2\;AB}\right) \\
=\int_{M}\left(\Sigma_{0}^{AB}\wedge\gamma_{1\;A}^{C}\wedge\gamma_{2\;CB}-2\gamma_{1}^{AB}\wedge\gamma_{2\;C(A}\wedge\Sigma_{0\;B)}^{C}\right) =0,
\end{multline*}
with the final equality following after re-arranging contractions on spinor indices.     $\Box$

\medskip

The expression \eqref{MHV1} provides a generating functional for the MHV amplitudes, but how do we actually extract a formula for the $n$-point amplitude?  In particular, we still need to perturbatively expand the SD background $M$ to pull out the `hidden' $n-2$ self-dual gravitons.  On a flat background, this was done by transforming the problem to twistor space, where the perturbative expansion can be achieved by making a suitable coordinate transformation on the spinor bundle \cite{Mason:2008jy}.  There are a variety of obstructions to doing this with a cosmological constant, including the fact that twistor space no longer fibers over $\P^{1}$.  Hence, we instead approach the problem via conformal gravity \emph{before} moving to twistor space.   

\subsubsection*{\textit{Relationship with conformal gravity}}

It is easy to see that the generating functional for scattering amplitudes in conformal gravity is given by the second term in \eqref{CGA3}.  Evaluated on-shell with Einstein scattering states, this is:
\be{CGGF}
\la h_{2}|h_{1}\ra^{\mathrm{CG}}=\frac{2i}{\varepsilon^{2}}\int_{M} \d\mu \; \psi_{1}^{ABCD}\psi_{2\;ABCD},
\ee
where $M$ is again the SD background which encodes the $n-2$ remaining gravitons.  By \eqref{CGB2}, this inner product should be equal to some constant multiple of \eqref{MHV1} on-shell (i.e., applying the equations of motion), and this is indeed the case.

\begin{propn}\label{CGDS}
On-shell, $\la h_{2}|h_{1}\ra=-\frac{3\varepsilon^{2}}{\Lambda\kappa^{2}}\la h_{2}|h_{1}\ra^{\mathrm{CG}}$.
\end{propn}
\proof  \eqref{CGGF} can be rewritten as
\begin{equation*}
\la h_{2}|h_{1}\ra^{\mathrm{CG}}=\frac{i}{\varepsilon^2}\int_{M}\psi_{1}^{ABCD}\Sigma_{0\;CD}\wedge\psi_{2\;ABEF}\Sigma_{0}^{EF}.
\end{equation*}
Using the linearized field equation \eqref{LFE2} for $h_{2}$, this becomes
\begin{equation*}
\la h_{2}|h_{1}\ra^{\mathrm{CG}}=\frac{i}{\varepsilon^2}\int_{M}\psi_{1}^{ABCD}\Sigma_{0\;CD}\wedge\left(\D_{0}\gamma_{2\;AB}-\frac{\Lambda}{3}\sigma_{2\;AB}\right).
\end{equation*}
Integrating by parts in the first term gives
\begin{equation*}
-\int_{M}\D_{0}\psi_{1}^{ABCD}\Sigma_{0\;CD}\wedge\gamma_{2\;AB}+\int_{\partial M}\psi_{1}^{ABCD}\Sigma_{0\;CD}\wedge\gamma_{2\;AB}=\int_{\partial M}\psi_{1}^{ABCD}\Sigma_{0\;CD}\wedge\gamma_{2\;AB},
\end{equation*}
using lemma \ref{ZRM}.  In the second term, a combination of both field equations \eqref{LFE2} for $h_{1}$ and \eqref{LFE1} for $h_{2}$ as well as integration by parts leaves
\begin{equation*}
-\frac{2\Lambda}{3}\int_{M}\gamma_{1}^{AB}\wedge\gamma_{2\;C(A}\wedge\Sigma_{0\;B)}^{C}+\frac{\Lambda^{2}}{9}\int_{M}\sigma_{1}^{AB}\wedge\sigma_{2\;AB}-\frac{\Lambda}{3}\int_{\partial M}\gamma_{1}^{AB}\wedge\sigma_{2\;AB}.
\end{equation*}
Combining both calculations gives:
\begin{multline*}
\la h_{2}|h_{1}\ra^{\mathrm{CG}}=\frac{i}{\varepsilon^2}\left(-\frac{2\Lambda}{3}\int_{M}\gamma_{1}^{AB}\wedge\gamma_{2\;C(A}\wedge\Sigma_{0\;B)}^{C}+\frac{\Lambda^{2}}{9}\int_{M}\sigma_{1}^{AB}\wedge\sigma_{2\;AB} \right) \\
-\frac{i}{\varepsilon^2}\left(\int_{\partial M}\psi_{1}^{ABCD}\Sigma_{0\;CD}\wedge\gamma_{2\;AB}-\frac{\Lambda}{3}\int_{\partial M}\gamma^{AB}_{1}\wedge\sigma_{2\;AB}\right) \\
=-\frac{\Lambda \;\kappa^{2}}{3\varepsilon}\la h_{2}|h_{1}\ra +\mbox{boundary terms}.
\end{multline*}

So the proof is complete if we can show that the boundary terms vanish.  Applying \eqref{LFE2} to the first of these terms leaves us
\begin{equation*}
\mbox{boundary terms} \sim \int_{\partial M}\D_{0}\gamma_{1}^{AB}\wedge\gamma_{2\;AB}-\frac{\Lambda}{3}\int_{\partial M}\gamma_{2}^{AB}\wedge\sigma_{1\;AB}-\frac{\Lambda}{3}\int_{\partial M}\gamma_{1}^{AB}\wedge\sigma_{2\;AB},
\end{equation*}
with the second and third terms cancelling due to skew symmetry in $h_{1},h_{2}$.  Finally, 
\begin{multline*}
\int_{\partial M}\D_{0}\gamma_{1}^{AB}\wedge\gamma_{2\;AB}=\int_{\scri^{+}}\D_{0}\gamma_{1}^{AB}\wedge\gamma_{2\;AB}-\int_{\scri^{-}}\D_{0}\gamma_{1}^{AB}\wedge\gamma_{2\;AB} \\
=-\int_{\scri^{+}}\gamma_{1}^{AB}\wedge\D_{0}\gamma_{2\;AB}-\int_{\scri^{-}}\D_{0}\gamma_{1}^{AB}\wedge\gamma_{2\;AB}=0,
\end{multline*}
by the fact that $h_{1}|_{\scri^{-}}\in V^{-}$ and $h_{2}|_{\scri^{+}}\in V^{-}$, as required.     $\Box$

\medskip

At this point, we have established that Einstein gravity MHV amplitudes can be computed via the conformal gravity generating functional, but we need a good theory for operationalizing this calculation.  It turns out that this is provided for us by translating the generating functional to twistor space.  For this, we need a twistor action.


\subsubsection{Remarks on $\cN=4$ conformal super-gravity}

Before proceeding directly to a discussion of the twistor action for conformal gravity, let us make some brief remarks about how the embedding of Einstein gravity into conformal gravity extends to the supersymmetric setting.  Analogues of conformal gravity with extended supersymmetry were first constructed in \cite{Bergshoeff:1980is}, and it is believed that these theories are well-defined for $\cN\leq 4$ (c.f., \cite{Ferrara:1977ij, deWit:1978pd}).  As we saw in our discussion of gauge theory, $\cN=4$ supersymmetry is most natural from our perspective since this results in a Calabi-Yau twistor space.  The maximally supersymmetric $\cN=4$ conformal supergravity (CSG) comes in two phenotypes: \emph{minimal} and \emph{non-minimal} based upon the presence of a certain global symmetry.  Einstein supergravity embeds into minimal CSG, but \emph{not} into the non-minimal models.

The field content of $\cN=4$ CSG consists of the spin-2 conformal gravitons along with bosonic fields $V^{a}_{\mu\;b}$, anti-self-dual tensors $T^{ab}_{\mu\nu}$, scalars $\{E_{ab}, D^{ab}_{cd}, \varphi\}$ and fermions $\{\psi^{a}_{\mu}, \chi^{a}_{bc}, \lambda_{a}\}$, where $a=1,\ldots,4$ is a $\SU(4)$ $R$-symmetry index.  \emph{Minimal} $\cN=4$ CSG is characterized by a global $\SU(1,1)$ symmetry acting non-linearly on the complex scalar $\varphi$ \cite{Bergshoeff:1980is}, and is related to the presence of $\cN=4$ Poincar\'e supergravity sitting inside the CSG \cite{Cremmer:1977tt}. 

This symmetry is manifested by replacing $\varphi$ with a doublet of complex scalars $\Phi_{\alpha}=(\Phi_{1},\Phi_{2})$ which transform under $\SU(1,1)\times\U(1)$ according to
\begin{equation*}
\Phi_{\alpha}\mapsto \mathsf{M}_{\alpha}^{\beta}\Phi_{\beta}, \qquad \Phi_{\alpha}\mapsto e^{i\lambda(x)}\Phi_{\alpha}, \qquad \mathsf{M}\in\SU(1,1),\;\lambda\in C^{\infty}(M,\C),
\end{equation*} 
subject to the constraint $\eta^{\alpha\beta}\overline{\Phi}_{\beta}\Phi_{\alpha}=1$ for $\eta$ the quadratic form on $\SU(1,1)$.  By gauge-fixing the local $\U(1)$ symmetry, one obtains the scalar $\varphi$ as a parametrization of the coset space $\SU(1,1)/\U(1)$, where $\U(1)$ is the diagonal subgroup.  

\begin{figure}
\centering
\includegraphics[width=3.25 in, height=1.25 in]{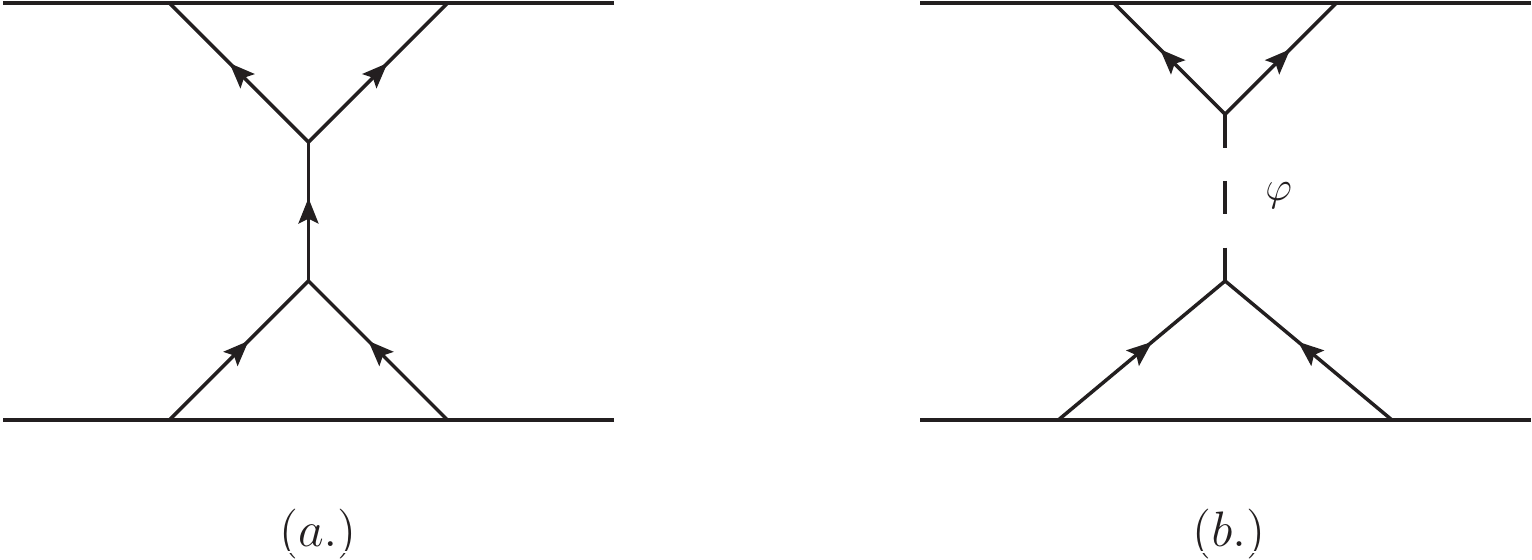}\caption{\textit{In minimal} $\cN=4$ \textit{CSG, external gravitons only couple to other gravitons in the bulk} (\emph{a}.); \textit{in the non-minimal model they can couple to the scalar} $\varphi$ (\emph{b}.).}\label{CSGs}
\end{figure}

Since we are interested in scattering processes with external states corresponding to conformal gravitons, it is particularly enlightening to consider the effects of this symmetry on the portions of the Lagrangian including the spin-2 fields and the scalar.  Clearly, the Lagrangian must contain the $(\mbox{Weyl})^2$ term of the $\cN=0$ action \eqref{CGA1}, but since $\varphi$ is charged under the global $\SU(1,1)$ symmetry, there can be no coupling between the conformal gravitons and this complex scalar.  Furthermore, the $\SU(4)_{R}$-symmetry of the remaining fields excludes any other couplings between bosonic or fermionic fields and the Weyl curvature.  This leads to a unique Lagrangian
\begin{equation*}
\cL^{\mathrm{min}}=C^{\mu\nu\rho\sigma}C_{\mu\nu\rho\sigma}+\varphi \Box^{2}\bar{\varphi} +\cdots\, ,
\end{equation*}
where the multitude of remaining kinetic and interaction terms will be irrelevant for our purposes.  Einstein supergravities at $\cN=4$ can be constructed from minimal CSG \cite{deRoo:1985jh} and so restricting to Einstein scattering states, Maldacena's argument should still apply and we can extract the tree-level Einstein gravity scattering amplitudes (see Figure \ref{CSGs} (\emph{a})).

Minimal CSG can be obtained as a gauge theory of the superconformal group $\SU(2,2|4)$.  A weaker version of the minimal Lagrangian can also be obtained by coupling abelian $\cN=4$ SYM to a $\cN=4$ CSG background \cite{deRoo:1984gd, deRoo:1985jh} and extracting the UV divergent portion of the partition function \cite{Liu:1998bu, Buchbinder:2012uh}.  It has also been shown that minimal $\cN=4$ CSG interacting with a $\SU(2)\times\U(1)$ $\cN=4$ SYM theory is finite and power-counting renormalizable \cite{Fradkin:1981jc, Fradkin:1985am}.

If we remove this $\SU(1,1)$ symmetry, then new interaction terms can appear in the Lagrangian resulting in \emph{non-minimal} $\cN=4$ CSG, which was first conjectured to exist in \cite{Fradkin:1983tg, Fradkin:1985am}.  Indeed, (local) conformal invariance still allows for terms such as
\begin{equation*}
\cL^{\mathrm{non-min}}=C^{\mu\nu\rho\sigma}C_{\mu\nu\rho\sigma}+\varphi \Box^{2}\bar{\varphi} + f(\varphi) C^{\mu\nu\rho\sigma}C_{\mu\nu\rho\sigma}+ig(\varphi)C^{\mu\nu\rho\sigma}C^{*}_{\mu\nu\rho\sigma} +\cdots,
\end{equation*}
for $f,g$ arbitrary real-analytic functions.  For generic choices of these functions, the scalar will provide a source for the Weyl curvature in the bulk, and \emph{vice versa}.  At the level of scattering amplitudes, conformal graviton states in the non-minimal theory can interact with the scalar in the bulk via three-point vertices of the form $\varphi (\mbox{Weyl})^2$, so there will be Feynman diagrams for which there is no analogue in Einstein gravity, as illustrated in Figure \ref{CSGs} (\emph{b}).  Without a consistent algorithm to subtract these diagrams, Maldacena's argument \emph{cannot} be applied to non-minimal CSG.  

Non-minimal $\cN=4$ CSG can be obtained by coupling non-abelian $\cN=4$ SYM to the $\cN=4$ CSG background and again extracting the UV divergent partition function.  While there is some doubt over whether non-minimal conformal supergravity is well-defined at the quantum level \cite{Romer:1985yg, Buchbinder:2012uh}, there is substantial evidence that the twistor-string theory of Berkovits and Witten corresponds to non-minimal $\cN=4$ CSG coupled to $\cN=4$ SYM \cite{Berkovits:2004jj}.  Indeed, spurious amplitudes related to the non-minimal coupling between conformal gravitons and scalars were found explicitly in \cite{Dolan:2008gc, Adamo:2012nn}.


\subsection{Twistor Action for Conformal Gravity}

In our study of $\cN=4$ SYM, the twistor action arose by translating the structure of the Chalmers-Siegel action to twistor space.  We now show that the same thing can be done with the conformal gravity action \eqref{CGA3}.  We give this action in Mason's original coordinate-free formulation as well as in a form using an explicit choice of background complex structure, and both constructions naturally generalize to $\cN=4$ to give minimal theories.  Restricting to the Einstein subsector gives us both a twistorial formulation for the MHV generating functional \eqref{CGGF}, as well as a candidate twistor action for Einstein gravity itself \cite{Adamo:2013tja}.


\subsubsection{$\cN=0$ action}

\subsubsection*{\textit{Coordinate-free approach}}

Let us begin by deriving a twistor action which avoids any explicit choices of coordinates or background complex structure.  The first term in the space-time action \eqref{CGGF} corresponds to the SD sector of solutions to the Bach equations.  By theorem \ref{NLG}, we know that this is equivalent to a twistor space $\CPT$ with integrable complex structure $J$, so the associated Nijenhuis tensor $N_{J}$ must vanish.  We can encode this requirement in an action functional by introducing a Lagrange multiplier field $\mathscr{G}\in\Omega^{3,0}(\CPT,\Omega^{1,1})$ and taking \cite{Mason:2005zm}
\be{CFTA1}
S_{1}[J,\mathscr{G}]=\int_{\CPT}N_{J}\lrcorner\mathscr{G},
\ee 
which has field equations
\begin{equation*}
N_{J}=0, \qquad \dbar^{J}\mathscr{G}=0.
\end{equation*}
This action is invariant under diffeomorphisms as well as $\mathscr{G}\rightarrow \mathscr{G}+\dbar^{J}\gamma$ for $\gamma\in\Omega^{3,0}(\CPT,\Omega^{1,0})$, so we can interpret $\mathscr{G}$ as a cohomology class.  The vanishing of $N_{J}$ corresponds on space-time to $\Psi_{ABCD}=0$, which is the first equation of \eqref{CGFE} when $\varepsilon=0$.  So to establish that \eqref{CFTA1} describes self-dual conformal gravity, we need to show that $\mathscr{G}$ corresponds to a space-time field satisfying the Bach equation.

Write $\mathscr{G}=g\otimes\D^{3}Z$, where $g\in\Omega^{1,1}(\CPT,\cO(-4))$ and $\D^{3}Z$ is the tautologically defined section of $\Omega^{3,0}(\CPT,\cO(4))$.  The field equations indicate that $g\in H^{0,1}(\CPT,\Omega^{1,0}(-4))$, so we can apply the Penrose transform.  Picking an arbitrary representative of the conformal class, we can construct an array of space-time fields from $g_{\alpha}\D Z^{\alpha}$:
\be{ASD1}
\Gamma_{\delta ABC}=\left(
\begin{array}{c}
G^{D}_{ABC} \\
\gamma_{D'ABC}
\end{array}\right) = \int_{X}\tau\wedge\lambda_{A}\lambda_{B}\lambda_{C}g_{\delta}, \qquad \nabla^{AA'}\Gamma_{\delta ABC}=0.
\ee
Recalling that $\nabla$ acts on $\Gamma_{\delta ABC}$ via the local twistor connection as in \eqref{ltc2},
\be{ASD2*}
\nabla^{AA'}\Gamma_{\delta ABC}=0 \leftrightarrow \left\{
\begin{array}{c}
\nabla^{AA'}G^{D}_{ABC}-i\gamma^{A'D}_{BC} = 0 \\
\nabla^{AA'}\gamma_{D'ABC}-i\Phi^{AA'}_{DD'}G^{D}_{ABC} = 0
\end{array}\right. .
\ee

Now, the Penrose transform of $Z^{\alpha}g_{\delta}$ is given by
\begin{equation*}
\int_{X}\tau \wedge Z^{\alpha}\lambda_{B}\lambda_{C}g_{\delta}=\left(
\begin{array}{c}
\Gamma_{\delta ABC} \\
0
\end{array}\right),
\end{equation*}
because the restriction to $X\subset\CPT$ implies that $X_{\alpha\beta}Z^{\alpha}=0$, where $X_{\alpha\beta}\in\T_{[\alpha\beta]}$ corresponds to the point $x\in M$ \cite{Mason:1990}.  Hence, imposing the usual local twistor gauge-fixing $Z^{\alpha}g_{\alpha}=0$ on twistor space has the consequence $G^{A}_{ABC}=0$ on space-time.  Therefore, $G_{DABC}=G_{(DABC)}$ and we can substitute the first z.r.m. equation into the second to obtain
\begin{equation*}
\nabla^{AA'}\nabla^{D}_{D'}G_{BCDA}+\Phi^{AA'}_{DD'}G^{D}_{ABC}=\left(\nabla^{AA'}\nabla^{D}_{D'}+\Phi^{ADA'}_{D'}\right)G_{ABCD}=0,
\end{equation*}
which is the required Bach equation.  This is precisely what is predicted for a linearized ASD field in conformal gravity on twistor space \cite{Mason:1987}.

Hence, \eqref{CFTA1} is indeed the twistor action for self-dual conformal gravity.  We still need to describe the ASD interactions, which are given by the second term in \eqref{CGA3}.  This is easy though, since we already know that $\mathscr{G}$ encodes the space-time Lagrange multiplier $G_{ABCD}$ in its $(1,1)$-form part $g$.  Indeed, since $g$ is a cohomology class,
\be{GPent}
G_{ABCD}(x)=\int_{X}\sigma_{A}\sigma_{B}\sigma_{C}\sigma_{D}\;g(Z(x,\sigma)),
\ee
and the interaction term on twistor space becomes:
\be{TCGInt}
S_{2}[J,\mathscr{G}]=\int_{\PS\times_{M}\PS}\d\mu\;(\sigma_{1}\sigma_{2})^{4}g_{1}\wedge g_{2},
\ee
where $\PS\times_{M}\PS\cong M\times X\times X$, $(\sigma_{1}\sigma_{2})$ is the $\SL(2,\C)$-invariant inner product between the homogeneous coordinates on $X$, and $\d\mu$ is a measure on the space of rational curves $X\subset\CPT$.  Of course, this integral must be performed over a \emph{real} four-dimensional slice determined by some reality conditions which single out $\CPT_{\R}$.  

Hence, we have the full conformal gravity twistor action:
\be{CFTA2}
S[J,\mathscr{G}]=S_{1}[J,\mathscr{G}]-\varepsilon^{2}S_{2}[J,\mathscr{G}].
\ee    
The following theorem ensures that this twistor action is as good as the one we used in our study of gauge theory:
\begin{thm}[Mason \cite{Mason:2005zm}]\label{MasThm}
The twistor action $S[J,\mathscr{G}]$ is classically equivalent to the conformal gravity action \eqref{CGA3} off-shell in the sense that solutions to its Euler-Lagrange equations are in one-to-one correspondence with solutions to the field equations \eqref{CGFE} up to space-time diffeomorphisms.  Additionally, upon fixing Woodhouse gauge and Euclidean reality conditions, $S[J,\mathscr{G}]$ is equal to the space-time action.
\end{thm}

\subsubsection*{\textit{Flat background complex structure}}

For the purposes of this review, it will actually be advantageous for us to work with the twistor action using an explicit choice of background complex structure.  In particular, we take the background complex structure to be $\dbar$ associated to the flat twistor space $\PT$ of Minkowski space; the complex structure on the curved twistor space is then given by $\dbar_{f}=\dbar+f$, which is integrable by \eqref{contact} whenever
\be{Nijenhuis}
\dbar^{2}_{f}=\left(\dbar f^{\alpha}+[f,f]^{\alpha}\right)\partial_{\alpha}\equiv N^{\alpha}\partial_{\alpha} =0.
\ee
Of course, this is not equal to the full Nijenhuis tensor (which is generally a non-polynomial object), but the vanishing of the two quantities is equivalent thanks to the Newlander-Nirenberg theorem in our chosen coordinate frame.

The action now becomes a functional of $f\in\Omega^{0,1}(\CPT,T^{1,0}_{\CPT})$ and $g\in\Omega^{1,1}(\CPT,\cO(-4))$, and the only change is with respect to the self-dual contribution:
\be{TCGSD}
S_{1}[g,f]=\int_{\CPT}\D^{3}Z\wedge g_{\alpha}\wedge N^{\alpha},
\ee
where $g_{\alpha}\in\Omega^{0,1}(\CPT, \cO(-5))$, subject to $Z^{\alpha}g_{\alpha}=0$.  The self-dual field equations are
\be{SDFES}
N^{\alpha}=0, \qquad \dbar_{f}g=0,
\ee
and the gauge freedom is \cite{Mason:2005zm}
\begin{equation*}
g_{\alpha}\rightarrow g_{\alpha}+\partial_{\alpha}\gamma +\dbar_{f}\chi_{\alpha},
\end{equation*}
for $\gamma\in\Omega^{0}(\CPT,\cO(-4))$, $\chi_{\alpha}\in\Omega^{0}(\CPT,\cO(-5))$.  As in the coordinate-free setting, this means that we can consider $g_{\alpha}$ to be a cohomology class on $\CPT$, and it again corresponds to the space-time field $G_{ABCD}$ via the Penrose transform.  The ASD interactions are still encoded by $S_{2}$ from \eqref{TCGInt}, but now with the understanding that the $\P^1$ fibers are holomorphic with respect to $\dbar_{f}$.  In other words, the rational curves in twistor space are constructed by the constraint \eqref{holomap}.

We denote this representation of the twistor action by
\be{TCG}
S[g,f]=S_{1}[g,f]-\varepsilon^{2}S_{2}[g,f],
\ee
and use it almost exclusively in what follows.  While this sacrifices some formal flexibility, it also enables us to be quite explicit with some calculations, as we will see when considering Einstein degrees of freedom.


\subsubsection{Minimal $\cN=4$ action}

The $\cN=0$ conformal gravity twistor action generalizes naturally to $\cN=4$ supersymmetry.  In this setting, the curved twistor space is topologically an open subset of $\P^{3|4}$, and points in the chiral complex space-time $M$ still correspond to rational curves $X\subset\CPT$.  The twistor map from $\PS$ to $\CPT$ is promoted to:
\be{sincid}
Z^{I}(x^{\mu},\theta^{Aa}, \sigma_{B})=\left(\lambda_{A}(x,\theta,\sigma), \mu^{A'}(x,\theta,\sigma), \chi^{a}(x,\theta,\sigma)\right),
\ee
and the canonical holomorphic section of the Berezinian is denoted by $\D^{3|4}Z$.  Considering the complex structure to be a finite deformation of the flat one on $\PT$, the data for the twistor action becomes
\be{sdata}
\dbar_f=\dbar+ f^I\frac\p{\p Z^I} \, , \qquad g=g_I \D Z^I\in \Omega^{1,1}(\CPT)\, , \quad \D Z^I=\rd Z^I-f^I\, .
\ee
This means that the holomorphic curves $X$ in twistor space are constructed by the supersymmetric analogue of \eqref{holomap} 
\be{sholomap}
\dbar|_{X} Z^{I}(x,\theta,\sigma)=f^{I}(Z).
\ee

With these structures in play, we can easily write down the $\cN=4$ generalization of the twistor action \eqref{TCG}:
\be{minTA1}
S_{1}[g,f]=\int_{\CPT}\D^{3|4}Z\wedge g_{I}\wedge \left(\dbar f^{I}+[f,f]^{I}\right),
\ee
\be{minTA2}
S_{2}[g,f]=\int\limits_{\PS\times_{M}\PS}\d\mu\wedge g_{1}\wedge g_{2},
\ee
where $\d\mu$ is now promoted to a measure on the $(4|8)$-dimensional space of curves $X\subset\CPT_{\R}$.  One interesting consequence of the $\cN=4$ supergeometry is that the conditions $\partial_{I}f^{I}=0$ and $Z^{I}g_{I}=0$ are not sufficient to fix the gauge freedoms
\begin{equation*}
f^{I}\rightarrow f^{I}+Z^{I}\alpha, \qquad g_{I}\rightarrow\partial_{I}\beta,
\end{equation*}
for $\alpha,\beta\in\Omega^{0,1}(\CPT,\cO)$.  This follows because $\beta$ has homogeneity zero, as opposed to the homogeneity $-4$ gauge transformations from the $\cN=0$ setting.  

On the $\cN=4$ twistor space, we can expand $g$ in the anti-commuting variables as
\begin{equation*}
g=g^{0}+\chi^{a}g^{-1}_{a}+\cdots \frac{\chi^{4}}{4!}g^{-4},
\end{equation*} 
where each $g^{k}\in\Omega^{0,1}(\CPT,\Omega^{1,0}(k))$.  Our calculations at $\cN=0$ already showed us that $g^{-4}$ corresponds to the ASD spinor field $G_{ABCD}$ which satisfies the Bach equation on space-time.  We can use the Penrose transform and local twistor formalism to investigate the other components of $g$ with $\cN=4$.

For instance, consider $g^{0}\in H^{0,1}(\CPT, \Omega^{1,0})$, whose Penrose transform of this object was first described in \cite{Mason:1990}.  Write $g^{0}=a_{\alpha}\D Z^{\alpha}$ for $a_{\alpha}\in H^{0,1}(\CPT, \cO(-1))$.  Choosing an arbitrary conformal frame, the Penrose transform gives:
\be{cptr1}
\Gamma_{\alpha B'}=\left(
\begin{array}{c}
\Psi^{A}_{B'} \\
\Phi_{A'B'}
\end{array} \right) =\int_{X}\tau\wedge\frac{\partial a_{\alpha}}{\partial \mu^{B'}}, \qquad \nabla^{BB'}\Gamma_{\alpha B'}=0.
\ee
Using the local twistor connection, the z.r.m. equations of \eqref{cptr1} can be written on space-time as:
\begin{equation*}
\left\{\begin{array}{ccc}
\nabla^{BB'}\Psi^{A}_{B'}-i\epsilon^{BA}\Phi^{B'}_{B'} & = & 0 \\
\nabla^{BB'}\Phi_{A'B'} & = & 0
\end{array} \right. ,
\end{equation*}
while the Penrose transform of $Z^{\alpha}a_{\alpha}$ gives the conditions $\nabla_{BB'}\Psi^{B}_{A'}-i\epsilon_{B'A'}\Phi^{B'}_{B'}=0$ and $\Phi^{B'}_{B'}=0$.  This means that we can write $\Psi_{AA'}=\Box\varphi$, and the content of \eqref{cptr1} is reduced to $\Box^{2}\varphi =0$, the z.r.m. equation for a conformal scalar. 

An identical procedure will give the following equations for the remaining components:
\begin{eqnarray}
g^{-1}_{a} \Rightarrow & \Box\nabla_{BB'}\psi_{a}^{B}-i\nabla_{AA'}\left(\Phi^{AA'}_{CB'}\psi_{a}^{C}\right) =0, \\
g^{-2}_{ab} \Rightarrow & \left(\nabla_{AA'}\nabla_{BB'}+\Phi_{ABA'B'}\right)T_{ab}^{AB} = 0, \\
g^{-3\;a} \Rightarrow & \left(\nabla_{BD'}\nabla^{AA'}\nabla_{CC'}+\Phi^{AA'}_{CC'}\nabla_{BD'}\right)\eta^{a\;D'}_{AC} = 0.
\end{eqnarray}
These correspond to the linearized spinor, ASD tensor, and conformal gravitino z.r.m. equations of $\cN=4$ CSG, so $g$ is the natural supersymmetric extension of the $\cN=0$ Lagrange multiplier field on twistor space.

This means that $g$ defines a chiral superfield on space-time:
\be{s-tfield}
\mathcal{G}(x,\theta)=\int_{X} g(Z(x,\theta,\sigma)),
\ee
where $\mathcal{G}$ has an expansion like:
\begin{equation*}
\mathcal{G}(x,\theta)=\varphi +\theta^{a}_{A}\psi_{a}^{A}+\cdots +\theta^{4\;ABCD}G_{ABCD}+\cdots ,
\end{equation*}
and the space-time translation of our $\cN=4$ twistor action will look like
\be{CSUGRAct}
S[\mathcal{W},\mathcal{G}]=\int_{M}\d\mu \left(\mathcal{W}(x,\theta)\;\mathcal{G}(x,\theta)-\varepsilon^{2}\mathcal{G}(x,\theta)^{2}\right) \rightarrow \frac{1}{\varepsilon^2}\int_{M}\d\mu \; \mathcal{W}(x,\theta)^2,
\ee
where $\mathcal{W}(x,\theta)$ is the a chiral superfield which, on-shell, is a Lorentz scalar encoding the ASD $\cN=4$ Weyl multiplet (c.f., \cite{Bergshoeff:1980is}).  It has been shown that this superfield action has the correct linear reduction for $\cN=4$ CSG \cite{Berkovits:2004jj} and must correspond to the \emph{minimal} theory since there are no cubic (or higher) couplings between the dilaton and Weyl curvature.  This is evident directly from twistor space as well, since we have a $\U(1)$-symmetry 
\begin{equation*}
g\rightarrow e^{4i\alpha}g, \qquad \chi^{a}\rightarrow e^{-i\alpha}\chi^{a},
\end{equation*}
which eliminates all $\varphi (\mbox{Weyl})^2$ couplings at the level of twistor representatives.\footnote{This actually corresponds to a degenerate limit of the $\SU(1,1)$ symmetry of minimal $\cN=4$ CSG; see \cite{Adamo:2013tja} for additional discussion.  This does not affect our ability to isolate Einstein degrees of freedom, since Einstein supergravity still forms a subsector of this degenerate theory \cite{Cremmer:1977tt}.}

Since all our considerations in this review will be at tree-level for gravity, there is no particularly compelling reason to consider the $\cN=4$ twistor action as opposed to the $\cN=0$ action.  However, the supersymmetric action is `cleaner' in the sense that $S_{2}[g,f]$ doesn't need an explicit weighting factor of $(1 2)^4$, and the Calabi-Yau nature of the twistor space is also advantageous.  Hence, we will often choose to work with the $\cN=4$ framework in the future when performing explicit calculations.


\subsection{Einstein gravity}

Given the twistor action for conformal gravity (or its minimal $\cN=4$ extension), we now want to extract the Einstein subsector using the Maldacena argument outlined earlier.  In particular, by restricting to Einstein degrees of freedom on a de Sitter background, the conformal gravity twistor action should encode the scattering amplitudes of general relativity.  We perform this reduction here explicitly, and show that it gives a twistorial expression for the MHV generating functional of proposition \ref{CGDS}.  Additionally, this reduction allows us to conjecture a form of the twistor action for Einstein gravity itself.


\subsubsection{Reduction to Einstein degrees of freedom}

The first step in reducing the degrees of freedom in the twistor action to Einstein gravity is to break conformal invariance.  This is accomplished by introducing an infinity twistor, just as in $\M$.  Since we work on a background with cosmological constant, the infinity twistor differs from \eqref{infty} by now having rank four:
\be{inftyCC}
I_{\alpha\beta}=\left(
\begin{array}{cc}
\epsilon^{AB} & 0 \\
0 & \Lambda \epsilon_{A'B'}
\end{array} \right), \qquad 
I^{\alpha\beta}=\left(
\begin{array}{cc}
\Lambda \epsilon_{AB} & 0 \\
0 & \epsilon^{A'B'}
\end{array}\right).
\ee  
These can be generalized easily to $\cN=4$ supersymmetry, with the fermionic components of the infinity twistor corresponding to a gauging of the R-symmetry \cite{Wolf:2007tx}.  Since we will not be concerned with this gauging, we leave these fermionic components implicit.  We will also adopt the notation:
\begin{equation*}
I_{IJ}A^{I}B^{J}\equiv\la A,B\ra, \qquad I^{IJ}A_{I} B_{J}\equiv [A,B],
\end{equation*}
for contractions with the infinity twistor (with identical conventions for the $\cN=0$ twistors).  

Theorem \ref{NLG} tells us that an Einstein solution corresponds to a weighted contact structure on $\CPT$ specified by the 1-form $\tau$.  The infinity twistor gives a canonical structure to $\tau$, and also defines a (weighted) Poisson structure and bracket on $\CPT$:
\be{infstruct}
\tau= \la Z,\D Z\ra, \qquad \Pi=I^{IJ}\partial_{I}\wedge\partial_{J}, \qquad \{f,g\}=[\partial f,\partial g].
\ee
The complex deformation $\dbar_{f}$ must now respect both the Poisson and contact structures; this means that $f$ must be Hamiltonian with respect to $\Pi$:
\begin{equation*}
\cL_{f}\Pi =0\: \Rightarrow f=[\partial h, \partial], \qquad h\in\Omega^{0,1}(\CPT,\cO(2)).
\end{equation*}
Note that if $h=\dbar\gamma$, then $f=\dbar(\Pi(\gamma))$ is pure gauge, so we can take $h$ to be a cohomology class.  

In the $\cN=0$ setting, the Penrose transform tells us that this will correspond to a graviton of helicity $+2$.  Feeding this into \eqref{TCGSD}, we obtain:
\begin{multline}\label{EinR1}
S_{1}[g,f]\rightarrow S_{1}[g,h]=\int_{\CPT}\D^{3}Z\wedge g_{\alpha}\wedge I^{\alpha\beta}\partial_{\beta}\left(\dbar h +\frac{1}{2} \left\{h,h\right\}\right) \\
=\int_{\CPT}\D^{3}Z\wedge I^{\alpha\beta}\partial_{\alpha}g_{\beta}\wedge\left(\dbar h +\frac{1}{2} \left\{h,h\right\}\right),
\end{multline}
with the last line following via integration by parts.  We can identify $I^{\alpha\beta}\partial_{\alpha}g_{\beta}$ as the other graviton in the Einstein reduction:
\begin{lemma}
For $g_{\alpha}\in H^{0,1}(\CPT, \cO(-5))$, the Penrose transform of $I^{\alpha\beta}\partial_{\alpha}g_{\beta}$ can be identified with a graviton of helicity $-2$.
\end{lemma}
\proof Recall the Penrose transform of $g_{\alpha}$ given by \eqref{ASD1}.  In the de Sitter conformal structure, the z.r.m. equations \eqref{ASD2*} become:
\be{ASD2}
\nabla^{AA'}\Gamma_{\delta ABC}=0 \leftrightarrow \left\{
\begin{array}{c}
\nabla^{AA'}G^{D}_{ABC}-i\gamma^{A'D}_{BC} = 0 \\
\nabla^{AA'}\gamma_{D'ABC}-i\Lambda\epsilon^{A'}_{D'}G^{A}_{ABC} = 0
\end{array}\right. .
\ee
Using the fact that $G_{ABCD}=G_{(ABCD)}$, we can immediately reduce these to
\begin{eqnarray}
\nabla^{AA'}G^{D}_{ABC}-i\gamma^{A'D}_{BC} & = & 0, \label{zrm1}\\     
\nabla^{AA'}\gamma_{D'ABC} & = & 0 . \label{zrm2}
\end{eqnarray}

We now want to lower the homogeneity of $g_{\alpha}$ by applying a twistorial derivative; the Penrose transform of such an operation obeys \cite{Mason:1990}:
\begin{multline*}
\partial_{\alpha}g_{\beta}\xrightarrow{\mathrm{Penrose}\;\mathrm{transform}}\Phi_{\alpha\beta CDEF}=\left(
\begin{array}{c}
3\epsilon^{A}_{(C}\Gamma_{\beta DEF)} \\
-i\nabla_{CA'}\Gamma_{\beta DEF}
\end{array}\right) \\
=\left(
\begin{array}{cc}
3\epsilon^{A}_{(C}G^{B}_{DEF)} & 3\epsilon^{A}_{(C}\gamma_{|B'|DEF)} \\
-i\nabla_{CA'}G^{B}_{DEF}+\epsilon^{B}_{C}\gamma_{A'DEF} & -i\nabla_{CA'}\gamma_{B'DEF}+\Lambda\epsilon_{A'B'}G_{CDEF}
\end{array}\right).
\end{multline*}
Using \eqref{inftyCC}, we can deduce:
\be{ASD3}
I^{\alpha\beta}\partial_{\alpha}g_{\beta}\xrightarrow{\mathrm{Penrose}\;\mathrm{transform}}-\Lambda G_{ABCD}-i\nabla_{AA'}\gamma^{A'}_{BCD}.
\ee

It suffices to show that this obeys the spin-2 z.r.m. field equation for an ASD field.  Using \eqref{zrm1}, we have:
\begin{equation*}
I^{\alpha\beta}\partial_{\alpha}g_{\beta}\xrightarrow{\mathrm{Penrose}\;\mathrm{transform}}\psi_{ABCD}\equiv -\Lambda G_{ABCD}-\nabla_{AA'}\nabla^{EA'} G_{BECD}.
\end{equation*}
Now, note that $\nabla_{AA'}\nabla^{EA'}=\frac{1}{2}\epsilon^{E}_{A}\Box+\epsilon^{EF}\Box_{AF}$, where $\Box_{AF}=\nabla_{A'(A}\nabla^{A'}_{F)}$.  This leaves us with
\begin{multline*}
\psi_{ABCD}=-\Lambda G_{ABCD}-\frac{1}{2}\Box G_{ABCD}-\epsilon^{EF}\Box_{AF} G_{BECD} \\
=-\Lambda G_{ABCD}-\frac{1}{2}\Box G_{ABCD}+\Lambda\left( G_{ABCD}-G_{ABCD}-2G_{ABCD}+G_{ABCD}+G_{ABCD}\right)\\
=-\Lambda G_{ABCD}-\frac{1}{2}\Box G_{ABCD}.
\end{multline*}

Using \eqref{zrm1}, it follows that
\begin{equation*}
\nabla^{AA'}\psi_{ABCD}=-i\left(\Lambda\gamma^{A'}_{BCD}+\frac{1}{2}\Box\gamma^{A'}_{BCD}\right).
\end{equation*}
Now, \eqref{zrm2} tells us that $\nabla^{AA'}\gamma_{D'ABC}=0$, so any higher derivatives will also vanish.  In particular,
\begin{equation*}
\nabla_{DA'}\nabla^{AA'}\gamma_{D'ABC}=\frac{1}{2}\Box\gamma_{D'DBC}+\Lambda\gamma_{D'DBC}=0,
\end{equation*} 
which immediately implies that $\nabla^{AA'}\psi_{ABCD}=0$.  Since this is the spin-2 ASD zero-rest-mass field equation, the proof is complete.    $\Box$

\medskip

The Penrose transform tells us that we can represent $I^{\alpha\beta}\partial_{\alpha}g_{\beta}$ by an element of $H^{0,1}(\CPT,\cO(-6))$.  But given any $\tilde{h}\in H^{0,1}(\CPT,\cO(-6))$ we can also write $g_{\alpha}=I_{\alpha\gamma}Z^{\gamma}\tilde{h}$, which obeys $g_{\alpha}\in H^{0,1}(\CPT,\cO(-5))$ and $Z^{\alpha}g_{\alpha}=0$.  Hence, \eqref{EinR1} becomes:
\begin{multline*}
S_{1}[g,h]\rightarrow S_{1}[\tilde{h},h]=\int_{\CPT}\D^{3}Z\wedge I^{\alpha\beta}\partial_{\alpha}\left(I_{\beta\gamma}Z^{\gamma}\tilde{h}\right)\wedge\left(\dbar h +\frac{1}{2} \left\{h,h\right\}\right) \\
= 2\Lambda \int_{\CPT}\D^{3}Z\wedge\tilde{h}\wedge\left(\dbar h +\frac{1}{2} \left\{h,h\right\}\right). 
\end{multline*}
This process goes through in exactly the same fashion for the self-dual $\cN=4$ action; the only difference is that $h$ now encodes the positive helicity graviton \emph{multiplet}, and $\tilde{h}\in H^{0,1}(\CPT,\cO(-2))$ now encodes the negative helicity multiplet.  The resulting action is
\be{EinR2}
S_{1}[g,f]\rightarrow S_{1}[\tilde{h},h]=2\Lambda\int_{\CPT}\D^{3|4}Z\wedge\tilde{h}\wedge\left(\dbar h +\frac{1}{2} \left\{h,h\right\}\right).
\ee

As expected, this is precisely the self-dual twistor action for Einstein gravity, up to the factor of $\Lambda$ required by \eqref{CGB2} \cite{Mason:2007ct, Adamo:2012nn}.  The corresponding reduction for the second term of the twistor action follows easily:
\be{EinR3}
S_{2}[g,f]\rightarrow S_{2}[\tilde{h},h]=\int\limits_{\PS\times_{M}\PS} \d\mu\wedge\tau_{1}\wedge\tau_{2}\wedge\tilde{h}_{1}\wedge \tilde{h}_{2},
\ee
using the $\cN=4$ formalism.  So the reduction of the conformal gravity twistor action to Einstein wavefunctions is simply
\be{EinCGTA}
S[\tilde{h},h]=S_{1}[\tilde{h},h]-\varepsilon^{2}S_{2}[\tilde{h},h].
\ee
The remaining diffeomorphism freedom on $\CPT$ is captured by the transformations:
\begin{equation*}
Z^{\alpha}\rightarrow Z^{\alpha}+\left\{Z^{\alpha},\chi\right\}, \qquad h\rightarrow h+\dbar\chi +\left\{h,\chi\right\},
\end{equation*}
for $\chi$ a weight $+2$ function.

Now observe that we have arrived at a twistorial expression for the generating functional of MHV amplitudes in Einstein gravity.  In particular, we know that the conformal gravity generating functional is given by $S_{2}[g,f]$, so proposition \ref{CGDS} tells us that
\be{TGenF}
\la \tilde{h}_{2}|\tilde{h}_{1}\ra = -\frac{3\varepsilon^{2}}{\Lambda\kappa^{2}}S_{2}[\tilde{h},h],
\ee
where $S_{2}$ is given by \eqref{EinR3} for $\cN=4$, or with an additional factor of $(12)^4$ for $\cN=0$.  The positive helicity gravitons of the amplitude are encoded by the non-linear SD background $M$, which serves as the space of rational curves $X$ in twistor space constructed by solving \eqref{sholomap} on Einstein states:
\begin{equation*}
\dbar|_{X} Z^{I}=I^{IJ}\partial_{J}h(Z).
\end{equation*}


\subsubsection{Einstein twistor actions}

We can go beyond simply using the reduction to Einstein states to write down the amplitude generating functional, though.  Dividing \eqref{EinCGTA} by a power of $\Lambda$ in accordance with \eqref{CGB2}, we arrive at a functional which is a candidate twistor action for Einstein gravity itself.  In particular, for $\cN=4$ we have \cite{Adamo:2013tja}:
\begin{multline}\label{EinTA4}
S^{\mathrm{Ein}}_{\cN=4}[\tilde{h},h]=\int_{\CPT}\D^{3|4}Z\wedge\tilde{h}\wedge\left(\dbar h+\frac{1}{2}\left\{h,h\right\}\right)  \\
-\frac{\varepsilon^{2}}{\Lambda \kappa^{2}}\int\limits_{\PS\times_{M}\PS}\d\mu\wedge\tau_{1}\wedge\tau_{2}\wedge\tilde{h}_{1}\wedge\tilde{h}_{2}.
\end{multline}
This action has the correct self-dual reduction when $\varepsilon=0$ \cite{Mason:2007ct}, is obtained directly from the embedding of Einstein gravity into conformal gravity, is well-defined off-shell in any gauge, and (as we show in Section \ref{Chapter6}) produces the correct MHV amplitudes.\footnote{The only prior proposal for such an action, given in \cite{Mason:2008jy}, is really a generating functional for flat-space MHV amplitudes of Einstein gravity in `BGK' form.  It does not extend to space-times with cosmological constant and may not even be diffeomorphism invariant.}

While all of these facts indicate that \eqref{EinTA4} is a correct proposal, there is currently no known analogue of theorem \ref{MasThm} for this Einstein action, which \emph{proves} that it corresponds to Einstein gravity.  The basic reason for this is that we arrived at \eqref{EinTA4} by using our explicit choice of background complex structure.  While this resulted in a well-defined action functional, the geometric meaning of the terms in the Einstein twistor action is no longer clear.  In particular, how are we to interpret the self-dual action?  It certainly does not contain a Nijenhuis tensor (even in some special coordinate frame).  This should be contrasted against \eqref{CFTA1} for conformal gravity, where the geometrical meaning is clear and no background complex structure has been chosen.

The proposed Einstein twistor action has a field equation analogous to $N_{J}=0$:
\be{GREOM1}
\D^{3|4}Z\wedge\left(\dbar h+\frac{1}{2}\left\{h,h\right\}\right)=\frac{\varepsilon^{2}}{\Lambda\kappa^{2}}\;\d\mu\wedge\tau\;\int_{X_1}\tilde{h}_{1}\wedge\tau_{1},
\ee
where $X_1$ is the rational curve in $\CPT_{\R}$ (fixed by the reality conditions) which contains $Z$.  If one could show that this was a consistent subset of the field equations of the $\cN=4$ CSG twistor action, then it would prove that \eqref{EinTA4} is correct by the Maldacena argument.  A related approach would be to show that the Feynman rules of the two twistor actions are consistent with respect to Maldacena's argument; this would show that \eqref{EinTA4} is correct perturbatively.

Finally, let us point out that our Einstein twistor action has natural generalizations which should account for supergravities with $\cN\leq 8$.  For $\cN=0$ general relativity, we have
\begin{multline}\label{EinTA0}
S^{\mathrm{Ein}}_{\cN=0}[\tilde{h},h]=\int_{\CPT}\D^{3}Z\wedge\tilde{h}\wedge\left(\dbar h+\frac{1}{2}\left\{h,h\right\}\right)  \\
-\frac{\varepsilon^{2}}{\Lambda \kappa^{2}}\int\limits_{\PS\times_{M}\PS}\d\mu\;(\sigma_{1}\sigma_{2})^{4}\;\tilde{h}_{1}\wedge\tau_{1}\wedge\tilde{h}_{2}\wedge\tau_{2}.
\end{multline}
For $\cN=8$ supersymmetry, twistor space is topologically $\P^{3|8}$ and the single graviton multiplet is encoded by $\mathcal{H}\in\Omega^{0,1}_{\CPT}(2)$, which incorporates the negative helicity graviton in the term $\chi^{8}\tilde{h}$.  This leads to an action:
\begin{multline}\label{EinTA8}
S^{\mathrm{Ein}}_{\cN=8}[\mathcal{H}]=\int_{\CPT}\D^{3|8}Z\wedge \mathcal{H}\wedge\left(\dbar \mathcal{H}+\frac{1}{3}\left\{\mathcal{H},\mathcal{H}\right\}\right)  \\
-\frac{\varepsilon^{2}}{\Lambda \kappa^{2}}\int\limits_{\PS\times_{M}\PS}\d\mu\;\frac{\mathcal{H}_{1}\wedge\tau_{1}\wedge \mathcal{H}_{2}\wedge\tau_{2}}{(\sigma_{1}\sigma_{2})^{4}}.
\end{multline}


\subsection{Non-minimal Twistor Actions}

Before proceeding to study the Einstein reduction of the conformal gravity twistor action, let us make some remarks on the possibility of formulating \emph{non-minimal} $\cN=4$ CSG in twistor space.  Since we cannot consistently embed Einstein supergravity in such a theory, such a twistor action won't be useful in obtaining Einstein amplitudes.  Hence, this subsection can be treated as a curiosity and simply skipped over by the reader who is not interested. 

We outline here a proposal for a twistor action describing a particular version of non-minimal $\cN=4$ CSG due to Berkovits and Witten \cite{Berkovits:2004jj}.  While not attempting to prove this proposal, we argue that its perturbation theory will produce all of the expected tree-level scattering amplitudes.  Of course, there are unresolved questions as to whether such a theory is well-defined at the quantum level \cite{Romer:1985yg, Buchbinder:2012uh}, but all of our considerations here will be classical. 

Non-minimal versions of $\cN=4$ CSG are highly non-unique: arbitrary analytic functions can couple the scalar $\varphi$ to the conformal gravitons of the theory.  This can also be captured at the level of a chiral superspace action.  In the minimal case, we saw that the action \eqref{CSUGRAct} served to define a chiral superspace action in terms of $\mathcal{W}$.  However, since $\mathcal{W}$ has conformal weight zero, an action of the form
\begin{equation*}
S[\mathcal{W}]=\int_{M}\d\mu\;F(\mathcal{W}) +\int_{\bar{M}}\d\bar{\mu}\;\overline{F(\mathcal{W})},
\end{equation*}
where $\bar{M}$ is the anti-chiral super-manifold, will be conformal and supersymmetric for \emph{any} holomorphic function $F$.  While $F(\mathcal{W})=\mathcal{W}^2$ corresponds to the minimal theory, other choices clearly lead to interactions between the scalars and conformal gravitons.  For instance, $F(\mathcal{W})=\mathcal{W}^3$ will clearly give a Lagrangian term $\varphi \Psi^{ABCD}\Psi_{ABCD}$.

The twistor-string theory of Berkovits and Witten appears to correspond to a very particular choice of non-minimal $\cN=4$ CSG, with holomorphic function $F(\mathcal{W})=e^{2\mathcal{W}}$ \cite{Berkovits:2004jj}.  We refer to this as Berkovits-Witten CSG, or BW-CSG for short.  As a classical $\cN=4$ theory, it is easy to distinguish BW-CSG from the minimal theory by looking at its scattering amplitudes.  In the twistor-string theory for BW-CSG one finds a degree zero three-point amplitude of the form \cite{Berkovits:2004jj, Dolan:2008gc, Adamo:2012nn}:
\be{BW3pt}
\int \D^{3|4}Z\wedge\left(\partial_{K}f^{I}_{1}\partial_{I}f_{2}^{J}\partial_{J}f^{K}_{3} -\partial_{J}f^{I}_{1}\partial_{K}f_{2}^{J}\partial_{I}f_{3}^{K}\right).
\ee
Applying the Penrose transform, it is easy to see that this amplitude corresponds to a term $\bar{\varphi} \widetilde{\Psi}^{A'B'C'D'}\widetilde{\Psi}_{A'B'C'D'}$ in the space-time action.

Similarly, at degree one, there are amplitudes with an arbitrary number of $g$-insertions; at three-points, this provides the parity conjugate of \eqref{BW3pt}.  The $n$-point version of this amplitude is clearly generated by the chiral part of the space-time action:
\begin{equation*}
\int_{M}\d\mu\;\exp\left(\mathcal{W}(x,\theta)\right)=\sum_{n=2}^{\infty}\int_{M^{0}}\d\mu^{0}\;\varphi^{n-2}\;\Psi^{ABCD}\Psi_{ABCD}+\cdots,
\end{equation*}
where $\d\mu^{0}$ denotes the measure on the bosonic body $M^0$.  Parity invariance demands that we therefore have $n$-point analogues of \eqref{BW3pt}, coming from the anti-chiral part of the space-time action.

Let us try to find a corresponding twistor action: our strategy is to proceed by requiring the twistorial theory to produce the tree-level scattering amplitudes of BW-CSG.  To begin, we note that BW-CSG still has an anti-MHV three point amplitude (like the minimal theory); this comes from the self-dual twistor action we had before:
\be{BW-SD}
S_{1}[g,f]=\int_{\CPT}\D^{3|4}Z\wedge g_{I}\wedge\left(\dbar f^{I}+[f,f]^{I}\right).
\ee
Similarly, the twistorial version of $\int \d\mu e^{\mathcal{W}}$ is an easy generalization of \eqref{minTA2}
\be{BWchiral}
S^{\mathrm{chiral}}[g,f]=\int_{M}\d\mu\;\exp\left(\int_{X} g\right).
\ee
If we expand in fermionic variables, it is clear that on space-time this is the chiral portion of the action
\begin{equation*}
S^{\mathrm{chiral}}\sim \int \d\mu\; \exp(\varphi)\;\Psi^{ABCD}\Psi_{ABCD}+\cdots,
\end{equation*}
as expected.  

We still need to obtain the parity conjugates of the amplitudes generated by \eqref{BWchiral}.  Consider a holomorphic Chern-Simons theory on the tangent bundle $T^{1,0}_{\CPT}$:
\be{hCS}
S^{\mathrm{hCS}}[g,f]=\int_{\CPT}\D^{3|4}Z\wedge\tr\left(f\wedge\dbar f+\frac{2}{3}f\wedge f\wedge f\right)
\ee
Clearly, the cubic term in this action leads to the three-point amplitude \eqref{BW3pt} of BW-CSG.  The quadratic term in \eqref{BW-SD} leads to a $g-f$-propagator, so we can tie any number of $\bar{\mbox{MHV}}$-vertices onto \eqref{BW3pt} to form a $n$-point amplitude which has all $f$ external states.  These all-$f$ amplitudes form the parity-conjugate set to the all-$g$ amplitudes generated by \eqref{BWchiral}.

Hence, we conjecture that the twistor action
\be{BWTA}
S^{\mathrm{BW-CSG}}[g,f]=S_{1}[g,f]+S^{\mathrm{hCS}}[g,f]-\varepsilon^{2}S^{\mathrm{chiral}}[g,f],
\ee
should be (classically) equivalent to the non-minimal $\cN=4$ CSG of Berkovits and Witten.  Of course, our argument relies entirely upon the fact that \eqref{BWTA} has the same tree amplitudes as BW-CSG.  Furthermore, it is rather unfortunate that the anti-chiral portion of the space-time action is encoded only implicitly (i.e., we do not have an explicit $\exp(\bar{\mathcal{W}})$ term on twistor space).  In a sense, this is to be expected because parity invariance is often obscured in twistor space \cite{Witten:2004cp}.


\section{Gravity Tree Amplitudes in Twistor Space}
\label{Chapter6}

In the previous section, we saw how the embedding of Einstein gravity into conformal gravity was manifested at the level of twistor actions.  Now we operationalize these insights to actually compute scattering amplitudes in Einstein gravity on both de Sitter and flat backgrounds.  Our particular focus will be on the MHV amplitude, with two negative helicity gravitons (or $\cN=4$ graviton multiplets) and the rest positive helicity.  

To proceed, we first develop the perturbation theory associated to our twistor actions, identifying the propagators and vertices just as we did in the Yang-Mills case.  The main difference from gauge theory arises in the complicated structure of the vertices, which require their own perturbative expansion in terms of a diagram calculus on $\P^1$.  Applying this formalism, we show that the vertices of the twistor actions (for both \eqref{EinCGTA} and \eqref{EinTA4}) correspond to the MHV amplitudes, for which we obtain an expression for any value of the cosmological constant.  In the flat-space limit, we show that this limits onto Hodges' formula for the MHV amplitude \cite{Hodges:2012ym}.  Finally, we provide an alternative formula for the MHV amplitude based on BCFW recursion in twistor space.


\subsection{Feynman Rules}
\label{CGPerT}

We have two routes open to us for computing Einstein gravity amplitudes on twistor space: via the conformal gravity action \eqref{minTA1}-\eqref{minTA2}, or via the proposed Einstein gravity action \eqref{EinTA4}.\footnote{For ease of notation, we will work with the $\cN=4$ formalism.  As ever, the $\cN=0$ content is can be extracted by a fermionic integral.}  In either case, we need to develop the Feynman rules associated to the twistor action.  

In our study of $\cN=4$ SYM, we saw that the CSW gauge (an axial gauge given by a reference twistor $Z_{*}$) was optimal for performing amplitude calculations.  For the conformal gravity twistor action, the CSW gauge is a choice of coordinates and gauge for $g$ such that one of the anti-holomorphic form components of $f$ and $g$ vanish in the direction of $Z_{*}$:  
\be{CGCSW-gauge}
\overline{Z_*\cdot\frac{\partial}{\partial Z}} \lrcorner f=0=\overline{Z_*\cdot\frac{\partial}{\partial Z}} \lrcorner g  \, ,
\ee 
with identical restrictions on $h$ and $\tilde h$ in the Einstein case.  As in the gauge theory case, this eliminates the cubic vertex from the self-dual portion of the twistor action, and leaves us with:
\be{GFCG}
S[g,f]=\int_{\CPT}\D^{3|4}Z\wedge g_{I}\dbar f^{I} -\varepsilon^{2}\int\limits_{\PS\times_{M}\PS}\d\mu\; g_{1}\wedge g_{2},
\ee
for $\cN=4$ CSG and
\be{GFEin}
S[\tilde{h},h]=\int_{\CPT}\D^{3|4}Z\wedge\tilde{h}\wedge\dbar h - \frac{\varepsilon^{2}}{\Lambda \kappa^{2}}\int\limits_{\PS\times_{M}\PS}\d\mu\wedge\tau_{1}\wedge\tau_{2}\wedge\tilde{h}_{1}\wedge\tilde{h}_{2},
\ee
for the Einstein action.

In each case, we see that the kinetic term is provided by the gauge-fixed portion of the self-dual action while the vertices must be generated by the remaining non-self-dual interactions.  Equation \ref{TGenF} tells us that (upon restricting to Einstein states and dividing by the appropriate power of $\Lambda$) this interaction term should be the generating functional of MHV amplitudes.  In other words, the second term in \eqref{GFCG} or \eqref{GFEin} plays the role of $\log\det(\dbar+\cA)$ from $\cN=4$ SYM.  Clearly, we need a method for perturbatively expanding these terms as generating functionals to obtain the vertices.  Since this structure is universal (i.e., arises for both actions), we address it after first discussing the propagator.


\subsubsection{Twistor propagators}

For the proposed Einstein gravity twistor action, the kinetic term is simply
\begin{equation*}
S^{\mathrm{kin}}[\tilde{h},h]=\int_{\CPT}\D^{3|4}Z\wedge\tilde{h}\wedge\dbar h,
\end{equation*}
so we know that the propagator will look like a distributional form $\bar{\delta}^{2|4}$ with appropriate weights.  Indeed, the correct propagator in CSW gauge is easily seen to be
\be{Einprop}
\Delta^{\mathrm{Ein}}(Z_{1},Z_{2})= \bar{\delta}^{2|4}_{2,0,-2}(Z_{1},*,Z_{2})=\int_{\C^2}\frac{\d s}{s}t\;\d t\;\bar{\delta}^{4|4}(Z_{1}+sZ_{*}+tZ_{2}),
\ee
where the subscript denotes the weights.  

In the conformal gravity twistor action, the kinetic term reads
\begin{equation*}
S^{\mathrm{kin}}[\tilde{h},h]=\int_{\CPT}\D^{3|4}Z\wedge g_{I}\wedge\dbar f^{I}, 
\end{equation*}
so the kinetic operator is once again $\dbar$, but now the propagator will have a tensor structure that must account for the freedom in $g_{I}$ and $f^{I}$.  As mentioned in the previous section, the $\cN=4$ geometry makes this situation somewhat ambiguous since the $\partial_{I}f^{I}=0$ and $Z^{I}g_{I}=0$ conditions do not fix the gauge freedom in $f,g$.  

Focusing on the $\cN=0$ representatives, we know that the twistor propagator must impose $\p_\alpha f^\alpha=0$ and $Z^\alpha g_\alpha=0$.  Since we are on a projective twistor space and the freedom in $f^{\alpha}$ corresponds to adding multiples of $Z^{\alpha}$, we only really need to deal with the condition on $g_{\beta}$.  This can be accounted for with the tensor structure of the propagator, leaving us with
\be{bCGprop}
\Delta^\alpha_\beta(Z_{1},Z_{2})=\delta^\alpha_\beta \bar{\delta}^{2}_{1,0,-5}(Z_{1},*,Z_{2})  -\frac{1}{4} Z_{1}^{\alpha}\frac{\partial}{\partial Z_{2}^{\beta}} \bar{\delta}^{2}_{0,0,-4}(Z_{1},*,Z_{2}),
\ee
so that $Z^{\prime \beta}\Delta_\beta^\alpha=0$ (up to an irrelevant anomaly proportional to the reference twistor).  This is then extended to each propagator component to build the full $\cN=4$ propagator.

Of course, restricting $\cN=4$ CSG to Einstein degrees of freedom sets $f^{I}=I^{IJ}\partial_{J}h$ and $g_{I}=I_{IJ}Z^{J}\tilde{h}$, which automatically fixes the gauge freedom.  Hence, for calculations in conformal gravity restricted to Einstein states (what we are ultimately interested in), we can always take the $\cN=4$ CSG propagator to be:
\be{CGprop}
\Delta^{I}_{J}(Z_{1},Z_{2})|_{\mathrm{Ein}}= \delta^{I}_{J}\bar{\delta}^{2|4}_{1,0,-1}(Z_{1},*,Z_{2})=\delta^{I}_{J}\int_{\C^2}\frac{\d s}{s}\;t\;\d t\;\bar{\delta}^{4|4}(Z_{1}+sZ_{*}+tZ_{2}).
\ee


\subsubsection{Vertices and tree diagrams}

In CSW gauge, the vertices of both twistor actions are generated by the interaction term $S_{2}$.  Upon restriction to Einstein states, this is generating functional is equivalent in both actions, and is given by
\be{CGGF2}
S_{2}[\tilde{h},h]=\int\limits_{\PS\times_{M}\PS}\d\mu\wedge\tau_{1}\wedge\tau_{2}\wedge\tilde{h}_{1}\wedge\tilde{h}_{2}.
\ee
In order to obtain vertices for conformal gravity itself, one could simply choose two \emph{independent} reference twistors $I^{IJ}$ and $\tilde{I}_{IJ}$.  This is equivalent to giving a basis for conformal gravity polarization states in terms of Einstein degrees of freedom \cite{Adamo:2013tja}.

Equation \eqref{TGenF} indicates that these vertices should correspond (on-shell) to MHV amplitudes.  Twistorially, we can see this by noting that the generating functional contains two negative helicity gravitons in $\tilde{h}_{1}$, $\tilde{h}_{2}$, and the self-dual background space-time $M$ encodes the positive helicity gravitons via theorem \ref{NLG}.  What we need is a way of systematically expanding this background to recover the $n-2$ individual positive helicity wavefunctions.  

\subsubsection*{\textit{Perturbative iteration and measure}}

The non-linear graviton construction tells us that $M$ is realized twistorially as the space of holomorphic curves $X\subset\CPT$ which are constructed by solving \eqref{sholomap}
\be{GCE}
\dbar Z^{I}(x,\sigma)=f^{I}(Z)\equiv I^{IJ}\partial_{J}h(Z),
\ee
where we abbreviate $\dbar=\dbar|_{X}$ and $(x,\sigma)=(x^{\mu}, \theta^{Aa}, \sigma_{B})$ from now on to lighten notation.  Generically, the functional form of $Z^{I}$ may be very complicated; to simplify the situation we look for a coordinate transformation on $\PS$ which trivializes the incidence relations \eqref{sincid}.  This provides us with a mechanism for perturbatively expanding the SD background $M$, which we later realize as a calculus of tree diagrams on $\P^1$.

\begin{figure}[t]
\centering
\includegraphics[width=3.20 in, height=2 in]{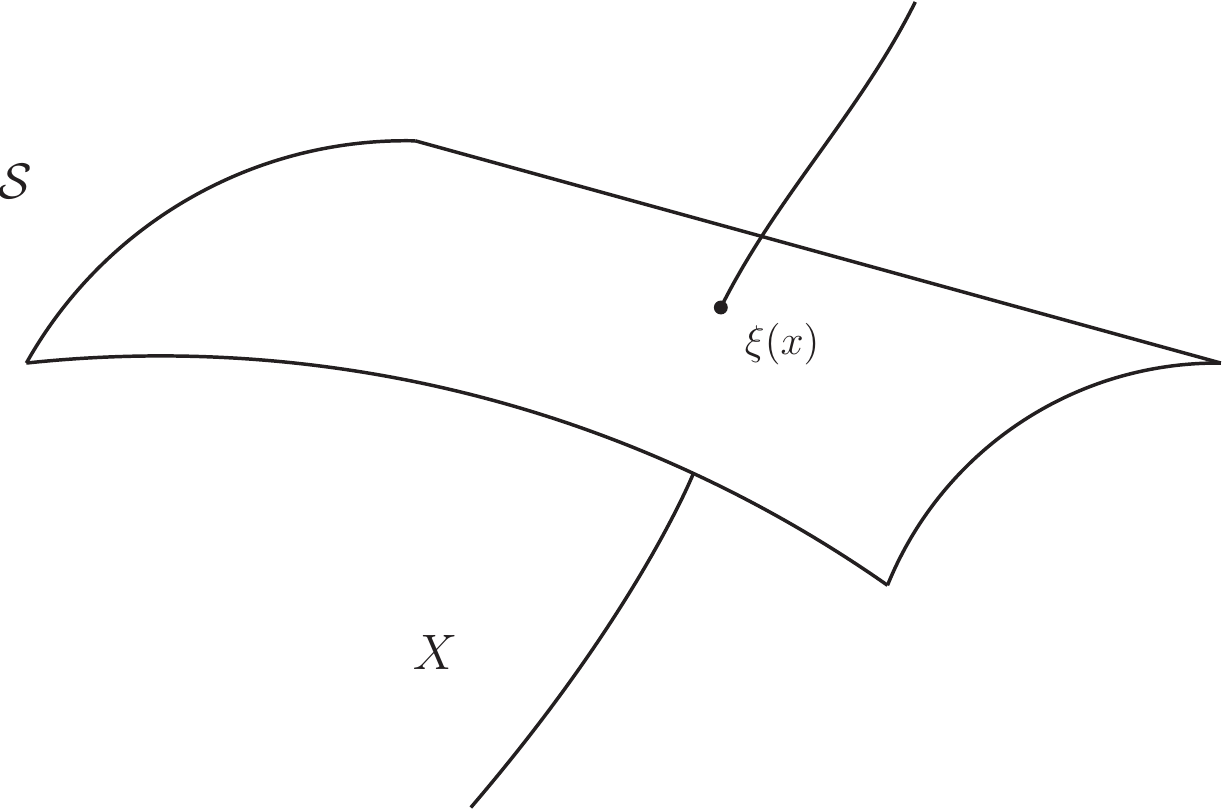}\caption{\textit{The surface $\mathcal{S}$ induces a choice for the coordinates $\cX$ on $\PS$.}}\label{Gauge1}
\end{figure}

By assumption, $Z^{I}(x,\sigma)$ is homogeneous of degree one in $\sigma$, so we can always find an array $\cX^{I A}(x,\sigma)$ which obeys
\be{Auxcord1}
\sigma_{A}\cX^{I A}(x,\sigma)=Z^{I}(x,\sigma).
\ee
\emph{A priori}, the $\cX$ are $(8|8)$ complex functions on $\PS$, but requiring them to be projective and obey \eqref{Auxcord1} reduces the number of independent components to $(5|8)$, so the $\cX$s act as a change of coordinates on $\PS$.  There is considerable freedom in the choice of $\cX^{I A}$; one viable choice is to take a surface $\mathcal{S}\subset\CPT$.  Each curve $X\subset\CPT$ will intersect $\mathcal{S}$ at a unique point $Z^{I}(x,\xi)$, as illustrated in Figure \ref{Gauge1}.  We then define:
\be{Auxcord2}
\cX^{I A}(x,\sigma)=\frac{Z^{I}(x,\sigma)\xi^{A}(x)-Z^{I}(x,\xi)\sigma^{A}}{(\sigma\xi)},
\ee
which clearly obeys \eqref{Auxcord1}.

Now, suppose $X^{I A}\sigma_{A}$ is the homogeneous solution to \eqref{holomap},
\begin{equation*}
X^{J A}=(\delta^{A}_{B}, x^{AB'}, \theta^{bA}).
\end{equation*}
We can (formally) solve \eqref{holomap} by a Picard-like iteration,\footnote{Note that the iteration here is a perturbative expansion for the amplitude generating functional, and differs from the (actual) Picard expansion which appears in \cite{Mason:2008jy}.  A translation between the two can be achieved, but only after some rather non-obvious applications of the Schouten identity (c.f., \cite{Nguyen:2009jk}).} with $Z^{I}_{0}=X^{I A}\sigma_{A}$ and 
\be{Picard1}
Z_{i}^{I}(x,\sigma)=X^{I A}\sigma_{A}+\dbar^{-1}\left(f^{I}(Z_{i-1})\right).
\ee
In the flat background limit, this iteration can be understood as perturbatively solving the good cut equation; in the space-time context, one would require a Green's function for the $\eth$-operator rather than $\dbar$ (c.f., \cite{Adamo:2009vu}).

Since $f^{I}$ is a form of weight $+1$, there is an ambiguity in what we mean by $\dbar^{-1}$.  It is natural to make a choice which is compatible with our coordinates $\cX$, so that the iteration becomes:
\be{Picard2}
Z_{i}^{I}(x,\sigma)=X^{I A}\sigma_{A}+\int_{\P^1}\frac{\D\sigma'}{(\sigma\sigma')}\frac{(\xi\sigma)^2}{(\xi\sigma')^2}f^{I}(Z_{i-1}(\sigma')).
\ee
Here, $\xi\in\P^{1}$ is an arbitrary point on the Riemann sphere reflecting the (two-fold) ambiguity in inverting the $\dbar$-operator; physical observables such as scattering amplitudes should be independent of $\xi$ at the end of our calculations.

This choice induces an expansion for the $\cX$s:
\be{Picard3}
\cX^{I A}(x,\sigma)=X^{I A}+\xi^{A}\int_{\P^1}\frac{\D\sigma'}{(\sigma\sigma')}\frac{(\xi\sigma)}{(\xi\sigma')^2}f^{I}(X\cdot\sigma')+\cdots.
\ee
Clearly, the $\cX$s are redundant coordinates depending on the choice of spin frame, but we can now read off their $\sigma$-dependence from \eqref{Picard3}:
\be{Picard4}
\dbar \cX^{I A}(x,\sigma)=\frac{\xi^{A}f^{I}}{(\xi\sigma)}.
\ee
This enables us to take the exterior derivative of $\cX$ with respect to the space-time coordinate $x$, finding
\be{Yder}
\dbar\left(\d_{x}\cX^{I A}(x,\sigma)\right)=\partial_{J} f^{I}\frac{\xi^{A}\sigma_{B}}{(\xi\sigma)}\d_{x}\cX^{J B}(x,\sigma).
\ee
Since $\partial_{I}f^{I}=0$, this means that the top-degree form $\d^{8|8}\cX$ is holomorphic in $\sigma$ and of weight zero; by Liouville's theorem, it is therefore independent of $\sigma$.  But this means that
\begin{equation*}
\d\mu = \frac{\d^{8|8}\cX}{\mathrm{vol}\;\GL(2,\C)}=\frac{\d^{8|8}X}{\mathrm{vol}\;\GL(2,\C)},
\end{equation*}
is an invariant volume form on the space-time $M$ itself.

Now we want to implement this iteration by perturbatively expanding \eqref{CGGF2}.  \emph{A priori}, the deformation of $Z^{I}$ given by \eqref{Picard1} can act on the wavefunctions $\tilde{h}_{1,2}$ (which have non-polynomial dependence on $Z$) or the contact structures $\tau_{1,2}$.  From \eqref{Picard2}, the action on a wavefunction is given by
\be{thdef}
\tilde{h}_{1}\rightarrow \int_{\P^1}\frac{\D\sigma_{i}\;(\xi\;1)^{2}}{(1\;i)(\xi\;i)^{2}}[\partial\tilde{h}_{1}, \partial h_{i}],
\ee
where we use the shorthand $[\partial\tilde{h}_{1}, \partial h_{i}] = I^{IJ}\partial_{1 I}\tilde{h}_{1}\partial_{i\;J}h_{i}$.  As for the contact structures, recall that each $\tau$ is quadratic in $Z$:
\begin{equation*}
\tau=\la Z, \partial Z\ra=I_{IJ}Z^{I}\;\partial Z^{J},
\end{equation*}
and can therefore absorb at most two deformations.

Furthermore, since the deformation always carries a power of the infinity twistor $I^{IJ}$, and this will contract with the $I_{JK}$ in $\tau$, such a deformation will always result in a power of $\Lambda$.  A bit of algebra shows that a single deformation of the contact structure (say, $\tau_{1}$) results in \cite{Adamo:2013tja}:
\be{tauprop1}
\psi^{1}_{i}=
\Lambda \frac{\D\sigma_i (\xi 1)^4}{(1i)^{2}(\xi i)^2} \rd_1\left( \frac{ (i1)}{(\xi 1)^2}Z^I_1\right)\partial_{iI} h_i 
=\Lambda \frac{\D\sigma_1 \D\sigma_i (\xi 1)}{(1i)^{2}(\xi i)^2}\left[(\xi i)\;Z^{I}_{1}+(1i)\;Z^{I}(\xi)\right]\partial_{iI} h_i \, ,
\ee
which is then integrated over the $\P^{1}$ corresponding to $\sigma_{i}$.  The second expression here uses the linearity of $Z^{I}(x,\sigma)$ in $\sigma$.  If a second deformation acts at the same contact structure, we can use the first expression to arrive at
\be{tauprop2}
\omega^{1}_{ij}=-\Lambda\frac{\D\sigma_1 \D\sigma_i \D\sigma_j (1\xi)^{4}(ij)}{(1i)^{2}(1j)^{2}(\xi i)^{2}(\xi j)^{2}}\left[\partial_i,\partial_j \right] h_i h_j \, ,
\ee 
which is now integrated over the additional $\P^{1}$ corresponding to $\sigma_{j}$.  

Inspecting the expression for $\psi^{1}_{i}$ in \eqref{tauprop1}, we can actually manipulate it into a format which is $\d_{i}$-exact:
\be{psip}
\psi^{1}_{i}=2\Lambda \D\sigma_{1}\; \sigma_{1A}\;\d_{i} \left(\frac{\sigma^{A}_{i}(\xi 1)h_{i}}{(1i)^{2}(\xi i)}\right).
\ee
As a result, we may be tempted to conclude that such deformations vanish.  If the new wavefunction $h_{i}$ is undeformed by any additional iterations, then this is indeed true \cite{Adamo:2013tja}.  However, we will see that for generic perturbative expansions, $h_{i}$ will also receive deformations and the contribution from $\psi^{1}_{i}$ is non-trivial.

\subsubsection*{\textit{Tree diagram calculus}}

At this point, the perturbative iteration gives us a method for expanding the SD background in \eqref{CGGF2} to extract the positive helicity gravitons of the vertex.  To compute the $n$-point vertex, we must expand to order $n-2$, since each iteration introduces a single new positive helicity graviton via \eqref{thdef}, \eqref{tauprop1}, or \eqref{tauprop2}.  There are many different ways in which this expansion can be performed.  For instance, when $n=3$ there are four distinct contributions, since the deformation of $Z^{I}$ can act at $\tau_{1,2}$ or $\tilde{h}_{1,2}$.  Clearly, the number of possible expansions grows dramatically with $n$, since we can also perturb the resulting wavefunctions $\{h_{i}\}$ at higher order.  

It turns out that we can represent each such iteration uniquely by a \emph{forest of tree graphs} on $n+2$ vertices.  Suppose we have perturbatively expanded the generating functional to order $n-2$ (so there are $n-2$ positive helicity wavefunctions $h_i$).  Then the graph associated to given expansion is constructed by:
\begin{itemize}
\item Draw a black vertex for the each negative helicity wavefunction $\tilde{h}_{1}$, $\tilde{h}_{2}$.

\item Draw a grey vertex for each contact structure $\tau_{1}$, $\tau_{2}$.

\item Draw a white vertex for each of the $n-2$ positive helicity wavefunctions $h_{i}$.

\item Draw an arrow connecting each white vertex to its source in the expansion.
\end{itemize}
\begin{figure}[h]
\centering
\includegraphics[width=2.1 in, height=0.35 in]{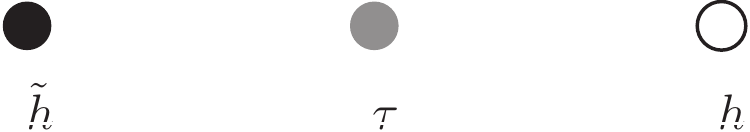}\caption{\small{\textit{Building blocks for Feynman diagrams}}}\label{FeynRules}
\end{figure}  

\begin{figure}[t]
\centering
\includegraphics[width=4.20 in, height=2 in]{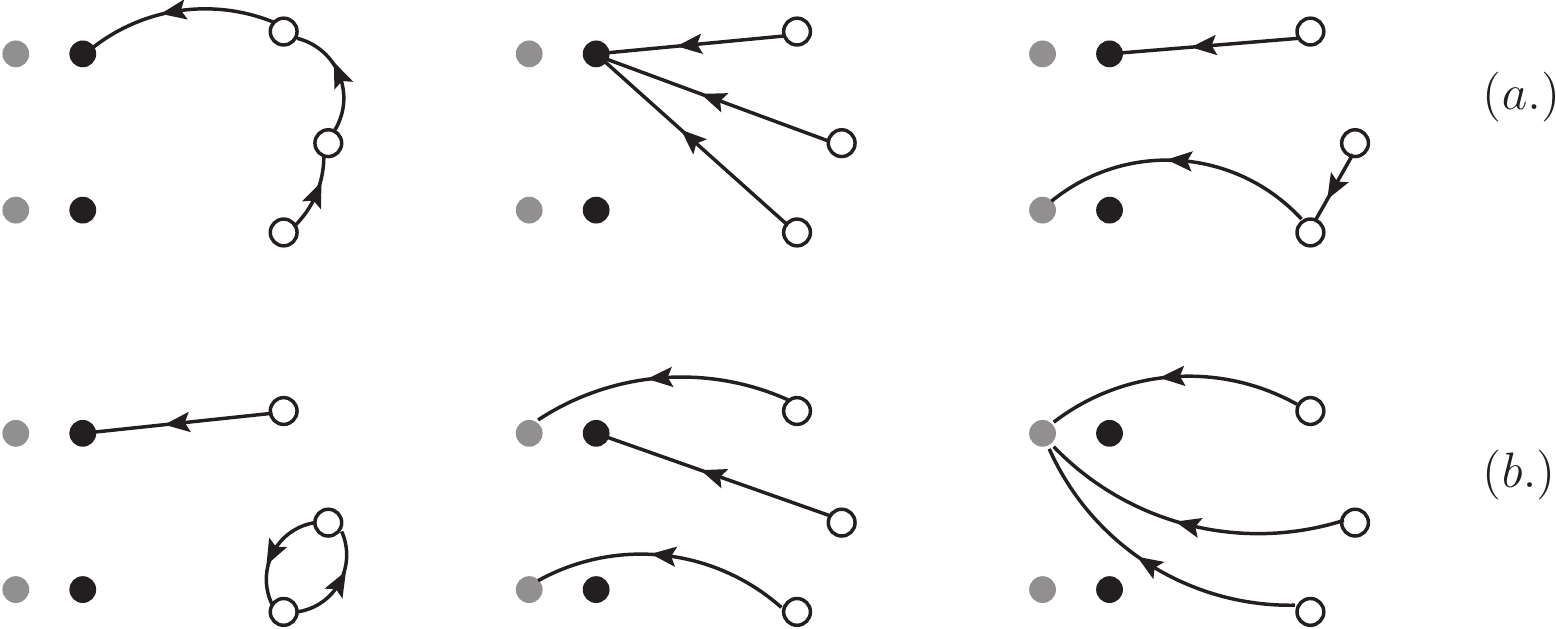}\caption{\small{\textit{Some diagrams for the 5-point vertex which have a non-vanishing \emph{(a.)}, or excluded/vanishing \emph{(b.)} contribution.}}}\label{Diags}
\end{figure}

By following the arrows through the diagram, one can trace each branch of the diagram back to its source, which must be one of the grey or black vertices representing the factors of the original generating function \eqref{CGGF2}.  So starting anywhere in the graph, if we follow the arrows then eventually we will wind up at a grey or black vertex.  In other words, each graph corresponds to a forest of trees each of which is \emph{rooted} at a grey or black vertex.  The fact that each connected component is a \emph{tree graph} follows from the fact that each step in the expansion produces a new positive helicity wavefunction (so we can't follow arrows and end up completing a loop).  Some examples of diagrams which either contribute (\emph{a}.) or are excluded from the contribution to the 5-point vertex are shown in Figure \ref{Diags}.

Of course, we want to associate each of these diagrams with some integrand on the moduli space $\CM_{n,1}$; summing all the contributions and integrating should give the vertex just like in the Yang-Mills story.  The computational dictionary required is simple and determined by the perturbative iteration itself, which endows the edges with weights.  An arrow from a white vertex $i$ to a white or black vertex $j$ corresponds to a weight
\be{piprop}
\HH_{ij}=\frac{I^{IJ}\;(\xi j)^{2}}{(ij)\;(\xi i)^2}\frac{\partial}{\partial Z^{I}(\sigma_i)}\frac{\partial}{\partial Z^{J}(\sigma_j)}=\frac{(\xi j)^{2}}{(ij)\;(\xi i)^2}[\partial_{i},\partial_{j}],
\ee
while an arrow from a white vertex to a grey vertex corresponds to $\psi^{1,2}_{i}$ from \eqref{tauprop1}.  If a grey vertex has two incoming arrows, say from white vertices $i$ and $j$, then we can account for both with the weight factor $\omega^{1,2}_{ij}$ from \eqref{tauprop2}.

This picture can be made precise by writing the vertex generating functional in a way that makes the construction of the SD background $M$ via \eqref{GCE} explicit.  Introducing a Lagrange multiplier $Y\in\Omega^{1,0}(\P^{1},T^{*}_{\CPT})$, this becomes \cite{Adamo:2013tja}:
\begin{equation*}
S_{2}[\tilde{h},h]=\int_{M}\d\mu \left[\int_{X}Y_{I}\left(\dbar Z^{I}-I^{IJ}\partial_{J}h\right)+\frac{\varepsilon^{2}}{\Lambda \kappa^{2}}\left(\int_{X} \tilde{h}\wedge\tau\right)^2 \right].
\end{equation*}
Integrating out $Y$ returns \eqref{GCE}, but keeping it in play allows us to perform the perturbative expansion of $S_{2}$ to order $n-2$ by using Feynman diagrams on $\P^1$ with vertices
\begin{equation*}
V_{\tilde{h}}=\int\limits_{X\times X}\tau_{1}\wedge\tau_{2}\wedge\tilde{h}_{1}\wedge\tilde{h}_{2}, \qquad V_{h_i}=\int_{X}[Y,\partial_{i}h_i],
\end{equation*}
for $i=3,\ldots,n$.  Contractions occur via the $\P^1$ propagator for the $YZ$-system (which arises in the similar context of the Berkovits-Witten twistor-string \cite{Berkovits:2004jj}):
\begin{equation*}
\left\la Y_{I}(\sigma_{i})\;Z^{J}(\sigma_{j})\right\ra = \delta^{J}_{I}\frac{\D\sigma_{i}}{(ij)}\frac{(\xi j)^{2}}{(\xi i)^2},
\end{equation*}
and working classically, this results in the the forests of trees we just described.

In sum, the vertices of the twistor actions are given by summing these weighted tree diagrams.  If we write the set of diagrams contributing to the $n$-point vertex as $\cF^{n}$, then this has a natural disjoint union splitting based on the number of arrows which are incoming at the grey vertices.  That is,
\begin{equation*}
\mathcal{F}^{n}=\bigsqcup_{k=0}^{4}\mathcal{F}^{n}_{k},
\end{equation*}
where each diagram $\Gamma\in\mathcal{F}^{n}_{k}$ is a forest on $n+2$ vertices which has $k$ arrows into the grey vertices (we cannot have $k>4$ because $\tau$ is only quadratic in $Z$).  We then write the $n$-point vertex of the twistor action--somewhat heuristically--as
\be{GVert1}
\mathcal{V}_{n}=\frac{1}{\Lambda}\sum_{k=0}^{4}\sum_{\Gamma\in\mathcal{F}^{n}_{k}}\int\limits_{M\times(\P^1)^n}\d\mu\;F_{\Gamma}\;\tau_{1}\tilde{h}_{1}\;\tau_{2}\tilde{h}_{2}\prod_{i=3}^{n}h_{i}\;\D\sigma_{i},
\ee
where $F_{\Gamma}$ encodes the contribution from diagram $\Gamma$ built out of the weights.\footnote{Here, we think of the integral over $\CM_{n,1}$ as being over $\M\times (\P^1)^n$.}  Below, we will turn this somewhat esoteric formula into a concrete expression for the MHV amplitude.

Before proceeding, one should note that these tree diagrams first arose in the context of a semi-classical connected tree formalism for the worldsheet CFT of Berkovits-Witten twistor-string theory \cite{Adamo:2012xe}.  In that arena, trees were needed to extract the minimal content from the non-minimal $\cN=4$ CSG in the twistor-string at MHV; the fact that we obtain the same formalism directly from the minimal twistor action proves that trees indeed isolate the minimal content.  The more puzzling question of why trees were required in the worldsheet CFT (which in principle should include all loop and disconnected diagrams) found its resolution in Skinner's twistor-string \cite{Skinner:2013xp}: there worldsheet supersymmetry suppresses the loops and the resulting tree diagrams lead directly to the flat-space amplitudes of Einstein gravity.


\subsection{The MHV Amplitude}
\label{MHVLambda}

The embedding of Einstein gravity into conformal gravity tells us that the vertices of the twistor action should correspond to MHV amplitudes on-shell.  We have just described how to obtain these vertices by summing weighted tree diagrams in \eqref{GVert1}, but we still need a concrete method for performing this sum.  It turns out that summing forests of tree graphs (with weights) is a natural operation in algebraic combinatorics; the key result in this area is the \emph{Matrix-Tree theorem} (an analogue of Kirchoff's theorem for directed graphs), which relates the counting of graphs with weights to a determinant of the Laplacian matrix of the graph (c.f., \cite{Stanley:1999, vanLint:2001, Stanley:2012} or \cite{Feng:2012sy} for an overview with direct connections to gravity amplitudes).

For an arbitrary oriented graph $G$ with $n$ vertices, let us denote the edge from vertex $i$ to vertex $j$ by $(i,j)\in\mathcal{E}$, where $\mathcal{E}$ denotes the set of edges in $G$.  If the edge $(i,j)$ is endowed with weight $w_{ij}\in\C$, then the weighted Laplacian matrix of $G$ is the $n\times n$ matrix with entries
\begin{equation*}
\mathcal{L}_{ij}(G)=\left\{
\begin{array}{c}
-w_{ij} \;\mathrm{if}\;i\neq j\;\mathrm{and}\;(i,j)\in\mathcal{E}\\
\sum_{(i,k)\in\mathcal{E}}w_{ik}\;\mathrm{if}\;i=j \\
0 \;\;\mathrm{otherwise}
\end{array}\right. .
\end{equation*}  
The Matrix-Tree theorem for rooted forests on the directed graph $G$ is then:
\begin{thm}[Weighted Matrix-Tree Theorem for Forests]\label{MTT}
Let $\mathcal{F}^{(i_1,\ldots i_r)}(G)$ be the set of forests of $G$ rooted at $\{i_1,\ldots, i_r\}$ and $\mathcal{L}(G)$ be the weighted Laplacian matrix of $G$.  For each $F\in\mathcal{F}^{(i_1,\ldots i_r)}(G)$, denote by $E_F\subset\mathcal{E}$ the set of edges in the forest.  Then
\be{MTT*}
\left|\mathcal{L}(G)^{i_{1}\cdots i_{r}}_{i_{1}\cdots i_{r}}\right|= \sum_{F\in\mathcal{F}^{(i_1,\ldots i_r)}(G)}\left(\prod_{(i,j)\in E_{F}}w_{ij}\right),
\ee
where $ \left|\mathcal{L}(G)^{a\cdots b}_{c\cdots d}\right|$ denotes the determinant of $\cL(G)$ with the rows $\{a,\ldots, b\}$ and columns $\{c,\ldots, d\}$ removed.
\end{thm}
A proof of this particular version of the matrix-tree theorem can be found in \cite{Feng:2012sy}.

Our goal is now to apply theorem \ref{MTT} to formula \eqref{GVert1} and obtain an explicit formula for the MHV amplitude.  This formula will be defined initially with $\Lambda\neq 0$, so we actually find a twistorial expression for the MHV `scattering amplitude' on a de Sitter (or anti-de Sitter) background.\footnote{Clearly, the asymptotically flat definition of the S-matrix no longer holds on de Sitter backgrounds.  While $\scri^{-}$ to $\scri^{+}$ scattering is still mathematically defined (i.e., a meta-observable in the sense of \cite{Witten:2001kn}), no physical observer can integrate over the whole space-time.  In AdS, the situation is improved and we can consider correlation functions in the boundary CFT.  We make some remarks about how to interpret the twistor MHV formula in these contexts in Section \ref{Chapter7}.}  Passing to the $\Lambda\rightarrow 0$ limit, we will recover Hodges' formula for the flat-space MHV amplitude.


\subsubsection{Summing diagrams and the vertex formula}  

\begin{figure}
\centering
\includegraphics[width=4 in, height=1 in]{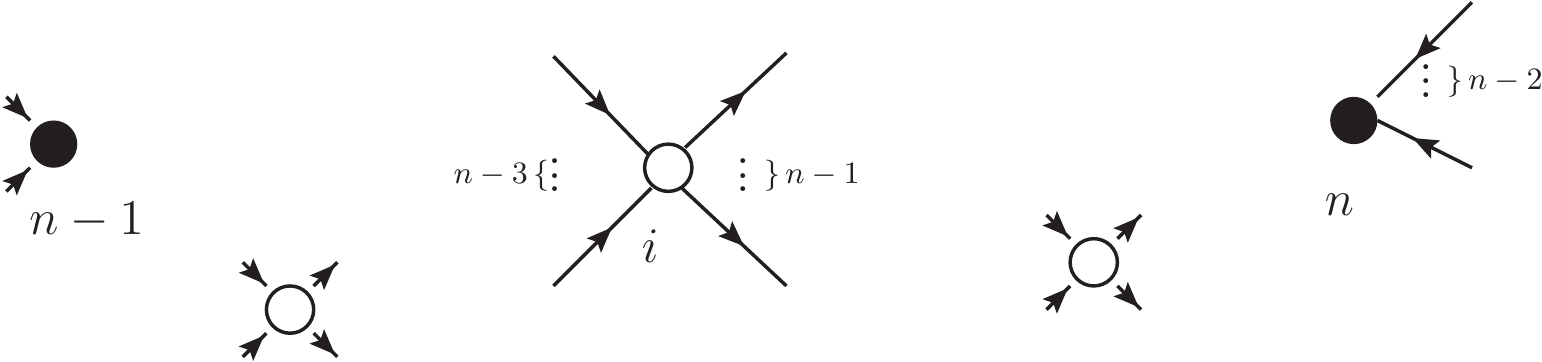}\caption{\textit{The graph $G^{n}_{0}$ features all possible edges which could contribute to $\cF_{0}^{n}$}}\label{G}
\end{figure}

In the case of the twistor action vertices, we can apply theorem \ref{MTT} to each subsector of $\cF^{n}=\sqcup_{k=0}^{4}\mathcal{F}^{n}_{k}$ successively.  Let us begin with $\cF^{n}_{0}$; this includes all diagrams which have no arrows into the grey vertices (i.e., there are no deformations of the contact structures $\tau_{1,2}$).  In this case, the directed graph $G^{n}_{0}$ which forms the input for the Matrix-Tree theorem can be built from $n-2$ white vertices and $2$ black vertices, since they grey vertices play no role.  The edges of $G^{n}_{0}$ feature all possible perturbative expansions which could produce the $n-2$ white vertices.  So each white vertex has $n-1$ outgoing edges (one to every other vertex) and $n-3$ incoming edges (one from every other white vertex), while each black vertex has $n-2$ incoming edges and no outgoing edges. See Figure \ref{G} for an illustration of this configuration.

Up to a rank-two error term and conjugation (both of which are irrelevant), the weighted Laplacian matrix associated to $G^{n}_{0}$ is given by \cite{Adamo:2012xe}:
\be{wLap1}
\mathcal{L}(G^{n}_{0})=\HH=\left(
\begin{array}{cccccc}
\HH_{11} & \HH_{12} & \cdots & \HH_{1n} \\
\HH_{21} & \ddots   &        & \HH_{2n} \\
\vdots   &          & \HH_{n-1\;n-1} & \HH_{n-1\;n} \\
\HH_{n1} & \cdots   &\HH_{n\;n-1} & \HH_{nn}
\end{array}\right), \qquad \HH_{ii}=-\sum_{j\neq i}\HH_{ij},
\ee
where the off-diagonal entries are precisely the weights $\HH_{ij}$ from \eqref{piprop}.  As there are no grey vertices in play, each forest of trees contributing to the vertex in the class $\cF^{n}_{0}$ must be rooted at one of the two black vertices.  Then theorem \ref{MTT} indicates that the required sum of weights is accomplished by taking the determinant $|\HH^{12}_{12}|$, with the rows and columns corresponding to $\tilde{h}_{1}$, $\tilde{h}_{2}$ removed from the weighted Laplacian \eqref{wLap1}.

We can also write the undeformed contact structures as
\begin{equation*}
\tau_{1}=\la Z_{1},\partial Z_{1}\ra= I_{IJ}X_{A}^{I}\sigma_{1}^{A}\;X^{J}_{B}\d\sigma_{1}^{B}=X^{2}\;\D\sigma_{1},
\end{equation*}
where we abbreviate $X^{2}=\la X_{A},X^{A}\ra$.  Combined with the Matrix-Tree theorem, this gives us the contribution to the vertex $\mathcal{V}_{n}$ from graphs in $\cF^{n}_{0}$:
\be{cont0}
\int \d\mu\;(X^{2})^{2}\;\left| \HH^{12}_{12}\right|\;\prod_{i=1}^{n}h_{i}\;\D\sigma_{i},
\ee
where we understand that $h_{1,2}\equiv\tilde{h}_{1,2}$.

Precisely the same process of considering the graph $G^{n}_{k}$ of all possible deformations, writing down the weighted Laplacian, and then applying the Matrix-Tree theorem allows us to account for the contributions from every other sector $\cF^{n}_{k>0}$.  For instance, in $\cF^{n}_{1}$ there is a single arrow incoming to one of the grey vertices, which can come from any of the $n-2$ white vertices in play, say $i$.  This gives an overall factor of $\psi^{1}_{i}$ or $\psi^{2}_{i}$, depending on which contact structure is deformed.  The remaining portions of the diagram are again accounted for by the weighted Laplacian $\HH$, but now we have to eliminate \emph{three} rows and columns when applying theorem \ref{MTT}, since the trees can also be rooted at vertex $i$ now.  The result for $\cF^{n}_{1}$ is therefore:
\be{cont1}
\sum_{\Gamma\in\mathcal{F}^{n}_{1}}\int \d^{4|8}x\;X^{2}\;F_{\Gamma}\;\prod_{i=1}^{n}h_{i}\;\D\sigma_{i} =\int \d\mu\;X^{2}\sum_{i=3}^{n}\psi^{1}_{i}\;\left| \HH^{12i}_{12i}\right|\; \prod_{j=1}^{n}h_{j}\;\D\sigma_{j}+(1\leftrightarrow 2),
\ee   
where the factor of $X^{2}$ is due to the undeformed contact structure still in play.

Proceeding in this fashion, we arrive at the formula for the $n$-point vertex \cite{Adamo:2013tja}:
\begin{multline}\label{MHVamp}
\mathcal{V}_{n}=\frac{1}{\Lambda}\int \d\mu\;\left[ (X^2)^2\left|\HH^{12}_{12}\right|+ X^{2} \sum_{i}\psi^{1}_{i}\left|\HH^{12i}_{12i}\right| +X^{2}\sum_{i,j}\omega^{1}_{ij}\left|\HH^{12ij}_{12ij}\right| \right. \\
\left. +\sum_{i,j}\psi^{1}_{i}\psi^{2}_{j}\left|\HH^{12ij}_{12ij}\right| +\sum_{i,j,k}\psi^{1}_{i}\omega^{2}_{jk}\left|\HH^{12ijk}_{12ijk}\right| +\sum_{i,j,k,l}\omega^{1}_{ij}\omega^{2}_{kl}\left|\HH^{12ijkl}_{12ijkl}\right|\right]\prod_{m=1}^{n}h_{m}\;\D\sigma_{m}\:+(1\leftrightarrow 2).
\end{multline}
In this expression, the sums are understood to run over all indices which are not excluded from the determinant, and also to symmetrize on those indices.  For instance, in the first term of the second line $\sum_{i,j}$ runs over all $i,j=3,\ldots n$ with $i\neq j$.  The expression \eqref{MHVamp} makes sense \emph{off-shell} (i.e., as a vertex of the twistor action) since no step in our derivation from $S_{2}[\tilde{h},h]$ assumed that the wavefunctions were $\dbar$-closed on twistor space.  

A non-trivial test on this formula for $\mathcal{V}_{n}$ is independence of the reference spinor $\xi\in\P^{1}$.  This entered the definition of the perturbative iteration due to the ambiguity in defining $\dbar^{-1}$ on $\P^1$.  From \eqref{Picard3}, we see that a change in $\xi$ induces a variation in $\cX$, which are coordinates on the projective spinor bundle $\PS$.  Hence, a variation in $\xi$ should correspond to a diffeomorphism on $\PS$ under which $\mathcal{V}_{n}$ should be invariant.  This can be checked explicitly by considering the infinitesimal variation generated by $\d_{\xi}=\d\xi^{A}\frac{\partial}{\partial\xi^{A}}$, and it can be shown that \cite{Adamo:2013tja}:
\be{grgauge}
\d_{\xi}\mathcal{V}_{n}=\int\frac{\d^{8|8}X}{\mathrm{vol}\;\GL(2,\C)}\frac{\partial}{\partial X^{IA}} V^{IA}=0,
\ee
where $V^{IA}$ are the components of a smooth vector field (roughly speaking, on $\CM_{n,1}$).\footnote{Of course, this argument is on-shell in nature; if $\mathcal{V}_{n}$ appears inside a Feynman diagram it need not be $\xi$-independent on its own.  Only the full amplitude being calculated needs to be independent of the reference spinor.}

So on-shell, \eqref{MHVamp} is a well-defined formula for the MHV amplitude with cosmological constant.  However, the on-shell condition actually allows us to simplify the expression substantially, as we will show now.


\subsubsection{The amplitude}

Equation \eqref{MHVamp} provides a perfectly valid representation of the MHV amplitude with $\Lambda\neq 0$; by inserting momentum eigenstates or some other on-shell wavefunctions, we pass from a vertex to the amplitude $\mathcal{V}_{n}\rightarrow\cM_{n,0}$.  It turns out that this formula can be simplified considerably on-shell, though.

To begin, note that all the weights which appear in \eqref{MHVamp} take the form of differential operators, given by \eqref{piprop}, \eqref{tauprop1}, and \eqref{tauprop2}.  These operators act on the wavefunctions $\{h_{i}\}$, and when these are chosen to be momentum eigenstates this action becomes rather complicated, involving derivatives of delta-functions due to the $\Lambda\neq0$ infinity twistor.  Clearly, things would be much simpler if we could treat things algebraically.

This can be accomplished by working with \emph{dual twistor} wavefunctions:
\be{dtwf}
h(Z(\sigma_{i}))=\int_{\C}\frac{\d t_{i}}{t_{i}^{1+w_{i}}}\exp\left(i t_{i}W_{i}\cdot Z(\sigma_{i})\right), \qquad w_{i}=\left\{
\begin{array}{cc}
-2 & \mbox{if}\;i=1, 2 \\
2 & \mbox{otherwise}
\end{array} \right. .
\ee
Here $W_{i\;I}=(\tilde{\mu}^{A}, \tilde{\lambda}_{i A'})$ are coordinates on $n$ copies of dual twistor space, $\PT^{\vee}$.  These wavefunctions have been used before in other contexts \cite{Mason:2009sa, Cachazo:2012pz}, and can be paired with momentum eigenstates in an appropriate manner to obtain functionals of momenta at the end of any calculation.  Furthermore, the scaling parameters $t_{i}$ can be absorbed into the worldsheet coordinates by defining a new set of non-homogeneous coordinates: $\sigma_{i}t_{i}\rightarrow\sigma_{i}$, $\d t_{i}\D\sigma_{i}\rightarrow\d^{2}\sigma_{i}$.

With \eqref{dtwf}, all the weights in our diagram calculus become purely algebraic.  In particular, we now have:
\begin{equation*}
\HH_{ij}=-\frac{[W_{i},W_{j}]}{(ij)},
\end{equation*}
while deformations of the contact structure are
\begin{equation*}
\psi^{1}_{i}=\Lambda\;i\frac{(\xi 1)\;W_{i\;I}}{(1i)^{2}(\xi i)^2}\left[(\xi i)\;Z^{I}(\sigma_{1})+(1i)\;Z^{I}(\xi)\right], \qquad \omega^{1}_{ij}=\Lambda \frac{[W_{i},W_{j}]\;(1\xi)^{4}(ij)}{(1i)^{2}(1j)^{2}(\xi i)^{2}(\xi j)^{2}}.
\end{equation*}
In this framework all the ingredients of \eqref{MHVamp} are transformed from differential operators to algebraic functions of the dual twistors.  Furthermore, the product of wavefunctions and $\P^1$ measures can also be expressed compactly and in a manner that uses a generalization of momentum to the dual twistor framework.  In particular, we have    
\begin{equation*}
\prod_{i=1}^{n}h_{i}\;\D\sigma_{i}=e^{i\mathcal{P}\cdot X}\;\d^{2}\sigma, \qquad \cP_{I}^{A}=\sum_{i=1}^{n}W_{i\;I}\sigma_{i}^{A}, \qquad \d^{2}\sigma\equiv \prod_{i=1}^{n}\d^{2}\sigma_{i},
\end{equation*}
with $\mathcal{P}$ playing the role of total `momentum' in this framework.  

Now, using the dual twistor wavefunctions \eqref{dtwf} note that the second term in the first line of \eqref{MHVamp} can be written as
\begin{multline*}
\int\d\mu\;X^{2}\sum_{i} \psi^{1}_{i}\left|\HH^{12i}_{12i}\right|\;e^{i\cP\cdot X}\d^{2}\sigma \\
=\Lambda \int \d\mu\;X^{2}\sum_{i}\left|\HH^{12i}_{12i}\right|\left(\frac{(\xi 1)(\xi i)\sigma_{1}^{A}+(\xi 1)(1i)\xi^{A}}{(1i)^{2}(\xi i)^{2}}\right)\frac{\partial e^{i\cP\cdot X}}{\partial\sigma_{i}^{A}}\d^{2}\sigma ,
\end{multline*}
using the dual twistor expression for $\psi^{1}_{i}$.  This is motivated by the alternative expression for $\psi^{1}_{i}$ given by \eqref{psip}, in which it is expressed as a derivative with respect to $\sigma_i$.   Hence, we can integrate by parts with respect to $\d^{2}\sigma_{i}$ to find:
\begin{multline*}
\int\d\mu\;X^{2}\sum_{i} \psi^{1}_{i}\left|\HH^{12i}_{12i}\right|\;e^{i\cP\cdot X}\d^{2}\sigma  \\
=-\Lambda\int\d\mu\;X^{2}e^{i\cP\cdot X}\sum_{i}\frac{\partial}{\partial\sigma_{i}^{A}}\left(\left|\HH^{12i}_{12i}\right|\frac{(\xi 1)(\xi i)\sigma_{1}^{A}+(\xi 1)(1i)\xi^{A}}{(1i)^2(\xi i)^2}\right) \d^{2}\sigma \\
=-\Lambda\int\d\mu\;X^{2}e^{i\cP\cdot X}\sum_{i,j}\left|\HH^{12ij}_{12ij}\right|\frac{[W_{i},W_{j}](1\xi)^{4}(ij)}{(1i)^{2}(1j)^{2}(\xi i)^{2}(\xi j)^{2}}\d^{2}\sigma \\
=-\int\d\mu\;X^{2}\sum_{i,j}\omega^{1}_{ij}\left|\HH^{12ij}_{12ij}\right| e^{i\cP\cdot X}\d^{2}\sigma .
\end{multline*}
with the third line following after symmetrizing over $(i\leftrightarrow j)$ and several applications of the Schouten identity.

Thus, we see that following an integration by parts the second term in \eqref{MHVamp} cancels the third term.  A similar calculation demonstrates that the fourth and fifth terms also cancel with each other, so the amplitude can be written much more compactly as:
\be{MHVamp2}
\cM_{n,0}=\frac{1}{\Lambda}\int\d\mu\;\left[ (X^2)^2\left|\HH^{12}_{12}\right| +\sum_{i,j,k,l}\omega^{1}_{ij}\omega^{2}_{kl}\left|\HH^{12ijkl}_{12ijkl}\right|\right]\prod_{m=1}^{n}h_{m}\;\D\sigma_{m}\:+(1\leftrightarrow 2),
\ee
where we have restored arbitrary twistor wavefunctions and homogeneous coordinates.  Clearly this formulation is an improvement over \eqref{MHVamp} in terms of simplicity, although the arguments which lead to it are on-shell in nature (i.e., based upon the dual-twistor wavefunctions) so we do not expect it to be suitable for use as a \emph{vertex} of the twistor action.


\subsubsection{Hodges' formula from the flat-space limit}

The formulae \eqref{MHVamp}, \eqref{MHVamp2} provide expressions for the MHV amplitude of Einstein gravity with non-vanishing cosmological constant.  While we have checked that these expressions are gauge-invariant with \eqref{grgauge}, another obvious check we should perform is the flat space-limit, where $\cM_{n,0}$ should reproduce the flat-space scattering amplitude.

While several forms of the flat-space MHV amplitude for gravity have been known for some time (e.g., the BGK or BCFW formulas \cite{Berends:1988zp, Mason:2008jy, Bedford:2005yy}), the optimal one was only recently discovered by Hodges \cite{Hodges:2012ym}.  In $\cN=4$ language, Hodges' formula reads:
\be{HForm}
\cM^{\mathrm{Hodges}}_{n,0}(\Lambda=0)=\int \d\mu\; \frac{(12)^{2}}{(1i)^{2}(2i)^{2}}\left|\HH^{12i}_{12i}\right|\prod_{j=1}^{n}h_{j}\;\D\sigma_{j},
\ee   
where the entries of the matrix $\HH$ are now built from the flat-space infinity twistor \eqref{infty}.  This is optimal (compared to all previous formulations) in the sense that it makes no reference to an ordering of the external gravitons, requires no explicit sum over permutations, and is the natural analogue of the Parke-Taylor amplitude \eqref{ParkeTaylor} from Yang-Mills theory.

In flat space, the weighted Laplacian $\HH$ has some nice properties, which we make note of here:
\begin{lemma}[Hodges \cite{Hodges:2012ym}]\label{Co-ranklem}
With $\Lambda=0$ and inserting momentum eigenstates into \eqref{HForm}, the matrix $\HH$ is independent of $\xi\in\P^{1}$, obeys
\be{cr3}
\sum_{j=1}^{n}\HH_{ij}\sigma_{j}^{A}\sigma_{j}^{B}=0,
\ee
and hence has co-rank 3.
\end{lemma}
\proof After partially integrating over $\d\mu$ in \eqref{HForm}, momentum conservation emerges using traditional momentum eigenstates in the form
\begin{equation*}
\sum_{j=1}^{n}\tilde{p}^{A'}_{j}\sigma^{A}_{j}=0.
\end{equation*}
Now, suppose we chose a second reference spinor $\zeta\in\P^{1}$.  The only change in $\HH$ is in the diagonal entries:
\begin{multline*}
\Delta\HH_{ii}=-\sum_{j\neq i}\frac{[ij]}{(ij)}\left(\frac{(\xi j)^{2}}{(\xi i)^{2}}-\frac{(\zeta j)^2}{(\zeta i)^2}\right) \\
=-\sum_{j\neq i}\frac{[ij]}{(ij)}\left(\frac{(\xi j)(\zeta i)-(\xi i)(\zeta j)}{(\xi i)(\zeta i)}\right)\left(\frac{(\xi j)}{(\xi i)}+\frac{(\zeta j)}{(\zeta i)}\right) \\
=\frac{(\zeta \xi)\tilde{p}_{A'\;i}}{(\xi i)(\zeta i)}\left(\frac{\xi_{A}}{(\xi i)}+\frac{\zeta_{A}}{(\zeta i)}\right)\sum_{j=1}^{n}\tilde{p}_{j}^{A'}\sigma_{j}^{A}=0,
\end{multline*}
so $\HH$ is independent of $\xi$.  Then we have
\begin{equation*}
\sum_{j=1}^{n}\HH_{ij}\sigma_{j}^{A}\sigma_{j}^{B}=\sum_{j\neq i}\frac{[ij]}{(ij)}\left[\sigma_{j}^{A}\sigma_{j}^{B}-\sigma_{i}^{A}\sigma_{i}^{B}\frac{(\xi j)^2}{(\xi i)^2}\right] =0,
\end{equation*}
due to independence of $\xi$.  Since the symmetric array $\sigma_{j}^{A}\sigma_{j}^{B}$ has three degrees of freedom, the matrix $\HH$ has co-rank 3 as claimed.     $\Box$

This lemma allows us to conclude that our formula for $\cM_{n,0}$ in \eqref{MHVamp2} would vanish as $\Lambda\rightarrow 0$ if we had not included the overall factor of $\Lambda^{-1}$, which appears due to the embedding of Einstein gravity inside conformal gravity.  Indeed, suppose we had not included this factor, then the flat-space limit would be 
\begin{equation*}
\lim_{\Lambda\rightarrow 0}\int\d\mu\;(X^2)^{2}\;\left|\HH^{12}_{12}\right|\;\prod_{i=1}^{n}h_{i}\;\D\sigma_{i},
\end{equation*}
since $\omega^{1}_{ij}\sim O(\Lambda)$ means that we can drop the second term in \eqref{MHVamp2} in the limit.  But when $\Lambda\rightarrow 0$, $X^{2}=1$ and lemma \ref{Co-ranklem} tells us that $\HH$ has co-rank three so the twice-reduced determinant vanishes after performing the $\d\mu$ integral against momentum eigenstates.

However, it still appears that our formula for $\cM_{n,0}$ (now with the correct factor of $\Lambda^{-1}$) is a long way off from Hodges' formula.  If we use dual twistor wavefunctions \eqref{dtwf}, then we are interested in
\be{3red1}
\lim_{\Lambda\rightarrow 0}\frac{1}{\Lambda}\int \frac{\d^{8|8}X}{\mathrm{vol}\;\GL(2,\C)}\;(X^2)^{2} \left|\HH^{12}_{12}\right|\;e^{i\cP\cdot X}\;\d^{2}\sigma.
\ee
Although this expression is finite as $\Lambda\rightarrow 0$ by lemma \ref{Co-ranklem}, it is based on a twice-reduced determinant.  How can we get to the thrice-reduced determinant which is the basis for Hodges' formula?

The answer is provided by noticing that we can represent each factor of $X^{2}$ in \eqref{3red1} by a differential `wave operator' acting on $e^{i\cP\cdot X}$:
\be{waveop}
X^{2}\rightarrow \Box :=\frac{I_{IJ}}{(12)}\frac{\partial}{\partial W_{1\;I}}\frac{\partial}{\partial W_{2\;J}}.
\ee
Doing this allows us to re-write the twice-reduced contribution to $\cM_{n,0}$ as
\be{3red2}
\frac{1}{\Lambda}\int \frac{\d^{8|8}X}{\mathrm{vol}\;\GL(2,\C)}\d^{2}\sigma\;\left|\HH^{12}_{12}\right|\;\Box^{2}e^{i\cP\cdot X} =\frac{1}{\Lambda}\int \frac{\d^{2}\sigma}{\mathrm{vol}\;\GL(2,\C)}\left|\HH^{12}_{12}\right|\;\Box^{2}\delta^{8|8}(\cP).
\ee
On the support of this delta-function, we know that $\HH$ has co-rank three by lemma \ref{Co-ranklem}, so we can integrate by parts once with respect to $\frac{\partial}{\partial W_{2}}$ to give
\begin{multline*}
-\frac{1}{\Lambda}\int \frac{\d^{2}\sigma}{\mathrm{vol}\;\GL(2,\C)}\frac{\partial}{\partial W_{2\;J}}\left|\HH^{12}_{12}\right|\frac{I_{IJ}}{(12)}\frac{\partial}{\partial W_{1\;I}}\Box\delta^{8|8}(\cP) \\
=-\int \frac{\d^{2}\sigma}{\mathrm{vol}\;\GL(2,\C)} \sum_{i}\frac{(\xi 2)^{2}}{(12)(i2)(\xi i)^{2}}\left|\HH^{12i}_{12i}\right|\;W_{i}\cdot\frac{\partial}{\partial W_{1}} \Box \delta^{8|8}(\cP).
\end{multline*}
Once again, the support of the delta-function indicates that we can take $W_{i}\cdot\frac{\partial}{\partial W_{1}}=\sigma_{1}\cdot\frac{\partial}{\partial\sigma_{i}}$, and then integrate by parts once again with respect to $\d^{2}\sigma_{i}$.  This leaves us with
\begin{multline}
\int \frac{\d^{2}\sigma}{\mathrm{vol}\;\GL(2,\C)} \sum_{i}\frac{(12)^{2}}{(1i)^{2}(2i)^{2}}\left|\HH^{12i}_{12i}\right|\;\Box\delta^{8|8}(\cP) \\
+\int \frac{\d^{2}\sigma}{\mathrm{vol}\;\GL(2,\C)} \sum_{i,j}\left(\frac{(\xi 2)^{2}(1\xi)(ji)+(\xi 2)^{2}(1j)(\xi i)}{(12)(i2)(ji)(\xi i)(\xi j)^{2}}\right)\HH_{ij}\;\left|\HH^{12ij}_{12ij}\right|\;\Box\delta^{8|8}(\cP).
\end{multline}

The contribution from the second line can be further simplified by noting that the summation entails symmetrization, term-by-term, in both $1\leftrightarrow 2$ and $i\leftrightarrow j$.  A straightforward calculation involving several applications of the Schouten identity allows us to reduce this to
\begin{equation*}
\int \frac{\d^{2}\sigma}{\mathrm{vol}\;\GL(2,\C)} \sum_{i,j}\left(\frac{(\xi 1)^{2}(i2)(j2)+(\xi 2)^{2}(i1)(j1)}{(1i)(2i)(1j)(2j)(\xi i)(\xi j)}\right)\HH_{ij}\;\left|\HH^{12ij}_{12ij}\right|\;\Box\delta^{8|8}(\cP).
\end{equation*}

Upon using the symmetry of $i\leftrightarrow j$ and the basic properties of determinants, we are finally left with an expression for the amplitude which has no overall factor of $\Lambda^{-1}$ and now features thrice-reduced determinants:
\begin{multline}\label{3red3}
\cM_{n,0}=\int \d\mu \left[\sum_{i,j}\left(\frac{(\xi 1)^{2}(i2)(j2)+(\xi 2)^{2}(i1)(j1)}{(1i)(2i)(1j)(2j)(\xi i)(\xi j)}\right)\left|\HH^{12i}_{12j}\right| \right. \\
\left. \sum_{i}\frac{(12)^{2}}{(1i)^{2}(2i)^{2}}\left|\HH^{12i}_{12i}\right| \right]\;\prod_{m=1}^{n}h_{m}\;\D\sigma_{m},
\end{multline}
where we have reverted to arbitrary twistor wavefunctions and taken all remaining $\Lambda$-dependence to zero.

Now, the summation in the second line of \eqref{3red3} only appearance of the reference spinor $\xi\in\P^1$ is in the diagonal entries of the matrix $\HH$.  But lemma \ref{Co-ranklem} tells us that $\HH$ is actually \emph{independent} of the choice of $\xi$ in the flat-space limit.  So the second line of \eqref{3red3} is independent of $\xi$.  The first line is also independent of $\xi$ on its own; this can be shown directly with a residue computation \cite{Adamo:2012xe}.  This means that we can freely set $\xi=\sigma_{1}$, leaving  
\be{Flat2}
\cM_{n,0}(\Lambda=0)=\int \d\mu\left[\sum_{i,j}\frac{(12)^{2}}{(1i)(1j)(2i)(2j)}\left|\HH^{12i}_{12j}\right| +\sum_{i}\frac{(12)^{2}}{(1i)^{2}(2i)^{2}}\left|\HH^{12i}_{12i}\right|\right]\;\prod_{k=1}^{n}h_{k}\;\D\sigma_{k}.
\ee

Finally, we note that on the support of overall momentum conservation, every term in \eqref{Flat2} is equivalent.  This follows from the basic properties of reduced determinants and is built into the Hodges' formula itself, which has many equivalent expressions \cite{Hodges:2012ym, Cachazo:2012kg}.  So up to an irrelevant integer constant (which can be accounted for with proper normalizations), we find:
\begin{equation*}
\lim_{\Lambda\rightarrow 0}\cM_{n,0}=\int \d\mu\; \frac{(12)^{2}}{(1i)^{2}(2i)^{2}}\left|\HH^{12i}_{12i}\right|\;\prod_{j=1}^{n}h_{j}\;\D\sigma_{j}= \cM^{\mathrm{Hodges}}_{n,0}(\Lambda=0),
\end{equation*}
as required.  


\subsubsection{Towards the MHV formalism}

The primary utility of the twistor action for $\cN=4$ SYM was that it naturally encoded the MHV formalism for gauge theory.  The allowed us to easily compute tree-level amplitudes directly, and also formed the basis for loop-level computations at the level of the integrand to all orders in perturbation theory.  On twistor space, the building blocks for the MHV formalism on twistor space were the twistor propagator and the vertices, which were easily seen to correspond on-shell to the Parke-Taylor amplitudes.  

It is easy to see that we have now built the same building blocks on twistor space for our gravitational actions.  For both the conformal and Einstein gravity actions, we now have an expression for the vertices given by \eqref{MHVamp}, and their respective propagators restricted to Einstein states are given by \eqref{CGprop} and \eqref{Einprop}.  Clearly, this is enough to define a MHV formalism on twistor space along the lines of the one we developed for gauge theory in Section \ref{Chapter3}.  If we could translate this prescription to momentum space (or even operationalize efficiently in twistor space) it would represent a major breakthrough, since traditional definitions of an MHV formalism for gravity break down \cite{BjerrumBohr:2005jr, Bianchi:2008pu}.  

Recall that on twistor space, the MHV degree $k$ of an amplitude is the count of the number of external $\tilde{h}$s in an amplitude minus 2.  Since each vertex of the twistor action contains two $\tilde{h}$s, and each propagator $\Delta(Z_{1},Z_{2})$ takes the place of one $\tilde{h}$, we have
\be{MHV-deg}
k= |\mathcal{V}|+l-1
\ee
where $|\mathcal{V}|$ is the number of vertex insertions and $l$ is the number of loops in the diagram.  So to compute a N$^k$MHV tree amplitude, we must sum diagrams with $k+1$ MHV vertices and $k$ propagators--just like we did for $\cN=4$ SYM.

Now, if our proposal for the Einstein gravity twistor action is correct then it should not matter whether we perform the computation with the Einstein action or the conformal gravity action restricted to Einstein states.  For instance, a NMHV diagram for the Einstein action would correspond to a contribution of the form
\be{grNMHV1}
\int \D^{3|4}Z_{1}\;\D^{3|4}Z_{2}\;\mathcal{V}(Z_{1},\ldots)\;\Delta^{\mathrm{Ein}}(Z_{1},Z_{2})\;\mathcal{V}(\ldots, Z_{2}),
\ee
where the vertex is given by \eqref{MHVamp} and the propagator by \eqref{Einprop}.  The analogous calculation in conformal gravity involves replacing $I_{IJ}Z^{J}_{1}\tilde{h}_1$ in one vertex and $I^{IJ}\partial_{2J}h_{2}$ in the other with a propagator and then dividing by the overall factor of $\Lambda$ required by the embedding of Einstein gravity.  In particular, \eqref{grNMHV1} should be equal to
\be{grNMHV2}
\frac{1}{\Lambda}\int \D^{3|4}Z_{1}\;\D^{3|4}Z_{2}\;\mathcal{V}^{J}(Z_{1},\ldots)\;\Delta^{I}_{J}(Z_{1},Z_{2})\;\mathcal{V}_{I}(\ldots, Z_{2}),
\ee
where the propagator is given by \eqref{CGprop} and the vertices now carry a twistor index.  Showing that \eqref{grNMHV1} is equal to \eqref{grNMHV2} would establish that the Einstein twistor action is correct at the level of perturbation theory, and initial calculations indicate that this is true.

Regardless of the validity of the Einstein twistor action, it is clear that conformal gravity induces a MHV formalism on twistor space.  However, the structure of this formalism is significantly different from previous momentum space proposals.  The functional form of the vertex $\mathcal{V}_{n}$ begins with a twice-reduced determinant, as in \eqref{MHVamp}.  This indicates that in flat space, the MHV formalism on twistor space will \emph{not} simply correspond to an off-shell extension of the Hodges formula linked with $p^{-2}$ propagators (or at least not in an obvious way).  A better idea of what is happening on momentum space could be had by translating the twistor propagator to momentum space as we did in the build up to proposition \ref{MHVpropn}, but its action on the vertices could be hard to deduce.  Obviously this is an important goal for future research.


\subsection{BCFW Formulae}

We conclude this section by presenting an alternative route to formulae for amplitudes with cosmological constant by using BCFW recursion. It is known that BCFW recursion holds for gravity scattering amplitudes on a flat background \cite{Bedford:2005yy, Cachazo:2005ca, Benincasa:2007qj}.  As we will see, this can be easily extended to backgrounds with cosmological constant.  By determining the three-point seed amplitudes (i.e., MHV and $\overline{\mbox{MHV}}$), we can in principle compute all tree-level amplitudes using momentum eigenstates, although we focus on the $n$-point MHV amplitude here.


\subsubsection{Three-point amplitudes}

\subsubsection*{\textit{Anti-MHV 3-point}}

The three-point $\overline{\mbox{MHV}}$ amplitude comes from the cubic vertex in $S_{1}[\tilde{h},h]$, where no Picard iteration is needed.  Using either the conformal gravity twistor action restricted to Einstein states or the Einstein action itself:
\be{MHVbar1}
\cM_{3,-1}(1,2,3;\Lambda)=\int_{\CPT}\D^{3|4}Z\wedge\tilde{h}_{1}\wedge\left\{h_{2},h_{3}\right\}.
\ee

To evaluate this, we must insert momentum eigenstates for the gravitons.  We use eigenstates with 4-momentum $p^{AA'}=p^{A}\tilde{p}^{A'}$ and fermionic momentum $\eta_{a}p_{A}$
\be{gmomeig}
\tilde{h}_{i}=\int_{\C}s_{i}\;\d s_{i}\;\bar{\delta}^{2}(s_{i}\lambda_{i}-p_{i})e^{s_{i}[[\mu_{i}\tilde{p}_{i}]]}, \qquad h_{i}=\int_{\C}\frac{\d s_{i}}{s^{3}_{i}}\;\bar{\delta}^{2}(s_{i}\lambda_{i}-p_{i})e^{s_{i}[[\mu_{i}\tilde{p}_{i}]]}.
\ee
From the point of view of de Sitter space, such eigenstates are rather un-natural since they are singular on a finite light cone and do not recognize infinity.  In other words, they are adapted to the affine patch \eqref{dSmetric2} rather than the Poincar\'{e} patch \eqref{dSmetric3}.  As we shall see, the pay-off for making this seemingly awkward choice is formulae that limit nicely as $\Lambda\rightarrow 0$.

Plugging this into \eqref{MHVbar1} gives:
\begin{multline*}
\int_{\CPT}\D^{3|4}Z\wedge\tilde{h}_{1}\wedge\left(\frac{\partial h_{2}}{\partial\mu_{2\;A'}}\frac{\partial h_{3}}{\partial\mu_{3}^{A'}}-\Lambda \frac{\partial h_{2}}{\partial\lambda_{2\;A}}\frac{\partial h_{3}}{\partial\lambda_{3}^{A}}\right) \\
=\int \D^{3|4}Z\frac{s_{1}\;[2\;3]}{s_{2}^{2}s_{3}^{2}}\prod_{i=1}^{3}\d s_{i}\bar{\delta}^{2}(s_{i}\lambda_{i}-p_{i})\;\left(1-\Lambda\Box_{p}\right) \exp\left(\sum_{i=1}^{3} s_{i}[[\mu_{i}\tilde{p}_{i}]]\right) \\
=\frac{[23]^{2}}{[12]^{2}[31]^{2}}\left(1-\Lambda\Box_{p}\right)\delta^{4|8}\left(\sum_{i=1}^{3}p_{i}\right).
\end{multline*}
Here, the second line follows by noting that $\frac{\partial}{\partial \lambda_{A}}$ can be re-expressed as a derivative with respect to $p_{A}$, which in turn leads to $\tilde{p}_{A'}\p/\p p_{AA'}$ when acting eventually on the momentum conserving delta function.  We therefore have:
\be{MHVbar2}
\cM_{3,-1}(1,2,3;\Lambda)=\frac{[23]^{2}}{[12]^{2}[31]^{2}}\left(1-\Lambda\Box_{p}\right)\delta^{4|8}\left(\sum_{i=1}^{3}p_{i}\right).
\ee
As claimed, this limits nicely to the flat-space result as $\Lambda\rightarrow 0$, and the $\Box_{p}$ manifests the breaking of Poincar\'{e} symmetry in de Sitter space.

\subsubsection*{\textit{MHV 3-point}}

The three-point MHV amplitude is the first non-trivial application of our perturbative iteration, which acts only once to produce a single positive helicity wavefunction $h$.  This can act at either contact structure or negative helicity wavefunction; however we know that the perturbation of a single contact structure is $\d$-exact by \eqref{psip}.  Since there are no further perturbations at three points, any deformation of the contact structures vanishes by Stokes' theorem \cite{Adamo:2012xe, Adamo:2013tja}.  

So the only deformations are of the form
\begin{equation*}
\tilde{h}_{i}\rightarrow \int_{\P^1}\frac{\D\sigma_{j}\;(\xi\;i)^{2}}{(i\;j)(\xi\;j)^{2}}I^{IJ}\partial_{I}\tilde{h}_{i}\partial_{J}h_{j}.
\end{equation*}
and the three-point amplitude becomes
\be{MHV3pt1}
\cM_{3,0}(1,2,3;\Lambda)=
\frac{1}{\Lambda}\int \d\mu\; (X^{2})^{2}\frac{(\xi 3)^{2}}{(13)\;(\xi 1)^{2}}[\partial_{1},\partial_{3}]\prod_{i=1}^{3}h_{i}\D\sigma_{i} + (2\leftrightarrow 3).
\ee
Since this exhausts the perturbative iteration at three points, the incidence relations are $Z^{I}=X^{I}_{A}\sigma^{A}$, allowing us to write
\begin{equation*}
\partial_{I}\tilde{h}_{3}=\frac{\sigma^{B}_{2}}{(32)}\frac{\partial \tilde{h}_{3}}{\partial X^{I B}}.
\end{equation*}
Inserting this into \eqref{MHV3pt1}, we can now integrate by parts with respect to $X$.  Our choice means that $\frac{\partial}{\partial X}$ annihilates $\tilde{h}_{2}$ as well as $I^{IJ}\partial_{J}h_{1}$, since this vector is divergence-free.  Hence, the only contribution is:
\begin{multline*}
\frac{4}{\Lambda}\int \d\mu\; X^{2} I_{IK}Z_{2}^{K} \frac{(\xi 3)^{2}}{(32)(13)(\xi 1)^{2}}I^{IJ}\tilde{h}_{3}\partial_{J}h_{1} \tilde{h}_{2}\prod_{i=1}^{3}\D\sigma_{i} +(2\leftrightarrow 3) \\
=-4 \int \d\mu\;X^{2}\frac{(\xi 3)^{2}}{(32)(13)(\xi 1)^{2}}\tilde{h}_{3}\;Z_{2}\cdot\partial_{1} h_{1}\tilde{h}_{2}\prod_{i=1}^{3}\D\sigma_{i}+(2\leftrightarrow 3).
\end{multline*}

As $Z^{I}(x,\sigma)$ is a degree one function in $\sigma$ by assumption, we can let the differential operator $Z_{2}\cdot\partial_{1}$ act as
\begin{equation*}
Z_{2}\cdot\partial_{1}\sim \sigma_{2}\cdot\frac{\partial}{\partial\sigma_{1}}.
\end{equation*}
This enables us to integrate by parts with respect to $\D\sigma_{1}$, obtaining
\begin{equation*}
4 \int \d\mu\;X^{2}\frac{(\xi 3)^{2}}{(32)(13)(\xi 1)^{2}}\left(\frac{(32)(\xi 1)-2(\xi 2)(13)}{(13)(\xi 1)}\right) h_{1}\tilde{h}_{2}\tilde{h}_{3}\prod_{i=1}^{3}\D\sigma_{i} +(2\leftrightarrow 3).
\end{equation*}
After two applications of the Schouten identity and inserting momentum eigenstates \eqref{gmomeig}, we have:
\begin{multline}\label{MHV3pt2}
\cM_{3,0}(1,2,3;\Lambda)=\int \d\mu\;X^{2}\frac{(23)^{2}}{(12)^{2}(31)^{2}}h_{1}\tilde{h}_{2}\tilde{h}_{3}\prod_{i=1}^{3}\D\sigma_{i} \\
=\frac{\la 23\ra^{2}}{\la 12\ra^{2} \la 31\ra^{2}}(1-\Lambda \Box_{p})\delta^{4|8}\left(\sum_{i=1}^{3}p_{i}\right),
\end{multline}
with the $\Box_{p}$ arising from the Fourier transformation of $X^{2}=1-\Lambda x^{2}$.  Once again, note that this has the correct $\Lambda\rightarrow 0$ limiting behavior.


\subsubsection{$\cN=8$ supergravity and BCFW}

These three-point amplitudes can be used to seed the tree-level BCFW recursion for Einstein gravity.  First we confirm that BCFW recursion indeed extends to gravitational scattering amplitudes on (anti-)de Sitter backgrounds \cite{Adamo:2012nn}.
\begin{lemma}\label{GBCFWlem}
BCFW recursion is valid for gravitational scattering amplitudes ($0\leq\cN\leq8$) on backgrounds with cosmological constant.
\end{lemma}
\proof  BCFW recursion is derived by picking two external momenta for a scattering amplitude and analytically continuing them with a complex variable $z$ while keeping them on-shell and maintaining overall momentum conservation.  The amplitude then becomes a complex function $\cM(z)$: it has simple poles wherever internal propagators go on-shell, and $\cM(0)$ is the original amplitude.  These simple poles correspond to the terms arising in the BCFW recursion, so provided $\cM(z\rightarrow\infty)$ vanishes, Cauchy's theorem implies the recursion.  In the $\Lambda=0$ case, it was proven that $\cM(z\rightarrow\infty)=0$ using a background field method in \cite{ArkaniHamed:2008yf}.  With $\Lambda\neq 0$, $\cM(z)$ still has simple poles corresponding to propagators going on-shell, so the only potential subtlety arises with the fall-off as $z\rightarrow\infty$, and it suffices to show that the methods of \cite{ArkaniHamed:2008yf} still work.  In the large $z$ regime, we are interested in quadratic fluctuations on a classical background, where the fluctuations correspond to the two shifted particles and the soft background looks like de Sitter space.  For our gravitational amplitudes, this entails inserting a metric $g_{\mu\nu}+h_{\mu\nu}$, and extracting the portion which is quadratic in $h$ \cite{Christensen:1979iy}:
\begin{multline*}
\cL_{\mathrm{quad}}=\sqrt{-g}\left[\frac{1}{4}\tilde{h}^{\mu\nu}(2R_{\mu\rho}g_{\mu\sigma}-2R_{\mu\rho\nu\sigma}-g_{\mu\rho}g_{\nu\sigma}\Box)h^{\rho\sigma}-\frac{1}{2}\nabla^{\rho}\tilde{h}_{\rho\mu}\nabla^{\sigma}\tilde{h}^{\mu}_{\sigma}\right. \\
\left. -\tilde{h}(R_{\rho\sigma}-\frac{1}{4}g_{\rho\sigma}R)h^{\sigma}_{\mu}-\frac{1}{2}\Lambda\tilde{h}^{\mu\nu}h_{\mu\nu}\right],
\end{multline*}  
where $\tilde{h}_{\mu\nu}=h_{\mu\nu}-\frac{1}{2}g_{\mu\nu}h$, and $h=g_{\mu\nu}h^{\mu\nu}$.  To this, we add the de Donder gauge-fixing term, as well as a Lagrangian density for a conformally-invariant scalar field, leaving us with:
\begin{multline*}
\cL_{\mathrm{quad}}=\sqrt{-g}\left[\frac{1}{4}\tilde{h}^{\mu\nu}(2R_{\mu\rho}g_{\mu\sigma}-2R_{\mu\rho\nu\sigma}-g_{\mu\rho}g_{\nu\sigma}\Box)h^{\rho\sigma}-\tilde{h}(R_{\rho\sigma}-\frac{1}{4}g_{\rho\sigma}R)h^{\sigma}_{\mu}\right. \\
\left. -\frac{1}{2}\Lambda\tilde{h}^{\mu\nu}h_{\mu\nu}+\frac{1}{2}g^{\mu\nu}\nabla_{\mu}\phi\nabla_{\nu}\phi -\Lambda\phi^{2}\right].
\end{multline*}

Now, we take our background metric $g_{\mu\nu}$ to be de Sitter, and implement the field re-definition used in \cite{Bern:1999ji}:
\begin{equation*}
h_{\mu\nu}\rightarrow h_{\mu\nu}+g_{\mu\nu}\phi, \qquad \phi\rightarrow \frac{h}{2}+\phi.
\end{equation*}
A bit of tensor algebra reveals that the quadratic Lagrangian transforms to become:
\begin{equation*}
\cL_{\mathrm{quad}}\rightarrow\sqrt{-g}\left[\frac{1}{4}g^{\mu\nu}\nabla_{\mu}h^{\sigma}_{\rho}\nabla_{\nu}h^{\rho}_{\sigma}-\frac{1}{2}h_{\mu\nu}h_{\rho\sigma}R^{\mu\rho\nu\sigma}+\frac{1}{2}g^{\mu\nu}\nabla_{\mu}\phi\nabla_{\nu}\phi -\Lambda\phi^{2}\right].
\end{equation*}
This transformation successfully eliminates all the trace terms, and after decoupling the re-defined scalar field, the Lagrangian is exactly the same as the one used in the flat background calculation.  From this point, the proof that $\cM(z\rightarrow\infty)$ vanishes follows in exactly the same fashion as in the $\Lambda=0$ case of \cite{ArkaniHamed:2008yf}, as desired.     $\Box$

\medskip 

BCFW recursion for Einstein gravity is most easily expressed for maximal (i.e., $\cN=8$) SUGRA, where there is only a single supermultiplet \cite{Cremmer:1978km}.  $\cN=8$ SUGRA is an interesting theory in its own right: it is obtained by dimensional reduction from 11-dimensions \cite{Cremmer:1979up}; has a valid `no-triangle' hypothesis \cite{BjerrumBohr:2008ji}; contains a non-linear global $E_{7(7)}$ symmetry \cite{Cremmer:1978ds}; possesses additional recursive-like relations (the so-called `bonus relations'); has an S-matrix which is well-defined everywhere in the moduli space; and may even be UV finite \cite{Bern:2006kd}.  However, we will simply use $\cN=8$ supersymmetry as a convenient calculational tool.

Supersymmetric BCFW recursion for gravity \cite{ArkaniHamed:2008gz} is still seeded by the three-point amplitudes, and its translation into twistor space is well-understood for a flat background \cite{Mason:2009sa}. Here we rewrite those formulae in a notation that is suggestive of twistor actions and twistor-string theory and extend them to $\Lambda\neq 0$.  We will be working directly with the Einstein amplitudes so the overall factor of $\Lambda$ present in the formulae above will be absent and we can take $\Lambda\rightarrow 0$ if desired.  

For the remainder of this section, we work in $\cN=8$ supertwistor space $\PT_{[8]}$ so that $Z^I=(Z^\alpha,\chi^a)$ where now $a=1,\ldots ,8$ and the corresponding holomorphic volume form $\D^{3|8}Z$ now has weight $-4$ (so $\PT_{[8]}$ is no longer Calabi-Yau). We can embed the graviton fields into the $\cN=8$ framework by setting
\be{4in8}
\cH= h+\cdots +\frac{\chi^{8}}{8!} \tilde h\, .
\ee
A generic $\cH$ of homogeneity degree two will encompass the full $\cN=8$ linear gravity supermultiplet in the same way that $\cA$ encoded the full $\cN=4$ SYM multiplet.

Of course, one may ask the natural question: is this operation well defined on a de Sitter background?  The immediate answer is `no,' simply because there is no unbroken unitary representation of supersymmetry in de Sitter space \cite{Witten:2001kn}.  For our purposes though, the $\cN=8$ supersymmetry is just a formal tool we use to encode both $\cN=0$ graviton helicities in the single field \eqref{4in8}.  After computing the amplitude in the $\cN=8$ formalism, we can truncate immediately to $\cN=0$ by performing fermionic integrations in the usual fashion.  Furthermore, from the perturbative point of view our amplitudes are just polynomials in $\Lambda$, and one can simply reverse the sign to consider a (perturbative) amplitude for gauged supergravity on anti-de Sitter space, where supersymmetry is unbroken. 

The BCFW recursion in \cite{Mason:2009sa} was based on a split signature framework in which the twistors are totally real and the $n$-point amplitude was represented as a distribution on $n$ copies of $\PT_{\R}\subset\RP^{3|8}$.  We begin by translating this setup to the complex setting adopted throughout this review.

In Chapter \ref{Chapter3}, we represented scattering amplitudes on twistor space in terms of their integral kernel.  With $\cN=4$ supersymmetry, this entailed picking twistor representatives in $\Omega^{0,2}(\PT,\cO)$.  With $\cN=8$ supersymmetry, the relevant pairing between twistor wavefunctions takes the form:
\begin{equation*}
\Omega^{0,1}(\PT_{[8]},\cO(k))\times\Omega^{0,2}_{c}(\PT_{[8]},\cO(4-k))\rightarrow\C , \qquad (\phi,\alpha)\mapsto\int_{\PT_{[8]}}\D^{3|8}Z\wedge\phi\wedge\alpha.
\end{equation*}
The twistor wavefunctions for $\cN=8$ SUGRA take values in $H^{0,1}(\PT_{[8]},\cO(2))$, so this means that we should represent our scattering states in the integral kernel as: 
\be{delta-fn-w}
\cH_{i}=\bar\delta^{3|8}_{2,2}(Z_i,Z(\sigma_{i}))=\int_\C \frac{\d s}{s^3}\bar\delta^{4|8}(Z_i+sZ(\sigma_{i}))\, .
\ee
As in Chapter \ref{Chapter3}, integration with respect to $\D\sigma_{i}$ in our calculations reduces this to a $(0,2)$-form of weight +2 as desired.

The recursion is seeded by the three point $\overline{\mbox{MHV}}$ and MHV amplitudes.  The formulae \eqref{MHVbar2}, \eqref{MHV3pt2} extend easily to $\cN=8$ SUGRA; removing the overall factor of $\Lambda$ gives
\be{MHV-bar-N=8}
\cM^{\cN =8}_{3,-1}(1,2,3)= \int_{\PT_{[8]}}  \D^{3|8}Z\wedge \cH_{3}\; \{\cH_{1},\cH_{2}\},
\ee
and
\be{MHV3ptN=8}
\cM^{\cN =8}_{3,0}(1,2,3)=\int\d\mu\;X^{2} \prod_{i=1}^3 \frac { \cH_{i}\; \D\sigma_i }{(\sigma_i\cdot\sigma_{i+1})^2},  
\ee
where we use the notation\footnote{The fermionic parts of the infinity twistor correspond to some gauging of the R-symmetry of supergravity \cite{Wolf:2007tx}; for our purposes we can let these components be zero.} 
\begin{equation*}
Z^{I}(\sigma)=X^{IA}\sigma_{A}.
\end{equation*}

From \eqref{BCFR3}, BCFW on twistor space becomes
\be{recursion}
\cM(Z_1,\ldots,Z_n)=\sum_{L
,R}\int_{\C\times \PT_{[8]}}\D^{3|8}Z \frac{\d z}z  \cM_L(Z_1, Z_2,\ldots,Z_i,Z)\;\cM_R(Z,Z_{i+1},\ldots,Z_n+z Z_1)
\ee
where the sum is over all $1<i<n-1$ and permutations fixing $1$ and $n$.  

In the solution to the recursion relations, a particularly important role is played by the contributions in which either $\cM_L$ or $\cM_R$ is a three point amplitude.   Up to various shifts, these are the main terms involved in solving the recursion relations inductively.  In these cases it emerges from three-particle kinematics that the contributions are only nontrivial when $\cM_L$ is MHV or $\cM_R$ is the $\overline{\mbox{MHV}}$.  The latter case is known as the `homogeneous term,' and when the amplitude under consideration is the $n$-point MHV amplitude, the homogeneous terms form the entire recursion.  The integrations for the homogeneous term were performed explicitly in \cite{Mason:2009sa} for the split signature, flat background case; we extend them to the complex $\Lambda\neq 0$ setting here.

Our starting point is the MHV recursion:
\be{homogeneous}
\cM_{n,0}(1,\ldots, n)=\sum \int \D^{3|8}Z\frac{\d z}{z}\cM(Z_{1},\ldots, Z_{n-1},Z)\;\cM_{3,-1}(Z, Z_{n-1},Z_{n}+zZ_{1}).
\ee
In the integral kernel formalism, the three point seed amplitude is obtained from \eqref{MHV-bar-N=8}:
\begin{multline*}
\cM_{3,-1}(Z, Z_{n-1},Z_{n}+zZ_{1})=\int \D^{3|8}Z'\;\bar{\delta}^{3|8}_{2,2}(Z_{n}+zZ_{1},Z')\left\{\bar{\delta}^{3|8}_{2,2}(Z,Z'), \bar{\delta}^{3|8}(Z_{n-1},Z')\right\} \\
=\left[\frac{\partial}{\partial Z}, \frac{\partial}{\partial Z_{n-1}}\right]\;\bar{\delta}^{3|8}(Z,Z_{n}+zZ_{1}) \; \bar{\delta}^{3|8}(Z_{n-1},Z_{n}+zZ_{1}),
\end{multline*}
simply using the properties of the complex distributional forms.  This leaves us with a recursion of the form
\begin{equation*}
\sum \int \D^{3|8}Z\frac{\d z}{z}\cM(Z_{1},\ldots, Z_{n-1},Z)\;\left[\partial, \partial_{n-1}\right]\; \bar{\delta}^{3|8}_{2,2}(Z,Z_{n}+zZ_{1}) \; \bar{\delta}_{2,2}^{3|8}(Z_{n-1},Z_{n}+zZ_{1}).
\end{equation*} 
Integrating by parts with respect to $\D^{3|8}Z$ and using the delta-function support leaves us with
\be{homr1}
-\sum I^{IJ}\int \frac{\d z}{z}\frac{\partial}{\partial Z_{n-1}^{I}}\cM(1,\ldots, n-1)\frac{\partial}{\partial Z^{J}_{n-1}}\bar{\delta}_{2,2}^{3|8}(Z_{n-1},Z_{n}+zZ_{1}).
\ee

Now, a simple calculation shows that
\begin{multline*}
\frac{\partial}{\partial Z^{J}_{n-1}}\bar{\delta}^{3|8}_{2,2}(Z_{n-1},Z_{n}+zZ_{1})=-\frac{\partial}{\partial Z^{J}_{n-1}}\bar{\delta}^{3|8}_{2,2}(Z_{n}+zZ_{1},Z_{n-1}) \\
=-\frac{\partial}{\partial Z^{J}_{n-1}}\int_{\C}\frac{\d s}{s^{3}} \bar{\delta}^{4|8}(Z_{n}+zZ_{1}+sZ_{n-1}) = \frac{\partial}{\partial Z_{n}^{J}}\int_{\C}\frac{\d s}{s^2}\bar{\delta}^{4|8}(Z_{n}+zZ_{1}+sZ_{n-1}),
\end{multline*}
so we can rewrite \eqref{homr1} as
\be{homr2}
-\sum I^{IJ}\int \frac{\d z}{z}\frac{\d s}{s^{2}}\frac{\partial}{\partial Z_{n-1}^{I}}\cM(1,\ldots, n-1)\frac{\partial}{\partial Z_{n}^{J}}\bar{\delta}^{4|8}(Z_{n}+zZ_{1}+sZ_{n-1}).
\ee
But at this point we can solve the recursion completely by noting that
\begin{equation*}
\int_{\C^{2}}\frac{\d z}{z}\frac{\d s}{s^{2}}\bar{\delta}^{4|8}(Z_{n}+zZ_{1}+sZ_{n-1})=\bar{\delta}^{2|8}_{0,1,3}(Z_{1},Z_{n-1},Z_{n}).
\end{equation*}

Taking into account the sum over BCFW decompositions, we find that the full homogeneous term in the $\cN=8$ recursion is:
\begin{equation}\label{hgs-term}
\cM_{n,0}=\sum_{i\neq 1,n}I^{IJ}\frac{\partial}{\partial Z_{n}^{I}}\bar{\delta}^{2|8}_{0,1,3}(Z_{1},Z_{i},Z_{n})\frac{\partial}{\partial Z_{i}^{J}} \cM_{n-1,0}(1,\ldots, i,\ldots, n-1).
\end{equation}
The final step is to show that this can be re-expressed in a manner which immediately allows us to obtain the integral kernel we are after for any value of $\Lambda$.

\begin{propn}\label{CCBCFW}
The $n$-point MHV amplitude for $\Lambda\neq0$ is given by BCFW recursion on twistor space as
\begin{multline}\label{grav-MHV-BCFW}
\cM_{n,0}(Z_1,\ldots, Z_n;\Lambda)=\int\d\mu\;\left(\prod_{i=4}^n 
\frac{[\p_{i}, \p_{i-1}]\; \cH_{i} \D\sigma_i }{(i\; i-1)}\right)\frac{\cH_3 \D\sigma_3 \; \cH_2 \D\sigma_2 \; \cH_1\tau_1}{(32)^2(21)^2(1n)^2}\\
 +\mathrm{Perms}(2,\ldots, n-1),
\end{multline}
where $\cH_i=\bar\delta^{3|8}_{2,2}(Z_i,Z(\sigma_i))$ and the terms in the product are ordered with increasing $i$ to the left.
\end{propn}

\proof It suffices to demonstrate that the first step of the recursion obeys this pattern, after which \eqref{grav-MHV-BCFW} follows inductively.  At four points, \eqref{hgs-term} gives
\begin{multline*}
\cM_{4,0}=I^{IJ}\frac{\partial}{\partial Z_{4}^{I}}\bar{\delta}^{2|8}_{0,1,3}(Z_{1},Z_{3},Z_{4})\frac{\partial}{\partial Z_{3}^{J}} M_{3,0}(1,2,3) \\
=\int\frac{\d s}{s^2}\frac{\d t}{t}I^{IJ}\partial_{4\;I}\bar{\delta}^{4|8}(Z_{4}+sZ_{3}+tZ_{1})\;\partial_{J\;3}\frac{\cH_3 \D\sigma_3 \; \cH_2 \D\sigma_2 \; \cH_1\tau_1}{(32)^2(21)^2(13)^2},
\end{multline*}
using \eqref{MHV3ptN=8}.  Now define $q\sigma_{4}=s\sigma_{3}+t\sigma_{1}$; this implies the following relations:
\begin{equation*}
s=q\frac{(14)}{(13)}, \qquad t=q\frac{(34)}{(31)}, \qquad qZ(\sigma_{4})=sZ(\sigma_{3})+tZ(\sigma_{4}).
\end{equation*}

From this, we obtain
\begin{equation*}
\int \d s\;\d t \frac{(13)}{q^{3}(43)}I^{IJ}\partial_{4\;I}\bar{\delta}^{4|8}(Z_{4}+qZ(\sigma_{4}))\partial_{J\;3}\frac{\cH_3 \D\sigma_3 \; \cH_2 \D\sigma_2 \; \cH_1\tau_1}{(32)^2(21)^2(14)^2},
\end{equation*}
using the support of the delta-functions in play.  Now, using the relations above, we have
\begin{multline*}
(31)\d s\wedge\d t=\frac{(31)}{2}\left(\d s\wedge\d t-\d t\wedge\d s\right)=\frac{(31)}{2}\left(\frac{(13)}{(14)}\d s\wedge\frac{s}{q}\d t-\frac{(31)}{(34)}\d t\wedge \frac{t}{q}\d s\right) \\
=\frac{(31)}{2}\d q\wedge \left(\frac{s}{q}\d t -\frac{t}{q}\d s\right)=\frac{\d q}{2}\wedge\D\sigma_{4},
\end{multline*}
neglecting terms which will wedge to zero in the overall expression.  Finally, this leaves us with:
\begin{multline*}
\int \frac{\D\sigma_{4}}{(43)}I^{IJ}\frac{\d q}{q^3}\partial_{I\;4}\bar{\delta}^{4|8}(Z_{4}+qZ(\sigma_{4}))\partial_{J\;3}\frac{\cH_3 \D\sigma_3 \; \cH_2 \D\sigma_2 \; \cH_1\tau_1}{(32)^2(21)^2(14)^2} \\
=\int \frac{\D\sigma_{4}}{(43)}I^{IJ}\partial_{I\;4}\cH_{4}\partial_{J\;3}\frac{\cH_3 \D\sigma_3 \; \cH_2 \D\sigma_2 \; \cH_1\tau_1}{(32)^2(21)^2(14)^2},
\end{multline*}
as required.     $\Box$ 

\medskip

Of course, we should still be free to take the $\Lambda\rightarrow 0$ limit of this expression, in which case it should be comparable to Hodges' formula derived in the previous section.  To do this, we must multiply by generic wave-functions and integrate out the $Z_i$; but at this point \eqref{grav-MHV-BCFW} seems to entail a sum over chains rather than trees.  We can reconcile these two pictures by invoking the arguments of \cite{Drummond:2009ge} to use a cyclically ordered version of the recursion in which we take just the one term in \eqref{hgs-term} and then sum the final result over all permutations of $2$ to $n-1$.

More explicitly, one can perform a derivation similar to the one given here, but using \emph{dual twistor} wavefunctions.  In the $\Lambda\rightarrow 0$ limit, this reproduces the recursion initially derived by Hodges from $\cN=7$ supersymmetry \cite{Hodges:2011wm}.  This indicates that \eqref{grav-MHV-BCFW} indeed has the correct flat-space limit.  It would be interesting to see if one can prove the equivalence between this formula and \eqref{MHVamp2}, though.


\section{Open Questions and Future Directions}
\label{Chapter7}

In this review, we have explored many facets of the twistor action approach to gauge theory and gravity.  For Yang-Mills theory, we have seen that the twistor action manifests the MHV formalism, computes the tree-level S-matrix, and even allows for some progress in the study of loop amplitudes.  Furthermore, we showed that local operators and null polygonal Wilson loops in gauge theory have a natural expression in twistor space; this led to proofs of several interesting correspondences to all levels in perturbation theory (at the level of the integrand).  

While the situation is a bit more complicated for gravity, we were still able to make significant progress by utilizing the embedding of Einstein gravity inside conformal gravity on a de Sitter background.  This enabled us to derive a formula for the MHV amplitude in the presence of a cosmological constant directly from the twistor action as well as using BCFW recursion.  We also arrived at a conjecture for the twistor action of Einstein gravity, which is supported by a correct self-dual reduction, the appropriate MHV amplitudes, and gauge invariance.

In many ways, these results raise more questions than they answer: Can general progress be made at loop-level in twistor theory?  Do other gauge theories in other dimensions, have a twistor action description?  Is there a sensible MHV formalism for gravity?  What applications--if any--do these results have in pure mathematics?  We conclude this review with a brief overview of some open problems and potential future directions for research in this field.  This is by no means an exhaustive list, and is heavily biased by the opinions and interests of the author.


\subsection{Gauge Theory}

The gauge theory under consideration in this review was maximally supersymmetric Yang-Mills theory; however, a twistor action description exists for \emph{all} Yang-Mills theories in four-dimensions.  For $\cN=0,1,2$, these are simply provided by more subtle versions of the $\cN=4$ action studied in Chapters \ref{Chapter3} and \ref{Chapter4} \cite{Mason:2005zm, Boels:2006ir}.  Even the $\cN=3$ theory (which is really equivalent to $\cN=4$) has a description in terms of ambitwistor space \cite{Mason:2005kn}.  Hence, the most exciting questions with respect to gauge theory are not about \emph{how} to provide perturbative descriptions of Yang-Mills theories, but rather what can be achieved with the descriptions we have, and what other theories can we hope to study?

\subsubsection*{\textit{Beyond tree-level and the planar limit}}

The treatment of Chapter \ref{Chapter3} was most complete at tree-level.  For loop amplitudes, we saw that a generic amplitude on twistor space will require some sort of regulation which accounts for the `0/0' behavior of the shifted R-invariants.  While the twistor action can be defined for the Coulomb branch of $\cN=4$ SYM, and even leads to the massive MHV formalism (see Appendix \ref{Appendix2}), it remains to be seen how--or if--this gives the proper regulating behavior on twistor space.  A similar issue arises in the context of the twistor Wilson loop for the planar sector of $\cN=4$ SYM: here, one can obtain the loop integrand but it is unclear how this can be evaluated correctly.  In the simplest case, the question becomes: how do we evaluate the `Kermit' integral for the one-loop MHV amplitude in twistor space?

Recently, Mason and Lipstein demonstrated that the Kermit integral can immediately be cast in $\d\log$-form in twistor space \cite{Lipstein:2012vs}, and using a suitable choice of contours, be properly integrated \cite{Lipstein:2013}.  This throws open the doorway to obtaining loop \emph{amplitudes} rather than just integrands using the twistor Wilson loop.  Furthermore, we saw in Chapter \ref{Chapter3} that a careful treatment of the integration contour was required to obtain the correct behavior of the two-point vertex for the Feynman rules of the twistor action itself.  Applying the methodology of \cite{Lipstein:2013} to the twistor action could lead to a method for isolating the correct IR behavior of loop amplitudes, and therefore extend the techniques reviewed here to generic loop amplitudes.

Furthermore, while the twistor Wilson loop only describes amplitudes in the planar limit of $\cN=4$ SYM (i.e., the scattering amplitude/Wilson loop duality only holds in the planar limit), there is no such restriction on studying the amplitudes of the twistor action itself.  Integrability techniques have always provided a substantial amount of power in the planar limit, and it appears that they can successfully be used to determine the planar S-matrix to \emph{all} values of the coupling \cite{Basso:2013vsa, Basso:2013aha}!  Hence, it seems natural for us to study twistor theory outside of the planar limit, where it may lead to new insights not accessible to the powerful integrability methods.

\subsubsection*{\textit{The Grassmannian approach}}

Pioneered by Arkani-Hamed and various collaborators \cite{ArkaniHamed:2009si, ArkaniHamed:2009dn, ArkaniHamed:2010kv}, the Grassmannian approach to scattering amplitudes aims (very roughly speaking) to associated a $n$-particle N$^{k}$MHV amplitude with a top-degree form on the Grassmannian $\mathrm{Gr}(k+2,n)$.  The power of this method lies primarily in its ability to manifest all the symmetries of the scattering amplitudes it is describing: in particular, both superconformal and \emph{dual} superconformal invariance can be manifested in the Grassmannian \cite{ArkaniHamed:2009vw}.  This Grassmannian formalism provides a description for the integrand which is manifestly in $\d\log$-form.  Additionally, there is an interesting correspondence between the Grassmannian formulae and bipartite graphs on planar Riemann surfaces; these graphs have become known as `on-shell diagrams,' and encode properties of scattering amplitudes such as BCFW factorization \cite{ArkaniHamed:2012nw}.  On-shell diagrams (and their generalizations) can also be used to represent classes of $\cN=1,2$ quiver gauge theories, where operations on the graphs get reinterpreted as dualities and limits of the field theory (e.g., Seiberg duality or Higgsing) \cite{Franco:2012mm, Xie:2012mr}.

However, the Grassmannian approach lacks the coherent organizing principle we usually associate to a `theory' in physics.  In particular, the entire approach is dictated by a set of symmetries which are used to specify the three-point amplitudes; everything else follows from BCFW recursion.  But where did these symmetries come from?  Presumably, they were inherited from a physical theory (defined in terms of a Lagrangian, or some other organizing principle) which is lurking off-stage.  In other words, the Grassmannian approach provides an efficient way for building scattering amplitudes and a representation that manifests many of their symmetries; it is \emph{not} a physical theory, though.

Hence, finding a theory which produces the Grassmannian formalism (or from which the formalism follows naturally) seems like an important goal, and twistor actions may provide the answer.  These obviously constitute an organizing principle, and throughout this review we have seen how they manifest symmetries which are obscured on space-time.  Furthermore, it has already been shown explicitly at 1-loop that the twistor Wilson loop provides the sought-after $\d\log$-form of the planar integrand \cite{Lipstein:2012vs}.  If this can be extended to an algorithm for all loops, then it should be clear that the twistor action can deliver the same sort of representations as the Grassmannian.  Indeed, there may even be a precise map between the two formulations which matches the MHV formalism of the twistor action to the BCFW foundations of the Grassmannian approach.

\subsubsection*{\textit{Instantons and non-perturbative data}}

Everything we have studied here lies within the realm of perturbative quantum field theory.  In a sense, this is dictated by the fact that our twistor actions (for gauge theory or conformal gravity) are based on a Chalmers-Siegel expansion for the space-time theory.  For Yang-Mills theory, the Chalmers-Siegel Lagrangian differs from the initial Yang-Mills Lagrangian by the topological term $\int \tr(F\wedge F)$, which does not affect the perturbation theory.  Unfortunately, this indicates that it will not be possible to compute non-perturbative quantities (such as a partition function) using the twistor action formulation as we currently understand it.

Nevertheless, there are hints that twistor theory could have something to say about gauge theoretic invariants.  It has been known for some time that Donaldson theory on 4-manifolds can be recast in terms `instanton counting' invariants of a topologically twisted $\cN=2$ Yang-Mills theory \cite{Witten:1988ze}.  On $\R^{4}$ or $S^4$, there is not sufficient topology to get interesting invariants in Donaldson-Witten theory; however, one can still define instanton counting invariants of the gauge theory itself by working equivariantly with respect to a subgroup of the Lorentz group.  This leads to Nekrasov's partition function, which can be thought of as counting instantons on the four-dimensional $\Omega$-background \cite{Nekrasov:2002qd}.  Twistor methods have always been well suited to describing instanton calculations, and the Coulomb branch of these gauge theories can also be described on twistor space (see Appendix \ref{Appendix2}).  It would be fascinating if the Ward correspondence could be adapted to this equivariant setting, perhaps leading to a twistorial method for computing Nekrasov's partition function and hence the instanton prepotential of Seiberg-Witten theory. 

\subsubsection*{\textit{Other gauge theories}}

An obvious question to ask about the twistor action program is whether it extends to other gauge theories in other dimensions.  In an arbitrary number of space-time dimensions, one can define twistors to be the pure spinors of the conformal group (in four-dimensions, the purity condition is trivial) and many results such as the Penrose transform can be proven, albeit often in more complicated forms.  Particularly interesting candidates are ABJM theory (a $\cN=6$ Chern-Simons theory with dual superconformal symmetry) or $\cN=8$ SYM in three-dimensions, and the elusive $\cN=(0,2)$ theory in six-dimensions.  Momentum twistor techniques have already been employed to study the scattering amplitudes of the three-dimensional theories (e.g., \cite{Huang:2010rn, Lipstein:2012kd}), and there has been some progress towards establishing the analogue of a Ward correspondence for the six-dimensional $\cN=(0,2)$ theory \cite{Mason:2011nw, Saemann:2011nb}.  However, definitions of a twistor action for any of these theories in the general non-abelian regime remain a long way off.

Finding such twistor actions could prove an important breakthrough in understanding these theories more generally.  As we have seen throughout, this could lead to efficient calculational tools like the MHV formalism, and in the case of the $\cN=(0,2)$ theory there is no known Lagrangian description at all.  For $\cN=8$ SYM, one might hope to proceed by `dimensional reduction' on the twistor action for $\cN=4$ SYM.  This could be accomplished by imposing an axial gauge defined by a time-like vector representing the dimensional reduction in space-time, although subtleties involving gauge invariance of any resulting MHV formalism may arise.  

For the six-dimensional theory, one requires a twistorial treatment of non-abelian, self-dual gerbes.  While twistor actions have been found for the linear fields of $\cN=(0,2)$, it remains to be seen whether this construction can be extended to the fully non-linear, non-abelian regime \cite{Mason:2012va}.  It appears that a six-dimensional superconformal theory containing a non-abelian tensor multiplet can be formulated at the level of equations of motion in twistor space, using the Penrose-Ward transform \cite{Saemann:2012uq, Saemann:2013pca}.  While this construction has yet to pass various tests associated with the $\cN=(0,2)$ theory (e.g., reduction to super-Yang-Mills in five dimensions), and its precise connection with M-theory is still unclear, it is nevertheless an exciting arena of research--one in which twistor actions may play an important clarifying role.     

\subsubsection*{\textit{Holomorphic linking and elliptic curves}} 

From the twistor Wilson loop of $\cN=4$ SYM, we know that scattering amplitudes can be interpreted as \emph{holomorphic linking} between irreducible components of a nodal elliptic curve.  This is the natural generalization of the Gauss linking number to the holomorphic category.  It may be possible to use holomorphic linking to provide an alternative definition for scattering amplitudes in terms of abstract objects and operations in algebraic geometry.  For an abelian holomorphic Chern-Simons theory, holomorphic linking can be understood entirely in terms of homological algebra \cite{Frenkel:2005qk}. Translating these ideas into physical language and then generalizing them to the non-abelian setting should define a set of `homological Feynman rules' which allows us to interpret scattering amplitudes as abstract (but well-defined) objects in algebraic geometry.

On a related note, one can ask: is it possible to define a holomorphic Wilson loop on \emph{arbitrary} elliptic curves which has the twistor Wilson loop as its limit when the curve degenerates?  This is more difficult than it sounds, because the moduli space of bundles on an elliptic curve is non-empty (recall that on each $\P^{1}$ component of the nodal curve, we could apply the Birkhoff-Grothendieck theorem).  It may be possible to proceed by working with bundles `close' to the trivial bundle and applying the results of Atiyah.  One could then consider the expectation values of these Wilson loops with respect to a holomorphic Chern-Simons theory, and hope to define holomorphic analogues of knot invariants \cite{Witten:1988hf}.

\subsubsection*{\textit{Hopf algebras}}

Hopf algebras are associative, coassociative bi-algebras equipped with a certain structure known as an `antipode.'  These algebras play a fascinating role in the description of renormalization in gauge theories (c.f., \cite{Connes:1999yr, Brown:2011pj}), but they are also lurking behind many of the loop-level structures in UV finite theories such as $\cN=4$ SYM.  For instance, the multiple zeta values and polylogarithms which appear in loop amplitudes have natural Hopf algebras associated with them \cite{Goncharov:2002}.

There is a natural structure in $\cN=4$ SYM which also seems highly amenable to a Hopf algebra description: the BCFW recursion relation.  It may be possible to understand BCFW recursion as a Hopf algebra structure, with factorization corresponding to a co-product and the 3-point seed amplitudes corresponding to the primitive elements.  This could in turn lead to a diagrammatic mechanism for computing information about the transcendental functions making up scattering amplitudes.  If it exists, such a structure will persist not just for $\cN=4$ SYM but indeed for \emph{any} gauge theory which obeys BCFW recursion.


\subsection{Gravity}

Much of what we were able to say about gravity in this review was accomplished in a rather roundabout fashion, by working via conformal gravity.  Hence, the most obvious open problem is to either prove the validity of the Einstein twistor action \cite{Adamo:2013tja}, or else find another proposal that works.  Skinner's discover of a twistor-string theory which describes the flat-space amplitudes of $\cN=8$ supergravity certainly indicates that a correct twistor action should exist, although the crucial presence of worldsheet supersymmetry in this theory may prove an obstacle.  With such a clear goal in mind, the remainder of this section is devoted to other interesting directions that research on twistor theory and gravity could take.

\subsubsection*{\textit{Twistor-string theory}}

Skinner's twistor-string theory \cite{Skinner:2013xp} is anomaly free with $\cN=8$ supersymmetry, includes explicitly the conformal symmetry-breaking infinity twistor, and produces the tree-level S-matrix of $\cN=8$ SUGRA on a flat background \cite{Cachazo:2012kg}.  While it seems the worldsheet theory is well-defined for $\Lambda\neq 0$, it is not known how to compute gauge-invariant correlators in this regime.  The problem is analogous to proving $\xi$-independence as encountered in Section \ref{Chapter6}: when one attempts to compute a worldsheet correlation function in the twistor-string, the answer is not independent of reference spinors or the location of picture changing operators.  This indicates that either something is missing from our description of the twistor-string (e.g., a new class of vertex operators which do not contribute in the $\Lambda\rightarrow 0$ limit), or else that the twistor-string fails to describe gravity in the presence of a cosmological constant.  In this regard, the twistor action approach (even via conformal gravity) appears to have an edge on the twistor-string as we currently understand it.  

In any case, the string theory is anomaly free for all values of the worldsheet genus.  This indicates that one should, in principle, be able to compute loop-level amplitudes for $\cN=8$ supergravity.  However, there are several features of the twistor-string which can be treated na\"ively at genus zero which will become more complicated on non-rational worldsheets (see section 5 of \cite{Skinner:2013xp} for a good overview of these issues).  Once again, there may be additional vertex operators which need to be taken into account, so determining the full spectrum of such operators is clearly important for both computing loops as well as with a cosmological constant.  Understanding how to overcome either of these issues will represent a breakthrough in our ability to apply twistor methods to a quantum theory of gravity.   

\subsubsection*{\textit{Beyond MHV}}

A major lesson from Sections \ref{Chapter5} and \ref{Chapter6} is that the conformal gravity twistor action is not so dissimilar from the twistor action for $\cN=4$ SYM.  Both have $\dbar$ as their kinetic operator, and both have vertices which correspond to MHV amplitudes.  In the gauge theory setting, we were able to extend these notions off-shell to derive the MHV formalism.  For Einstein gravity, the traditional definition of the MHV formalism by a Risager shift fails \cite{BjerrumBohr:2005jr, Bianchi:2008pu}, but it is easy to see that this is not the unique definition for such a formalism.  

This raises the intriguing possibility that a MHV formalism for Einstein gravity could be defined by extending the Feynman rules for the conformal gravity twistor action off-shell and then restricting to Einstein states for the external legs as proposed in \cite{Adamo:2013tja}.  It may also be possible to approach Einstein gravity directly, either by working with Skinner's twistor-string or by developing a twistor action for general relativity directly.  It is worth noting that an MHV-like expansion for the Einstein sector has been developed and checked numerically \cite{Penante:2012wd} by `relaxing delta-functions' in a Grassmannian representation of the S-matrix \cite{He:2012er, Cachazo:2012pz}, but it remains to be seen if this can be translated into a compact prescription or checked analytically.
  
\subsubsection*{\textit{Graviton non-gaussianities}}

An important goal for future research is to translate formulae for `scattering amplitudes' with cosmological constant (such as \eqref{MHVamp2} or \eqref{grav-MHV-BCFW}) into momentum expressions which make physical sense.  The methods reviewed here are directed towards obtaining answers which limit to scattering amplitudes as $\Lambda\rightarrow 0$; however, for computations relevant to cosmological observables in a de Sitter background one must utilize the `in-in formalism.' (c.f., \cite{Maldacena:2002vr}).  In this picture one works on the Poincar\'e patch of de Sitter space, and uses Bunch-Davies vacua for the wavefunctions.  These states are then integrated from the horizon to the operator insertion point, and then back to the horizon rather than out to $\scri$.  

The computation of such graviton correlators with a cosmological constant is of substantial interest in both cosmology \cite{Maldacena:2002vr, Maldacena:2011nz} and the AdS/CFT correspondence \cite{Raju:2012zr}.  In particular, from the cosmological point of view, the three-point correlators represent the first deviation from the Gaussian spectrum of background fluctuations predicted by single field inflationary models of the universe.  If we can translate our three-point formulae into this framework, it would demonstrate that twistor methods can be applied to these issues.  Furthermore, the current state-of-the-art for these computations in the AdS/CFT setting is $n=4$ \cite{Raju:2012zs}; equations \eqref{MHVamp2} or \eqref{grav-MHV-BCFW} should give the MHV correlator for all $n$ though!

Of course, translating this expression into something that will prove useful from the cosmology or AdS/CFT perspectives remains a non-trivial task.  Choosing a contour in $\CM_{n,1}$ corresponding to the Poincar\'{e} slicing is easy; one just needs to line up with the usual contour integral (c.f., \cite{Maldacena:2002vr}).  More difficult is finding twistor representatives for the Bunch-Davies vacua (i.e., the scattering states) and appropriately fixing the $\GL(2,\C)$ freedom in a way that respects the de Sitter group.

\acknowledgments

This review is adapted from my D.Phil. thesis at the University of Oxford.  As such, thanks must go first and foremost to Lionel Mason for being an excellent supervisor and collaborator, as well as to Mat Bullimore and David Skinner for collaboration on various projects and numerous interesting conversations over the years.  I have also benefited at various times from the insight, interest, and encouragement of Fernando Alday, Philip Candelas, Rob Clancy, Michael Green, Michael Gr\"ochenig, Keith Hannabuss, Andrew Hodges, Frances Kirwan, Arthur Lipstein, Xenia de la Ossa, Roger Penrose, Ron Reid-Edwards, Markus R\"oser, James Sparks, George Sparling, Arkady Tseytlin, and Pierre Vanhove.

This work was supported primarily by the National Science Foundation (Graduate Research Fellowship 1038995), as well as the Clarendon Scholarship and Balliol College.

\appendix


\section{Superconnections for $\cN=4$ Super-Yang-Mills}
\label{Appendix1}

In this appendix, we review the construction of superconnections on chiral Minkowski superspace $\M$ for $\cN=4$ SYM.  In particular, we wish to demonstrate that given the field equations and some constraints, the superconnection can be determined order-by-order in $\theta$.  We begin by reviewing the \emph{non}-chiral construction of superconnections via dimensional reduction, and then study the restriction to the chiral setting for abelian and $\SU(N)$ gauge groups.

\subsection{Non-Chiral Constraints and Field Equations}

We have seen that $\cN=4$ SYM is a chiral theory, so its natural setting is the $(4|8)$-dimensional supermanifold $\M$, charted with coordinates $(x^{AA'},\theta^{aA})$.  There is a corresponding anti-chiral space $\tilde{\M}$, also of dimension $(4|8)$, and these two spaces combine to yield the full non-chiral space $\M^{\mathrm{nc}}$.  More formally, the chiral and anti-chiral super-spaces are given by $\M=G(2|0,\C^{4|4})$ and $\tilde{\M}=G(2|4,\C^{4|4})$ respectively.  $\M^{\mathrm{nc}}$ is then given by the flag manifold $F(2|0,2|4,\C^{4|4})$, with chiral and anti-chiral projections
\begin{equation*}
\xymatrix{
 & \M^{\mathrm{nc}} \ar[dl]_{\pi} \ar[dr]^{\tilde{\pi}} & \\
 \M & & \tilde{\M} }
\end{equation*}
It is not hard to see that both $\pi$ and $\tilde{\pi}$ are submersions of relative super-dimension $(0|8)$, so it follows that $\M^{nc}$ is a supermanifold of dimension $(4|16)$ \cite{Manin:1997}.

In this section, we review some old results which establish a link between constraints on superconnections and the field equations of gauge theories with different amounts of supersymmetry depending on the space-time dimension.  We begin by studying $\cN=1$ Yang-Mills theory in (complex) ten-dimensional Minkowski space-time; solutions to the corresponding field equations were shown to be equivalent to a set of constraints on superconnections for a bundle $\widehat{E}\rightarrow\widehat{\M}$ (where $\widehat{\M}$ is the trivial complex $(10|16)$-dimensional supermanifold) by Witten, Harnad and Shnider \cite{Witten:1985nt, Harnad:1985bc}.  We then discuss how these results are equivalent, via a dimensional reduction procedure, to a similar correspondence between the field equations of $\cN=4$ SYM in four-dimensions and constraint equations on superconnections for a bundle $E\rightarrow\M^{\mathrm{nc}}$.

$\widehat{\M}$ is charted by coordinates $(x^{\mu},\theta^{i})$, for $\mu=0,\ldots, 9$ and $i=1,\ldots, 16$.  The $\theta^{i}$ are anti-commuting Grassmann coordinates, and the R-symmetry $i$-index lives in the fundamental representation $\mathbf{16}$ of $\Spin(10,\C)$; the Clifford algebra splits as $\mathbf{32}=\mathbf{16}+\bar{\mathbf{16}}$, so downstairs indices live in the anti-fundamental representation $\bar{\mathbf{16}}$.  Although one can show that these two representations are inequivalent (c.f., \cite{Brink:1976bc}), the $\gamma$-matrices map $\mathbf{16}$ and $\bar{\mathbf{16}}$ into each other, and take the form:
\begin{equation*}
\gamma^{\mu}=\left(
\begin{array}{c c}
 0 & \gamma^{\mu\;kl} \\
\gamma^{\mu}_{ij} & 0 
\end{array} \right).
\end{equation*}
We can then define the translation operators $\{\partial_{\mu},\;q_{i}\}$ by
\begin{equation*}
\partial_{\mu}=\frac{\partial}{\partial x^{\mu}}, \qquad q_{i}=\frac{\partial}{\partial\theta^{i}}+\gamma^{\mu}_{ij}\theta^{j}\frac{\partial}{\partial x^{\mu}}.
\end{equation*}
It is easy to see that these translation operators generate the $\cN=1$ SUSY algebra:
\be{1susy}
\{q_{i},q_{j}\}=2\gamma^{\mu}_{ij}\partial_{\mu}.
\ee

Now consider a $G$-bundle $\widehat{E}\rightarrow\widehat{\M}$ with connection $\nabla= \mathrm{d} +\CA$, where $\CA$ is the connection 1-form taking values in $\mathfrak{g}^{\C}$.  The components of $\nabla$ and $\CA$ can be obtained by contracting with the translation operators (now interpreted as forming a basis of $T^{1,0}\widehat{\M}$):
\begin{eqnarray*}
\Gamma_{\mu}=\partial_{\mu}\lrcorner\CA, \qquad \Gamma_{i}=q_{i}\lrcorner\CA ; \\
\nabla_{\mu}=\partial_{\mu}\Gamma_{\mu}, \qquad \nabla_{i}=q_{i}+\Gamma_{i}.
\end{eqnarray*}
Witten demonstrated that a system of constraints on $\nabla$ corresponds to the $\cN=1$ SYM field equations in ten dimensions; this is enabled by the useful fact:
\begin{lemma}[Witten \cite{Witten:1978xx}]
The following statements are equivalent:
\begin{enumerate}
\item The constraints 
\be{1const}
\left\{\nabla_{i},\nabla_{j}\right\}=2\gamma^{\mu}_{ij}\nabla_{\mu},
\ee
hold;

\item the connection $\nabla$ is flat when restricted to null lines in $\widehat{\M}$;

\item the equations for a covariantly constant section $s\in\Gamma(\widehat{E})$,
\begin{equation*}
\lambda^{\mu}\nabla_{\mu}s=0, \qquad \lambda^{\mu}\gamma_{\mu}^{ij}s=0,
\end{equation*}
are integrable.
\end{enumerate}
\end{lemma}
Since null lines in the complex space-time $\widehat{\M}$ correspond to points in ambitwistor space, it is clear that these constraints are twistorial in nature; and ($3$.) justifies the interpretation of constraints on the connection as integrability conditions.

These constraints can be written in a more useful form by defining the curvatures
\be{curvs}
\cF_{\mu\nu}=[\nabla_{\mu},\nabla_{\nu}], \qquad \cF^{j}=\frac{1}{10}\gamma^{\mu\;ij}[\nabla_{\mu},\nabla_{i}],
\ee
which are bosonic and fermionic respectively.  A bit of algebra then shows that \eqref{1const} are equal to the following three equations, which we again interpret as constraints on $\nabla$:
\begin{eqnarray}
[\nabla_{i},\nabla_{\mu}] & = & -\gamma_{\mu\;ij}\cF^{j} \nonumber \\
\nabla_{i}\cF^{j} & = & \frac{1}{2}\Sigma^{\mu\nu\;j}_{i}\cF_{\mu\nu} \nonumber \\
\nabla_{i}\cF_{\mu\nu} & = & \gamma_{\mu\;ij}\nabla_{\nu}\cF^{j}-\gamma_{\nu\;ij}\nabla_{\mu}\cF^{j} \label{1const2}
\end{eqnarray}
where $\Sigma^{\mu\nu\;j}_{i}=\gamma^{jk\;[\mu}\gamma^{\nu]}_{ki}$ are the quadric elements of the Clifford algebra.  We then have:
\begin{thm}[Witten \cite{Witten:1985nt}]
The constraint equations \eqref{1const2} imply the $\cN=1$ Yang-Mills equations in ten dimensions:
\begin{eqnarray}
\gamma^{\mu}_{ij}\nabla_{\mu}\cF^{j} & = & 0, \label{1Dirac} \\
\nabla^{\mu}\cF_{\mu\nu}+\frac{1}{2}\gamma_{\nu\;ij}\left\{\cF^{i},\cF^{j}\right\} & = & 0. \label{1YM}
\end{eqnarray}
\end{thm}

The next step is to rephrase the constraint conditions in a manner that allows us to reconstruct the full superfields and superconnection from initial (i.e., $O(\theta^{0})$) data; to do this we must eliminate the fermionic gauge freedom of the superconnection.  The simplest way to do this is to introduce the fermionic Euler vector field
\be{euler}
\Upsilon\equiv \theta^{i}\frac{\partial}{\partial\theta^{i}},
\ee
and require that the 1-form $\CA$ obey the radial gauge condition:
\be{radial}
\Upsilon\lrcorner\CA = \theta^{i}\Gamma_{i}=0.
\ee
It is easy to see that $\Upsilon$ makes sense covariantly (i.e., $\Upsilon=\theta^{i}\nabla_{i}=\theta^{i}\partial_{i}$).  We can use this to derive an equivalent series of constraints which provide a recursive way to determine all the components of the superconnection and superfields from initial data.  For instance, consider $\Upsilon\Gamma_{i}$; using \eqref{1susy} and \eqref{1const} we find
\begin{multline*}
\Upsilon\Gamma_{i}=\theta^{j}\left(\{\nabla_{j},\nabla_{i}\}-\{\nabla_{j},q_{i}\}\right)=\theta^{j}\left(2\gamma^{\mu}_{ij}\nabla_{\mu}-\{q_{j},q_{i}\}-\{\Gamma_{j},q_{i}\}\right) \\
=2\theta^{j}\gamma^{\mu}_{ij}\Gamma_{\mu}-\Gamma_{i}.
\end{multline*}

Proceeding in this way, we obtain the following system \cite{Harnad:1985bc}:
\begin{eqnarray}
(1+\Upsilon)\Gamma_{i} & = & 2\theta^{j}\gamma^{\mu}_{ij}\Gamma_{\mu}, \nonumber \\
\Upsilon\Gamma_{\mu} & = & -\theta^{i}\gamma_{\mu\;ij}\cF^{j}, \nonumber \\
\Upsilon \cF^{j} & = & \frac{1}{2}\theta^{i}\Sigma^{\mu\nu\;j}_{i}\cF_{\mu\nu}, \nonumber \\
\Upsilon \cF_{\mu\nu} & = & \theta^{i}\gamma_{\mu\;ij}\nabla_{\nu}\cF^{j}-\theta^{i}\gamma_{\nu\;ij}\nabla_{\mu}\cF^{j} \label{1const3}
\end{eqnarray}
From a set of initial data $\{\Gamma^{(0)}_{\mu},\cF^{(0)\;i}\}$, these equations uniquely determine $\{\Gamma_{\mu},\Gamma_{i},\cF^{i}\}$ to all orders in $\theta$.  Furthermore, this result goes the other way:
\begin{thm}[Harnad \& Shnider \cite{Harnad:1985bc}]\label{superThm1}
If $\{\Gamma^{(0)}_{\mu},\cF^{(0)\;i}\}$ satisfy the field equations, then so do the full fields $\{\Gamma_{\mu},\Gamma_{i},\cF^{i}\}$ defined by \eqref{1const3}.  Furthermore, the field equations imply the constraints \eqref{1const2} from which the recursive relations were derived.
\end{thm}

A corollary of these two theorems is that we have a three-way equivalence between solutions to the constraint equations \eqref{1const} or \eqref{1const2}, the field equations in ten dimensions \eqref{1Dirac}-\eqref{1YM}, and the recursive relations \eqref{1const3}.  We now want to reduce this equivalence from ten to four space-time dimensions.

The dimensional reduction is performed by writing $\widehat{\M}=\M^{\mathrm{nc}}\oplus\E_{6}$, given by the decompositions of the complex isometry and spin groups \cite{Harnad:1985bc}:
\begin{equation*}
\mathrm{O}(10,\C)\supset \mathrm{O}(4,\C)\times \mathrm{O}(6,\C), \qquad \Spin(10,\C)\supset [\SL(2,\C)\times\tilde{\SL}(2,\C)]\times \SL(4,\C),
\end{equation*}
where we assume that the real slice of $\M^{\mathrm{nc}}$ has Minkowski signature while the real slice of $\E_{6}$ is Euclidean.  Under this splitting, the fermionic variables can be written
\begin{equation*}
\left\{\theta^{i}\right\}_{i=1,\ldots, 16}=\left\{\theta^{aA},\theta^{A'}_{b}\right\}^{A,A'=0,1}_{a,b=1,\ldots, 4},
\end{equation*}
so $A,A'$ are now the Weyl 2-spinor indices of $\SL(2,\C)$ and $\tilde{\SL}(2,\C)$ and $a,b$ are the $\SU(4)$ R-symmetry indices.  The bosonic portion of $\M^{\mathrm{nc}}$ can be charted by coordinates in the vector representation, $\{x^{AA'}\}$, while the bosonic portion of $\E_{6}$ is charted with $\{y^{ab}\}$ for $y^{ab}=-y^{ba}$.

In this new basis, the translation operators $\{\partial_{\mu},q_{i}\}$ become $\{\partial_{AA'},\partial_{ab},q_{aA},q_{A'}^{b}\}$, given explicitly by
\begin{eqnarray*}
q_{aA} & = & \frac{\partial}{\partial\theta^{aA}}+\theta^{A'}_{a}\frac{\partial}{\partial x^{AA'}}-\theta_{A}^{b}\frac{\partial}{\partial y^{ab}}, \\
q_{A'}^{a} & = & \frac{\partial}{\partial\theta^{A'}_{a}}+\theta^{aA}\frac{\partial}{\partial x^{AA'}}-\frac{1}{2}\epsilon^{abcd}\theta_{bB'}\frac{\partial}{\partial y^{cd}}, \\
 & & \partial_{AA'}=\frac{\partial}{\partial x^{AA'}}, \qquad \partial_{ab}=\frac{\partial}{\partial y^{ab}}.
\end{eqnarray*}
These translate the SUSY algebra relations of \eqref{1susy} to
\be{2susy}
\left\{q_{aA},q_{bB}\right\}=2\epsilon_{AB}\partial_{ab}, \qquad \left\{q_{A'}^{a},q_{B'}^{b}\right\}=\epsilon_{A'B'}\epsilon^{abcd}\partial_{cd}, \qquad \left\{q_{aA},q_{B'}^{b}\right\}=2\delta^{b}_{a}\partial_{AB'}.
\ee
Finally, we can decompose the superconnection and its 1-form in this new basis to obtain four components:
\begin{eqnarray}
\nabla_{aA}=q_{aA}+\Gamma_{aA}, & \nabla_{A'}^{a}=q_{A'}^{a}+\Gamma_{A'}^{a}, \\
\nabla_{AA'}=\partial_{AA'}+\Gamma_{AA'}, & \nabla_{ab}=\partial_{ab}+\Gamma_{ab}.
\end{eqnarray}

We then find a direct translation of the constraints, field equations, and recursive relations.  In particular, we write the curvatures from the ten-dimensional setting as
\begin{equation*}
\cF_{\mu\nu}=(\cF_{AB},\cF_{A'B'},\cF_{abAA'},\cF_{abcd}), \qquad \cF^{i}=(\tilde{\cF}^{aA}, \tilde{\cF}^{A'}_{b}),
\end{equation*}
and make the useful identifications
\be{fcurvs}
\cF_{aA'}=\frac{1}{2}\epsilon_{A'B'}\tilde{\cF}^{B'}_{a}, \qquad \cF_{A}^{a}=\frac{1}{2}\epsilon_{AB}\tilde{\cF}^{aB}.
\ee
Then the constraint equations become \cite{Harnad:1985bc}:
\begin{eqnarray}
\left\{\nabla_{aA},\nabla_{bB}\right\} & = & 2\epsilon_{AB}\nabla_{ab}, \nonumber \\
\left\{\nabla_{A'}^{a},\nabla_{B'}^{b}\right\} & = & \epsilon_{A'B'}\epsilon^{abcd}\nabla_{cd}, \nonumber \\
\left\{\nabla_{aA},\nabla_{B'}^{b}\right\} & = &2\delta^{b}_{a}\nabla_{AB'}, \label{2const1}
\end{eqnarray}

\begin{eqnarray}
\left[\nabla_{aA},\nabla_{BC'}\right] & = & \epsilon_{AB}\cF_{aC'} , \nonumber \\
\left[\nabla_{A'}^{a},\nabla_{BC'}\right] & = & \epsilon_{A'C'}\cF_{B}^{a}, \nonumber \\
\left[\nabla_{aA}, \nabla_{bc}\right] & = & \epsilon_{abcd}\cF_{A}^{d}, \nonumber \\
\left[\nabla_{A'}^{a}, \nabla_{bc}\right] & = & 2\delta_{[b}^{a}\cF_{c]A'}, \label{2const2}
\end{eqnarray}

\begin{eqnarray}
\nabla_{aA}\cF_{bA'} & = & 2\cF_{abAA'}, \nonumber \\
\nabla_{A'}^{a}\cF_{B}^{b} & = & \epsilon^{abcd}\cF_{cdBA'}, \nonumber \\
\nabla_{aA}\cF_{B}^{b} & = & 2\delta^{b}_{a}\cF_{AB}+\frac{1}{2}\epsilon_{AB}\epsilon^{bcde}\cF_{deac}, \nonumber \\
\nabla_{A'}^{a}\cF_{bB'} & = & 2\delta^{a}_{b} \cF_{A'B'}-\frac{1}{2}\epsilon_{A'B'}\epsilon^{acde}\cF_{debc}; \label{2const3}
\end{eqnarray}

The recursion relations under the radial gauge (which now reads $\theta^{aA}\Gamma_{aA}+\theta^{A'}_{a}\Gamma_{A'}^{a}=0$) become:
\begin{eqnarray}
(1+\Upsilon)\Gamma_{aA} & = & 2\theta^{bB}\epsilon_{AB}\Gamma_{ab}+2\theta^{B'}_{a}\Gamma_{AB'}, \nonumber \\
(1+\Upsilon)\Gamma_{A'}^{a} & = & 2\theta^{aB}\Gamma_{BA'}+\epsilon_{A'B'}\epsilon^{abcd}\theta^{B'}_{b}\Gamma_{cd}, \nonumber \\
\Upsilon\Gamma_{AA'} & = & \epsilon_{BA}\theta^{aB}\cF_{aA'}+\epsilon_{B'A'}\theta^{B'}_{a}\cF_{A}^{a}, \nonumber \\
\Upsilon\Gamma_{ab} & = & \epsilon_{abcd}\theta^{cA}\cF_{A}^{d}+2\theta^{A'}_{[a}\cF_{b]A'}, \nonumber \\
\Upsilon \cF_{aA'} & = & 2\theta^{bA}\cF_{baAA'}+2\theta^{B'}_{a}\cF_{A'B'}-\frac{1}{2}\epsilon^{bcde}\theta_{bA'}\cF_{deac}, \nonumber \\
\Upsilon \cF_{A}^{a} & = & 2\theta^{aB}\cF_{AB}+\frac{1}{2}\epsilon^{acde}\theta_{A}^{b}\cF_{debc}+\epsilon^{bacd}\theta^{A'}_{b}\cF_{cdAA'}. \label{recursion1}
\end{eqnarray}

At this point, these equations hold for the full 10-dimensional super-space, written in a particularly suggestive basis.  If we assume that the superconnection and all superfunctions involved are invariant with respect to the translations in the bosonic portion of $\E_{6}$, then we obtain a dimensional reduction to $\M^{\mathrm{nc}}$.\footnote{The dimensional reduction to a $d=6$, $\cN=2$ theory on a complex space-time with Euclidean real slice is obtained by assuming invariance with respect to $\partial_{AA'}$.}  In this case, the portion of the superconnection $\nabla_{ab}$ is replaced by a anti-symmetric array of six complex scalars $\cF_{ab}$ in the following fashion:
\begin{equation*}
\nabla_{ab}\rightarrow \left[\cF_{ab}, \cdot\right], \qquad \cF_{abAA'}\rightarrow\nabla_{AA'}\cF_{ab}, \qquad \cF_{abcd}\rightarrow\left[\cF_{ab},\cF_{cd}\right].
\end{equation*}
This completes the reduction to four dimensions. 


\subsection{Reduction to Chiral Super-space}
\label{chiral}

We have now established that a superconnection $\CA$ over the complex non-chiral Minkowski space $\M^{\mathrm{nc}}$ can be built from consistency conditions which are compatible with the $\cN=4$ SYM field equations.  In other words, we a can establish a precise correspondence between solutions to the Yang-Mills equations and the fully non-chiral superconnection.  The question is, can a similar correspondence be deduced on the $(4|8)$-dimensional chiral Minkowski space $\M$?

To begin, we perform the reduction by setting $\bar{\theta}^{aA}=\theta^{A'}_{a}=0$, and investigate the consequences.  We are interested in expressing the constraints in terms of recursive relations and Bianchi identities, so note that the radial gauge condition becomes:
\be{radial2}
\theta^{aA}\Gamma_{aA}=0.
\ee

The recursive relations \eqref{recursion1} now become (for the reduced fermionic Euler vector $\Upsilon=\theta^{aA}\partial_{aA}$):
\begin{eqnarray}
(1+\Upsilon)\Gamma_{aA} & = & 2\theta_{A}^{b}\cF_{ba}, \nonumber \\
(1+\Upsilon)\Gamma_{A'}^{a} & = & 2\theta^{aB}\Gamma_{BA'}, \nonumber \\
\Upsilon\Gamma_{AA'} & = & \theta_{A}^{a}\cF_{aA'}, \nonumber \\
\Upsilon \cF_{ab} & = & \epsilon_{abcd}\theta^{cA}\cF_{A}^{d}, \nonumber \\
\Upsilon \cF_{aA'} & = & 2\theta^{bA}\nabla_{AA'}\cF_{ba}, \nonumber \\
\Upsilon \cF_{A}^{a} & = & 2\theta^{aB}\cF_{AB}. \label{4recursion}
\end{eqnarray}
By theorem \ref{superThm1}, these define a set of superfields satisfying the $\cN=4$ SYM equations, albeit independent of the anti-chiral coordinate $\theta^{A'}_{a}$.  However, there are of course solutions to the field equations which are not compatible with such chiral superconnections.  So although this reduction ensures the existence of a chiral superconnection, we lose the uniqueness which gives us the `if and only if' statements established in the non-chiral setting.  We now investigate what is required to close the chiral constraint equations and determine the superconnection uniquely.

By performing our chiral reduction, we have eliminated $\cF_{A'B'}$ and $\cF_{A}^{a}$, $\Gamma_{A'}^{a}$ become auxiliary fields since they can be defined, order-by-order in $\theta$, by the other fields using \eqref{4recursion}.  Additionally, we have `forgotten' about the six-dimensional constraint equations on $\E_{6}$ which arose via dimensional reduction.  While this was fine in the non-chiral case, the chiral reduction could make some of the constraint equations non-integrable since we leave out the additional equations on $\E_{6}$.

The curvature constraint equations of interest are
\be{const4}
\left\{\nabla_{a(A},\nabla_{B)b}\right\}=0, \qquad \left[\nabla_{a(A},\nabla_{B)B'}\right]=0,
\ee
which define the curvatures $\cF_{ab}$ and $\cF_{aB'}$ respectively.  If we choose these as our sole constraint equations, it is clear that our formalism will no longer hold totally off-shell, since there are additional constraints in \eqref{2const1}-\eqref{2const3}.  We will see that additional constraints must be imposed which are precisely the field equations that are non-trivial in the SD sector (i.e., those equations with no factors of the 't Hooft coupling).

To see this, note that the first of the constraints in \eqref{const4} yields the Bianchi identity
\be{Bianchi1}
\nabla_{aA}\cF_{bc}=\nabla_{A[a}\cF_{bc]}.
\ee
The following lemma confirms that this is enough to determine $\cF_{ab}$, $\cF_{A}^{a}$, and $\cF_{AB}$ without imposing any of the $\cN=4$ SYM field equations.  We prove this for the lowest order components, and observe that it holds for the higher components by using the recursion relations.

\begin{lemma}\label{lemma: odd}
Let $\nabla_{aA}=\partial_{aA}+\Gamma_{aA}$ be an odd superconnection on a $G$-bundle $E'\rightarrow\C^{0|8}$, subject to $\left\{\nabla_{a(A},\nabla_{B)b}\right\}=0$.  Then there is a one-to-one correspondence between such connections and the field multiplet with lowest component $\{\Phi_{ab}, \Psi_{A}^{a}, G_{AB}\}$, where these fields are understood to take values in $\End(E)\cong\mathfrak{g}^{\C}$.
\end{lemma}
\proof  As we are concerned with the purely odd superconnection, we restrict our attention to a bundle $E'$ over the totally anti-commuting spacetime $\C^{0|8}$, charted by the $\theta^{aA}$.  In this case, the twistor space is reduced to $\P^{1|4}$ with homogeneous coordinates $(\lambda_{A},\chi^{a})$, and the reduced incidence relation
\begin{equation*}
\chi^{a}=\theta^{aA}\lambda_{A}.
\end{equation*}
Hence, any point $\theta\in\C^{0|8}$ corresponds to a line $X_{\theta}\cong\P^{1}\subset\P^{1|4}$, while any $(\lambda,\chi)\in\P^{1|4}$ corresponds to $\alpha_{(\lambda,\chi)}$, which is the `super' portion of an $\alpha$-plane in the chiral super-space (or equivalently, the `super'-part of a null geodesic in $\M^{4|8}$).  

Define a bundle $E\rightarrow\P^{1|4}$ with connection $\nabla_{aA}$ pulled back to the reduced twistor space from $E'\rightarrow\C^{0|8}$ with fibers
\begin{equation*}
E_{(\lambda,\chi)}\equiv\left\{ s\in\Gamma(\alpha_{(\lambda,\chi)},E'):\;\lambda^{A}\nabla_{aA}s=0\right\}.
\end{equation*}
The constraint $\left\{\nabla_{a(A},\nabla_{B)b}\right\}=0$ is the flatness condition on the $\alpha$-planes, so that fibers of $E$ have the same dimension as fibers of $E'$, and $\End(E)=\End(E')\cong\mathfrak{g}^{\C}$.  If $E$ is a holomorphic vector bundle, this determines $\nabla_{aA}$, so we must classify holomorphic vector bundles over $\P^{1|4}$.  Such bundles are determined by their $\dbar$-operator, $\dbar+\cA$, where $\cA\in\Omega^{0,1}(\P^{1|4},\mathfrak{g}^{\C}\otimes\cO)$ and is holomorphic in the $\chi^{a}$ with no components in the $\chi^{a}$-directions.

Explicitly, we can find a solution $H$ to $\lambda^{A}\nabla_{aA} H=0$ which takes values in $\mathfrak{g}^{\C}$ and is holomorphic in $\theta$ (i.e., $H$ is a frame of $E$).  We then define
\begin{equation*}
\cA=H^{-1}\dbar H,
\end{equation*}
which means that $\lambda^{A}\partial_{aA}\cA=0$, so $\cA$ defines a genuine $\dbar$-operator on $E$.  The gauge freedom in $H$ is right multiplication by a $\mathfrak{g}^{\C}$-valued function $\gamma(\lambda,\bar{\lambda},\chi)$, so $\cA$ transforms as
\begin{equation*}
\cA\rightarrow \gamma^{-1}\cA \gamma +\gamma^{-1}\dbar \gamma,
\end{equation*}
as required for a $(0,1)$-connection.  Now, as there are no $(0,2)$-forms on $\P^{1|4}$, it follows immediately that $\dbar\cA=0$, so we can expand in fermions:
\begin{equation*}
\cA=a+\chi^{a}\tilde{\psi}_{a}+\frac{1}{2}\chi^{a}\chi^{a}\phi_{ab}+\frac{\epsilon_{abcd}}{3!}\chi^{a}\chi^{b}\chi^{c}\psi^{d}+\frac{\chi^{4}}{4!}g,
\end{equation*}
where each bosonic contribution is a $(0,1)$-form on $\P^{1}$ taking values in the proper homogeneity bundle, and by the $\dbar$-closure of $\cA$, each lies in $H^{0,1}(\P^{1},\cO(-n))$ for $n=0,\ldots,4$.

We can use gauge freedom to put $\cA$ into the Woodhouse gauge, where all its bosonic components define holomorphic $(0,1)$-forms when restricted to the $X_{\theta}\cong\P^{1}$-fibers of the twistor space over $\C^{0|8}$.  As $H^{0,1}(\P^{1},\cO)=H^{0,1}(\P^{1},\cO(-1))=0$, this means that $a=\tilde{\psi}^{a}=0$, and the remaining fields can be given explicitly by choosing Euclidean reality conditions and applying theorem \ref{Wrep1}:
\begin{equation*}
\phi_{ab}=\Phi_{ab}\frac{\la\hat{\lambda} \mathrm{d}\hat{\lambda}\ra}{\la\lambda\hat{\lambda}\ra^{2}}, \qquad \psi^{a}=\Psi_{A}^{a}\hat{\lambda}^{A}\frac{\la\hat{\lambda} \mathrm{d}\hat{\lambda}\ra}{\la\lambda\hat{\lambda}\ra^{3}}, \qquad g=G_{AB}\hat{\lambda}^{A}\hat{\lambda}^{B}\frac{\la\hat{\lambda} \mathrm{d}\hat{\lambda}\ra}{\la\lambda\hat{\lambda}\ra^{4}}.
\end{equation*}
Upon restoring $x$-dependence, the $\Phi_{ab}$, $\Psi_{A}^{a}$, and $G_{AB}$ are z.r.m. fields on $\M$, and gauge freedom is reduced to that of space-time gauge transformations.  This completes the proof.     $\Box$

\medskip

We must still account for the fermionic curvature $\cF_{aA'}$; a Bianchi identity for this portion of the curvature is read off from the fifth relation in \eqref{4recursion}, giving:
\be{Bianchi2}
\nabla_{aA}\cF_{bA'} = 2\nabla_{AA'}\cF_{ab}.
\ee
Since there is mixing between the fields on both sides of this Bianchi identity and the auxiliary fields $\cF_{A}^{a}$, $\Gamma_{A'}^{a}$ in the recursive relations, we need to append additional equations for this constraint to be consistent.  The only possibilities are the field equations of $\cN=4$ SYM itself.  Lemma \ref{lemma: odd} is a statement about the SD sector of the theory, so the required the field equations are precisely those which are non-trivial in the instanton sector (i.e., $\lambda=0$) for a general non-abelian gauge group.  In terms of leading order field components, these are \eqref{FE1}, \eqref{FE3}, and \eqref{FE4}.  This amounts to a proof of the following chiral version of theorem \ref{superThm1}:

\begin{propn}\label{spropn}
Let $\nabla=\d+\CA$ be a chiral superconnection on the $G$-bundle $E\rightarrow\M$, with $\CA=\Gamma_{AA'}\d x^{AA'}+\Gamma_{aA}\d\theta^{aA}$ a $\mathfrak{g}^{\C}$-valued $1$-form, and curvatures $\cF_{ab}$, $\cF_{aA'}$ as above.  If $\{\Gamma^{(0)}_{AA'},\Gamma^{(0)}_{aA},\cF^{(0)}_{ab},\cF^{(0)}_{aA'}\}$ satisfy the field equations, and $\nabla$ obeys
\be{constraints}
\left\{\nabla^{a}_{(A},\nabla^{b}_{B)}\right\}=0, \qquad \left[\nabla^{a}_{(A},\nabla_{B)B'}\right]=0,
\ee
\be{Bianchi3}
\nabla_{aA}\cF_{bc}=\nabla_{A[a}\cF_{bc]} \qquad \nabla_{aA}\cF_{bA'} = 2\nabla_{AA'}\cF_{ab},
\ee
\be{recursions}
(1+\Upsilon)\Gamma_{aA}=2\theta_{A}^{b}\cF_{ab}, \qquad \Upsilon\Gamma_{AA'}=\theta_{A}^{b}\cF_{bA'},
\ee
subject to:
\begin{eqnarray}
D^{B}_{A'}G_{AB} -\left\{\tilde{\Psi}_{a\;A'},\Psi^{a}_{A}\right\}+\frac{1}{2}\left[\Phi_{ab},D_{AA'}\bar{\Phi}^{ab}\right] & = & 0, \label{fe1*} \\
D^{AA'}\Psi^{a}_{A} - \left[\tilde{\Psi}_{b\;A'},\bar{\Phi}^{ab}\right] & = & 0, \label{fe2*} \\
\Box \Phi_{ab}-\left\{\tilde{\Psi}_{[b}^{A'},\tilde{\Psi}_{a]\;A'}\right\}-\lambda\epsilon_{abcd}\left\{\Psi^{c}_{A},\Psi_{d\;A}\right\}-\lambda\left[\Phi_{c[a},[\bar{\Phi}^{cd},\Phi_{b]d}]\right] & = & 0 \label{fe3*},
\end{eqnarray}
then so do the resulting superfields $\{\Gamma_{AA'},\Gamma_{aA},\cF_{ab},\cF_{aA'}\}$.  Furthermore, the field equations imply the constraints from which the recursive equations \eqref{recursions} are derived.
\end{propn}


\subsection{Explicit Superconnections}

We now apply proposition \ref{spropn} to explicitly calculate the form of the superconnection; this entails integrating the Bianchi identities \eqref{Bianchi3} order-by-order in $\theta$. For an abelian gauge group, this is possible to all orders in $\theta$ quite easily; for a $\SU(N)$ gauge group we present this up to $O(\theta^4)$.  At each order in $\theta$, there are a finite number of irreducible expressions in the fermionic variables; these will enable us to compactly represent superconnection components and are given by the following table:

\begin{center}
\begin{tabular}{c r}
Order of $\theta$ & Irreducibles \\
\hline
$O(\theta)$ & $\theta^{Aa}$ \\
\hline
$O(\theta^{2})$ & $\theta^{2\;ab}_{AB}=\theta_{(A}^{[a}\theta_{B)}^{b]}$ \\
                & $\theta^{2\;ab}=\theta^{A(a}\theta^{b)}_{A}$ \\
\hline
$O(\theta^{3})$ & $\theta^{3\;abc}_{ABC}=\theta_{(A}^{[a}\theta_{B}^{b}\theta_{C)}^{c]}$ \\
                & $\theta^{3\;Ac}_{ab}=\theta^{2\;AB}_{ab}\theta_{B}^{c}$ \\
\hline
$O(\theta^{4})$ & $\theta_{4\;ABCD}^{abcd}=\theta_{(A}^{[a}\theta_{B}^{b}\theta_{C}^{c}\theta_{D)}^{d]}$ \\
                & $\theta^{4\;ABa}_{b}=\theta^{3\;ABC}_{b}\theta^{a}_{C}=\theta^{3\;ABCc}_{bc}\theta^{a}_{C}$ \\
                & $\theta^{4\;abcd}=\theta^{2\;ABab}\theta^{2\;cd}_{AB}$\\
\hline
$O(\theta^{5})$ & $\theta^{5\;ABC}_{abc}=\theta^{5\;(ABC)}_{[abc]}$ \\
                & $\theta^{5\;Aab}_{c}=\theta^{4\;ABa}_{c}\theta^{b}_{B}$ \\
\hline
$O(\theta^{6})$ & $\theta^{6\;AB}_{ab}=\theta^{6\;(AB)}_{[ab]}$ \\
                & $\theta^{6\;ab}=\theta^{5\;Ac(a}_{c}\theta^{b)}_{A}$ \\
\hline
$O(\theta^{7})$ & $\theta^{7\;A}_{a}$ \\
\hline
$O(\theta^{8})$ & $\theta^{8}$
\end{tabular}
\end{center}

It is equally important for us to know how arbitrary products of the $\theta$s decompose in terms of these irreducibles.  Some useful examples are:
\begin{eqnarray*}
\theta^{Aa}\theta^{Bb} & = & \theta^{2\;ABab}-\frac{1}{2}\epsilon^{AB}\theta^{2\;ab} \\
\theta^{Aa}\theta^{2\;BCbc} & = & \theta^{3\;ABCabc}+\frac{\epsilon^{AB}}{3}\theta^{3\;Cbca}+\frac{\epsilon^{AC}}{3}\theta^{3\;Bbca} \\
\theta^{Aa}\theta^{2\;bc} & = & \frac{4}{3}\theta^{3\;Aa(bc)} \\
\theta^{Aa}\theta^{3\;BCDbcd} & = & \theta^{4\;ABCDabcd}+\frac{1}{12}\epsilon^{A(B}\theta^{4\;CD)a}_{e}\epsilon^{ebcd} \\
\theta^{Aa}\theta^{3\;Bbcd} & = & \frac{1}{8}\theta^{4\;ABd}_{e}\epsilon^{ebca}+\frac{1}{2}\epsilon^{AB}\theta^{4\;bcad}
\end{eqnarray*}

\subsubsection*{\textit{Abelian gauge group}}

If we suppose that our fields live in the abelian gauge group $\U(1)$, then solving for the superconnection is relatively easy.  Using the radial gauge condition and the fact that $\cF_{ab}=\cF_{[ab]}$, it follows immediately that we can set our initial data to be
\be{eqn: zero}
\cF^{(0)}_{ab}=\Phi_{ab}, \qquad \Gamma^{(0)}_{aA}=0, \qquad \cF^{(0)}_{bA'}=\tilde{\Psi}_{bA'}, \qquad \Gamma^{(0)}_{AA'}=A_{AA'}.
\ee

For abelian gauge group all commutators vanish, so the $\nabla_{aA}$ appearing in \eqref{Bianchi3} can be replaced by $\partial_{aA}$.  This means that rather than integrate the Bianchi identity order-by order for $\cF_{ab}$, we can simply note that
\begin{equation*}
\partial_{dD}\partial_{eE}\partial_{aA}\cF_{bc}=\partial_{dD}\partial_{eE}\partial_{A[a}\cF_{bc]}.
\end{equation*}
As $\partial_{dD}\partial_{eE}$ anti-commute, it is clear that this entire expression is skew over all five R-symmetry indices: $[deabc]$.  But as the R-symmetry only ranges from $1,\ldots,4$ this implies that
\begin{equation*}
\partial_{dD}\partial_{eE}\partial_{aA}\cF_{bc}=0.
\end{equation*}
Combined with anti-symmetry of the indices on the curvature itself, we can easily see the correct expansion for $\cF^{ab}$,
\be{cab1}
\cF_{ab}=\Phi_{ab}+\frac{3}{2}\Psi_{abdA}\theta^{dA}+2G_{AB}\theta^{2\;AB}_{ab},
\ee
where $\Psi_{abdA}\equiv\epsilon_{abcd}\Psi^{c}_{A}$.  Feeding this into \eqref{recursions}, we find
\be{abcon1}
\Gamma_{aA}=\Phi_{ab}\theta^{b}_{A}+\Psi_{abdB}\theta^{2\;Bdb}_{A}+G_{BC}\theta^{3\;BC}_{aA}.
\ee

For the even portion of the superconnection, the initial data immediately determines
\begin{equation*}
\Gamma^{(1)}_{AA'}=\theta^{b}_{A}\tilde{\Psi}_{bA'}.
\end{equation*}
The Bianchi identity at $O(\theta^{0})$ is simply
\begin{eqnarray*}
\partial_{aA}\cF^{(1)}_{bA'} & = & 2 \partial_{AA'}\cF^{(0)}_{ab}=2\partial_{AA'}\Phi_{ab} \\
\Rightarrow \cF^{(1)}_{bA'} & = & 2\partial_{BA'}\Phi_{ab}\theta^{aB}.
\end{eqnarray*}
The recursion relation then gives
\begin{equation*}
\Gamma^{(2)}_{AA'}=\partial_{BA'}\Phi_{ab}\theta^{2\;Bba}_{A}.
\end{equation*}

At $O(\theta)$, the Bianchi identity reads
\begin{equation*}
\partial_{aA}\cF^{(2)}_{bA'}=3\partial_{AA'}\Psi_{abdB}\theta^{dB},
\end{equation*}
which is only integrable (by reference to our algebra of super-space coordinates) if $\partial_{AA'}\Psi_{abdB}=\partial_{A'(A}\Psi_{B)abd}$.  This integrability condition is satisfied precisely when we apply the $\cN=4$ field equation \eqref{fe2*}, which allows the first order Bianchi identity to be integrated:
\begin{equation*}
\cF^{(2)}_{bA'}=\frac{3}{2}\partial_{DA'}\Psi_{abdB}\theta^{2\;DBad}.
\end{equation*}
This in turn provides us with the third-order portion of the even connection:
\begin{equation*}
\Gamma^{(3)}_{AA'}=-\frac{1}{2}\partial_{DA'}\Psi_{abdB}\theta^{3\;DBbad}_{A}-\frac{1}{6}\partial_{AA'}\Psi_{abdB}\theta^{3\;Badb}-\frac{1}{6}\partial_{DA'}\Psi_{abdA}\theta^{3\;Dadb}.
\end{equation*}

Finally, the Bianchi identity at $O(\theta^{2})$ reads
\begin{equation*}
\partial_{aA}\cF^{(3)}_{bA'}=4\partial_{AA'}G_{BC}\theta^{2\;BC}_{ab},
\end{equation*}
which also comes with an integrability condition: $\partial_{DA'}G_{BC}=\partial_{A'(D}G_{BC)}$.  This is again satisfied upon recourse to the field equation \eqref{fe1*}, so we can obtain the curvature at third order and consequently the final (fourth-order) contribution to the connection:
\begin{equation*}
\cF^{(3)}_{bA'}=-\frac{4}{3}\partial_{DA'}G_{BC}\theta^{3\;DBC}_{b}, \qquad \Gamma^{(4)}_{AA'}=-\frac{1}{3}\partial_{DA'}G_{BC}\theta^{4\;BCD}_{A}.
\end{equation*}
This provides the total expression for the even portion of the super-connection:
\begin{multline}\label{abcon2}
\Gamma_{AA'}=A_{AA'}+\theta^{b}_{A}\tilde{\Psi}_{bA'}+ \partial_{BA'}\Phi_{ab}\theta^{2\;Bba}_{A}-\frac{1}{2}\partial_{DA'}\Psi_{abdB}\theta^{3\;DBbad}_{A}-\frac{1}{6}\partial_{AA'}\Psi_{abdB}\theta^{3\;Badb} \\
 -\frac{1}{6}\partial_{DA'}\Psi_{abdA}\theta^{3\;Dadb}-\frac{1}{3}\partial_{DA'}G_{BC}\theta^{4\;BCD}_{A}.
\end{multline}
So the full abelian superconnection is given by
\begin{equation*}
\CA^{\U(1)}=\Gamma_{AA'}\d x^{AA'}+\Gamma_{aA}\d\theta^{aA},
\end{equation*}
with the components given by \eqref{abcon1}-\eqref{abcon2}.

\subsubsection*{$\SU(N)$ \textit{gauge group}}

For a non-abelian gauge group $\SU(N)$, integrating the Bianchi identities order-by-order is a rather cumbersome task.  We present some partial results for both the odd ($\Gamma_{aA}$) and even ($\Gamma_{AA'}$) portions of the superconnection.  Again, we begin with the initial ansatz (which follows itself from the choice of radial gauge and integrability condition):
\begin{equation*}
\cF^{(0)}_{ab}=\Phi_{ab}, \qquad \Gamma^{(0)}_{aA}=0.
\end{equation*}
From this, we can readily use the recursion equations to get
\be{odd1}
\Gamma^{(1)}_{aA}=\Phi_{ab}\theta^{b}_{A}.
\ee

We now write down the Bianchi identity at zeroth-order, which reads:
\begin{equation*}
\partial_{aA}\cF^{(1)}_{bc}=\partial_{A[a}\cF^{(1)}_{bc]},
\end{equation*}
whose solution we can pull directly from the abelian case:
\be{odd2}
\cF^{(1)}_{bc}=\frac{3}{2}\Psi_{abdA}\theta^{dA}, \qquad  \Gamma^{(2)}_{aA}=\Psi_{abdB}\theta^{2\;dbB}_{A},
\ee
as before.

The first meaningful difference from the abelian theory appears when we write down the Bianchi identity at order one:
\begin{equation*}
\partial_{aA}\cF^{(2)}_{bc}+\left[\Gamma^{(1)}_{aA},\cF^{(0)}_{bc}\right]=\partial_{A[a}\cF^{(2)}_{bc]}+\left[\Gamma^{(1)}_{A[a},\cF^{(0)}_{bc]}\right].
\end{equation*}
Now, we can use our prior results to fill in the known quantities in this equation, leaving
\begin{equation*}
\partial_{aA}\cF^{(2)}_{bc}=\partial_{A[a}\cF^{(2)}_{bc]}+\theta^{d}_{A}\left(\frac{2}{3}[\Phi_{da},\Phi_{bc}]-\frac{1}{3}[\Phi_{bd},\Phi_{ac}]+\frac{1}{3}[\Phi_{cd},\Phi_{ab}]\right).
\end{equation*}
We know that $\cF^{(2)}_{bc}$ can contain a term from the abelian case, but we must now consider additional terms which solve the fully non-abelian Bianchi identity.  A bit of intuition shows that the correct ansatz is $\kappa [\Phi_{b(a},\Phi_{d)c}]\theta^{2\;ad}$, and plugging this into the above equation allows us to fix $\kappa=-1$.  Hence, we obtain:
\begin{equation*}
\cF^{(2)}_{ab}=2G_{AB}\theta^{2\;AB}_{ab}-\left[\Phi_{a(c},\Phi_{d)b}\right]\theta^{2\;cd},
\end{equation*}
from which we find
\be{cono3}
\Gamma^{(3)}_{aA}=G_{DB}\theta^{3\;DB}_{Aa}-\frac{2}{3}\left[\Phi_{a(c},\Phi_{d)b}\right]\theta^{3\;b(cd)}_{A}.
\ee

The situation becomes increasingly more complex as we consider the Bianchi identity at second order.  A lengthy but straightforward calculation along the lines of the $O(\theta)$ case eventually reveals:
\begin{equation*}
\cF^{(3)}_{bc}=\left(3\left[\Phi_{d[c},\Psi_{b]aeB}\right]-2\left[\Psi_{Bde[c},\Phi_{b]a}\right]\right)\left(\theta^{3\;Be(da)}-\theta^{3\;Beda}\right)+\frac{3}{4}\left[\Phi_{d[c},\Psi_{b]aeB}\right]\theta^{3\;Ba(ed)},
\end{equation*}
leading to
\begin{multline}\label{cono4}
\Gamma^{(4)}_{aA} = \left(\frac{3}{5}\left[\Phi_{d[b},\Psi_{a]feA}\right]-\frac{2}{5}\left[\Psi_{Ade[b},\Phi_{a]f}\right]\right)\theta^{4\;e[df]b}-\frac{3}{20}\left[\Phi_{d[b},\Psi\_{a]feA}\right]\theta^{4\;f(ed)b} \\
-\left(\frac{3}{20}\left[\Phi_{d[b},\Psi_{a]feB}\right]-\frac{1}{10}\left[\Psi_{Bde[b},\Phi_{a]f}\right]\right)\theta^{4\;B[f}_{A\;c}\epsilon^{d]ceb}+\frac{3}{80}\left[\Phi_{d[b},\Psi_{a]feB}\right]\theta^{4\;B(d}_{A\;c}\epsilon^{e)cfb}.
\end{multline}
Note that this is the first superconnection component which is totally new in the non-abelian case; the abelian superconnection terminates at third order \eqref{abcon1}.

For the even portion of the superconnection the initial data
\be{ev0}
\cF^{(0)}_{aA'}=\tilde{\Psi}_{aA'}, \qquad \Gamma^{(0)}_{AA'}=A_{AA'}
\ee
immediately gives us the even connection to first order in $\theta$ using the recursive relations \eqref{recursions}:
\be{ev1}
\Gamma^{(1)}_{AA'}=\theta^{b}_{A}\tilde{\Psi}_{bA'}.
\ee
Now, the relevant Bianchi identity at $O(\theta^{0})$ reads
\begin{equation*}
\partial_{aA}\cF^{(1)}_{bA'}=2D_{AA'}\Phi^{ab},
\end{equation*}
which is easily integrated to give
\be{ev2}
\cF^{(1)}_{bA'}=2D_{AA'}\Phi_{ab}\theta^{aA}, \qquad \Gamma^{(2)}_{AA'}=D_{BA'}\Phi_{ab}\theta^{2\;Bba}_{A}.
\ee

Writing out the Bianchi identity at $O(\theta)$ gives:
\begin{equation*}
\partial_{aA}\cF^{(2)}_{bA}=3D_{AA'}\Psi_{abdB}\theta^{dB}+\theta^{d}_{A}\left(2\left[\tilde{\Psi}_{dA'},\Phi_{ab}\right]-\left[\Phi_{ad},\tilde{\Psi}_{bA'}\right]\right).
\end{equation*}
As in the abelian case, we will need the field equation \eqref{fe2*} to make the Bianchi identity integrable due to our chiral reduction.  A straightforward calculation eventually gives
\begin{equation*}
\cF^{(2)}_{bA'}=\frac{3}{2}D_{A'(A}\Psi_{B)abd}\theta^{2\;ABad}+\theta^{2\;ad}\left[\Phi_{b(a},\tilde{\Psi}_{d)A'}\right].
\end{equation*}
Feeding this into the recursion relations and using the algebra for fermionic coordinates yields the third-order portion of the connection:
\begin{multline}\label{ev3}
\Gamma^{(3)}_{AA'}=\frac{1}{2}D_{A'(B}\Psi_{C)abd}\theta^{3\;BCabd}_{A}-\frac{1}{6}D_{A'(A}\Psi_{C)abd}\theta^{3\;Cadb}-\frac{1}{6}D_{A'(B}\Psi_{A)abd}\theta^{3\;Badb} \\
+\frac{4}{9}\theta^{3\;b(ad)}_{A}\left[\Phi_{b(a},\tilde{\Psi}_{d)A'}\right].
\end{multline}

Additional connection components can be calculated up to $O(\theta^8)$ for the both the odd and even superconnections, but the calculations are rather long and tedious so we stop here.


\section{Coulomb Branch on Twistor Space}
\label{Appendix2}

In Section \ref{Chapter3}, we saw that the twistor action for $\cN=4$ SYM provided an efficient mechanism for computing tree-level and finite loop-level scattering amplitudes via the MHV formalism.  To truly capture all loop amplitudes though, we require a regularization mechanism on twistor space to deal with IR divergences.  The mass regularization scheme of \cite{Alday:2009zm} seems like the best candidate; in order for this to work we need to access the Coulomb branch on twistor space.  In this appendix, we show that this is possible and leads to Kiermaier's massive MHV formalism \cite{Kiermaier:2011cr, Elvang:2011ub}, which can be seen as a generalization of the massive MHV rules of Boels and Schwinn (c.f., \cite{Boels:2007pj, Boels:2008ef}).  In principle, this provides a mechanism for regularizing divergent quantities on twistor space, although we are still far from being able to implement this regularization in a practical fashion that is self-contained in twistor space.


\subsection{Coulomb Branch and Mass Regularization}

$\cN=4$ SYM has a quantum moduli space obtained by giving vacuum expectation values to the six scalar fields $\Phi_{ab}$. The scalars transform under supersymmetry in the same multiplet as the gauge fields and hence the moduli space is called the Coulomb branch of the theory.

We begin with $\cN=4$ SYM at the origin of the moduli space with unbroken gauge group $\U(N)$ and R-symmetry group $\SU(4)_R$. We then move onto the Coulomb branch by giving the following vacuum expectation value to the scalar fields
\be{gcoulomb}
\la \Phi_{ab} \ra = \epsilon_{ab} \,
\mathrm{diag}(\, \upsilon_1\mathbb{I}_{N_1}\, ,\, \upsilon_2\mathbb{I}_{N_2}\, ,\, \ldots\, ) \, .
\ee
This breaks the gauge group spontaneously to the product $\prod_r \U(N_r)$ and the R-symmetry group to the subgroup $\Sp(4)_R$ that leaves invariant the symplectic form $\epsilon_{ab}$. Following \cite{Craig:2011ws} we choose
\be{sympstruct}
\epsilon_{ab} =  \begin{pmatrix}
  i\sigma_2 & 0  \\
  0 & i\sigma_2 
 \end{pmatrix},
\ee
manifesting a $\SU(2)_R\times \SU(2)_R$ subgroup that will play an important role in classifying the structure of scattering amplitudes on the Coulomb branch.

The spectrum now consists of massless and massive particles. There are massless supermultiplets in the adjoint of the unbroken gauge group $\prod_r \U(N_r)$ containing the familiar massless gluons $g^{\pm}$, fermions $\widetilde{\Psi}_{a\;A'}$, $\Psi_{A}^{a}$ in the fundamental $\mathbf{4}$, and scalars $\Phi_{ab}$ in the antisymmetric tensor $\mathbf{6}$ of an unbroken $\SU(4)_R$ symmetry. In addition, there are now massive supermultiplets in the bifundamentals of the gauge groups $\U(N_i)\times \U(N_j)$ with masses $m_{ij} = \upsilon_i-\upsilon_j$. These contain massive vector bosons $W^{\pm}$ with longitudinal component $W^L = \frac{1}{\sqrt{2}}(w_{12}+w_{34})$ arising from scalars in the direction of the vacuum expectation value, fermions $\omega_a$ in the fundamental $\mathbf{4}$, and the remaining scalar fields $\{\frac{1}{\sqrt{2}}(w_{12}-w_{34}),w_{13},w_{23},w_{14},w_{24}\}$ transforming in the $\mathbf{5}$ representation of the remaining unbroken $\Sp(4)_R$ symmetry.  In terms of the motivating picture from AdS geometry, the gauge group $\U(N_i)$ represents a stack of $N_i$ D3-branes located a distance $\upsilon_i$ from the $AdS_5$-boundary, and the massive bifundamental $m_{ij}$ is given by a string stretching between two such stacks.

\begin{figure}
\centering
\includegraphics[width=1.5 in, height=1.5 in]{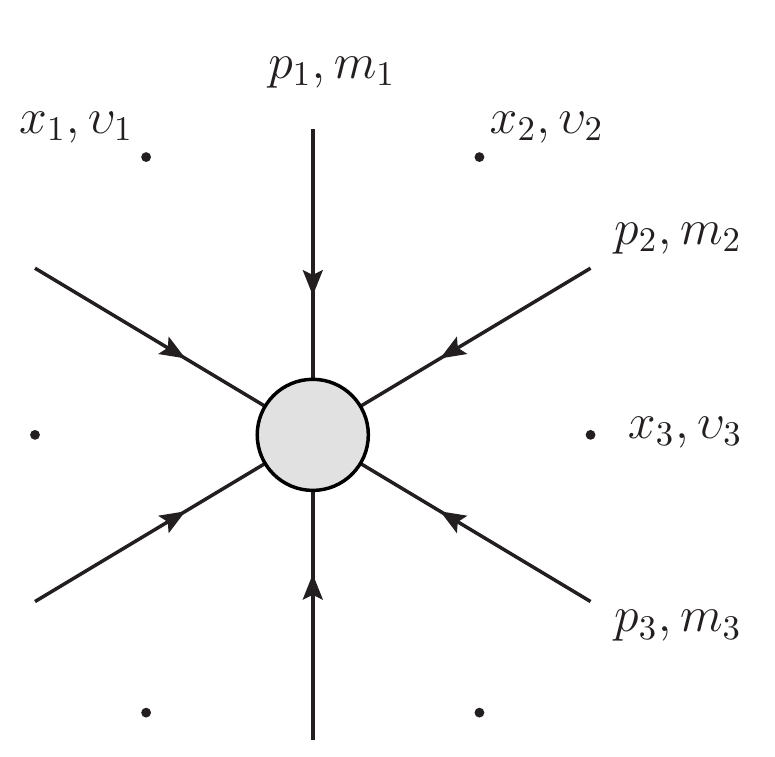}\caption{\textit{The assignment of VEVs to external regions and masses to external particles}}\label{Regions}
\end{figure}

For planar amplitudes on the Coulomb branch with incoming momenta $\{p_1,\ldots,p_n\}$, momentum conservation is made manifest by introducing region coordinates $\{x_1,\ldots,x_n\}$ forming a definite polygon 
\begin{equation*}
(x_{i+1}-x_i)^2=m_i^2
\end{equation*}
where $(m_1,\ldots,m_n)$ are masses and the incoming momenta are identified as $p_i = x_{i+1}-x_i$. Incoming particles transform in bifundamentals, so to have non-vanishing planar amplitudes, we assign gauge groups $ \U(N_1),\ldots,\U(N_n)$ with the vacuum expectation values $\{\upsilon_1,\ldots,\upsilon_n\}$ to external regions as shown in Figure \ref{Regions}. The masses $m_i = \upsilon_{i+1}-\upsilon_i$ then automatically satisfy the condition $\sum_im_i = 0$. 

For loop amplitudes we may also assign gauge groups and vacuum expectation values to all internal regions. Following \cite{Alday:2009zm, Kiermaier:2011cr}, we assign the same $\U(N)$ with vacuum expectation value $\upsilon$ to all internal regions and then use the translational freedom in $(\upsilon,\upsilon_1,\cdots,\upsilon_n)$ to choose $\upsilon = 0$. The planar limit where $N\gg N_i$ ensures that the leading contribution to such amplitudes are planar diagrams with all internal lines massless.

\begin{figure}
\centering
\includegraphics[width=3 in, height=1.5 in]{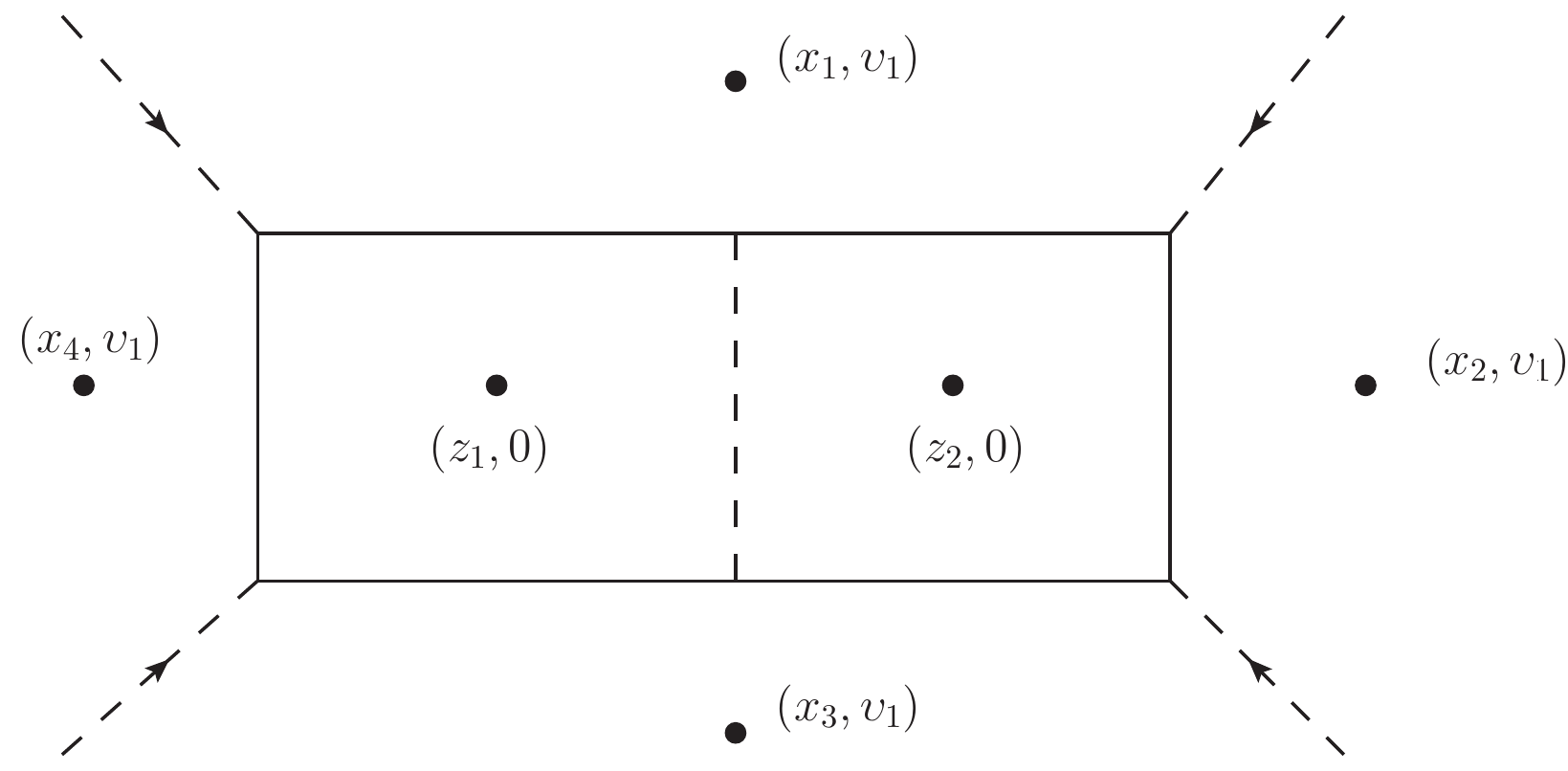}\caption{\textit{A 4-particle, 2-loop example of the massive regularization: external particles and full internal lines are massless, while legs on the outside of the loop have mass} $\upsilon_{1}$.}\label{Regions2}
\end{figure}

Furthermore, when all external regions are assigned the same gauge group with vacuum expectation value $\upsilon_1$, the incoming particles are all massless and the above framework becomes the AdS regulation scheme for amplitudes at the origin of the moduli space first proposed by \cite{Alday:2009zm} (see \cite{Henn:2011xk} for a review).  It has been shown that these mass-regularized amplitudes enjoy an exact dual superconformal symmetry, hence providing a regularization for the theory which does not obscure its Yangian symmetry algebra (in contrast to dimensional regularization).  A simple example is illustrated in Figure \ref{Regions2}.

We expand scattering amplitudes in the vacuum expectation values $\{\upsilon_1,\ldots,\upsilon_n\}$, or equivalently in the masses $\{m_1,\ldots,m_n\}$ subject to the condition $\sum_im_i=0$. Hence, we would like to decompose the incoming momenta $p_i$ into an $O(0)$ massless momentum and an $O(m_i^2)$ correction. This is achieved using an auxiliary null vector $q^\mu$ to decompose the momentum
\be{decomp}
p_i = p_i^{\perp} + \frac{m_i^2}{2p_i\cdot q}q\, .
\ee
The massless perpendicular component may then be expressed in terms of two-component spinors as $p^{\perp\,AA'} = |i^\perp\ra^{A}\,[ i^\perp|^{A'}$. The polarization vectors of massive on-shell states are defined using the auxiliary null vector $q^{AA'} = |q\ra^A[q|^{A'}$ and hence scattering amplitudes will depend explicitly on spinors $|q\ra^A$ and $|q]^{A'}$.

The supersymmetry algebra on the Coulomb branch is extended to include the central charge $Z_{ab} = \epsilon_{ab}$ in the direction of the vacuum expectation value. The on-shell massive supermultiplets containing the vector boson states $W^+$, $W^-$, $W^L=\frac{1}{\sqrt{2}}(w_{12}+w_{34})$, fermions $\omega_a$, and scalars $\{w_{13}$, $w_{14}$, $w_{23}$, $w_{24}$, $(w_{12}-w_{34})\}$, then transform under extended supersymmetry in a BPS supermultiplet. The extended supersymmetry algebra is realised on-shell by chiral superfields
\be{supermultiplet}
X(\eta) = W^+ + \eta^a \omega_a + \frac{1}{2!} \eta^a\eta^b w_{ab} + \frac{\epsilon_{abcd}}{3!}\eta^a\eta^b\eta^c \bar\omega^d + \frac{\epsilon_{abcd}}{4!}\eta^a\eta^b\eta^c\eta^d W^-\, .
\ee
The scattering amplitudes on the Coulomb Branch are then combined into superamplitudes $\la X_1\ldots X_n\ra$.

The MHV formalism at the origin of the moduli space \cite{Cachazo:2004kj} provides a perturbative expansion of scattering amplitudes in gauge theories that is drastically more efficient than Feynman diagram methods, and as we have seen, has a natural expression in twistor space.  Kiermaier proposed an extension of the MHV formalism for scattering amplitudes on the Coulomb branch of $\cN=4$ SYM \cite{Kiermaier:2011cr}. In this \emph{massive MHV formalism}, there are three classes of vertices at $O(m_i^0)$, $O(m_i)$ and $O(m_i^2)$ in the particle masses which are holomorphic in the perpendicular spinors $|i^\perp\ra$ associated with the reference null vector $q^{AA'}$. The three classes of vertex are given explicitly by the formulae\footnote{Here and elsewhere we suppress an overall bosonic momentum-conserving delta-function $\delta^8\left(\sum_{i}|i^\perp\ra^{A}\,[ i^\perp|^{A'}\right)$.}:
\be{vm0}
\frac{\delta^{(8)}\left(\sum\limits_i |i^\perp\ra\eta_{i}\right)}{\la 1^{\perp}2^{\perp}\ra\cdots\la n^{\perp}1^{\perp}\ra}, 
\ee
\be{vm1}
\left(\sum_{i}\frac{m_{i}\la 1^{\perp}i^{\perp}\ra}{\la1^{\perp}q\ra\la i^{\perp}q\ra}\right) \frac{\delta^{(4)}_{12}\left(\sum\limits_{i}|i^\perp\ra\eta_{i}\right)\delta^{(2)}_{34}\left(\sum\limits_{i}\la q i^{\perp}\ra\eta_{i}\right) + \left\{12\leftrightarrow 34\right\}}{\la 1^{\perp}2^{\perp}\ra\cdots\la n^{\perp}1^{\perp}\ra}, 
\ee
\be{vm2}
\left(\sum_{i}\frac{m_{i}\la 1^{\perp}i^{\perp}\ra}{\la1^{\perp}q\ra\la i^{\perp}q\ra}\right)^{2}\frac{\delta^{(4)}_{1234}\left(\sum\limits_{i}\la qi^{\perp}\ra\eta_{i}\right)}{\la 1^{\perp}2^{\perp}\ra\cdots\la n^{\perp}1^{\perp}\ra},
\ee
where the superscripts denote overall Grassmann degree and the subscripts indicate which components of the $\SU(2)_R\times \SU(2)_R$ broken R-symmetry the fermionic delta-functions occupy. 

The propagators are now massive scalar Feynman propagators of the form $1/(p^2-m^2)$. For a propagator bounding two regions $x_1$ and $x_2$ (which may be either internal or external) and associated vacuum expectation values $\upsilon_1$ and $\upsilon_2$, we have the propagator
\begin{equation*}
\frac{1}{(x_1-x_2)^2 - (\upsilon_1-\upsilon_2)^2}\, .
\end{equation*}
The propagators are then assigned holomorphic spinors $|p\ra^{A} = p^{AA'}|q]_{A'}$ using the anti-holomorphic component of the reference vector $q^{AA'} = q^Aq^{A'}$ in the same fashion as the MHV formalism at the origin of the moduli space.  These massive propagators arise from the resummation of the massless scalar propagator corrected by $O(m_{i}^2)$ two-point vertex insertions ($O(m_{i}^0)$ and $O(m_{i}^1)$ 2-point vertices vanish for kinematic reasons) \cite{Kiermaier:2011cr}. This massive MHV formalism is known to be correct via recursive arguments \cite{Elvang:2011ub}, but it can also be derived organically from the twistor action.


\subsection{Coulomb Branch Twistor Action}

In order to probe the Coulomb branch of $\cN=4$ SYM, we must first understand how to represent to scalar VEV on twistor space.  This amounts to constructing cohomological representatives for a constant scalar field on space-time. We then expand the twistor action for $\cN=4$ SYM around such background fields, and hence provide a derivation of the massive MHV rules.

\begin{figure}
\centering
\includegraphics[width=3 in, height=1 in]{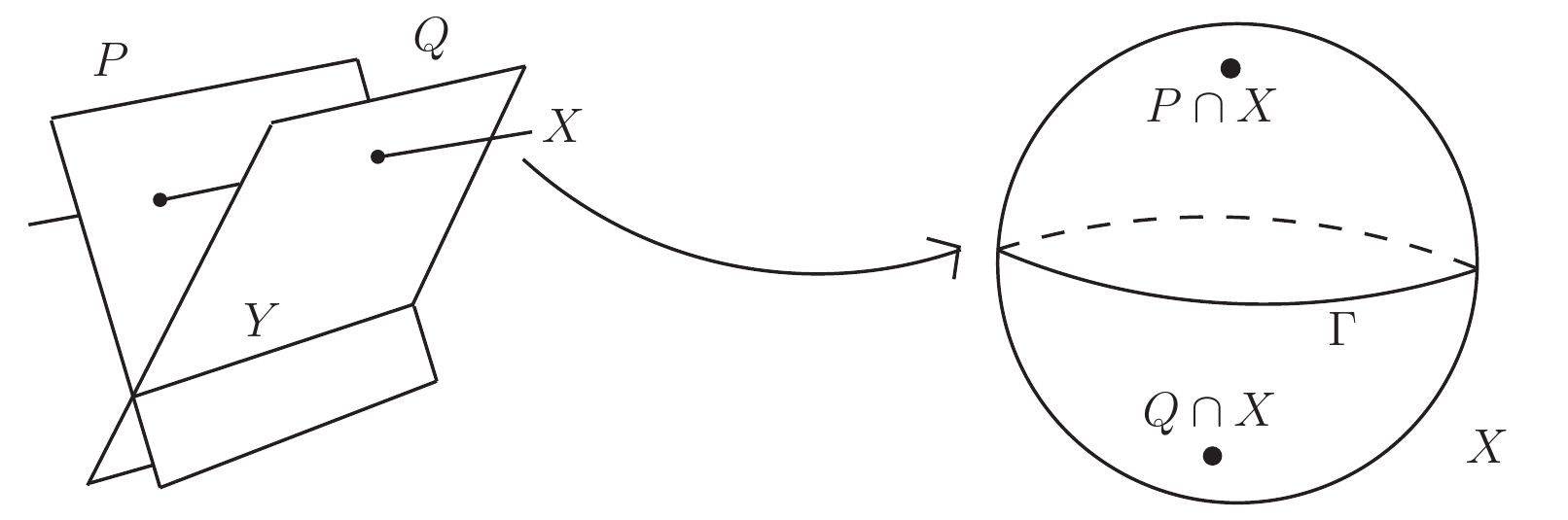}\caption{\textit{Geometric representation of the Penrose transform \eqref{PenroseExample}}}\label{PT}
\end{figure}

First consider a free massless scalar $\Phi(x)$. Solutions of the wave equation $\Box\Phi(x)=0$ may be constructed from a cohomology class on (bosonic) twistor space $\phi\in H^1(\PT_{b},\cO(-2))$ via the Penrose transform.  If we work in the \v{C}ech representation, the simplest twistor function of homogeneity -2 is the elementary state (c.f., \cite{Penrose:1986ca})
\be{elem1}
\phi(Z) =\frac{I^{\alpha\beta}P_\alpha Q_\beta}{(P_\alpha Z^\alpha)(Q_\beta Z^\beta)}\, ,
\ee
which becomes singular on the planes defined by 
\begin{equation*}
P=\{Z^{\alpha} |\;P_\alpha Z^\alpha=0\} \qquad \mathrm{and} \qquad Q=\{Z^{\alpha} |\;Q_\alpha Z^\alpha=0\}\, .
\end{equation*}
Therefore the pullback $\phi(\lambda_A,ix^{AA'}\lambda_A)$ to any line $X$ has simple poles on the intersection points $X\cap P$ and $X\cap Q$, and the compact contour $\Gamma$ in the Penrose transform is taken to surround one pole or the other, as illustrated in figure~\ref{PT}. Denoting the spinor components of the auxiliary dual twistors by $P_\alpha=(p^A,p_{A'})$ and $Q_\alpha=(q^A,q_{A'})$, the Penrose transform becomes the contour integral
\be{PenroseExample}
\Phi(x) =\frac{1}{2\pi i } \oint\limits_\Gamma \frac{p_A q^A}{\lambda_A( p^A - ix^{AA'}p_{A'})\; \lambda_B(q^B - ix^{BB'}q_{B'})} \D\lambda \, .
\ee
The contour integral is easily evaluated with the result
\be{elem2}
\Phi(x) = \frac{y^2}{(x-y)^2},
\ee
where the space-time point
\begin{equation*}
y^{AA'} = \frac{p^Aq^{A'} - q^Ap^{A'}}{\la p\, q\ra}
\end{equation*}
corresponds to the line $Y = P\cap Q$ in twistor space. The contour integral is well defined when the poles at $X\cap P$ and $X\cap Q$ do not coincide, or equivalently when the twistor lines $X$ and $Y$ are skew. If the lines $X$ and $Y$ intersect, then $(x-y)^2=0$ and the solution becomes singular. 

Now, moving $y^{AA'}$ towards infinity in space-time, the solution~\eqref{PenroseExample} tends towards unity at all finite distances. This requires that the intersection $P\cap Q$ becomes the line at infinity $I$ and hence that the dual twistors take the form $P_\alpha = I_{\alpha\beta}P^\beta$ and $Q_\alpha = I_{\alpha\beta}Q^\beta$, which have components $P_\alpha = (p^A,0)$ and $Q_\alpha = (q^A,0)$. This removes any position dependence from the Penrose transform and leads to a constant solution:
\begin{equation*}
\Phi(x) = \frac{1}{2\pi i } \oint\limits_\Gamma \frac{ p_A q^A}{ (p_A \lambda^A)(q_B\lambda^B)}\D\lambda \;  = 1\, .
\end{equation*}
Here we see clearly how the vacuum expectation value requires the introduction of the infinity twistor $I^{\alpha\beta}$ into the formalism, explicitly breaking the conformal symmetry.

Thus we have found a class of \v{C}ech cohomology representatives for the vacuum expectation value of a scalar field:
\be{Cech}
\tilde{\phi}(Z) = \frac{I_{\alpha\beta}P^\alpha Q^\beta}{(I_{\alpha\beta}P^\alpha Z^\beta)(I_{\alpha\beta}Q^\alpha Z^\beta)}.
\ee
An equivalent representative can also be found for the Dolbeault representative of the VEV, which takes the obvious form:
\be{Dolbeault}
\tilde{\phi}(Z)=\frac{I_{\alpha\beta}P^{\alpha}Q^{\beta}}{I_{\alpha\beta}P^{\alpha}Z^{\beta}}\dbar\left(\frac{1}{I_{\alpha\beta}Q^{\alpha}Z^{\beta}}\right).
\ee
This background field is very convenient for computations of scattering amplitudes in axial gauge and in this context, the spinor $q^A$ becomes that used to define the polarisations of massive on-shell states.  Note also that the representatives \eqref{Cech}, \eqref{Dolbeault} are independent of the choice of space-time signature or reality conditions on $\PT$.

If we fix Euclidean reality conditions, then yet another representation of the constant scalar VEV is available.  Using the basis \eqref{bforms}, any representative of a cohomology class in $H^{1}(\PT_{b},\cO(-2))$ can be written as
\begin{equation*}
\tilde{\phi}= \tilde{\phi}_{0}\hat{e}^{0}+\tilde{\phi}_{A'}\hat{e}^{A'}.
\end{equation*}
Fixing Woodhouse gauge, compactness and dimensionality of the $\P^{1}$ fibers gives $\tilde{\phi}_{0}=\tilde{\phi}_{0}(x)=1$.  So we may choose any Euclidean signature Dolbeault representative for the VEV of the form: 
\be{Cwrep}
\tilde{\phi}(Z)=\frac{\langle\hat{\lambda}\d\hat{\lambda}\rangle}{\langle\lambda\hat{\lambda}\rangle^2}+\tilde{\phi}_{A'}\bar{e}^{A'}.
\ee
By comparison with the signature-independent representatives of \eqref{Cech} or \eqref{Dolbeault}, we see that a choice is always implicit in the definition of the twistor representative.  In the signature-independent case, it is the choice of spinors $p^{A}$ and $q^{A}$, while in the Euclidean case it is the choice of anti-holomorphic involution. 

Finally, we must extend the twistor background field corresponding to the scalar VEV to the context of the Coulomb branch of $\cN=4$ SYM.  Recall that the full supermultiplet at the origin of the moduli space is encoded in the $(0,1)$-form $\cA$ of \eqref{superfield1}.  To include the scalar VEV, we introduce the background expectation value
\be{vev1}
\la \phi_{ab} \ra = \epsilon_{ab}\mathsf{U},
\ee
where
\begin{equation*}
\mathsf{U} = \left(\begin{array}{ccc}
\upsilon_{1}\mathbb{I}_{N_{1}} & 0 & \cdots \\
0 & \upsilon_{2}\mathbb{I}_{N_2} & \cdots \\
\vdots &  & \ddots 
\end{array} \right)\, .
\end{equation*}
Hence, we simply introduce the twistor space background field
\be{supertwistorvev}
\cU(Z) = \epsilon_{ab}\chi^a\chi^b \tilde{\phi}(Z) \mathsf{U},
\ee
where $\tilde{\phi}(Z)$ is one of the representatives \eqref{Cech}, \eqref{Dolbeault} or \eqref{Cwrep} discussed above and $\epsilon_{ab}$ is the R-symmetry breaking symplectic form \eqref{sympstruct}.

Now, recall our formulation of the twistor action for $\cN=4$ SYM at the origin of the moduli space \eqref{TwistorAction}.  By theorem \ref{BMSthm}, we know that this is (at least perturbatively) equivalent to the space-time Chalmers-Siegel action.  We now expand the twistor action around a new background connection by replacing $\cA \rightarrow \cU + \cA$, where $\cU$ is the twistor VEV of \eqref{supertwistorvev}.

We begin with holomorphic Chern-Simons theory on twistor space. Expanding around the background connection with respect to $\cU$ we find that
\begin{multline}
S_1[\cU+\cA] = S_1[\cA] \\ 
+ \frac{i}{2\pi}\int_{\PT} \D^{3|4}Z \wedge\tr\left(\cU\wedge\dbar\cA+\cA\wedge\dbar\cU+2\cU\wedge\cA\wedge\cA+2\cA\wedge\cU\wedge\cU\right).
\end{multline}
The interaction term in the twistor action is now
\begin{equation}
S_{2}[\cU+\cA]=\int_{\M_{\R}}\d^{4|8}X \log\det\left(\dbar+\cU+\cA\right)|_{X},
\end{equation}
and from this we can define a \emph{Coulomb branch twistor action}:
\be{cta}
S[\cA;\cU]=S_{1}[\cU+\cA]+\lambda\;S_{2}[\cU+\cA].
\ee
This nomenclature is motivated by the following fact:
\begin{propn}\label{CBProp}
The Coulomb branch twistor action is classically equivalent to the space-time Coulomb branch action of $\cN=4$ in the sense that solutions to its Euler-Lagrange equations are in one-to-one correspondence with solutions to the field equations up to space-time gauge transformations.  Additionally, upon fixing Woodhouse gauge and Euclidean reality conditions, \eqref{cta} is equal to the space-time action.
\end{propn}

\proof After implementing the Coulomb branch Higgs mechanism $\Phi_{ab}\rightarrow\Phi_{ab}+\la\Phi_{ab}\ra$ which breaks the gauge group from $\U(N)\rightarrow\prod_{i=1}^{r}\U(N_i)$, the space-time $\cN=4$ action picks up a correction:
\begin{equation*}
S^{C}=S^{C}_{\mathrm{SD}}[A,\widetilde{\Psi},\Phi]+\lambda\; I^{C}[\Phi,\Psi],
\end{equation*}
where
\be{mact1}
S^{C}_{\mathrm{SD}}=\int_{\M}\d^{4}x\,\tr\left(\widetilde{\Psi}_{aA'}\widetilde{\Psi}^{A'}_{b}\tilde{\epsilon}^{ab}-\frac{1}{2}\left[A_{AA'},\tilde{\epsilon}_{ab}\right]D^{AA'}\bar{\Phi}^{ab}-\frac{1}{4}\left[A_{AA'},\tilde{\epsilon}_{ab}\right]\left[A^{AA'},\tilde{\epsilon}^{ab}\right]\right),
\ee
\be{mact2}
I^{C}=\int_{\M}\d^{4}x\,\tr\left(2\Psi^{a}_{A}\Psi^{bA}\tilde{\epsilon}_{ab}+\frac{1}{4}\left[\tilde{\epsilon}^{ab},\bar{\Phi}^{cd}\right]\left[\Phi_{ab},\Phi_{cd}\right]+\frac{1}{8}\left[\tilde{\epsilon}^{ab},\bar{\Phi}^{cd}\right]\left[\Phi_{ab},\tilde{\epsilon}_{cd}\right]\right).
\ee
Here, we have made a notational simplification by combining the symplectic form and VEV structure generator:
\begin{equation*}
\tilde{\epsilon}_{ab}\equiv\epsilon_{ab}\mathsf{U}.
\end{equation*}

We follow the method of \cite{Boels:2006ir}, and will ignore irrelevant numerical factors.  As previously noted, the Woodhouse gauge condition automatically fixes the remaining gauge freedom to that of space-time, so it suffices for us to demonstrate that the massive corrections in \eqref{cta} are equal to \eqref{mact1}-\eqref{mact2}.  Let us first deal with $S_{1}[\cU+\cA]$.  With Euclidean reality conditions, we make use of the basis \eqref{bforms} to expand $\cA=\cA_{0}\bar{e}^{0}+\cA_{A'}\bar{e}^{A'}$, so the Woodhouse gauge condition is simply:
\begin{equation*}
\dbar^{*}|_{X}\cA_{0}=0.
\end{equation*}
Using the fact that $H^{1}(\P^{1},\cO)=H^{1}(\P^{1},\cO(-1))=0$ and theorems \ref{Wrep1} and \ref{Wrep2}, we can write the bosonic components of $\cA$ in this gauge as:
\begin{eqnarray*}
a=a_{A'}\hat{e}^{A'}, \qquad \tilde{\psi}_{a}=\tilde{\psi}_{aA'}\hat{e}^{A'}, \qquad \phi_{ab}=\Phi_{ab}\hat{e}^{0}+\phi_{abA'}\hat{e}^{A'}, \\
\psi^{a}=2\frac{\Psi^{a}_{A}\hat{\lambda}^{A}}{\langle\lambda\hat{\lambda}\rangle}\hat{e}^{0}+\psi^{a}_{A'}\hat{e}^{A'}, \qquad g=3\frac{G_{AB}\hat{\lambda}^{A}\hat{\lambda}^{B}}{\langle\lambda\hat{\lambda}\rangle^{2}}\hat{e}^{0}+g_{A'}\hat{e}^{A'}.
\end{eqnarray*}
Here lowercase (Greek or Roman) coefficients are potentially functions of $x$, $\lambda$ and $\hat{\lambda}$, while $\Phi$, $\Psi$, and $G$ are all the obvious z.r.m. fields on space-time.  Using \eqref{Cwrep}, we take
\begin{equation*}
\tilde{\phi}=\tilde{\phi}_{0}\hat{e}^{0}+\tilde{\phi}_{A'}\hat{e}^{A'}=\frac{\langle\hat{\lambda}\d\hat{\lambda}\rangle}{\langle\lambda\hat{\lambda}\rangle^2}+\frac{\hat{\lambda}^{A}A_{AA'}}{\langle\lambda\hat{\lambda}\rangle}\hat{e}^{A'},
\end{equation*}
for the scalar VEV.

After performing the fermionic integrals in the massive portion of $S_{1}$, we obtain
\begin{multline*}
S_{1}= \int_{\PT_{b}}\frac{\D^{3}Z\wedge\D^{3}\bar{Z}}{\la\lambda\hat{\lambda}\ra^4} \tr\left[\frac{\epsilon^{abcd}}{4}\left(\tilde{\epsilon}_{ab}(\lambda^{B}\partial_{BA'}\phi_{cd}^{A'}+[a_{A'},\phi_{cdA'}]) \right. \right. \\
\left. \left. +\Phi_{ab}(\lambda^{B}\partial_{BA'}\tilde{\phi}^{A'}\tilde{\epsilon}_{cd}+[a^{A'},\tilde{\phi}_{A'}\tilde{\epsilon}_{cd}]) +\phi_{abA'}[a^{A'},\tilde{\epsilon}_{cd}]\right)\right. \\
\left. \epsilon^{abcd}\left(\tilde{\epsilon}_{ab}\tilde{\psi}_{cA'}\tilde{\psi}^{A'}_{d}-\frac{1}{4}\phi_{abA'}\dbar_{0}\tilde{\phi}^{A'}\tilde{\epsilon}_{cd}- a_{A'}[\tilde{\epsilon}_{ab},\tilde{\phi}^{A'}\tilde{\epsilon}_{cd}]\right)\right].
\end{multline*}
Following \cite{Boels:2006ir}, we integrate out Lagrange multipliers in the holomorphic Chern-Simons portion of the action that depends only on $\cA$.  This gives:
\begin{equation*}
a_{A'}=\lambda^{A}A_{AA'}, \qquad \tilde{\psi}_{aA'}=\widetilde{\Psi}_{aA'}, \qquad \phi_{abA'}=\frac{\hat{\lambda}^{A}}{\la\lambda\hat{\lambda}\ra}D_{AA'}\Phi_{ab}.
\end{equation*}
These restrictions allow us to write $S_{1}$ entirely in terms of space-time fields and derivatives:
\begin{multline*}
S_{1}=\int_{\PT_{b}}\frac{\D^{3}Z\wedge\D^{3}\bar{Z}}{\la\lambda\hat{\lambda}\ra^4}  \tr\left(2\tilde{\epsilon}_{ab}\tilde{\Psi}^{a}_{aA'}\tilde{\Psi}_{b}^{A'}-\epsilon^{abcd}\left[\frac{1}{2}\frac{\hat{\lambda}^{A}\lambda^{B}}{\la\lambda\hat{\lambda}\ra}[A_{B}^{A'},\tilde{\epsilon}_{cd}]D_{AA'}\Phi_{ab} \right. \right. \\
\left. \left. -\frac{\lambda^{A}\hat{\lambda}^{B}}{\la\lambda\hat{\lambda}\ra}[A_{AA'}, \tilde{\epsilon}_{ab}][A_{B}^{A'}, \tilde{\epsilon}_{cd}]+\frac{1}{4}\frac{\lambda^{B}\hat{\lambda}^{C}}{\la\lambda\hat{\lambda}\ra}D_{BA'}D_{C}^{A'}[\tilde{\epsilon}_{ab},\Phi_{cd}]\right]\right). \\
\end{multline*}
Finally, we use the fact that in Euclidean space-time signature $\PT_{b}\cong\M\times\P^{1}$ to integrate out all $\lambda$-dependence, leaving
\begin{equation*}
S_{1}=2S^{C}_{\mathrm{SD}}+\int_{\M}\d^{4}x\,\tr\left(\frac{1}{2}\Box [\tilde{\epsilon}^{ab},\Phi_{ab}]\right).
\end{equation*}
The final term in this expression is an exact derivative and vanishes, as required.

For the non-local portion of the Coulomb branch twistor action, we note that the massive portion of $S_{2}$ has the perturbative expansion
\begin{multline*}
S_{2}=\int_{\M_{\R}}\d^{4|8}X\;\log\det\left(\dbar+\cU+\cA\right)|_{X}= \\
\int_{\M}\d^{4|8}X\;\tr\left(\log\dbar|_{X}+\sum_{n=1}^{\infty}\frac{1}{n}\left(\frac{-1}{2\pi i}\right)^{n}\int_{X^n}\frac{\D\lambda_{1}}{\langle\lambda_{n}\lambda_{1}\rangle}(\cA_{1}+\cU_{1})\cdots\frac{\D\lambda_{n}}{\langle\lambda_{n-1}\lambda_{n}\rangle}(\cA_{n}+\cU_{n})\right).
\end{multline*}
Here $\cA_{i}$, $\cU_{i}$ indicate the respective fields pulled back to the curve $X_{i}\cong\P^{1}$.  Consequently, we only need to consider $\cA_{0}$ and $\cM_{0}$ in our basis of $(0,1)$-forms $\{\hat{e}^{0},\hat{e}^{A'}\}$.  Due to the fermionic integral $\d^{8}\theta$ in $\d^{4|8}X$, we keep only those terms which are proportional to $\theta^8$.  In Woodhouse gauge, $\cA_{0}\sim O(\chi^{2})$, and the $\mathrm{Sp}(4)$ R-symmetry structure of $\cU$ constrains this sum considerably. There can be no terms higher than $O(\cU^2)$ and the series truncates at $O(\cA^4)$.  The $O(\cU^0)$ contribution simply reproduces $S_{2}[\cA]$ as expected; let us focus on the first $O(\theta^8)$, $O(\cU)$ term.

This goes as:
\begin{multline*}
\left(\frac{1}{24\pi^{3} i}\right)\int_{\M_{\R}}\d^{4|8}X \int_{X^3}\prod_{i=1}^{3}\frac{K_{i}}{\langle\lambda_{i}\lambda_{i+1}\rangle}\tr\left(\frac{2}{(3!)^2}\frac{\Psi^{a}_{A}\hat{\lambda}_{1}^{A}}{\langle\lambda_{1}\hat{\lambda}_{1}\rangle}\frac{\Psi^{b}_{B}\hat{\lambda}_{2}^{B}}{\langle\lambda_{2}\hat{\lambda}_{2}\rangle}\tilde{\epsilon}_{cd} \right. \\
\left. \lambda^{C}_{1}\lambda^{D}_{1}\lambda^{E}_{1}\lambda^{F}_{2}\lambda^{G}_{2}\lambda^{H}_{2}\lambda^{I}_{3}\lambda^{J}_{3}\epsilon_{efga} \epsilon_{hkjb}\theta_{C}^{e}\theta_{D}^{f}\theta_{E}^{g} \theta_{F}^{h}\theta_{G}^{k}\theta_{H}^{j}\theta_{I}^{c}\theta_{J}^{d}\right),
\end{multline*}
where $K_{i}$ is the canonical homogeneous volume form on $\P^{1}$:
\begin{equation*}
K_{i}=\frac{\langle\lambda_{i}\d\lambda_{i}\rangle\wedge\langle\hat{\lambda}_{i}\d\hat{\lambda}_{i}\rangle}{\langle\lambda_{i}\hat{\lambda}_{i}\rangle^{2}}.
\end{equation*}
We can perform the integration along the $X_{3}$ fiber using the relation:
\begin{equation*}
\int_{X_3}K_{3}\frac{\lambda^{A}_{3}\lambda^{B}_{3}\theta_{A}^{c}\theta_{B}^{d}}{\langle\lambda_{2}\lambda_{3}\rangle\langle\lambda_{3}\lambda_{1}\rangle}=-4\pi i\left(\frac{\lambda^{(A}_{1}\lambda^{B)}_{2}}{\langle\lambda_{1}\lambda_{2}\rangle^2}\right)\theta_{A}^{c}\theta_{B}^{d},
\end{equation*}
and after performing the $\d^{8}\theta$ integral, we are left with:
\begin{multline*}
\frac{-1}{3(3!)^2\pi^{2}}\int_{\M_{\R}}\d^{4}x \int_{X^2}K_{1}\wedge K_{2}\frac{\langle\lambda_{1}\lambda_{2}\rangle}{\langle\lambda_{1}\hat{\lambda}_{1}\rangle\langle\lambda_{2}\hat{\lambda}_{2}\rangle} \tr\left(\Psi^{a}_{A}\hat{\lambda}_{1}^{A}\Psi^{b}_{B}\hat{\lambda}_{2}^{B}\tilde{\epsilon}_{ab}\right) \\
= \frac{1}{24}\int_{\M}\d^{4}x\; \tr\left(\Psi^{a}_{A}\Psi^{bA}\tilde{\epsilon}_{ab}\right).
\end{multline*}

Up to a normalization constant, this is the first term from $I^{C}$ in \eqref{mact2}.  The remaining $O(\theta^8)$ terms are of the form $\cU\Phi^{3}$ and $\cU^{2}\Phi^{2}$ and yield the two remaining terms in a similar fashion.     $\Box$


\subsection{Massive MHV Formalism}
\label{MMHVForm}

Finally, we follow the analogy of the twistor action at the origin of the moduli space, and consider the Feynman rules of the Coulomb branch twistor action \eqref{cta} in CSW gauge.  We understand only how to describe massless external states in twistors space and therefore all mass terms will be treated perturbatively. The complete massive MHV rules are only obtained once the two-point amplitudes are resummed in momentum space.

We choose an axial gauge with reference twistor $Z^{I}_{*}=(0, q^{A'},0)$, and impose this gauge with respect to the new connection:
\be{CCSW}
\overline{Z_{*}^{I}\frac{\partial}{\partial Z^{I}}}\lrcorner (\cU+\cA) =0. 
\ee
As this removes one component from the connection, the cubic Chern-Simons term in $S_{1}[\cU+\cA]$ vanishes, leaving us with:
\begin{equation*}
\frac{i}{2\pi}\int_{\PT}\D^{3|4}Z\wedge\tr\left(\cA\wedge\dbar\cA+\cU\wedge\dbar\cA+\cA\wedge\dbar\cU\right).
\end{equation*}
Hence, the only portion of the action which is quadratic with respect to the gauge field $\cA$ is the abelian Chern-Simons term $\cA\wedge\dbar\cA$,\footnote{We treat quadratic corrections in the log-det term as interactions, just as we did at the origin of the moduli space.} so the twistor propagator is the same as for the theory at the origin of the moduli space, which we know from \eqref{TAprop}:
\be{CTAprop}
\Delta(Z_{1},Z_{2})=\bar{\delta}^{2|4}(Z_{1},Z_{*},Z_{2}),
\ee
The remaining linear contributions from the Chern-Simons term can be treated as interaction vertices, which vanish for kinematical reasons.

We now consider the vertices generated by expanding the non-local part of the Coulomb branch twistor action.  All such vertices will clearly be generated by the non-local $S_{2}[\cU+\cA]$ portion of \eqref{cta}, so we will make use of the perturbative expansion of log-det.  Pulling back the Dolbeault twistor background field $\cU$ of \eqref{Dolbeault} to the line $X\subset\PT$, we find
\begin{eqnarray}
\cU|_X(\lambda) &=&  \frac{\la p\, q\ra}{\la\lambda\,p \ra}\bar{\delta}^{1}(\la q\lambda\ra) \, \epsilon^{ab}\la \theta_a\lambda\ra\la \theta_b\lambda\ra\, \mathsf{U} \nonumber\\
&=& \frac{\la p\, q\ra}{\la\lambda\,p \ra} \bar{\delta}^{1}(\la q\lambda\ra)\, \left( \la \theta_1\lambda\ra\la \theta_2\lambda\ra  + \la \theta_3\lambda\ra\la \theta_4\lambda\ra \right)\, \mathsf{U},
\end{eqnarray}
where $\bar{\delta}^{1}$ is the homogeneity $-1$ delta $(0,1)$-form on $\P^1$ and in the second line we have used the symplectic form. The spinors $p^A$ and $q^A$ are the homogeneous coordinates on $X\cong\P^1$ of the intersection points $P\cap X$ and $Q\cap X$.  We can now insert this expression into the perturbative expansion of $\log\det(\dbar+\cU_\cA)|_{X}$ to read off the interaction vertices.  The reduced R-symmetry associated with the symplectic form means that all terms in this expansion of order greater than two in $\cU$ must vanish, so we are left with three classes of vertices: 

\subsubsection*{$O(\upsilon_{i}^{0})$ \textit{contributions}}

The first contributions from the non-local term contain no insertions of the background VEV and are identical to the vertices at the origin of the moduli space:
\be{o0}
\int_{\M}\d^{4|8}X\sum\limits_{n=2}^\infty \frac{1}{n}\left(\frac{1}{2\pi i}\right)^{n}  \int\limits_{X^n} \frac{\D\lambda_{1} \cdots \D\lambda_{n}}{\la \lambda_1\lambda_2\ra \cdots \la \lambda_n\lambda_1\ra} \tr\left(  \cA(\lambda_1)\cdots \cA(\lambda_n)\right), 
\ee
where $\cA(\lambda)$ denotes the pullback to the line $X$.  As we established in Chapter \ref{Chapter3}, each term in this expression is the $n$-particle MHV vertex of $\cN=4$ SYM. We will ignore the numerical factors of $1/n$ and $(2\pi i)^{-1}$ appearing in front of all vertices.

\subsubsection*{$O(\upsilon_{i}^{1})$ \textit{contributions}}

Now consider terms with a single insertion of the background field $\cU$. The terms containing $(n-1)$ further insertions of the twistor field $\cA$ are 
\begin{multline*}
\int_{\M}\d^{4|8}X \int\limits_{X^n}\frac{\D\lambda_{1}\cdots\D\lambda_{n}}{\la\lambda_{1}\lambda_{2}\ra\cdots\la\lambda_{n}\lambda_1\ra}\, \sum\limits_{j=1}^{n}\,\tr\left(\cA(\lambda_1)\cdots\cU(\lambda_{j})\cdots\cA(\lambda_{n})\right) \\
=\int_{\M}\d^{4|8}X \int\limits_{X^{n}}\frac{\D\lambda_{1}\cdots\D\lambda_{n}}{\la\lambda_{1}\lambda_{2}\ra\cdots\la\lambda_{n}\lambda_1\ra}   \sum\limits_{j=1}^{n}  \frac{\la p\,q\ra}{\la \lambda_j\, p\ra}\bar{\delta}^{1}(\la q\, \lambda_j\ra)\epsilon^{ab} \la\theta_{a}\lambda_{j}\ra\la\theta_{b}\lambda_{j}\ra \\
\times\tr\left(\cA(\lambda_1)\cdots\cA(\lambda_{j-1})\, \mathsf{U}\, \cA(\lambda_{j+1})\cdots\cA(\lambda_n)\right).
\end{multline*}
Now performing the $\lambda_j$ integral against the delta function and relabelling indices, we find new vertices
\begin{multline}\label{o1}
\int_{\M}\d^{4|8}X\, \epsilon_{ab} \la \theta^a q \ra\la\theta^b q\ra\, \int_{X^{n}}\frac{\D\lambda_{1}\cdots\D\lambda_{n}}{\la\lambda_{1}\lambda_{2}\ra\cdots\la\lambda_{n}\lambda_1\ra} \\
\times \sum\limits_{j=1}^{n}  \frac{\la \lambda_{j}\lambda_{j+1}\ra}{\la \lambda_j q\ra\la q\lambda_{j+1}\ra}\; 
\, \tr\left(\cA(\lambda_1) \cdots \mathsf{U}_j \cdots \cA(\lambda_n)\right) ,
\end{multline}
where the notation $\mathsf{U}_j$ indicates that the generator $\mathsf{U}$ is inserted in between the generators of $\cA(\lambda_j)$ and $\cA(\lambda_{j+1})$ in the colour trace.

\subsubsection*{$O(\upsilon_{i}^2)$ \textit{contributions}}

We can apply the same analysis when there are two insertions of the background field $\cU$. Since the VEV generates an abelian subalgebra (i.e., $[\mathsf{U},\mathsf{U}]=0$), we find vertices of the form:
\begin{multline}\label{o2}
\int_{\M}\d^{4|8}X\, \left (\epsilon_{ab} \la \theta^a q \ra\la\theta^b q\ra \right)^2 \int\limits_{X^{n}}\frac{\D\lambda_{1}\cdots\D\lambda_{n}}{\la\lambda_{1}\lambda_{2}\ra\cdots\la\lambda_{n}\lambda_1\ra} \\
\times\sum_{i=1}^{n} \sum_{j=1}^{n}  \frac{\la \lambda_{i}\lambda_{i+1}\ra}{\la \lambda_i q\ra\la q\lambda_{i+1}\ra}\frac{\la \lambda_{j}\lambda_{j+1}\ra}{\la \lambda_j q\ra\la q\lambda_{j+1}\ra}\; \tr\left( \cA(\lambda_1)\cdots \mathsf{U}_i \cdots \mathsf{U}_j \ldots \cA(\lambda_n)\right).
\end{multline}

\medskip

In the bra-ket notation employed in this appendix, the momentum eigenstates \eqref{YMeig} for the massless on-shell $\cN=4$ supermultiplet are 
\be{Eig2}
\cA_{i} =\int_{\C}\frac{\d s}{s}\exp\left[ s\;( [\mu\, i^\perp]+\chi\cdot\eta_i)\right]\bar{\delta}^{2}\left(s|\lambda\ra-|i^\perp\ra\right) \mathsf{T}_{i} ,
\ee
which describes a state with momentum $p_i^\perp = |i^\perp\ra[i^\perp|$ and supermomentum $|i^\perp\ra\eta_i$ and in the direction of the generator $\mathsf{T}_{i}$. Pulling this field back to the line $X$ we find
\be{pbeigh2}
\cA(\lambda_i) =\int_{\C}\frac{\d s}{s}\exp \left[s\;(i x\cdot |i^\perp\ra[i^\perp|+ \theta\cdot |i^\perp\ra\eta_i)\right] \bar{\delta}^{2}\left(s|\lambda\ra-|i^\perp\ra\right) \mathsf{T}_{i}
\ee
We now insert these momentum eigenstates into the vertices of the Coulomb branch twistor action.

After performing the $\d s$ parameter integrals, the $\lambda_j$-integrals are performed against the remaining $\bar{\delta}^1$-functions, replacing $|\lambda_j\ra$ with the perpendicular spinors $|j^\perp\ra$, and the bosonic space-time integral generates a momentum conserving delta-function via Nair's lemma:
\begin{equation*}
\int \d^4 x\, \exp\left( ix\cdot \sum\limits_{i=1}^n |i^\perp\ra[i^\perp| \right)= \delta^4\left( \sum\limits_{i=1}^n |i^\perp\ra\la i^\perp| \right)\, .
\end{equation*}

For vertices \eqref{o0} with no background field insertions, the fermionic space-time integral produces the corresponding supermomentum conserving delta-function. However, vertices with background field insertions depend on the fermionic coordinates, and the computation is altered. In the case of a single background field insertion for vertices \eqref{o1}, we have
\begin{multline}
\int \d^8\theta\, \left( \la\theta^1q\ra\la\theta^2q\ra+ \la\theta^3q\ra\la\theta^4q\ra\right) \exp\left( \theta \cdot \sum_i |i^\perp\ra\eta_i \right)\\
= \delta^{(4)}_{12}\left(\sum\limits_{i}|i^\perp\ra\eta_{i}\right)\delta^{(2)}_{34}\left(\sum\limits_{i}\la q i^{\perp}\ra\eta_{i}\right) + \delta^{(4)}_{34}\left(\sum\limits_{i}|i^\perp\ra\eta_{i}\right)\delta^{(2)}_{12}\left(\sum\limits_{i}\la q i^{\perp}\ra\eta_{i}\right),
\end{multline}
and similarly when there are are two background field insertions as in \eqref{o2},
\be{cdelta}
\int \d^8\theta\, \la\theta^1q\ra\la\theta^2q\ra\la\theta^3q\ra\la\theta^4q\ra \exp\left( \theta \cdot \sum_i |i^\perp\ra\eta_i \right)\\
= \delta^{(4)}_{1234}\left(\sum\limits_{i}\la q i^\perp\ra\eta_{i}\right)\, .
\ee

Now consider the colour structure of the vertices. The generator $\mathsf{T}_j$ lives in the bifundamental of $\U(N_j)\times\U(N_{j+1})$ and corresponds to the on-shell multiplet with momentum $p^\perp_j$ and mass $m_j = \upsilon_{j+1}-\upsilon_j$. Since the vacuum generator takes the form
\begin{equation*}
\mathsf{U} = \left(\begin{array}{ccc}
\upsilon_{1}\mathbb{I}_{N_{1}} & 0 & \cdots \\
0 & \upsilon_{2}\mathbb{I}_{N_2} & \cdots \\
\vdots &  & \ddots 
\end{array} \right)
\end{equation*}
we find that 
\begin{equation*}
\tr(\ldots\mathsf{T_{j-1}}\, \mathsf{U}\, \mathsf{T}_{j}\ldots) = \upsilon_j\,  \tr( \ldots \mathsf{T}_{j-1}\, \mathsf{T}_{j}\ldots)
\end{equation*}
and similarly
\begin{equation*}
\tr(\ldots\mathsf{T_{i-1}}\, \mathsf{U}\, \mathsf{T}_{i}\ldots\mathsf{T_{j-1}}\, \mathsf{U}\, \mathsf{T}_{j}\ldots) = \upsilon_i\upsilon_j\,  \tr( \ldots \mathsf{T}_{i-1}\, \mathsf{T}_{i} \ldots \mathsf{T}_{j-1}\, \mathsf{T}_{j}\ldots)
\end{equation*}
which holds for all values of $i$ and $j$ including when $i=j$. Hence, all $n$-particle vertices contain an overall factor of the colour trace $\tr(\mathsf{T}_1\ldots\mathsf{T}_n)$.

So omitting the momentum conserving delta-function and planar colour trace common to all three classes of vertices \eqref{o0}-\eqref{o2}, the insertion of the momentum eigenstates yields the following three series of vertices:
\be{cbv1}
\frac{\delta^{(8)}\left(\sum\limits_i |i^\perp\ra\eta_{i}\right)}{\la 1^{\perp}2^{\perp}\ra\cdots\la n^{\perp}1^{\perp}\ra} 
\ee
\be{cbv2}
\left(\sum_{i}\frac{\upsilon_i\, \la i\!-\!1^{\perp}i^{\perp}\ra}{\la i\!-\!1^{\perp}q\ra\la i^{\perp}q\ra}\right) \frac{\delta^{(4)}_{12}\left(\sum\limits_{i}|i^\perp\ra\eta_{i}\right)\delta^{(2)}_{34}\left(\sum\limits_{i}\la q i^{\perp}\ra\eta_{i}\right) + \left\{12\leftrightarrow 34\right\}}{\la 1^{\perp}2^{\perp}\ra\cdots\la n^{\perp}1^{\perp}\ra} 
\ee
\be{cbv3}
\left(\sum_{i}\frac{\upsilon_i\, \la i\!-\!1^{\perp}i^{\perp}\ra}{\la i\!-\!1^{\perp}q\ra\la i^{\perp}q\ra} \right)^{2}\frac{\delta^{(4)}_{1234}\left(\sum\limits_{i}\la qi^{\perp}\ra\eta_{i}\right)}{\la 1^{\perp}2^{\perp}\ra\cdots\la n^{\perp}1^{\perp}\ra}
\ee
A simple application of the Schouten identity then shows
\be{cbschouten}
\sum_{i=1}^n\frac{\upsilon_i\, \la i\!-\!1^{\perp}i^{\perp}\ra}{\la i\!-\!1^{\perp}q\ra\la i^{\perp}q\ra} = \sum\limits_{i=1}^n \frac{m_i \la \kappa i^\perp\ra}{\la \kappa q \ra\la i^\perp q\ra}
\ee
for any spinor $\kappa^{A}$.  In particular, setting $|\kappa\ra=|1^{\perp}\ra$ we recover precisely the Coulomb branch MHV vertices \eqref{vm0}-\eqref{vm2}.

The final step in recovering the full massive MHV rules is to derive the massive propagator.  In the original derivation, this arises by re-summing the $O(\upsilon_{i}^{2})$ 2-point corrections to the massless scalar propagator $1/p^2$.  On twistor space, such corrections arise in precisely the same fashion.  That is to say, although the $O(\upsilon_{i}^0)$ and $O(\upsilon_{i}^{1})$ 2-point corrections to $\Delta(Z_{1},Z_{2})$ vanish for kinematic reasons, the $O(\upsilon_{i}^{2})$ corrections do not.  A calculation reveals that upon inserting such a 2-point vertex on twistor space and then transforming to momentum space (see \cite{Adamo:2011cb} for details), the propagator is corrected to give
\begin{equation*}
-\frac{m_{I}^{2}}{p^4},
\end{equation*}
where $m_{I}$ represents the sum of masses flowing into the 2-point correction.  The infinite sum of such corrections then forms a geometric series which is re-summed to give
\be{massiveprop}
\frac{1}{p^{2}-m_{I}^{2}},
\ee
in precise accordance with the massive MHV rules of \cite{Kiermaier:2011cr}.

\bibliography{trefs}
\bibliographystyle{JHEP}

\end{document}